%% file: thesis.tex
\documentclass[ieeetr, cs,dissertation]{puthesis}

\usepackage{amsmath}
\usepackage{verbatim}
\usepackage{amssymb}
\usepackage{enumerate}
\usepackage{graphicx}
\usepackage{mdwlist}
\usepackage{multirow,colortbl,color,array}
\usepackage{algpseudocode}
\usepackage{hyperref}
\usepackage{booktabs}
\usepackage{multicol}
\usepackage{subfigure}
\nochapterblankpages

\title{%
  An Algorithmic Pipeline for Analyzing\\
  Multi-parametric Flow Cytometry Data%
}

\author{Md Ariful Azad}{Azad, Ariful}

\pudegree{Doctor of Philosophy}{Ph.D.}{May}{2014}

\majorprof{Alex Pothen}

\campus{West Lafayette}

\begin{document}
\sloppy  
\volume

\include{front}
\include{introduction}

\include{ch2-variance-stabilization}
\include{ch3-clustering}
\include{ch4-matching}

\include{ch5-template}

\include{ch6-template-based-classification}

\include{aml-classification}


\include{ch7-conclusions}




\bibliographystyle{natbib}
\bibliography{thesis}


\include{vita}

\end{document}

%% file: front.tex
%
%
%
%
%

\begin{dedication}

  To my mother Ayesha  Begum, \\eldest brother Md. Aslam hossain, \\ \& wife Rubaya Pervin, \\
  thank you for enlightening  my spirit.
 \end{dedication}

\begin{acknowledgments}
All praises go to the Almighty Allah for giving me the opportunity and strength to achieve a doctoral degree.
I remember my father who passed away at my young age. May Allah keep him in the eternal peace.

Foremost, I convey my sincere gratitude to my advisor Prof. Alex Pothen for his guidance and support in my Ph.D study.
He patiently nurtured me throughout the doctoral study and enthusiastically helped me in the research and writing the thesis.
The hospitality of Alex and his family made my stay at Purdue enjoyable.
He also advised me in many non-academic matters whenever needed.
I could not have imagined a better advisor and mentor.

I am grateful to Dr. Bartek Rajwa and Dr. Saumyadipta Pyne for helping me with their immense knowledge.
Without their help, it would be difficult for me to understand the biological systems for which I developed algorithms in my dissertation.
I would like to thank Prof. Olga Vitek and Prof. Ananth Grama for serving in my Ph.D advising committee and guiding me in various aspects of my research.

My sincere appreciation also goes to Dr. Mahantesh Halapannavar and Dr. John Feo for mentoring me in the summer internship at Pacific Northwest National Lab where I gained valuable experience from exciting projects.
I thank my fellow colleagues and friends: Dr. Assefaw Gebremedhin, Arif Khan, Mu Wang, Yu-hong Yeung, and Baharak Saberidokht for stimulating discussions and creating a friendly work atmosphere.
I am thankful to Johannes Langguth and Md. Mostofa Ali Patwary who visited from university of Bergen and discussed interesting problems.
I would like to give my deepest thanks to Bangladeshi students at Purdue for making my stay as comfortable as home.
Especially, I am grateful to my friends Imrul Hossain and Alim al Islam for their affection and generous help.

I am blessed to be born in a lovingly family.
As the youngest child, each member of my family raised me with affection and instilled values in me.
Especially my mother Ayesha Begum and the eldest brother Md. Aslam Hossain made me passionate about research and inspired me to pursue higher education abroad.
Finally, I was not able to finish this journey without the love and encouragement from my wife Rubaya Pervin.
Her cooperation and support in every aspect of life drive me forward.

It has been a great privilege to pursue my doctoral study in the Department of Computer Science at Purdue University.
I would like to thank the university and the department for providing me an excellent environment for learning and research.
I will always cherish my memory at Purdue.
\end{acknowledgments}


\tableofcontents

\listoftables

\listoffigures

\begin{abbreviations}
  AML & Acute Myeloid Leukemia\cr
  ANOVA & ANalysis Of VAriance \cr
  CD & Cluster of Differentiation \cr
  CI & Confidence Interval \cr
  EM & Expectation Maximization \cr
  EMD & Earth Mover's Distance \cr
  FC& Flow Cytometry \cr
  FS& Forward Scatter\cr
  GMM & Gaussian mixture model\cr
  HD & Healthy Donor\cr
  HIV & Human Immunodeficiency Virus\cr
  HM\&M & Hierarchical Matching and Merging\cr
  MANOVA & Multivariate ANalysis Of VAriance \cr
  ML &Maximum Likelihood \cr
  MEC & Mixed Edge Cover\cr
  MFI& Mean Fluorescence Intensity\cr
  PAM& Partitioning Around Medoids\cr
  PBMC & Peripheral Blood Mononuclear Cells\cr
  RCS & Relative Cluster Separation\cr
  SS& Side Scatter\cr
  TCP& T Cell Phosphorylation\cr
  UPGMA&Unweighted Pair Group Method with Arithmetic Mean\cr
  VS& Variance Stabilization\cr
\end{abbreviations}

\makeatletter \def\@chapter[#1]#2{%
\ifnum \c@secnumdepth >\m@ne \refstepcounter{chapter}%
\typeout{\@chapapp\space\thechapter.}%
\addcontentsline{toc}{chapter}{\protect\numberline{\thechapter}\uppercase{#1}} \fi \chaptermark{#1}%
\@makechapterhead{#2} \@afterheading \ifthen{\not \boolean{@@inchapters}} { \pagenumbering{arabic}%
\@@inchapterstrue } } \makeatother

\begin{abstract}
Flow cytometry (FC) is a single-cell profiling platform for measuring the phenotypes (protein expressions) of individual cells from millions of cells in biological samples. In the last several years, FC has begun to employ high-throughput technologies, and to generate high-dimensional data, and hence algorithms for analyzing the data represent a  bottleneck. 
This dissertation addresses several computational challenges arising in modern cytometry while mining information from high-dimensional and high-content biological data. A  collection of combinatorial and statistical algorithms for locating, matching, prototyping, and classifying cellular populations from multi-parametric flow cytometry data is developed. 

The algorithms developed in this dissertation are assembled into a data analysis pipeline called flowMatch.
This pipeline consists of five well-defined algorithmic modules for
(1) transforming data  to stabilize within-population variance, 
(2) identifying phenotypic cell populations by robust clustering algorithms,
(3) registering cell populations across samples,
(4) encapsulating a class of samples with templates, and 
(5) classifying samples based on their similarity with the templates.
Each module of flowMatch is designed to perform a specific task independent of other modules of the pipeline.
However, they can also be employed sequentially in the order described above to perform the complete data analysis.

The flowMatch pipeline is made available as an open-source R package in Bioconductor (http://www.bioconductor.org/).  
I have employed flowMatch for classifying leukemia samples, evaluating the phosphorylation effects on T cells, classifying healthy immune profiles, comparing the impact of two treatments for Multiple Sclerosis, and classifying the vaccination status of HIV patients. 
In these analyses, the pipeline is able to reach biologically meaningful conclusions quickly and efficiently with the automated algorithms.
The algorithms included in flowMatch  can also be applied to problems outside of flow cytometry such as in microarray data analysis and image recognition.
Therefore, this dissertation contributes to the solution of fundamental problems in computational cytometry and related domains.
\end{abstract}

%% file: introduction.tex
\chapter{Introduction}
\section{Diversity of the cellular systems}
The immune system of a living organism consists of numerous components performing in a coordinated
manner in order to provide protection against diseases as well as abnormally behaving host cells.
The immune system must be able to detect a wide variety of virulent agents, distinguish normal host cells from dis-regulated pre-cancerous cells, as well as learn and store information about various external perturbants \cite{coico2009immunology}. 
In order to decipher the operation and the regulatory networks of the immune system, it is crucial to obtain the quantitative description of its cellular components.

\begin{figure}[t]
   \centering
    	 \includegraphics[scale=.21]{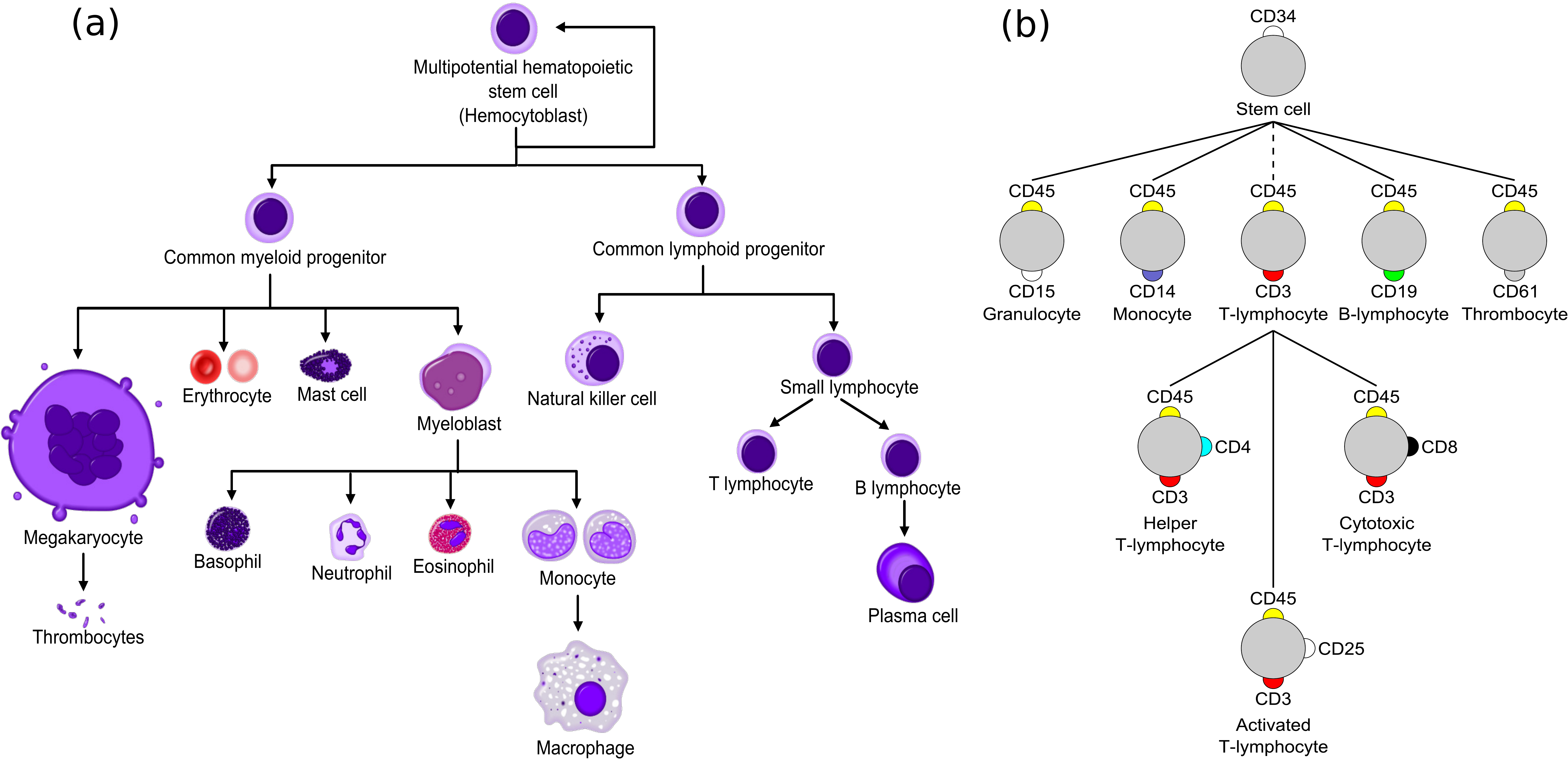}  
\vspace{-.2 cm}
   \caption[A simplified cellular hierarchy of human immune cells]{Simplified diagrams showing the cellular hierarchy of human immune cells. (a) Heterogeneous immune cells develop from haematopoietic stem cell (HSC). Different types of cells have distinctive morphological features such as size, shape, granularity etc. (b) Cluster of differentiation (CD) protein markers are shown for the common sub-types of white blood cells (leukocytes).  (Images are modified from Wikimedia Commons under Creative Commons license.)}
   \label{fig:blood_cell_hierarchy}
   \vspace{-.5 cm}
   \end{figure}

Different subsets of immune cells can be distinguished on a basis of their phenotypes, which in turn determine their functions and roles. 
The phenotypes of each cell are typically defined by a combination of morphological features such as size, shape, granularity etc. and the abundances of surface and intracellular markers such as the cluster of differentiation (CD) proteins. 
The immune cells can be organized in a hierarchy where cells performing general functions are positioned at the top, while cells performing specific functions are placed at the bottom of the hierarchy. 
I display a simplified diagram of the hierarchy of immune cells in Figure \ref{fig:blood_cell_hierarchy}, where morphological features are highlighted in the left subfigure and CD protein markers expressed by common sub-types of white blood cells (leukocytes) are shown in the right subfigure.

An established way to classify different phenotypes relies on the presence of specific proteins on the surface of cells. 
The surface molecules are assigned a CD (cluster of differentiation) number related to the type of specific monoclonal antibodies (mAb) that are shown to bind to that epitope.
For example, the mature helper T cells (also known as CD4$^+$ T cells) express CD45, CD3 and CD4 proteins. 
They are called CD45$^+$CD3$^+$CD4$^+$ cells in a common notation. 
Here,  `$+$' and  `high' indicate higher abundances of a marker, and `$-$' and `low' indicate lower levels of it.
Identifying the CD4$^+$ cell subset is pivotal in AIDS prognosis since the HIV virus infects this cell type and significantly reduces the number of functional CD4$^+$ T cells  towards the end of an HIV-1 infection \cite{mellors1997plasma}. 
Similar to the characterization of T cells, other cell types can be identified, described, and isolated on the basis of their morphology, physiology, and CD based immunophenotyping patterns.
The phenotypic patterns of cells are used to study their roles, interactions with other cell groups, and responses to various stimuli in healthy or diseased systems~\cite{maecker2012standardizing, van2012euroflow}. 
For these purposes, flow cytometry (FC), a profiling method measuring the phenotypic responses of the immune system at the resolution of single cells, is commonly employed.

\begin{figure}[t]
   \centering
    	 \includegraphics[scale=.4]{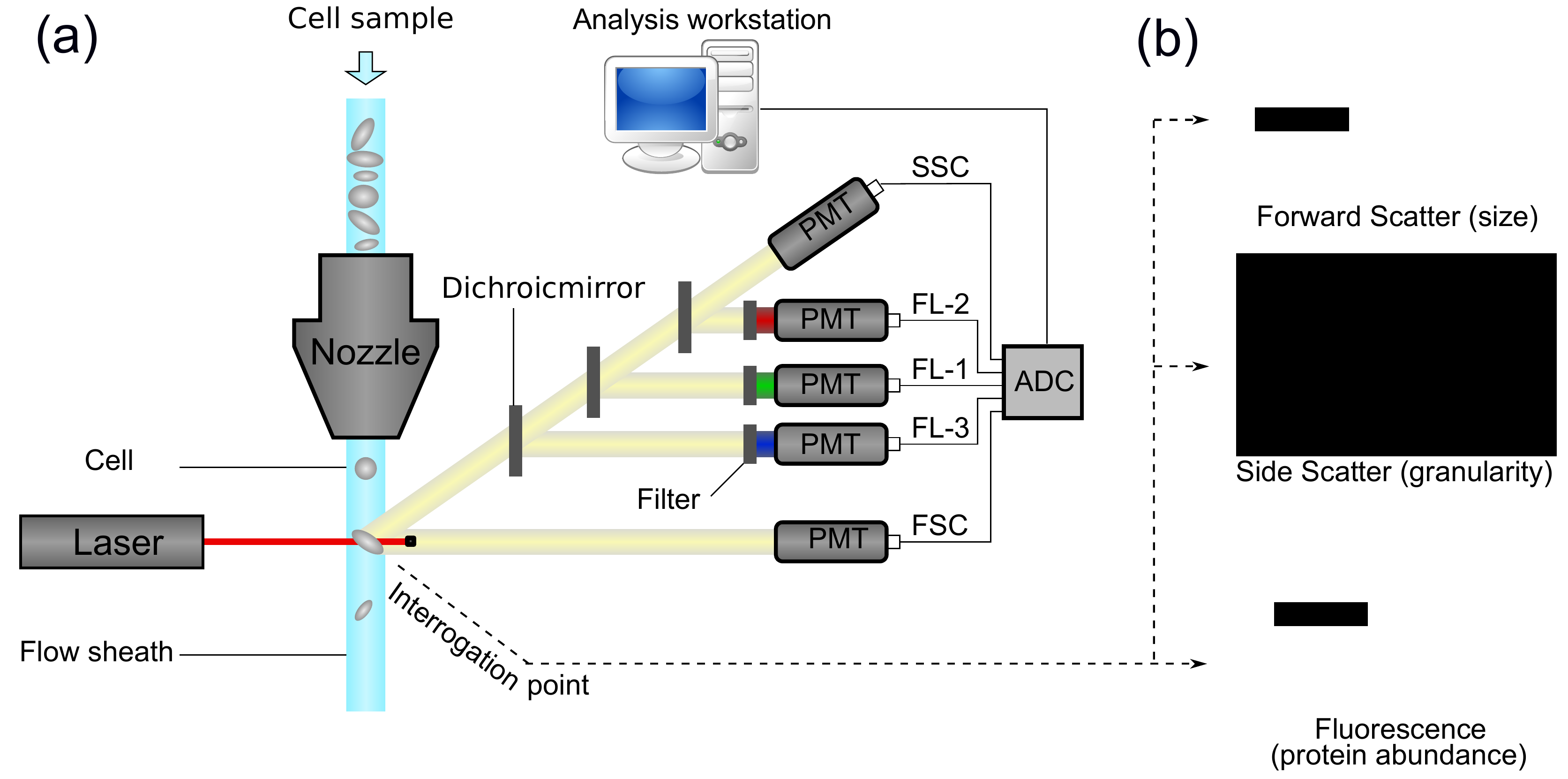}  
	 \vspace{-.2 cm}
	   \caption[A schematic diagram of a flow cytometer]{(a) A schematic diagram of a flow cytometer showing the hydrodynamic focusing of the fluid sheath, laser, optics (in simplified form, omitting focusing), photomultiplier tubes (PMTs), analogue-to-digital converter, and analysis workstation. (b) At the interrogation point, the size, granularity and expression of markers for a cell are measured by using forward scatter, side scatter and fluorescence intensities respectively. (Images are modified from Wikimedia Commons under Creative Commons license.)}
   \label{fig:cytometer}
   \vspace{-.5 cm}
\end{figure}

\section{ Flow cytometry (FC)}
Flow cytometry (FC) is a platform for measuring various features of individual cells from millions of cells in a sample.
The cellular features measured by FC include several morphological features such as size, shape, granularity etc. and the abundance of a number of proteins expressed by the cell. 
In typical FC experiments, cells in a suspension (or other biological particles) flow in a stream of fluid passing the region where a laser beam interrogates each cell. 
The light scattered by a cell in the forward and perpendicular directions -- known as Forward Scatter (FS) and Side Scatter (SS) channels respectively -- reveal the size, shape and granularity of the cell. 
The identity and abundance of the proteins are measured by the fluorescence signals emitted by the laser-excited, fluorophore-conjugated antibodies bound to the target proteins in a cell~\cite{shapiro2005practical}.
I show a schematic diagram of a flow cytometer and how it measures different features of a cell in Figure \ref{fig:cytometer}.
In a flow cytometer, the mixture of fluorescence signals are roughly separated by several band-pass filters and each signal is collected into a separate fluorescence (FL) channel.
In this context, I often use the terms ``a fluorescence channel" and ``a protein marker" interchangeably because we infer information about the latter through the signal collected at the former. 

Current fluorescence-based technology supports the measurements of up to 20 proteins simultaneously in each cell from a sample containing millions of cells~\cite{lugli2010data}.
Although emerging atomic mass cytometry systems such as CyTOF~\cite{bendall2011single} can measure more than 40 markers per cell, fluorescence detection is still the most used tool for single-cell measurements. 
FC is employed to study the complexity of the immune system and the changes in its components upon exposure to various external perturbants such as pathogens, chemical compounds (drugs), or vaccination, as well as events such as aging, autoimmune reactions, or presence of cancer. 
FC is now routinely employed to illustrate immune cells development and maturation~\cite{perfetto2004seventeen}, to study immune responses in the presence of a pathogen, to diagnose diseases of the immune system~\cite{peters2011leukemia}, and to develop novel vaccines (e.g., against HIV)~\cite{seder2008t}.

\section{Flow cytometry data analysis}
\begin{figure}[t]
   \centering
    	 \includegraphics[scale=.5]{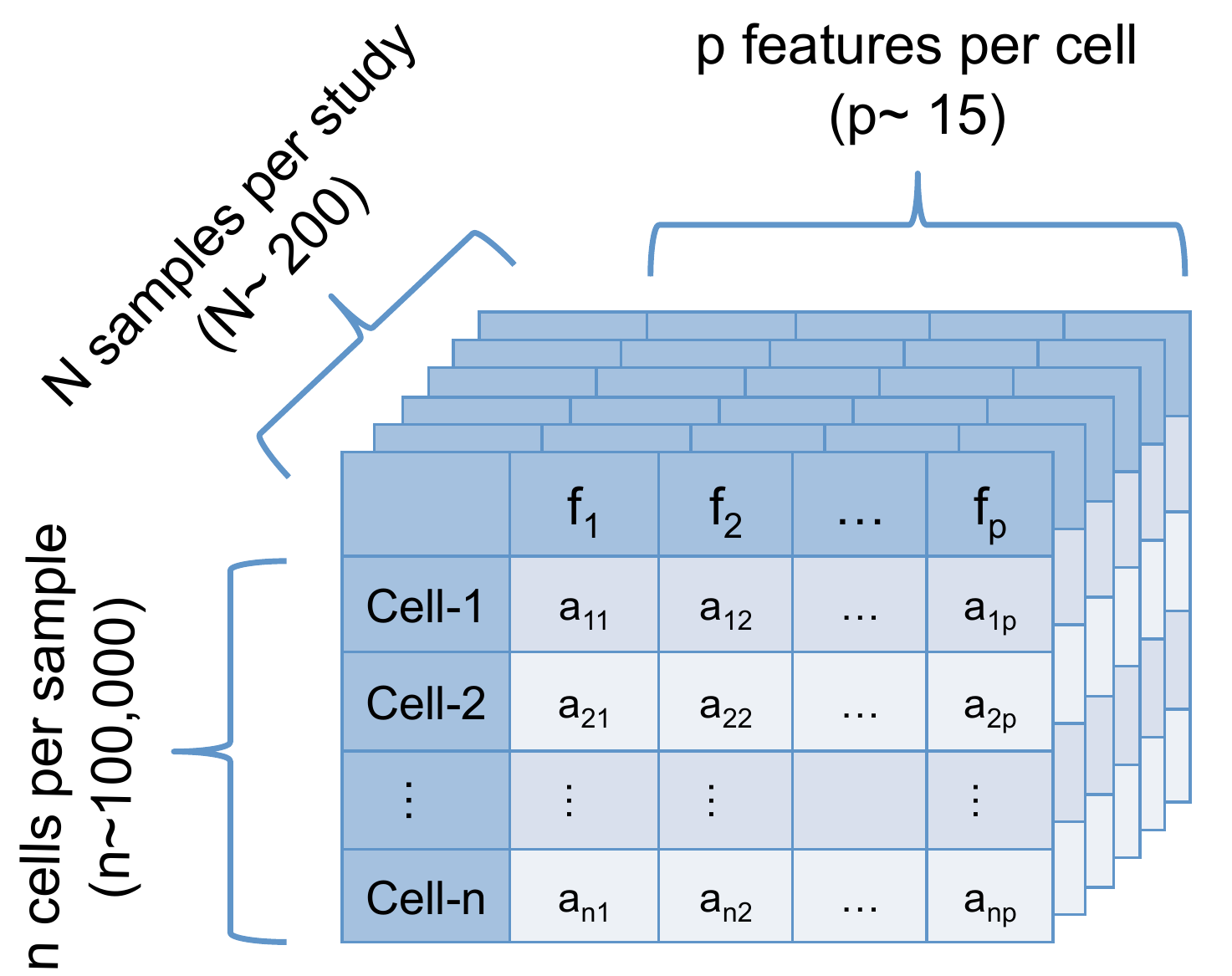}  
	 \vspace{-.2 cm}
	   \caption[A schematic view of flow cytometry (FC) samples]{A schematic view of flow cytometry samples from a typical experiment.}
	   \vspace{-.5 cm}
   \label{fig:FC_samples}
\end{figure}

An FC  sample measuring $p$  features (known as parameters in FC) from $n$ cells is represented by an $n\times p$ matrix $A$.
The matrix element $A(i,j)$ represents the measurements of the $j^{th}$ feature in the $i^{th}$ cell.
To this end, a cell is  represented by a $p$-dimensional feature vector capturing the morphology and protein expression profile of the cell, which are effectively measured by the light scatter and fluorescence signals.
In a typical experiment, we measure hundreds of samples with each sample measuring multi-dimensional features for up to millions of cells.
I show a schematic view of a data set generated in a typical FC experiment in Figure \ref{fig:FC_samples}. 
However, the actual numbers of features, cells, and samples vary from one experiment to another. 

Analysis of a collection of FC samples can lead to diagnosis of various diseases, monitoring of immune responses in the presence of a pathogen, development of novel vaccines etc.
However, before making useful biological conclusions, the raw FC measurements mixed with various sources of noise need to pass through several analysis steps in a systematic order.
These analysis steps are often guided by specific aspects of FC data and are customized to address the biological needs. 
I present a list of attributes arising in a typical multi-parametric FC experiment and relate each attribute to a necessary data analysis step in Table~\ref{tab:flow_data_properties}.
This list is not exhaustive and is subject to change depending on the design of experiment and the type of biological questions answered by a particular experiment.   
I briefly discuss several analysis steps in the next few paragraphs. 

\begin{table}[t]
   \centering
     \vspace{-.7 cm}
      \caption{Different analysis steps of FC data (right column) are originated from FC data attributes (left column).}
   \begin{tabular}{@{} lr @{}} 
      \toprule
      FC  data attributes  & Data analysis steps \\
      \midrule
      debris, dead cells, doublets & preprocessing \& quality control \\
      multiple fluorescence channels ($2$-$20$) & spectral unmixing (compensation)  \\
      variance increases with mean &  variance stabilization \\
      multiple cellular features ($4$-$20$) & multidimensional  distribution\\
      many observations per sample ($10^4$-$10^5$) & down-sampling  \\
      several discrete cell subsets ($2$-$50$)  &  clustering\\
      similar cell populations across samples & cluster matching or labeling \\
      many samples per cohort &  meta-clustering \& templates \\
      multiple classes of samples &  classification \\
      \bottomrule
   \end{tabular}
   \label{tab:flow_data_properties}
\end{table}

\subsection{Removing unintended cells}
\begin{figure}[t]
   \centering
   \includegraphics[scale=.48]{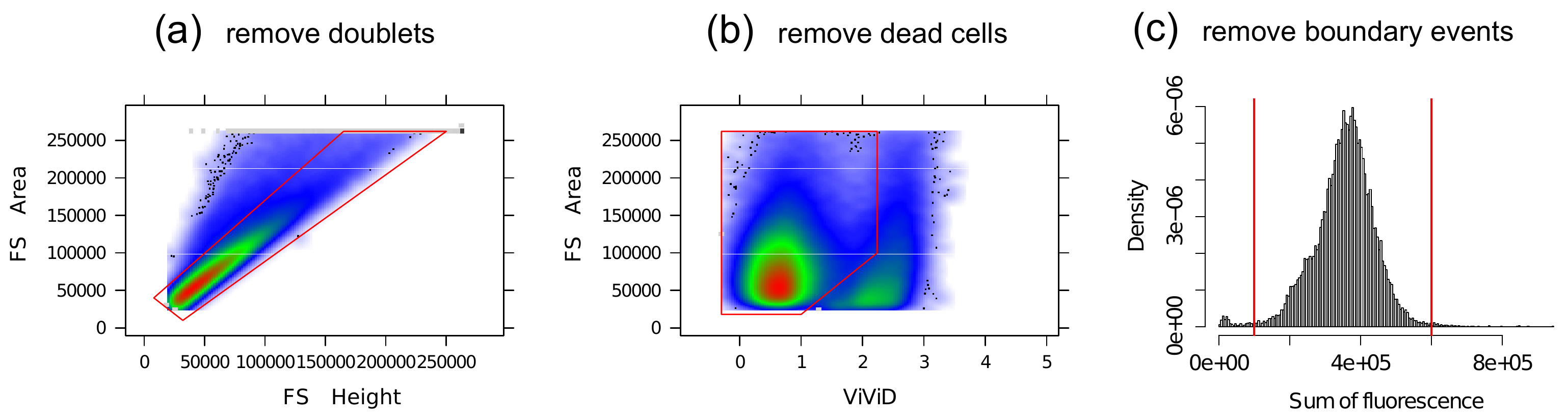} 
   \vspace{-.2 cm}
   \caption[Removing unintended events from an FC sample]{Removing unintended events from an FC sample. (a) Single intact cells (inside the red polygon gate) are separated from the doublets (outside of the red polygon gate). (b) A viability marker (ViViD) is used to remove dead cells (outside of the red polygon gate). (c) Cells emitting very low or very high fluorescence signals (outside of the read vertical lines) are removed as potential outlying events.}
   \label{fig:preprocessing}
   \vspace{-.5 cm}
\end{figure}

In the preprocessing phase, various unintended events such as doublets, dead cells, debris, etc. are removed from the FC data.
Figure \ref{fig:preprocessing} shows several preprocessing and quality control steps used in a typical FC data analysis.
A ``doublet" is a pair of attached cells, which has larger area but smaller height in the forward scatter (FS) channel relative to the single intact cells. 
Figure \ref{fig:preprocessing}(a) shows how we can separate single cells (inside the red polygon gate) from the doublets (outside of the red polygon gate).
Cell viability dyes, e.g., amine reactive viability dyes ViViD and Aqua Blue, are often used to separate dead cells from live cells~\cite{perfetto2006amine}.
Live cells are shown with a red polygon gate in Fig.~\ref{fig:preprocessing}(b) and dead cells outside of the gate are discarded from further analysis.
Furthermore, boundary events can also be removed from the histogram of total fluorescence as shown in Figure \ref{fig:preprocessing}(c). 
Cells outside of the read vertical lines are either too dim or too bright in terms of the total fluorescence emission and can be removed as outliers. 
Several other preprocessing steps are occasionally performed as part of quality control, for example see the discussion in~\cite{le2007data}.

\subsection{Spectral unmixing (compensation)} 
Flow cytometry measures the abundance of protein markers in a cell by the fluorescence intensities of the fluorophore-conjugated antibodies bound to the target proteins.
Because of the overlap of florescence spectra emitted by different fluorophores, a detector intended for a particular marker also captures partial emissions from other fluorophores.
The correct signal at each detector is then recovered by a process called the \emph{Spectral unmixing} or \emph{compensation} \cite{roederer2001spectral, Bagwell1993}. 

To see how a simple spectral unmixing method works, consider an FC system measuring the emission of $p$ fluorophores with $p$ detectors.
Then the general form of the compensation system is given by the following equation:
\begin{equation}
\mathbf{o} = \text{M}\mathbf{s} + \mathbf{a},
\end{equation}
where,
\begin{itemize}
  \item[] $\mathbf{o}$ = vector of the observed signals at $p$ detectors.
  \item[] M = $p\times p$ spillover matrix. The off-diagonal element $\text{M}[i,j]$ denotes the contribution of the $j^{th}$ fluorochrome to the detector of the $i^{th}$ fluorochrome. The diagonal elements represent the fraction of signal found in the appropriate channel. Each column of the matrix adds to unity.
  \item[] $\mathbf{s}$ = vector of original signal emitted from the $p$ fluorochromes.
  \item[] $\mathbf{a}$ = autofluorescence vector of length $p$ measuring the amount of background fluorescence.
\end{itemize}

\begin{figure}[t]
   \centering
   \includegraphics[scale=.6]{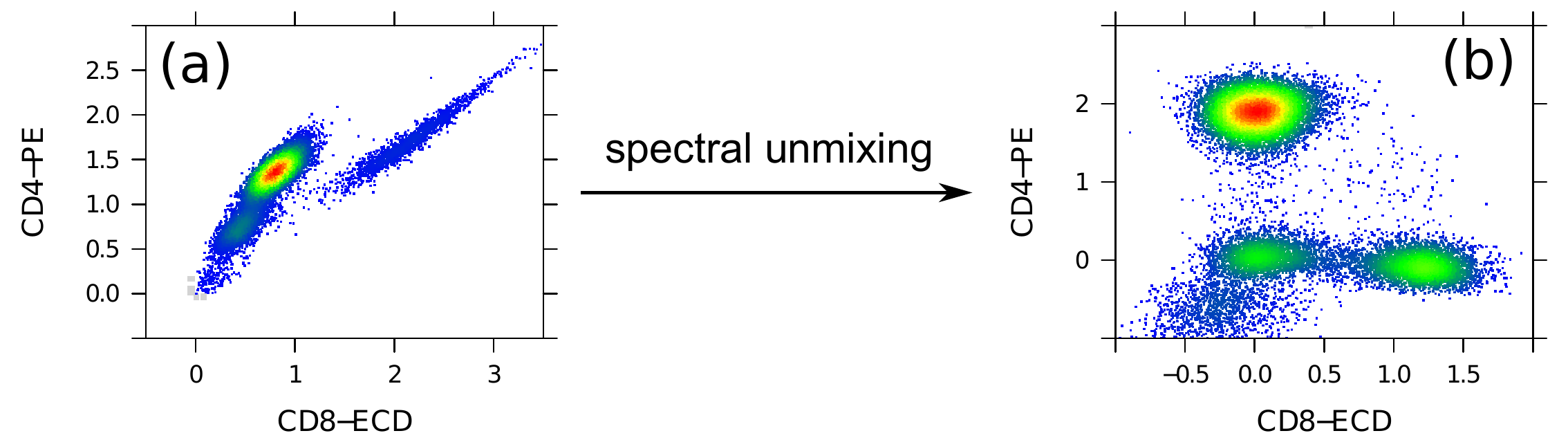} 
   \vspace{-.2 cm}
   \caption[Spectral unmixing in a pair of overlapped fluorescence channels]{(a) A pair of correlated channels due to the overlap of the corresponding fluorochrome spectra. (b) The same pair of channels after signals are unmixed from each other.}
   \vspace{-.5 cm}
   \label{fig:spectral_unmix}
\end{figure}

The goal of the spectral unmixing is to calculate the actual signal vector $\mathbf{s}$.
The simplest and widely used algorithm performing compensation is a straightforward application of linear algebra that requires the solution of a linear system of equations involving the spillover matrix $M$:  \cite{Bagwell1993, Novo2013}:
\begin{equation}
\mathbf{s} = \text{M}^{-1} \times (\mathbf{o} - \mathbf{a})
\label{eq:spectral_unmix}
\end{equation}
For example, Fig.~\ref{fig:spectral_unmix}(a) shows a pair of correlated channels due to the overlap of the corresponding fluorochrome spectra. 
The correlation is removed after the signals are unmixed from each other by Eq.~\ref{eq:spectral_unmix}, as illustrated in Fig.~\ref{fig:spectral_unmix}(b).
The accuracy of the reconstructed signal, however, depends on the nature of errors generated by the fluorescence emission process and photo-electric circuitry of the flow cytometer.
The error model can be approximated by a mixture of Poisson and Gaussian noise~\cite{Novo2013, snow2004flow}, a more accurate compensation scheme is discussed  by Novo et al.~\cite{Novo2013}.

\subsection{Data transformation and variance stabilization}
After performing initial pre-processing, FC data is often transformed for proper visualization and subsequent analysis.
Various nonlinear functions are typically used for this purpose, such as logarithm, hyperlog, biexponential, asinh transformations~\cite{bagwell2005hyperlog, parks2006new, dvorak2005modified, novo2008flow}.
These transformations are aimed primarily to resolve cell populations uniformly and to allow unimpeded visual interpretation especially by using a series of 2-D FC scatter plots.
However, owing to the nature of photon-counting statistics, variance of a fluorescence signal monotonically increases with the average signal intensity (average protein expression).
This signal-variance dependence creates problem in uniform feature extraction and comparing cell populations with different levels of marker expressions.
To remove the correlation between signal and variance, variance stabilization (VS) is performed.

Variance stabilization (VS)~\cite{efron1982transformation, huber2002variance} is a process that decouples signal from the variance such that the variance become approximately constant along the complete range of fluorescence intensities. 
In FC, VS can be achieved by transforming data with properly tuned transformation parameters.
Several past works followed this approach of parameter optimization, but mostly with an objective to maximize the likelihood of individual cell populations being generated by a mixture of multivariate-normal distributions on the transformed scale \cite{finak2010optimizing, ray2012computational}. 
To remove the correlation between signal and variance, I developed a variance stabilization (VS) method called \emph{flowVS} that decouples signal from the variance by an inverse hyperbolic sine (asinh) transformation function whose parameters are estimated by a likelihood-ratio test. 
I will discuss variance stabilization for FC samples in more detail in Chapter~\ref{chap:variance_stabilization}.

\subsection{Identifying cell populations (gating or clustering)}
For a given set of markers, a cell \emph{population} or cell cluster is a subset of cells in a sample with similar physical and fluorescence characteristics, and thus biologically similar to other cells within the subset but distinct from those outside the subset. 
In conventional FC practice, trained operators identify cell populations by visualizing cells in a collection of two-dimensional scatter plots. 
This is traditionally a manual process known as ``gating" in cytometry.
However, with the ability to monitor large numbers of cellular features simultaneously, the visualization-based approach is inadequate to identify high-dimensional populations. 
To address this problem, a number of researchers have proposed several customized clustering algorithms to identify cell populations in FC samples.
These algorithms include model-based clustering \cite{chan2008statistical, lo2008automated, pyne2009automated}, density based clustering\cite{qian2010elucidation, walther2009automatic}, spectral clustering \cite{zare2010data} and other non-parametric approaches \cite{aghaeepour2011rapid}.
The FlowCAP consortium (http://flowcap.flowsite.org/) designed a set of challenges with an aim to identify the best clustering algorithms for different FC datasets and Aghaeepour et. al. \cite{aghaeepour2013critical} provides a state-of-the-art summary of the field.

Automated population identification is a pivotal step in the FC data analysis.
Given the availability of a large collection of software packages for solving the clustering problem, it is not always trivial to select the best option to analyze a particular dataset, especially when no ``ground truth" or ``gold standard" is available.
In this situation various cluster validation methods can be used to evaluate the quality of different clustering solutions from which the best option can be picked~\cite{jain1999data, halkidi2001clustering}.
However, it has been shown that an ensemble clustering computed from a consensus of a collection of simple clustering solutions often outperforms any individual algorithm for FC data \cite{aghaeepour2013critical}.
Finding an optimal ensemble clustering is an NP-hard problem and different heuristics are often used  to obtain an approximate solution~\cite{hornik2008hard}.
Considering the importance of the clustering problem, I describe a new consensus clustering algorithm that maintains a hierarchy of clustering solutions from different algorithms.
This algorithm provides a new perspective of ensemble clustering to the FC community.
I will discuss different clustering algorithms and a novel consensus clustering algorithm in Chapter~\ref{chap:clustering}.

\subsection{Registering cell populations across samples}
 Phenotypically distinct cell populations often respond differently upon perturbation or change of biological conditions.
 This changed response results in increased protein expression in each cell, or in changes in the fraction of cells belonging to a cell type. 
To identify the population-specific changes, we register cell clusters across different samples~\cite{pyne2009automated, azad2012matching}.
The population registration problem can be solved by computing a similarity measure between each pair of cell clusters across samples and matching clusters with high similarity.
Additionally, when biologically meaningful labels for clusters are known for a sample, a population matching algorithm can label clusters from another sample, thus solving the population labeling problem as well. 

In conventional FC practice, population registration is often performed by mapping 2-D projections of clusters.
However, the visual mapping is inadequate for high-dimensional data.
Recently two different types of algorithms have been proposed in order to solve the population registration problem automatically and efficiently. 
In the first approach, the centers of different clusters are ``meta-clustered" (cluster of clusters) and the clusters whose centers fall into same meta-cluster are marked with same label \cite{finak2010optimizing}.
The second approach computes a biologically meaningful distance between each pair of clusters across samples and then matches similar clusters by using a combinatorial matching algorithm \cite{pyne2009automated, azad2010identifying}. 
I developed an algorithm of the second type, called  the mixed edge cover (MEC) algorithm~\cite{azad2010identifying, azad2012matching}.
The MEC algorithm uses a robust graph-theoretic framework to match a cluster from a sample to zero or more clusters in another sample and thus solves the problem of missing or splitting populations as well.
I discuss more about the cluster matching algorithms in Chapter~\ref{chap:matching}.

\subsection{Meta-clustering and templates}
A flow cytometry dataset often consists of samples belonging to a few representative classes representing multiple experimental conditions, disease or vaccination status, time points etc.
Towards this end, I assume that samples belonging to a particular class are more similar (given a biologically meaningful similarity measure) among themselves than samples from other classes.
In this scenario it is more efficient to summarize a collection of samples with a few representative \emph{templates}, each template representing samples from a particular class \cite{pyne2009automated, finak2010optimizing, azad2012matching}.
Thereby, overall changes across multiple conditions can be determined by comparing the cleaner and fewer class templates rather than the large number of noisy samples themselves.

A \emph{template} is usually constructed by matching similar cell clusters across samples and combining matched clusters into \emph{meta-clusters}. 
Clusters in a meta-cluster represent the same type of cells and thus have overlapping distributions in the marker space.
Therefore, the meta-clusters represent generic cell populations that appear in most samples with some sample-specific variation.
A template is a collection of relatively homogeneous meta-clusters commonly shared across samples of a given class, thus describing the key immunophenotypes of an overall class of samples in a formal, yet robust, manner.

In my dissertation, I have developed a hierarchical matching-and-merging (HM\&M) algorithm that builds templates from a collection of samples by repeatedly merging the most similar pair of samples or partial templates obtained by the algorithm thus far~\cite{azad2012matching}. 
A meta-cluster within a template represents a homogeneous collection of cell populations and acts as a blueprint for a particular type of cell.
However, traditional null hypothesis based significance testing (e.g., F-test or paired t-test) often has a high probability of making a Type I error when used to evaluate the homogeneity of a meta-cluster because of large cluster sizes.
Hence, to evaluate meta-cluster homogeneity, I propose the ratio of  between-cluster to within-cluster variance (relative cluster separation, $\phi$), in a MANOVA model, as an alternative method to evaluate the homogeneity of a meta-cluster.
I will discuss meta-clustering and template construction algorithm in detail in Chapter~\ref{chap:template}.

\subsection{Classifying FC samples}
Besides their use in high-level visualization and cross-class comparisons, templates can be employed to classify new samples with unknown status. 
Templates work as prototypes of different biological classes, e.g., disease status, time points etc.,  by emphasizing the common properties of the class while omitting sample-specific details. 
A new sample with unknown class label is predicted to come from a class whose template the sample is most similar to. 
The template-based classification is robust and efficient because it compares samples to cleaner and fewer class templates rather than the large number of noisy samples themselves.
While classifying new samples, the templates can be dynamically updated to incorporate the information gained from the classification of the new samples. 
This approach makes it possible to summarize the data from each laboratory using templates for each class, and then to merge the templates and template-trees across various laboratories, as the data is being continuously collected and classified.  
I will discuss the template-based classification algorithms in detail in Chapter~\ref{chap:template-based-classification}.

\section{The need for automation in FC data analysis}
FC data is large, continuous, high-dimensional, and perturbed by Poisson and Gaussian noise as we measure various features of individual cells for millions of intact cells in a sample. 
Traditionally, after some semi-automated preprocessing, cell populations are identified by visualizing cells as a collection of two-dimensional scatter plots. 
The reader will find it helpful to refer the Figure \ref{fig:clustering_hd}(b) in Chapter~\ref{chap:clustering} for an example where four types of immune cells are identified using five cluster of differentiation (CD) proteins.
A trained operator then visually compares these biaxial plots from different samples, registers populations across samples from multiple conditions (healthy vs. disease for example) and studies the differences across conditions to extract necessary biological information.

Recent advances in FC technology pose challenges to the traditional manual analysis in three aspects: (1) high-dimensionality of data, (2) large volume of data, and (3) compute-intensive analysis.
First, cell populations defined in higher dimension are difficult to locate in 2-D projections.
Furthermore, biaxial plots assume that the axes are orthogonal to each other; however, often there are correlations between these protein markers, and these are difficult to analyze using biaxial projections.
Second, it is laborious and often infeasible to manually analyze an FC dataset with hundreds of samples, each with millions of multi-parametric cells.
Third, several analysis steps require numerical calculations such as optimization of nonlinear functions and solutions of matrix equations.
However, a collection of tools designed to perform specific functions but not to work together does not improve the overall analysis of FC data, 
because a significant amount of time is spent in finding appropriate tools for each step, in identifying optimum parameters for the selected tools and, in processing data between different steps.
Therefore, to prevent the data analysis from being the bottleneck in scientific discovery based on cytometry, an automated and systematic cytomics pipeline is necessary.

Even though a rich set of computational tools is available for other fluorescence-based technologies such as the microarrays \cite{allison2006microarray}, they can not be directly applied -- as a black box -- to analyze FC data because of the technological differences.
Microarrays measure the expression of a large number of genes under different conditions, whereas FC measures a smaller number of proteins characteristic of a few immunophenotypes, across a large number of samples. 
As a result, the objectives of various analysis steps are often different for these two technologies.
For example, between-samples variances across a large number of genes are stabilized in microarray, whereas I stabilize within-population variances across a collection of samples in FC.
Another example of their difference is in data clustering, where multi-parametric cells are clustered to identify functional cell populations in FC; in contrast, genes are clustered based on their expression patterns in microarray.  
In summary, the nature of the data, its pre-processing, statistical treatment, and algorithms for downstream analysis are all substantially different for FC and other fluorescence-based technologies. 
Therefore, a customized pipeline adapted to the properties of FC data is necessary, especially to keep up with the ever increasing dimensionally and large number of samples generated routinely in FC experiments.
To address this need, recently there has been an influx of automated tools for analyzing FC data  \cite{aghaeepour2013critical, pyne2009automated, spidlen2013genepattern, kotecha2010web}.
The algorithms and software developed in my PhD work are the latest additions to the attempt of automating FC data analysis.

\section{Contributions}
This dissertation has four major contributions to computational cytometry: (a) algorithms, (b) software, (c) biological applications, and (d) publications. 

\emph{Algorithms:} This dissertation addresses the need to analyze the large volume of multi-parametric FC data by developing an algorithmic pipeline called \emph{flowMatch}~\cite{azad2014flowMatch}.
The pipeline contains six algorithmic modules performing six steps of FC data analysis:
(1) spectral unmixing to remove the effect of overlapping fluorescence channels, 
(2) stabilizing variance to decouple signals from noise, 
(3) robust clustering of cells to identify phenotypic populations, 
(4) registering cell clusters across samples from multiple conditions,  
(5) constructing templates by preserving the common expression patterns across samples, and 
(6) classifying samples based on the measured phenotypes. 
In addition to these six steps, the pipeline has several helper functions for preprocessing and visualizing FC data.

\begin{figure}[!t]
   \centering
   \includegraphics[scale=.65]{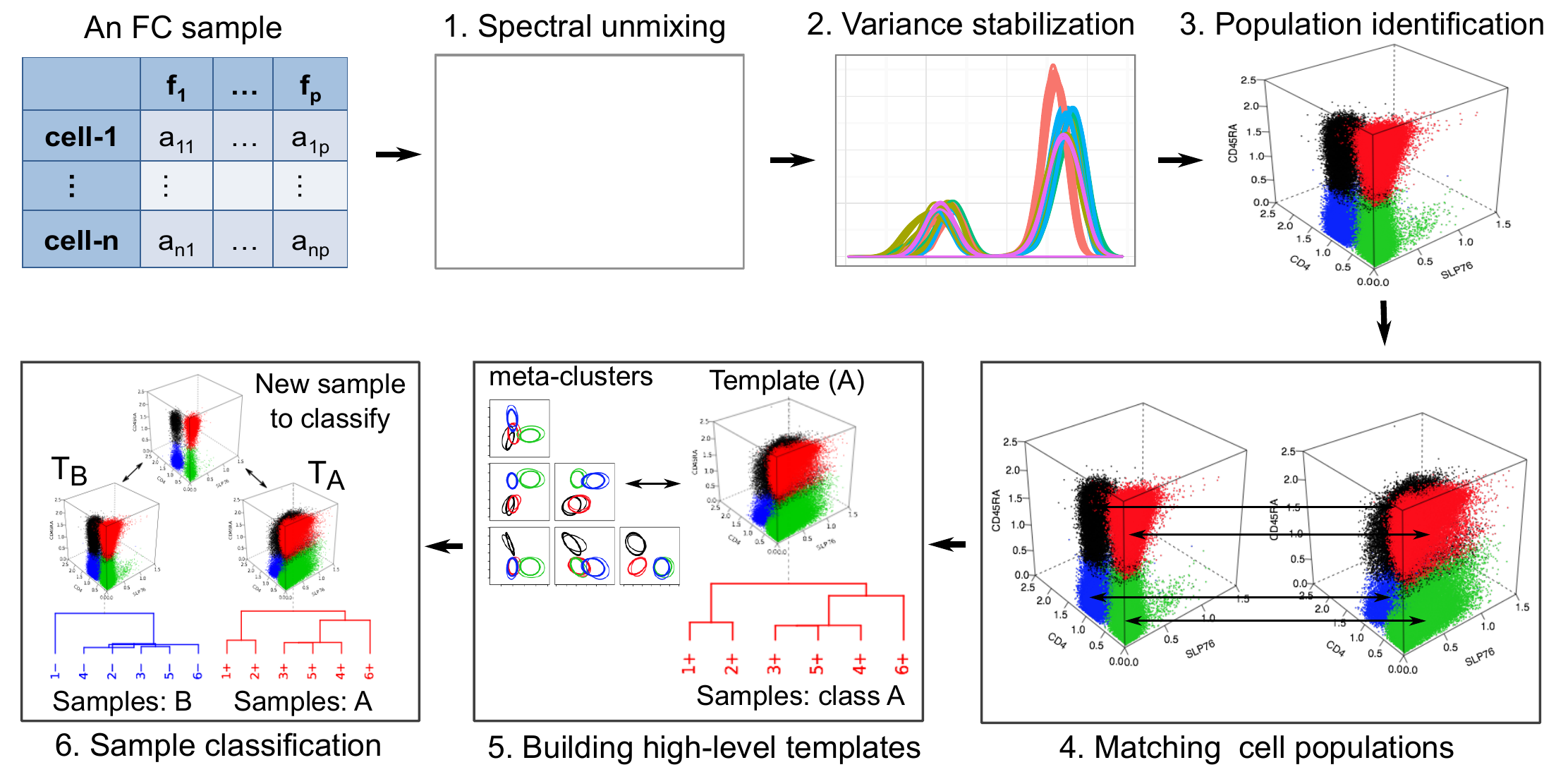} 
   \vspace{-.2 cm}
   \caption[An FC data analysis pipeline]{An FC data analysis pipeline. (1) unmixing of spectra to remove the effect of overlapping fluorescence channels, (2) stabilizing variances by transformation and normalization, (3) identifying cell populations by  automated clustering algorithm in a high-dimensional marker space, (4) registering cell populations across samples with a matching algorithm, (5) representing a class of samples with high-level templates, and (6) classifying samples by using the templates.}
   \label{fig:pipeline}
   \vspace{-.5 cm}
\end{figure}

Figure \ref{fig:pipeline} displays the schematic view of six functional modules of the flowMatch pipeline.
An FC sample is represented by an $n\times p$ matrix, where $n$ is the number of cells and $p$ is the number of features (physiological properties and expression levels of markers) measured in each cell.
Subfig.~\ref{fig:pipeline}(1) shows the overlap of two spectra (green and yellow) emitted by two fluorochromes. 
The yellow band pass filter captures a significant amount of signal from the green spectrum, which must be compensated (e.g., by Eq.~\ref{eq:spectral_unmix}) to correctly reconstruct the signal form the yellow fluorochrome.   
Subfig.~\ref{fig:pipeline}(2) displays the density plots of a single marker from several samples of a dataset after stabilizing the variance. 
Observe that the variances (width) of the density peaks (both positive and negative peaks across samples) are nicely stabilized.
Subfig.~\ref{fig:pipeline}(3) shows the application of a clustering algorithm to a three dimensional sample, where different colors denote four phenotypically distinct cell populations.
In Subfig.~\ref{fig:pipeline}(4), I show how functionally similar cell clusters are matched by a population registration algorithm, where same colors are used to mark the matched clusters.
The next Subfig.~\ref{fig:pipeline}(5) illustrates how a template is created by a hierarchical algorithm from six samples belonging to the same class. 
Finally, the classification of a new sample based on its similarity with two class templates is described in Subfig.~\ref{fig:pipeline}(6).

\emph{Software:} I have developed algorithms for the last five steps of the pipeline except the spectral unmixing step.
I included spectral unmixing and other related preprocessing tools from related packages to make this pipeline as complete as possible.
I have developed an R package {\tt flowMatch}~\cite{manual_flowMatch, azad2014flowMatch} by implementing the steps discussed in Fig.~\ref{fig:pipeline} in both R and C++ programming languages. 
This package has been made available through Bioconductor~\cite{gentleman2004bioconductor} (http://www.bioconductor.org/). 
I hope that flowMatch will be a useful addition to the available FC data analysis tools and can contribute to faster analysis of large FC datasets.

\emph{Biological applications:} The \emph{flowMatch} pipeline contains several combinatorial and statistical algorithms used to perform different steps of data analysis. 
The complete analysis is usually divided into a series of encapsulated sub-problems and each of them is solved by a module of the pipeline.
I demonstrated the application of different steps of the pipeline with three data sets: healthy donor data, T cell phosphorylation data, and  acute myeloid leukemia (AML) data. 
The healthy donor (HD) dataset consists of 65 five-dimensional samples from five healthy individuals who donated bloods on different days. 
I use this dataset to demonstrate that by stabilizing within-cluster variance we are able to construct homogenous meta-clusters despite the presence of biological, temporal, and technical variations.
The T cell phosphorylation (TCP) dataset consists of 29 pairs of samples before and after stimulation with an anti-CD3 antibody~\cite{maier2007allelic}.
I analyzed the data in order to demonstrate that templates can be created from multiple conditions and the meta-clusters can be matched across experimental conditions (before and after stimulation) to assess the overall changes in phosphorylation experiment.
The acute myeloid leukemia (AML) dataset consists of samples from 43 AML patients and 217 healthy individuals.
I used flowMatch pipeline to build healthy and AML templates, to identify AML  markers, and to classify AML samples by comparing test samples against the two templates\cite{azad2014immunophenotypes}.
I describe the datasets in Section\ref{sec:intro_dataset} in more detail.
In the past, I have also tested different components of flowMatch to solve problems in evaluating HIV vaccination success, and detecting correlations among multiple sclerosis (MS) treatments.

\emph{Publications:} The algorithms with their biological applications developed in this dissertation have been published to several peer-reviewed journals~\cite{aghaeepour2013critical, azad2012matching} (Nature Methods, BMC Bioinformatics) and conferences~\cite{azad2010identifying, azad2013classifying} (e.g., WABI, ACM BCB, GLBIO, SIAM~LS, Cyto, etc.).
Several papers are submitted for publication~\cite{azad2014immunophenotypes,azad2014flowMatch}.
The \emph{flowMatch} pipeline was one of the top performers to solve several challenges designed by the FlowCAP consortium at National Institute of Healthy (NIH).
I have developed several multithreaded algorithms for computing the maximum cardinality matching in large graphs, which are published in several conference proceedings (SC, IPDPS, IPDPSW)~\cite{azad2012ipdps, azad2011computing, azad2012multithreaded}.
Aside from this dissertation, I have also developed a robust variant of residual resampling technique for computing the uncertainty in evolutionary trees and illustrated its use with an analysis of genome-scale alignments of yeast~\cite{waddell2009resampling, waddell2010resampling}.

\emph{Difference from related work:} \emph{flowMatch} is similar to the GenePattern Flow Cytometry Suite in terms of the coverage of distinct algorithms for different analysis steps.
However, these two pipelines are significantly different from each other on how each step of the pipeline is performed. 
For example, GenePattern normalizes FC samples by aligning density peaks on each channel as described by Hahne et al. \cite{hahne2010per}. 
In contrast, flowMatch stabilizes variances of the density peaks on each channel without shifting the signals. 
Likewise, other steps of the flowMatch pipeline are significantly different from the corresponding steps in the GenePattern Flow Cytometry Suite.

The primary difference between flowMatch and other FC data analysis tools (discussed in Section~\ref{sec:intro_related_work}) is that I consider a collection of FC sample related to each other and analyze them collectively.
For example, I characterize a group of similar samples with representative templates and use templates in between-class comparison, classification,  and other overall biological conclusions.
In contrast, most of the existing tools analyze samples individually and only make sample-specific conclusions. 
Finally, flowMatch is not an exhaustive pipeline and I plan to include other functionalities into the pipeline in future.

\section{Datasets used in this thesis}
\label{sec:intro_dataset}
\subsection{Healthy donor (HD) dataset}
\label{sec:hd_data_description}
\begin{figure}[!ht]
   \centering
   \includegraphics[scale=.67]{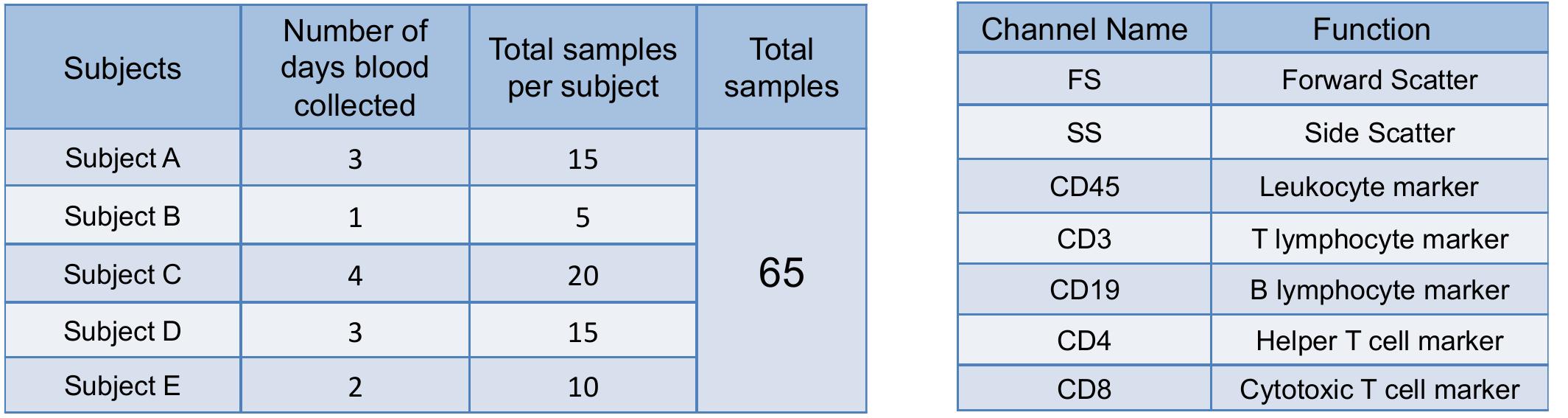} 
   \vspace{-.2 cm}
   \caption[Description of the healthy donor (HD) dataset]{ Description of the healthy donor (HD) dataset. The distribution of samples across subject is shown in the left and the functions of scatter and fluorescence channels are described in the right.}
   \label{fig:HD_data_desc}
   \vspace{-.5 cm}
\end{figure}

The healthy donor (HD) dataset represents a ``biological simulation" where peripheral blood mononuclear cells (PBMC) were collected from five healthy individuals on up to four different days.
Each sample was divided into five parts and analyzed through a flow cytometer at Purdue's Bindley Biosciences Center. 
Thus, we have five technical replicates for each sample from a subject, and each replicate was stained using labeled antibodies against CD45, CD3, CD4, CD8, and CD19 protein markers.
I show a summary of the HD dataset in Fig.~\ref{fig:HD_data_desc}.

The HD dataset includes three sources of variations: (1) technical or instrumental variation among replicates of the same sample, (2) within-subject temporal (day-to-day) variation, and (3) between-subject natural or biological variation. 
The dataset contains 13 replicated groups, with each group containing five copies of the same sample. 
Hence, the variation among five replicates within each replicated group originates from the technical variation in sample preparation, instrument calibration and sample measurement in the flow cytometer.
The within-subject temporal variation reflects the environmental impact  on the immune system on different days when the blood was drawn from a subject.
Finally, samples from different subjects have natural between-subject variations.
Different sources of variations present in the HD dataset mate it an ideal case to demonstrate the functionality of the \emph{flowMatch} pipeline.

\subsection{T cell phosphorylation (TCP) dataset}
\label{sec:tcp_data_description}
I reanalyze a  T cell phosphorylation (TCP) data set from Maier et al.~\cite{maier2007allelic} to determine differences in phosphorylation events downstream of T cell receptor activation in naive and memory T cells. 
For each of the 29 subjects in this study, whole blood was stained using labeled antibodies against CD4, CD45RA, SLP-76 (pY128), and ZAP-70 (pY292) protein markers before stimulation with an anti-CD3 antibody, and another aliquot underwent the same staining procedure five minutes after stimulation.
The first two markers (CD4, CD45RA) are expressed on the surface of different T cell subsets and the last two (SLP-76 and ZAP-70) are highly expressed after T cells are phosphorylated~\cite{maier2007allelic}.

During the stimulation anti-CD3 antibody binds with T cell receptors (TCR) and activates the T cells,
initiating the adaptive immune response. 
The binding with TCR induces  dramatic changes in  gene expression and cell morphology, 
and induces the formation of a phosphorylation-dependent signaling network via 
multi-protein complexes. 
ZAP-70 is a kinase that phosphorylates tyrosine in a trans-membrane protein called LAT, 
and LAT and SLP-76 are part of a platform that assembles the signaling proteins~\cite{Brockmeyer+:phosphorylation}.  
I reanalyzed this dataset in order to demonstrate that computationally meta-clusters are preserved across experimental conditions (before and after stimulation) when the homogeneity of meta-clusters is preserved.

The AML dataset is discussed in Chapter~\ref{chap:aml-classification}.

\section{Related work}
\label{sec:intro_related_work}
Recent advances in FC technology have led to an influx of automated tools to analyze FC data  \cite{aghaeepour2013critical, pyne2009automated, spidlen2013genepattern, kotecha2010web}.
However, many of the automated tools solve a small subset of the problems arising in FC data analysis.
For example, spectral unmixing (compensation) is discussed in \cite{roederer2001spectral, Bagwell1993, Novo2013} and various data transformation methods are discussed in \cite{bagwell2005hyperlog, parks2006new, dvorak2005modified, novo2008flow, finak2010optimizing, ray2012computational}.
Automated gating or clustering is arguably the most discussed problem in FC data analysis \cite{aghaeepour2013critical, pyne2009automated, lo2008automated, aghaeepour2011rapid} and Aghaeepour et al. \cite{aghaeepour2013critical} provides a state-of-the-art summary of the field.
Cluster matching and meta-clustering have been studied only recently by \cite{pyne2009automated, finak2010optimizing}, but have received relatively less attention from the researchers.
Other general-purpose and problem-specific methods are also discussed in the literature \cite{aghaeepour2012early, qiu2011extracting, robinson2012computational, maecker2012new, bashashati2009survey}.
A number of the aforementioned tools are implemented as free, open-source R packages such as flowCore \cite{hahne2009flowcore}, flowViz \cite{sarkar2008using}, flowClust \cite{lo2009flowclust}, flowTrans \cite{finak2010optimizing}, flowStats \cite{hahne2010flowstats}, flowType \cite{aghaeepour2012early} and other packages available  through Bioconductor \cite{gentleman2004bioconductor}.

Recently  Spidlen et al. introduced a web-based FC data analysis pipeline called ``GenePattern Flow Cytometry Suite"  (http://www.genepattern.org) that includes 34 open-source modules performing preprocessing, normalization, gating, cluster matching and other post-processing steps \cite{spidlen2013genepattern}. 
\emph{flowMatch} is similar to the GenePattern Flow Cytometry Suite in terms of the problem it solves, but significantly different on how each step is performed. 
For example, GenePattern normalizes FC samples by aligning density peaks on each channel as described by Hahne et al. \cite{hahne2010per}. 
In contrast, I stabilize variances of the density peaks on each channel without shifting the signals. 
Likewise, other steps of the flowMatch pipeline are significantly different from the corresponding steps in the GenePattern Flow Cytometry Suite.

To the best of my knowledge, there is no other software tool that would provide a comprehensive pipeline for FC data analysis. 
However, a few software tools, most of them commercial, integrate one or two steps of the complete analysis pipeline. 
For example, the Immunology Database and Analysis Portal (ImmPort, https://immport.niaid.nih.gov) integrates the FLOCK software that performs normalization and density based clustering \cite{scheuermann2009immport}. 
Cytobank \cite{kotecha2010web} is a web-based application focusing mainly on the organization and storage of cytometry data. 
Finally, most of the major commercial third party software vendors, including Tree Star, De Novo Software, and Verity Software House, offer computational tools for certain steps of the FC data analysis. 
However, all these tools solve only a few steps of the overall computational analysis with limited support for the other steps. 
I hope that  flowMatch will be a useful addition to the available FC data analysis tools and can contribute to faster analysis of FC datasets.

\section{Outline of the thesis}
This thesis describes various computational aspects of analyzing flow cytometry data.
I have developed a systematic pipeline, \emph{flowMatch} containing five functionally distinct modules for analyzing large volume of multi-parametric FC data.
These five algorithmic modules are discussed in five chapters of this thesis.
In Chapter~\ref{chap:variance_stabilization}, I describe \emph{flowVS}, a method for stabilizing variance in FC data by transforming each channel with nonlinear functions.
Chapter~\ref{chap:clustering} discusses several basic and two consensus clustering algorithms and explains how multiple cluster validation indices can simultaneously be optimized to select the ``best" clustering algorithm and algorithmic parameters. 
Chapter~\ref{chap:matching} discusses a robust mixed edge cover (MEC) algorithm for registering (labeling) cell clusters across samples.
The next Chapter~\ref{chap:template} discusses a hierarchical matching-and-merging (HM\&M) algorithm that summarizes a collection of similar samples with templates consisting of several homogeneous meta-clusters.
I discuss how the templates are used to classify new samples in a dynamic fashion in Chapter~\ref{chap:template-based-classification}. 
Chapter \ref{chap:aml-classification} discusses algorithms for classifying and immunophenotyping the acute myeloid leukemia (AML). 
Concluding remarks and future directions of research are presented in Chapter~\ref{chap:conclusions-future-work}.

%% file: ch2-variance-stabilization.tex
\chapter{Variance stabilization in flow cytometry}
\label{chap:variance_stabilization}
\section{Introduction}
In this chapter, I discuss \emph{flowVS} -- a novel method for stabilizing within-population variances across flow cytometry (FC) samples.
\emph{flowVS} stabilizes variance by transforming FC samples with the inverse hyperbolic sine (asinh) function whose parameters are optimized to homogenize the within-population variances.
Variance stabilization (VS) is a data-transformation process for dissociating data variability from the mean values.
In flow cytometry, the purpose of VS is to decouple biological signals (usually measured by average marker expressions of cell populations) from different sources of variations and noise so that only biologically significant effects are detected. 

Variance inhomogeneity is an inherent problem in fluorescence-based FC measurements and is a bottleneck for automated multi-sample comparisons.
The origin of the problem is the fluorescence signal formation and the detection process that monotonically increases the variance of the fluorescence signal with the average signal intensity \cite{snow2004flow, Novo2013}.
Due to such signal-variance dependence, a cell population (a cluster of cells with similar marker expressions) with higher level of protein expressions (i.e., higher fluorescence emission) has higher variance than another population with relatively low level of protein expressions (i.e., low fluorescence emission). 
This inhomogeneity of within-population variance creates problems with extracting features uniformly and comparing cell populations with different levels of marker expressions.
Furthermore, detecting statistically significant changes among populations, such as in an analysis of variance (ANOVA) model, explicitly requires that variance be approximately stabilized in populations.
By removing mean-variance dependence from FC data, VS makes it possible to detect biologically and statistically meaningful changes across populations from different samples.

VS is an explicit preprocessing step for other fluorescence-based technologies such as the microarrays~\cite{schena1995quantitative, chen1997ratio, durbin2002variance, huber2002variance}.
However, unlike microarray data, explicit VS is often not performed in FC data analysis. 
Traditionally, FC data is transformed with nonlinear functions to project cell populations with normally distributed clusters -- a choice that usually simplifies subsequent visual analysis \cite{lo2008automated, bagwell2005hyperlog, parks2006new, novo2008flow, dvorak2005modified, finak2010optimizing}. 
Recently Finak et al. used the maximum likelihood approach \cite{finak2010optimizing} with different transformations to explicitly satisfy normality of the cell populations.
Ray et al.~\cite{ray2012computational} transformed each channel  with the asinh function whose parameters are optimally selected by the Jarque-Bera test of normality (a goodness-of-fit test of whether sample data have the skewness and kurtosis matching a normal distribution). 
While these transformations approximately normalize FC data, they might not stabilize the variance.

The VS problem in FC, however, cannot be solved directly by applying the mature VS techniques from the microarray literature.
In microarrays, each gene is measured multiple times (possibly under multiple conditions) and the between-sample variance for each gene is stabilized with respect to the average expression of the gene across samples.
In contrast, variance is defined by within-population, cell-to-cell variation in FC and  this within-population variance is stabilized with respect to the average expression of markers within each population.
These contrasting objectives prevent us from applying VS methods from microarray literature directly to flow cytometry.

I address the need for explicit VS in FC with a maximum likelihood (ML) based method, \emph{flowVS}, which is built on top of a commonly used asinh transformation. 
The choice of asinh function is motivated by its success as a variance stabilizer for microarray data~\cite{durbin2002variance, huber2002variance}.
 \emph{flowVS} stabilizes the within-population variances separately for each marker (fluorescence channel) $z$ across a collection of $N$ samples.
After transforming $z$ by asinh($z/c$) ($c$ is a normalization cofactor), \emph{flowVS} identifies one-dimensional clusters (density peaks) in the transformed channel. 
Assume that a total of $m$ 1-D clusters are identified from $N$ samples with the $i^{th}$ cluster having variance $\sigma^2_i$.
Then the asinh($z/c$) transformation works as a variance stabilizer if the variances of the 1-D clusters are approximately equal, i.e., $\sigma^2_1 \sim \sigma^2_2\sim ... \sim \sigma^2_m$. 
To evaluate the homogeneity of variance (also known as homoskedasticity), I use  Bartlett's likelihood-ratio test\cite{Bartlett1937}.
From a sequence of cofactors used with the asinh function, \emph{flowVS} selects one with the ``best" VS quality computed by Bartlett's test. 
\emph{flowVS} is therefore an explicit VS method that stabilizes within-cluster variances in each marker/channel by evaluating the homoskedasticity of clusters with a Likelihood-ratio test.

I show, with a five-dimensional healthy dataset, that \emph{flowVS} removes the mean-variance dependence from raw FC data and makes the within-population variance relatively homogeneous.
Such variance homogeneity is especially useful to build meta-clusters from a collection of phenotypically similar cell populations across samples.
Previous studies (Hahne et al. \cite{hahne2010per}, for example) shifted the distribution of each fluorescence channel to ensure homogeneity in meta-clusters, but such artificial shifting may hide useful biological signals.
By contrast, \emph{flowVS} builds homogeneous meta-clusters from variance-stabilized clusters without losing the biological differences among samples.
I will discuss the impact VS on comparisons among samples (with related concepts of meta-clusters and templates) in Chapter~\ref{chap:template}.

The rest of the chapter is organized as follows. 
I start with a short Section \ref{sec:vs_related_work} on related work and the current transformation techniques employed in FC.
In Section \ref{sec:flowVS} I describe \emph{flowVS}, a method for stabilizing variance of FC data.
The next Section \ref{sec:vs_results} describes applications of this VS technique to an FC dataset and a microarray dataset. 
I conclude this Chapter in Section \ref{sec:vs_conclusions} by discussing limitations of \emph{flowVS} and possible future work.

\section{Related work}
\label{sec:vs_related_work}
From the beginning of the twentieth century, VS has been widely studied for its central role in making heteroskedastic data easily tractable by standard methods.
Heteroskedasticity appears in various datasets mostly because the data follows a distribution with correlated mean and variance, e.g., the Poisson distribution.
For well-known distribution families, VS is usually performed by transforming data with an analytically chosen function $f$.
For example, $f(z)=\sqrt{z+3/8}$ works as a good (asymptotic) stabilizer for a random variable $z$ following the Poisson distribution \cite{anscombe1948transformation}.
Variance stabilizers for several well-known distribution families are described in \cite{anscombe1948transformation, bar1988classical}.
For unknown distributions, heuristic and data-driven stabilizers are often used, e.g., see \cite{bartlett1936square, efron1982transformation, tibshirani1988estimating}.

However, traditional transformations are often inadequate for low-count (photon limited) signals~\cite{zhang2008wavelets, huber2002variance} because of unknown error patterns in fluorescence data.
Hence various \emph{ad hoc} variance stabilization schemes have been developed for different types of fluorescence data.
In microarrays, the VS problem has been addressed by various non-linear transformations~\cite{schena1995quantitative, chen1997ratio, durbin2002variance, huber2002variance}. 
Most notably, the widely used approach by Huber et al. \cite{huber2002variance} uses the inverse hyperbolic sine (asinh) transformation whose parameters are selected by a maximum-likelihood (ML) estimation.  

For flow cytometry data, researchers have used various non-linear transformations, such as the logarithm, hyperlog, generalized Box-Cox, and biexponential (e.g., logicle and generalized arcsinh) functions~\cite{lo2008automated, bagwell2005hyperlog, parks2006new, novo2008flow, dvorak2005modified, finak2010optimizing}.
In the past work, the parameters of these transformations were adjusted in a data-driven manner to maximize the likelihood (\emph{flowTrans} by Finak et al.~\cite{finak2010optimizing}), to satisfy the normality (\emph{flowScape} by Ray et al.~\cite{ray2012computational}), and to comply with simulations (\emph{FCSTrans} by Qian et al.~\cite{qian2012fcstrans}).
\emph {flowTrans} estimates transformation parameters for each sample by maximizing the likelihood of data being generated by a multivariate-normal distributions on the transformed scale.
\emph{flowScape} optimizes the normalization factor of asinh transformation by the Jarque-Bera test of normality.
\emph{FCSTrans} selects the parameters of the linear, logarithm, and logicle transformations with an extensive set of simulations. 
In contrast to these approaches that transform a single sample, \emph{flowVS} transforms a collection of samples together for stabilizing within-population variations.
Note that normalizing data may not necessarily stabilize its variance, e.g., for a Poisson variable $z$, $\sqrt{z+3/8}$ is an approximate variance-stabilizer, whereas $z^{2/3}$ is a normalizer~\cite{efron1982transformation}.

\section{Variance stabilization for flow cytometry data}
\label{sec:flowVS}
\subsection{The goal of variance stabilization}
The aim of variance stabilization (VS) in FC  is to make within-population variances of different cell populations approximately equal and thereby independent of the average protein expressed by populations.
Recall that the expression of a protein is measured by the intensity of a channel capturing the fluorescence of a particular wavelength.
VS therefore stabilizes the within-population fluorescence variance and makes it independent of the mean fluorescence intensities (MFI) of the cell populations.
In this context, I use the terms ``a fluorescence channel" and ``a protein marker" equivalently because we infer information about the latter through the signal collected at the former. 
However, I will use ``fluorescence channel" more frequently because the nature of fluorescence emissions -- not the protein expressions -- dominates the mean-variance relationship in FC data.
I do not stabilize variance on the scatter channels because, as pointed out by Finak et al.~\cite{finak2010optimizing}, there are few benefits to transforming forward and side scatter channels.

\subsection{Channel-specific variance stabilization}
I assume that compensated fluorescence channels are independent and stabilize variance on each channel (a column of the data matrix) separately. 
Besides being simple, one-dimensional VS prevents unnecessary correlation among transformed channels incurred by multi-dimensional VS.
Note that the correlations among fluorescence channels due to spectral overlap are compensated before we stabilize variance.
Even though the protein expressions can still be correlated \cite{parks2006new}, I do not include such correlations in the VS process because the nature of such problem-specific correlation is difficult to model.

Since the actual error model of FC data is unknown, it is not trivial to select a function to transform this data.
Even though researchers have studied a number of normalization functions, they are often selected arbitrarily~\cite{finak2010optimizing, ray2012computational}.
Considering the similarity in fluorescence-based data collections between FC and microarrays, I decided to use an inverse hyperbolic sine (asinh) function that has been shown to successfully stabilize variance in fluorescence readouts from microarray data  \cite{durbin2002variance, huber2002variance}.
This choice of asinh function is also motivated by its success in FC data visualization and normalization \cite{finak2010optimizing, ray2012computational}.
Stabilizing variance with other transformations can be performed using the same \emph{flowVS} framework but is not discussed here. 

To transform a fluorescence channel $z$, I use the asinh transformation with a single parameter $c$:
\begin{equation} 
\operatorname{asinh} (z/c) = \ln(z/c + \sqrt{(z/c)^2 + 1}).
 \label{eq:asinh}
\end{equation}
In this transformation, $c$ is called the normalization cofactor whose value is optimally selected to stabilize variance in channel $z$.
Let $\sigma^2_{ij}$ be the variance of the $i^{th}$ cluster defined in channel/marker $z$ in the $j^{th}$ sample. 
My objective is to select a cofactor for the asinh transformation such that after transforming channel $z$ in each sample, variance $\sigma^2_{ij}$ of  every cluster  becomes approximately homogeneous. 

In a more general form, asinh transformation is expressed with three parameters, $a * \text{asinh}(b+z/c)$, where in addition to the cofactor $c$, $a$ denotes a scaling after transformation, and $b$ denotes a translation before transformation.
I set $a=1$ since other values do not affect downstream analysis, and $b=0$ to avoid shifting of cell populations.
Hence I am left with a single parameter, $c$, that I aim to set in order to stabilize the variance.
Note that other transformations such as logarithmic, hyperlog~\cite{bagwell2005hyperlog}, logicle~\cite{parks2006new}, Box-Cox \cite{lo2008automated} etc., may also stabilize the variance but a detailed analysis covering all these transformations is out of the scope of this study. 

\subsection{\emph{flowVS}: an algorithm for per-channel variance stabilization}
\label{sec:vs_algorithm}
Given a collection of $N$ FC samples, the \emph{flowVS} algorithm stabilizes the variance in a fluorescence channel $z$ with the following steps.
\begin{enumerate}
\item {\em  Selecting a sequence of cofactors}: I select an evenly spaced increasing sequence of cofactors $c_1, c_2, ..., c_l$ to be used with the asinh transformation.
The start and end values ($c_1$ and $c_l$) are empirically selected so that the sequence includes a variance stabilizing cofactor.

\item {\em  Transforming data and evaluating homoskedasticity for each cofactor}:
For each cofactor $c_q \in \{c_1, c_2, ..., c_l\}$,  the following steps are performed.
\begin{enumerate}
\item {\em  Transforming channel/marker $z$ in each sample}: Let  $z_j$ be a vector denoting the selected channel in the $j^{th}$ sample where $1\leq j \leq N$. 
The algorithm transforms $z_j$  by the asinh function: $z'_j = \operatorname{asinh} (z_j/c_q)$, where $z'_j$ is the same channel after the transformation. 

\item {\em  Detecting 1-D density peaks (1-D clusters)}: I estimate the density of each transformed fluorescence vector $z'_j$ by a kernel density estimation method. 
The peaks in the density of $z'_j$ are identified as regions of high local density and significant curvature (also called landmarks in \cite{hahne2010per}).
To identify the 1-D density peaks,  \emph{flowVS} uses the {\tt curv1Filter} function from the {\tt flowCore} package~\cite{hahne2009flowcore} in R. 
Here a density peak represents a 1-D cluster of cells and therefore can also be identified by a clustering algorithm~\cite{aghaeepour2012early}.
Let $P_j$ be the collection of density peaks identified from $z'_j$.   

\item {\em  Collecting density peaks from all sample}: Density peaks from all samples are collected into a set $P$ such that $P = \cup_{1\leq j \leq N} P_j $. Let $P$ contain a total of $m$ density peaks where the $i^{th}$ peak contains $n_i$ cells with mean $\mu_i$ and variance $\sigma^2_i$.

\item {\em  Testing homoskedasticity}: The performance of the asinh transformation with cofactor $c_q$ is evaluated by a test of variance homogeneity (homoskedasticity).
For this purpose, I use Bartlett's test \cite{Bartlett1937}, a well-known likelihood-ratio test for homoskedasticity. 
Assuming $n=\sum_{1\leq i \leq m} n_i$ to be the total number of cells in $P$ and $\sigma^2_p$ to be the pooled variance of $m$ density peaks, I compute Bartlett's statistics as follows:

\begin{equation}
B(c_q) = \frac{(n-m)\: \ln(\sigma_p^2) - \sum_{i=1}^m (n_i-1)\: \ln(\sigma_i^2)}{1 + \frac{1}{3(m-1)} \left( \sum_{i=1}^{m} \frac{1}{n_i-1} - \frac{1}{n-m}\right)}.
\label{eq:bartlett}
\end{equation}
Bartlett's statistics $B(c_q)$ computes the degree of homogeneity across all 1-D clusters in channel $z$  after it is transformed by $\operatorname{asinh} (z_j/c_q)$.
\end{enumerate}

\item {\em  Finding a cofactor for optimum VS}: The optimum variance stabilization is achieved by a transformation giving minimum value of Bartlett's statistics.
Therefore, the asinh transformation with the cofactor 
\begin{equation}
c_{q^*} = \arg\!\min_{q=1}^l B(c_q),
\label{eq:bartlett}
\end{equation}
gives the optimum VS for the channel/marker $z$.
Channel $z_j$ in each sample is then transformed by $\operatorname{asinh} (z_j/c_{q^*})$ and used in subsequent analysis.
\end{enumerate}


\section{Results}
\label{sec:vs_results}
I demonstrate how the  \emph{flowVS} algorithm stabilizes variance with the HD dataset described in Section~\ref{sec:hd_data_description}.
Recall that the HD dataset consists of 65 samples from five healthy individuals who donated blood samples on different days. 
Each sample was divided into five replicates and each replicate was stained using labeled antibodies against CD45, CD3, CD4, CD8, and CD19 protein markers.
Before variance stabilization each sample is preprocessed, compensated for spectral overlap, and gated on FS/SS channels to identify lymphocytes.
Since lymphocytes always express CD45 protein (CD45 is a common leukocyte marker), I omit this marker from the rest of the discussion.

 \begin{figure}[!t]
   \centering
   \includegraphics[scale=.4]{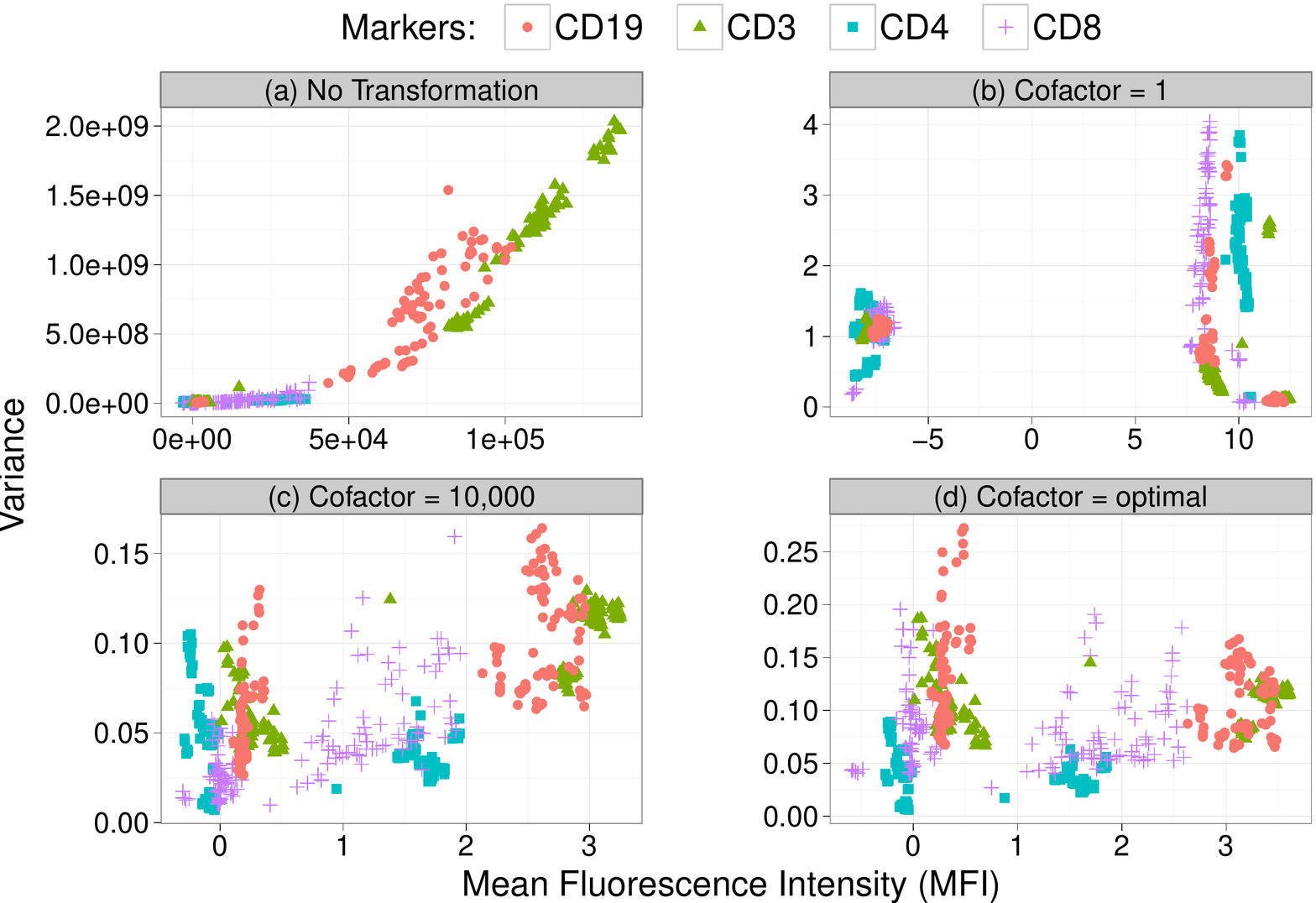} 
   \vspace{-.2 cm}
   \caption[Mean-variance relationship in samples from the HD dataset before and after variance stabilization]{Mean Fluorescence Intensity (MFI) of each 1-D cluster is plotted against the variance of the cluster for different choice of cofactors used with the asinh transformation.  Clusters in each marker are shown with the same symbol and color. (a) no transformation: variance increases monotonically with the mean, (b) small cofactor: VS is not achieved, (c) reasonable value of cofactor (same for all channels): mean-variance dependence remains except for CD4 channel/marker, and (d) optimal values of cofactors: variance is approximately stabilized for each marker/channel. } 
   \label{fig:VS_HD}
   \vspace{-.5 cm}
\end{figure}

\subsection{Selecting the optimum cofactors for the asinh transformation}
For each sample of the HD dataset, the \emph{flowVS} algorithm identifies density peaks in CD3, CD4, CD8, and CD19 markers/channels by following step 2(b) in Section~\ref{sec:vs_algorithm}.
In each of these four channels, the algorithm identifies two density peaks representing high-expressing (``positive") and low-expressing (``negative") cell populations (e.g., CD3$^-$, CD3$^+$ clusters in the CD3 marker/channel).
Therefore, from the 65 samples in the HD dataset, I obtain 130 ($65\times2$) 1-D clusters for each marker, giving a total of 520 ($65\times2\times4$) 1-D clusters in CD3, CD4, CD8, and CD19 channels.

For each of these 520 clusters, I compute the within-cluster variance and average marker expression (Mean Fluorescence Intensity, MFI).
Figure~\ref{fig:VS_HD} plots the mean-variance relationship for every cluster (density peak) before transforming the channels and after they are transformed by asinh function with different cofactors.
In this figure, clusters identified in channels transformed with different cofactors are shown in different panels and in each panel, clusters in a marker are shown with the same symbol and color.
Subfig.~\ref{fig:VS_HD}(a) reveals the non-linear correlation between cluster variance and mean, which is typically observed in raw FC data before applying any transformation.  
The mean-variance dependence  is not alleviated after transforming the data by asinh function with arbitrary cofactors as can be seen in Subfig.~\ref{fig:VS_HD}(b,c).
Finally, Subfig.~\ref{fig:VS_HD}(d) shows that the variances of the 1-D clusters become approximately stabilized and independent of the means after the optimum variance stabilization is performed.
On CD3 marker, for example, CD3$^-$, CD3$^+$ clusters (shown by green triangles in Subfig.~\ref{fig:VS_HD}(d)) have approximately stable variances and no visible correlation exists between the variances and the cluster means.

\begin{figure}[!t]
   \centering
    \includegraphics[scale=.4]{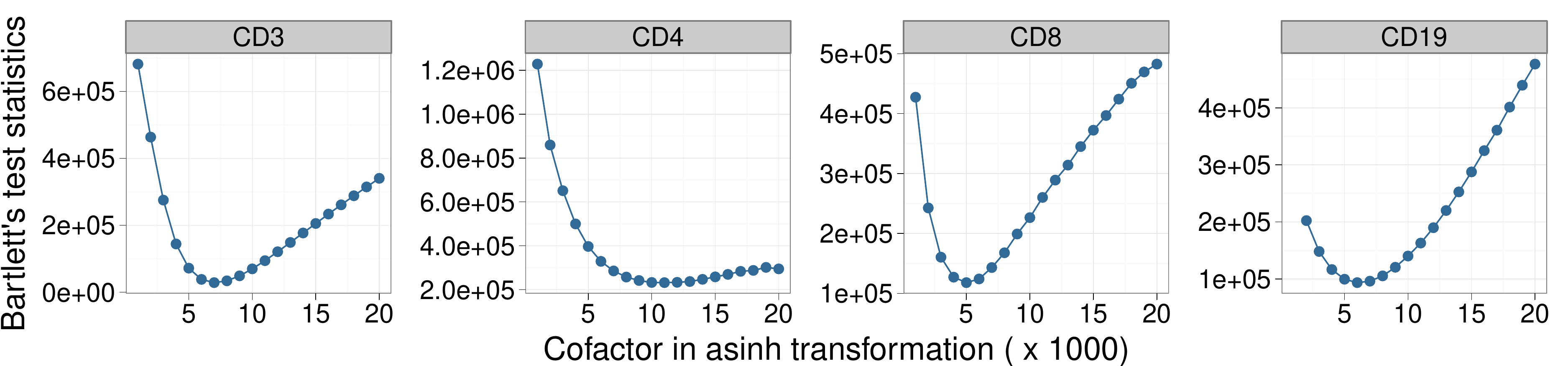}
     \vspace{-.8 cm}
   \caption[Variance stabilizing cofactors used with the asinh transformation for the HD dataset]{Finding the optimum values of the cofactors in asinh transformation for each fluorescence channel in the HD dataset. } 
    \label{fig:bartlett_minima}
     \vspace{-.5 cm}
\end{figure}

The optimum cofactor for the asinh transformation is selected by minimizing Bartlett's statistics separately for each of the four markers.
Figure~\ref{fig:bartlett_minima} shows Bartlett's statistics computed in each channel after it has been transformed by asinh function with a sequence of cofactors. 
A minimum is observed for every channel and the cofactor is set to the value of the minimizer of Bartlett's statistics. 
The variance stabilizing cofactors for different markers are: (a) 5,000 for CD8, (b) 6,000 for CD19, (c) 7,000 for CD3, and (d) 10,000 for CD4.
Therefore, the channels of the HD dataset are transformed by asinh function with the optimum cofactors.
As shown in Subfig.~\ref{fig:VS_HD}(d), the optimum transformations approximately stabilize variances in different channels.

\begin{figure}[!t]
   \centering
    \includegraphics[scale=.4]{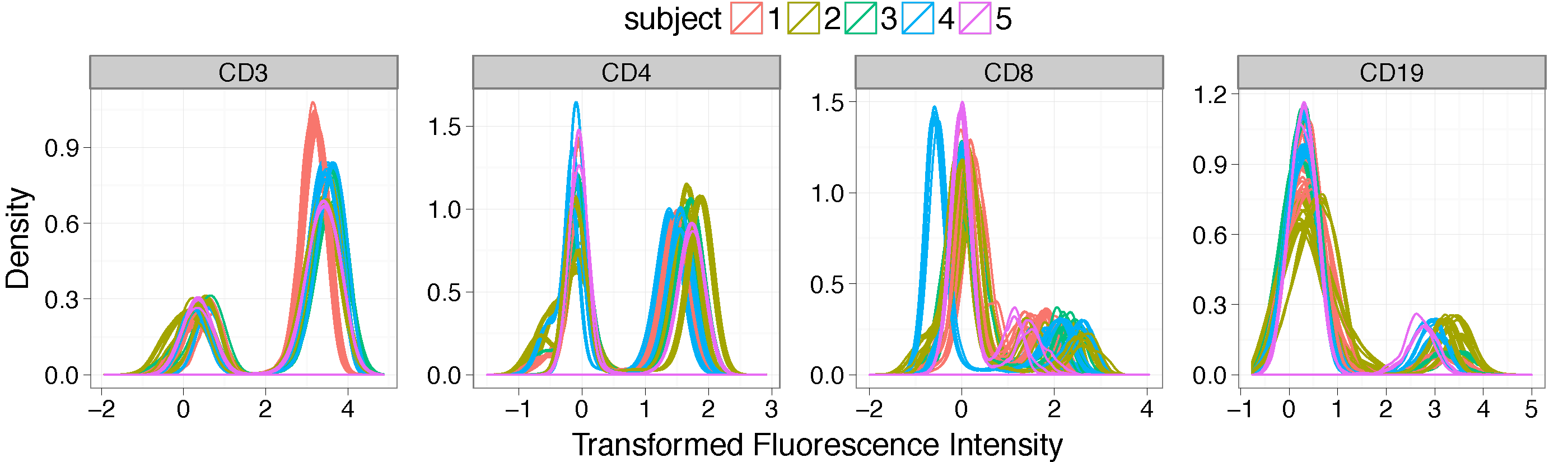}
         \vspace{-.2 cm}
   \caption[Density plot of the variance stabilized channels from samples in the HD dataset]{Density plot of the variance stabilized channels across all samples in the HD dataset. Different colors used for samples from different subjects. 
   } 
     \label{fig:density_plot_HD}	
      \vspace{-.5 cm}
\end{figure}

I plot the density of the variance stabilized channels in Fig.~\ref{fig:density_plot_HD}, where different colors are used to denote the samples from five different subjects.
In Fig.~\ref{fig:density_plot_HD}, both positive and negative density peaks (clusters with high or low marker expression)  spread approximately equally across all samples, confirming the homogeneity of variances in one dimensional clusters.
For this dataset, the density curves from the same subjects are tightly grouped together as expected.
However, clusters across subjects may not be well aligned due to the between-subject variations.
Aligning density peaks across samples -- as was done by Hahne et al. \cite{hahne2010per} -- is not an objective of the VS algorithm, because such shifting of density may potentially eclipse true biological signals across samples.

\subsection{Normality of the variance-stabilized clusters}
\begin{figure}[!t]
   \centering
    \includegraphics[scale=.5]{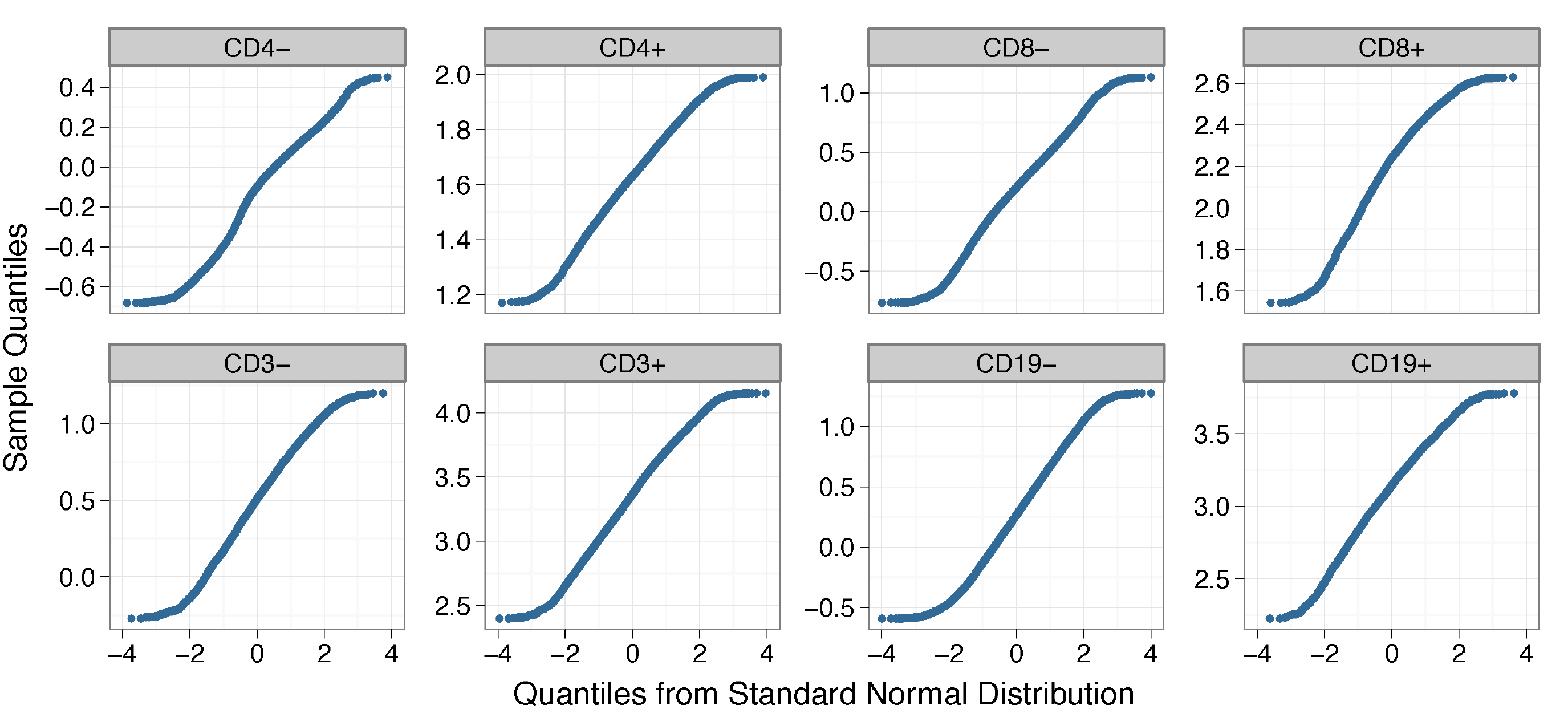}
          \vspace{-.2 cm}
   \caption[The normality of the one dimensional clusters after variance stabilization]{The Q-Q plots for the eight 1-D clusters obtained from a representative sample in the HD dataset. Every Q-Q plot  shows linearity in the central part, except for a little deviation at the end, indicating that the clusters approximately follow a normal distributions with heavier tails.} 
     \label{fig:qqplot}	
      \vspace{-.5 cm}
\end{figure}

In addition to stabilizing the variances, the transformed channels approximately follow the normal distribution.
To see this, I draw the Quantile-Quantile plots (Q-Q plots) \cite{wilk1968probability} for eight 1-D clusters obtained from a representative sample in the HD dataset in Figure \ref{fig:qqplot}.
In each Q-Q plot, the distribution of a 1-D cluster is compared with the standard normal distribution by plotting their quantiles against each other.
If a cluster is normally distributed (i.e., linearly related to the standard normal distribution), the points in the Q-Q plot approximately lie on a straight line.
All eight Q-Q plots in Fig.~\ref{fig:qqplot} show linearity in their central parts, except small deviations at the ends, indicating that the 1-D clusters approximately follow normal distributions with heavier tails.
As I will discuss in Section~\ref{sec:homogeneity_mc}, the normality of clusters is a desired condition for the analysis of variance (ANOVA) model used to evaluate the homogeneity of a collection of similar clusters, also known as a meta-cluster.

\subsection{Application to microarray data}
The VS approach based on optimizing Bartlett's statistics can also be used to stabilize variance in microarray data.  
However, the initial steps of \emph{flowVS} need to be adapted for microarrays.
Assume that the expressions of $m$ genes are measured from $N$ samples in a microarray experiment. 
After transforming the data by the asinh function, the mean $\mu_i$ and variance $\sigma^2_i$ of the $i^{th}$ gene $g_i$ are computed from the expressions of $g_i$ in all samples.  
\emph{flowVS} then stabilizes the variances of the genes by transforming data using the asinh function with an optimum choice of cofactor.
Unlike FC, a single cofactor is selected for all genes in the microarray data.

\begin{figure}[!t]
   \centering	
\subfigure[Selecting optimum cofactor for flowVS]{
    \includegraphics[scale=.5]{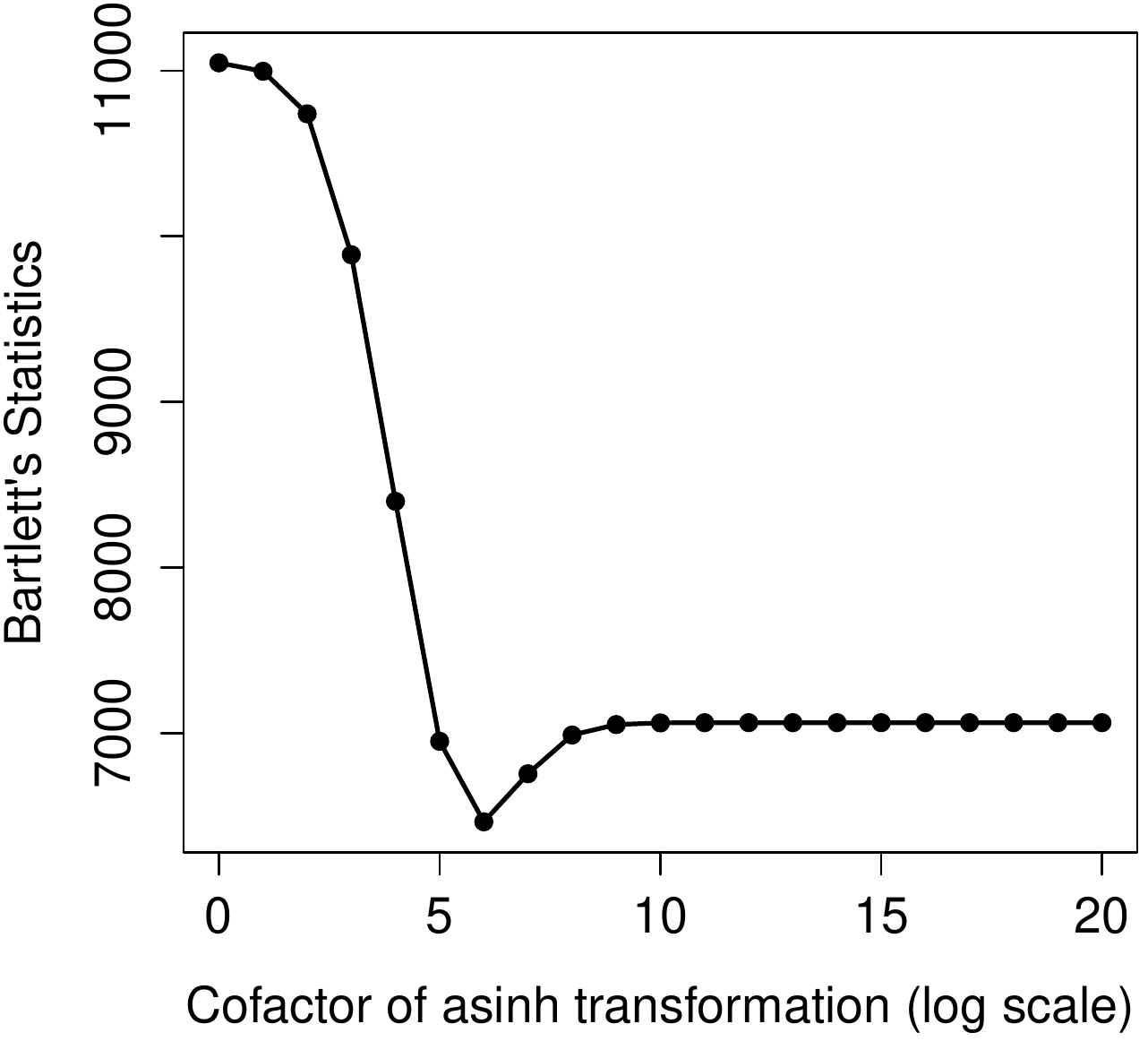}
    \label{fig:kidney_bartlett}
	}
	\hspace{.7 cm}
\subfigure[VS by different methods]{
    \includegraphics[scale=.5]{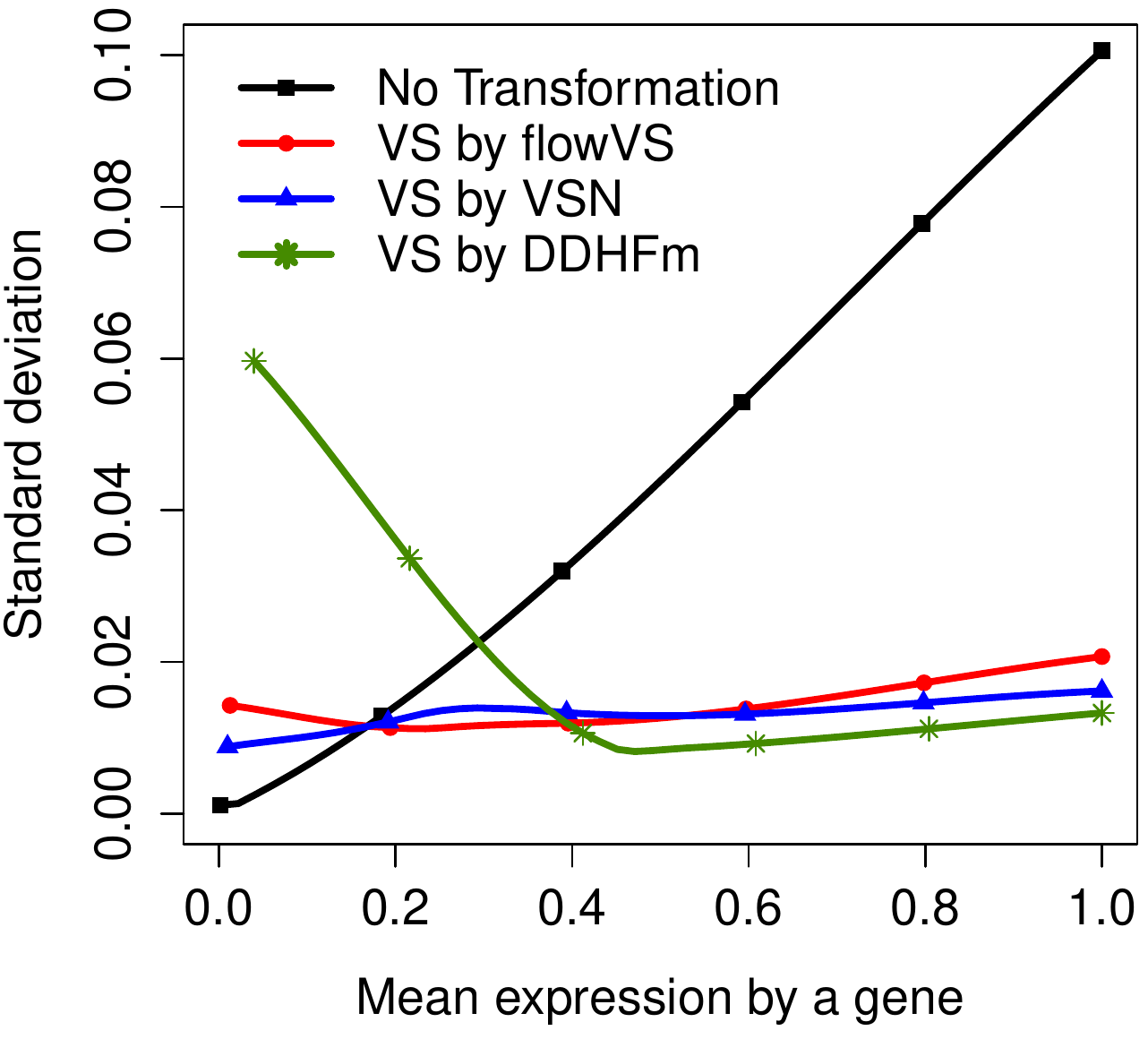}
    \label{fig:kidney_all_methods}	
  }
  \label{fig:kidney_vs}
  \vspace{-.2 cm}
  \caption[Three variance stabilization approaches applied to the Kidney microarray data]{(a) For Kidney microarray data~\cite{huber2002variance}, \emph{flowVS} selects the optimum cofactor for the asinh transformation by minimizing Bartlett's statistics. 
  The cofactors are shown in the natural logarithm scale.
  (b) The standard deviation and mean of each gene from the Kidney data are plotted before transformation and after variance stabilization by \emph{flowVS}, {\tt VSN}, and {\tt DDHFm}. Loess regression is used to smoothen the curves.  
   }
  \vspace{-.5 cm}
 \end{figure}

I have applied the modified \emph{flowVS} to the publicly available Kidney microarray data provided by Huber et al.~\cite{huber2002variance}.
The Kidney data reports the expression of 8704 genes from two neighboring parts of a kidney tumor by using cDNA microarray technology. 
For different values of the cofactor, \emph{flowVS} transforms the Kidney data with the asinh function and identifies the optimum cofactor by minimizing Bartlett's statistics.
Subfig.~\ref{fig:kidney_bartlett} shows that a minimum value of Bartlett's statistics is obtained when the cofactor is set to $\exp(6)$ ($\sim400$). 
The optimum cofactor is then used with the asinh function to transform the Kidney data.

I compare the VS performance of  \emph{flowVS} with two software packages, {\tt VSN} by Huber et al.~\cite{huber2002variance} and  {\tt DDHFm} by Motakis et al.~\cite{motakis2006variance}.
Similar to  \emph{flowVS}, {\tt VSN} uses asinh transformation whose parameters are optimized my maximizing a likelihood function~\cite{huber2002variance}.
\texttt{DDHFm}  applies a data-driven Haar-Fisz transformation (HFT)\cite{fryzlewicz2005data, motakis2006variance} to stabilize the variance.
Both  {\tt VSN} and \texttt{DDHFm} are developed for stabilizing variance in microarray data and can not be applied to flow cytometry data.

\begin{figure}[!t]
   \centering
    \includegraphics[scale=.48]{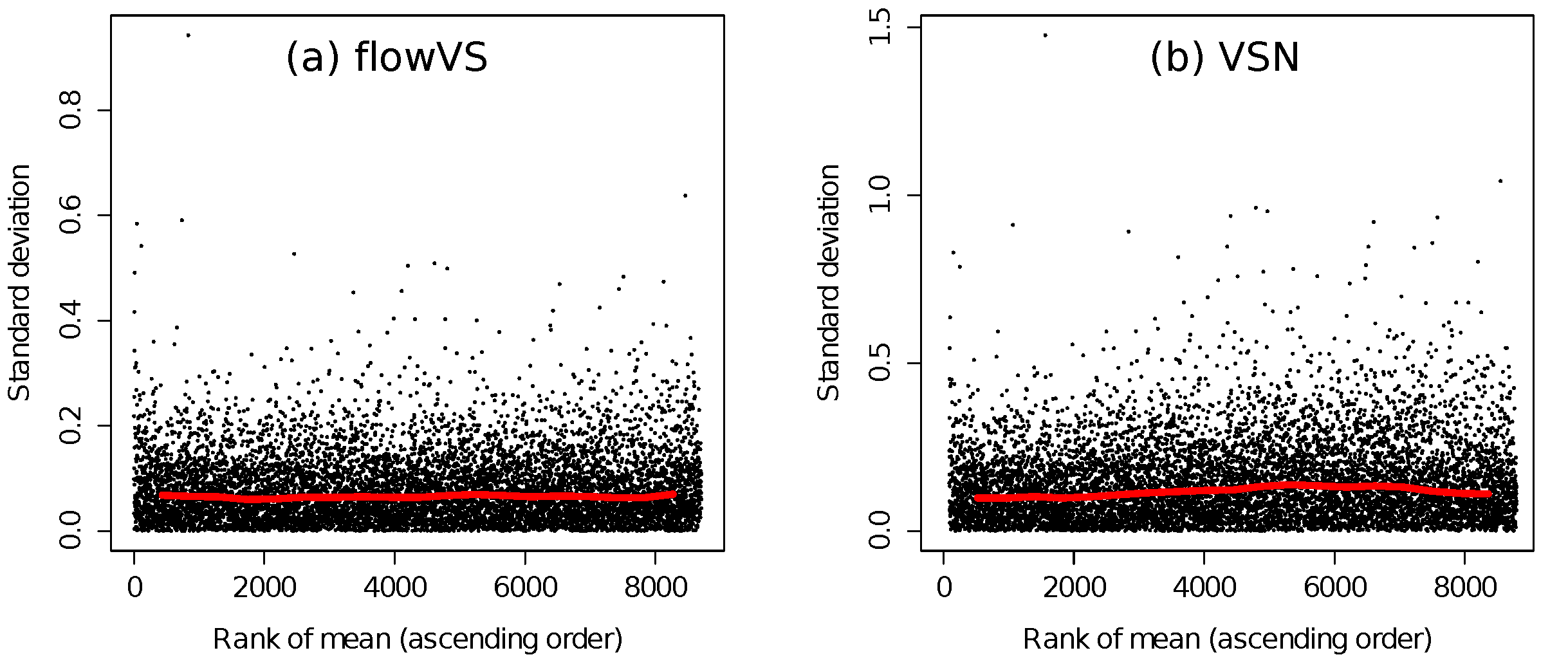}
          \vspace{-.5 cm}
   \caption[Variance stabilization of the Kidney microarray data]{Variance stabilization of the Kidney microarray data~\cite{huber2002variance} by (a) \emph{flowVS} and (b) VSN~\cite{huber2002variance}. 
   Each black dot plots the standard deviation of a gene against the rank of its mean. 
   The red lines depict the running median estimator. 
   If there is no mean-variance dependence, then the red lines should be approximately horizontal.
}
     \label{fig:kidney_flowVS_VSN}	
      \vspace{-.5 cm}
\end{figure}

Before transforming the Kidney data and after transforming it by \emph{flowVS}, {\tt VSN}, and \texttt{DDHFm}, I plot the mean and standard deviation of every gene in Subfig.~\ref{fig:kidney_all_methods}. 
In this figure, I have applied a loess regression to obtain smooth average curves.
We observe in Subfig.~\ref{fig:kidney_all_methods} that the standard deviation of the untransformed Kidney data increases monotonically with the mean. 
Both VSN and \emph{flowVS} approaches stabilize the variance approximately for all genes in this data.
However, the Haar-Fisz transformation achieves good VS properties only for genes with higher levels of expressions. 

To take a closer look at the transformed data by  \emph{flowVS} and {\tt VSN}, I plot the variances of the genes against the ranks of the means with two subfigures in Fig.~\ref{fig:kidney_flowVS_VSN}.
These figures are generated by {\tt meanSdPlot} function from the {\tt VSN} package. 
Here, the ranks of the means distribute the data evenly along the $x$-axis and thus make it easy to visualize the homogeneity of variances.
Both VSN and \emph{flowVS} remove the mean-variance dependence since the red lines are approximately horizontal in both Subfig.~\ref{fig:kidney_flowVS_VSN}(a) and Subfig.~\ref{fig:kidney_flowVS_VSN}(b).
Therefore, \emph{flowVS} performs equally well compared to the state-of-the-art approach developed for the microarray data.

\section{Conclusions}
\label{sec:vs_conclusions}
I describe a variance stabilization framework, \emph{flowVS}, that removes the mean-variance correlations observed in cell populations from flow cytometry samples.
This framework transforms each marker/channel by the asinh function whose normalization cofactor is optimally selected by Bartlett's likelihood-ratio test.
I show, with a five-dimensional healthy dataset consisting of 65 FC samples, that \emph{flowVS} removes the non-linear mean-variance dependence from raw FC data and makes the within-population variances relatively homogeneous across all populations.
\emph{flowVS} also performs comparably to the state-of-art variance stabilization approaches for the microarray data. 

Variance homogeneity (homoskedasticity) is a desirable property for comparing populations across conditions, building meta-clusters from phenotypically similar populations, and analyzing meta-clusters in an ANOVA model.
However, unlike the earlier approach by Hahne et al. \cite{hahne2010per}, \emph{flowVS} does not artificially shift populations to align them in the marker space.
By stabilizing the variances, \emph{flowVS} homogenizes similar cell populations and establishes the foundation of biologically meaningful meta-clusters and templates as will be discussed in Chapter~\ref{chap:template}.

The VS framework presented here has several limitations.
First, \emph{flowVS} stabilizes variance separately in each marker/channel.
This approach is inadequate to stabilize covariances across multiple channels, which is necessary when channels are correlated.
Second, \emph{flowVS} repeatedly identifies 1-D cell clusters (density peaks) and evaluates the homogeneity of clusters by the likelihood-ratio test.
Therefore, this framework does not perform well when cell clusters are not easily identifiable.
Third, \emph{flowVS} stabilizes variance more accurately when more samples are simultaneously passed to the framework.
Hence, the approach is not suitable for normalizing a single sample or stabilizing variances of sequentially arriving samples.
Note that we stabilize between-sample variances in  microarray data; therefore, VS can not performed on a single microarray sample.
Finally, Bartlett's test used in  \emph{flowVS} assumes that the deviation from normality is relatively modest. 
If data deviates significantly from normality, other likelihood ratio tests can be employed, such as LeveneÕs test~\cite{levene1960robust} or the Brown-Forsythe test~\cite{brown1974robust}.
However, I tested \emph{flowVS} with several FC datasets and in every case Bartlett's test outperforms the LeveneÕs and Brown-Forsythe tests in selecting variance stabilizing trnasformations.

\emph{flowVS} operates as an independent module in the FC data analysis pipeline.
It does not depend on the preprocessing algorithms applied before VS nor on the post-analysis methods such as matching, meta-clustering, and classification algorithms.
Hence, \emph{flowVS} is capable of working  with other downstream algorithms developed by other researchers, such as FLAME by Pyne et al.~\cite{pyne2009automated} and flowTrans by Finak et al.~\cite{finak2010optimizing}.
  

%% file: ch3-clustering.tex
\chapter{Population identification by clustering algorithms}
\label{chap:clustering}

\section{Introduction}
In this chapter, I discuss clustering algorithms for identifying phenotypically distinct cell clusters in a flow cytometry (FC) sample.
These algorithms partition a multidimensional sample into several phenotypic clusters such that cells within a cluster are biologically similar to each other but distinct from those outside the cluster. 
A phenotypic cell cluster usually represents a particular sub-type of cells with specific biological function and often called a \emph{cell population} in flow cytometry.
Analyzing cellular functions at the level of populations instead of individual cells conveys more robust biological information because of the natural variations within cells of the same type.  
Therefore, identifying cell populations from the mixture of different cell-types turns into an important step in any FC data analysis pipeline.

Traditionally, cell populations are identified by a manual process known as ``gating", where cell clusters are recognized in a collection of two-dimensional scatter plots (see Fig.~\ref{fig:clustering_hd}(b) for an example).
However, with the ability to monitor a large number of protein markers simultaneously and to process a large number of samples with a robotic arm, manual gating is not feasible for high-dimensional or high-throughput FC data.
To address the gating problem, researchers have proposed several automated clustering algorithms, such as {\tt CDP}~\cite{chan2008statistical}, {\tt FLAME}~\cite{pyne2009automated}, {\tt FLOCK}~\cite{qian2010elucidation}, {\tt flowClust/Merge}~\cite{lo2008automated, finak2009merging}, {\tt flowMeans}~\cite{aghaeepour2011rapid}, {\tt MM}~\cite{sugar2010misty}, {\tt curvHDR}~\cite{naumann2010curvhdr}, {\tt SamSPECTRAL}~\cite{zare2010data}, {\tt SWIFT}~\cite{naim2010swift}, {\tt RadialSVM}~\cite{quinn2007statistical}, etc.
For a summary of these algorithms, see Table 1 of \cite{aghaeepour2013critical}.
Aghaeepour et al. \cite{aghaeepour2013critical} provides a state-of-the-art summary of the field.

Different algorithms perform better for different FC datasets as was observed in a set of challenges organized by the FlowCAP consortium ({\tt http://flowcap.flowsite.org/})~\cite{aghaeepour2013critical}.
Given the large number of algorithmic options, it is often difficult to select the best algorithm for a particular dataset.
When the ``ground truth" or ``gold standard" about the clustering pattern is unavailable, we can evaluate the quality of a clustering solution with the cluster validation methods~\cite{jain1999data, halkidi2001clustering}.
The validation methods evaluate how well a given partition captures the natural structure of the data based on information intrinsic to the data alone.
They can be used in selecting the optimum parameters for a clustering algorithm (e.g., the optimum number of clusters), as well as choosing the best algorithm for a dataset.

I describe several cluster validation methods that can evaluate the quality of the clustering solution given by an algorithm.
In general terms, there are three major cluster validation approaches based on \emph{external}, \emph{internal}, and \emph{relative} validation criteria \cite{jain1999data, halkidi2001clustering}.
The internal validation techniques evaluate how well a given partition captures the natural structure of the data based on information intrinsic to the data, and therefore suitable to select optimum parameters for a clustering algorithm.
I discuss five popular internal cluster validation methods and show that they can be simultaneously optimized to select the algorithm with the best performance on an FC sample as well as the parameters with the algorithm.

If an agreement among the validation methods can not be reached when choosing an algorithm, we use a consensus of several clustering solutions~\cite{hornik2008clue, aghaeepour2013critical}.
I present two heuristic algorithms that compute consensus clusterings from a collection of partitions (cluster ensembles).
The first heuristic approach is called {\tt Clue} developed by Hornik et al.~\cite{hornik2008hard, hornik2008clue}, while the second heuristic approach is {\tt flowMatch} developed by myself~\cite{azad2012matching}.
Using a representative sample, I show that consensus clustering performs better than the individual algorithms under different cluster validation methods.
The superior performance of consensus clustering was also observed in~\cite{aghaeepour2013critical}, where the consensus clustering by {\tt Clue}~\cite{hornik2008clue} outperformed the ``best performing" algorithm based on their similarities with the manual (visual) gating (when evaluated by an external validation method called the F-measure).

The rest of the chapter is organized as follows.
In Section~\ref{sec:clust_algorithms}, I discuss several simple clustering algorithms that can be used as off-the-shelf tools to cluster FC data.
Section~\ref{sec:cluster_val_methods} discusses different cluster validation methods.
The next Section~\ref{sec:consensus_clust} discusses the two heuristic algorithms for constructing a consensus clustering from a collection of partitions (cluster ensembles).
In Section~\ref{sec:clustering_results}, I demonstrate different aspects of clustering algorithms with two FC samples from two separate datasets.
I conclude the chapter in Section~\ref{sec:clustering_conclusions}.

\section{Clustering algorithms} 
\label{sec:clust_algorithms}
Clustering (automated gating) is arguably the most researched topic in computational cytometry~\cite{aghaeepour2013critical, pyne2009automated, bashashati2009survey, chan2008statistical,  qian2010elucidation, lo2008automated, finak2009merging, aghaeepour2011rapid, sugar2010misty, naumann2010curvhdr, SamSPECTRAL, naim2010swift, quinn2007statistical}.
A state-of-the-art summary of this field is discussed by Aghaeepour et al.~\cite{aghaeepour2013critical}.
These algorithms often take few parameters whose values must be set appropriately to obtain good clustering solutions.
For example, the spectral clustering package {\tt SamSPECTRAL}~\cite{SamSPECTRAL} takes two arguments (in addition to other arguments), (a) {\tt normal.sigma} -- a scaling parameter that determines the``resolution" in the spectral clustering stage and  (b) {\tt separation.factor} -- a threshold controlling clusters separation.
To obtain a good clustering solution from {\tt SamSPECTRAL}, these two parameters need to be selected carefully for every dataset.
As a second example, the simple and fast algorithm {\tt flowMeans}~\cite{aghaeepour2011rapid} also depends on a  parameter {\tt MaxN} that influences the quality of clustering.
I will show how the selection of parameter influences clustering solution, with an example in Section~\ref{sec:clustering_hd}.

The algorithms discussed in FC literature are built on top of a few fundamental algorithms, such as the k-means, Gaussian mixture modeling, spectral clustering, hierarchical clustering, etc.~\cite{jain1999data}.
Here, I move one step back to see how the basic algorithms perform, in a completely unsupervised setting, in clustering FC samples.
For this purpose, I selected six popular algorithms whose efficient implementations are available as open-source packages.
The algorithms, their computational complexity, and the R packages that implement them are listed in Table~\ref{tab:clustering_algorithms}.
K-means is a popular algorithm that partitions $n$ observations into $k$ clusters in which each observation belongs to the cluster with the nearest mean~\cite{hartigan1979algorithm, lloyd1982least}.
Clara (Clustering LARge Applications) is a sampling based algorithm that creates a collection of $m$ subsamples of size $s$ from $n$ observations.
Once $k$ representative objects have been selected from the sub-dataset, each observation of the entire dataset is assigned to the nearest medoid~\cite{kaufman2009finding}.
Hclust is a hierarchical clustering algorithm that builds a hierarchy of data points into $k$ clusters.
Here, I used the Unweighted Pair Group Method with Arithmetic Mean (UPGMA) method~\cite{jain1999data, fastcluster}.
GMM is the Gaussian mixture modeling algorithm that models a sample with a mixture of $k$ normal distributions and assigns observations into $k$ Gaussian components by the Expectation-Maximization (EM) algorithm~\cite{bishop2006pattern, lebret2012rmixmod} .
The Self Organizing Tree Algorithm (SOTA) uses an unsupervised neural network with a binary tree topology and recursively divides  clusters with the largest diversity until $k$ clusters remain~\cite{herrero2001sota}. 
The spectral clustering works by embedding the data points into the subspace of the $k$ largest eigenvectors of a normalized affinity/kernel matrix~\cite{von2007tutorial, yan2009fastSpec}.
An informative discussion about different clustering algorithms can be found in~\cite{gan2007clusteringBook}.

\begin{table}[!t]
   \centering
      \vspace{-.7 cm}
   \caption[List of clustering algorithms used in the FC data analysis]{Several clustering algorithms are listed along with their complexity and the R packages implementing them. In the complexity,  $n$ denotes the number of cells in a sample and $p$ idenotes the number of features (dimension) measured pre cell. The parameter $l$ is the number of iterations required before an algorithm converges. For Clara, $m$ denotes the number of subsamples and $s$ denotes the number of items in each subsample. }
   \label{tab:clustering_algorithms}
   \begin{tabular}{lcl} 
         \toprule
         Clustering algorithm    & Computational complexity &  R package\\
         \toprule
      
      K-means~\cite{hartigan1979algorithm, lloyd1982least}      & $O(nl)$ & {\tt stats}\cite{R_kmeans} \\ 
      Clara (PAM)~\cite{kaufman2009finding}       & $O(n + ms^2)$ & {\tt cluster}\cite{cluster_manual} \\ 
      Hclust~\cite{jain1999data}       & $O(n^2)$  & {\tt fastcluster}\cite{fastcluster}  \\ 
        GMM~\cite{bishop2006pattern}       & $O(nl)$  & {\tt Rmixmod}\cite{lebret2012rmixmod}   \\ 
      SOTA~\cite{herrero2001sota}      & $O(n\log n)$  & {\tt clValid}\cite{brock2011clvalid}   \\ 
      Spectral~\cite{von2007tutorial, yan2009fastSpec}       & $O(n^3)$  & {\tt kernlab}\cite{karatzoglou2004kernlab}   \\
        \bottomrule
   \end{tabular}

\end{table}

I apply each of the six algorithms to cluster an FC sample.
For slower algorithms, such as the spectral clustering, I downsample the data into a smaller size so that the algorithms finish in a reasonable amount of time.
I always use the default arguments of these algorithms except the number of clusters $k$.
However, I often perform a second phase to estimate the statistical parameters of the clusters identified by the clustering algorithms.
For this purpose, I characterize an FC sample with a mixture of $k$ normally distributed clusters of $p$-dimensional points.
Each cluster is characterized by a multi-dimensional normal distribution and is represented by two parameters $ \boldsymbol {\mu}$, the $p$-dimensional mean vector, and $\Sigma$, the $p\times p$ covariance matrix \cite{ lo2008automated}.
To estimate the parameters of $k$ clusters, I use the Expectation-Maximization (EM) algorithm~\cite{bishop2006pattern}.

\section{Cluster validation methods}
\label{sec:cluster_val_methods}
Clustering is an unsupervised process of identifying phenotypically similar cell populations from an FC sample.
When clustering patterns and the number of clusters are not known \emph{a priori}, the clustering solution is evaluated for quality assurance\cite{ramze1998new, jain1999data, halkidi2001clustering}.
Cluster validation usually answers questions like ``how well does the resulting partition fit the data?", ``how many clusters are there in a sample?", and ``from a collection of algorithms, which algorithm provides the best partition?".

In general terms, there are three major cluster validation approaches based on \emph{external}, \emph{internal}, and \emph{relative} validation criteria \cite{jain1999data, halkidi2001clustering}.
External validation methods evaluate a clustering result based on an \emph{a priori}  knowledge of correct class labels or the ``gold standard".
Internal validation techniques evaluate how well a given partition captures the natural structure of the data based on information intrinsic to the data alone.
Relative validation methods compare two partitions to identify their relative differences.
While external and relative validation techniques depend on some benchmarking or ``gold standard", internal validation methods do not depend on any prior knowledge.
Given the limited number of ``gold standards" in flow cytometry, internal validation is the most practical approach for most FC dataset, and it is the focus of my discussion. 
However, occasionally prior human interpretation (e.g., manual gating) is available for a subset of samples of an FC dataset~\cite{aghaeepour2013critical}.
In this scenario, I discuss the application of external cluster validation methods briefly in the latter part of this section. 
The following discussion is limited to validating hard clustering where each data item is assigned to exactly one cluster.
Evaluation methods for fuzzy clustering are discussed in~\cite{pal1995cluster, ramze1998new, xie1991validity}.

\subsection{Internal cluster validation methods}
Consider an $n \times p$ data matrix $A$ storing $n$ $p$-dimensional observations.
In flow cytometry, $A$ represents a biological sample measuring $p$ features from $n$ cells.
Let $d(\mathbf x,\mathbf y)$ be the Euclidean distance between two data items, $\mathbf x$ and $\mathbf y$.
I use $d^2(\mathbf x,\mathbf y)$ to denote the squared Euclidean distance between $\mathbf x$ and $\mathbf y$.
Assume that $A$ is partitioned into $k$ clusters $C = \{c_1$, $c_2$, ..., $c_k\}$, where the $i^{th}$ cluster $c_i$ contains $|c_i|$ data items with mean $ \boldsymbol\mu_i$. 

Most internal cluster validation methods use the concepts of within-cluster \emph{compactness} and between-cluster \emph{separation}.
The former concept minimizes the intra-cluster distances, while the latter maximizes the between-cluster  distances.
In the discussion on internal validation methods, I use $\Delta(c_i)$ to denote the intra-cluster distance of a cluster $c_i$ and $D(c_i, c_j)$ to denote the between-cluster distance from $c_i$ to $c_j$.
Different cluster validation methods define the intra- and inter-cluster distances differently.
Based on these definitions, I discuss five popular internal cluster validation methods that I use to evaluate automated gating of FC samples.

{\em  Calinski-Harabasz (CH) Index:} 
Calinski and Harabasz~\cite{calinski1974dendrite} defined  the intra-cluster distance $\Delta(c_i)$ of a cluster $c_i$ as the sum of squared distances from each data item $\mathbf x\in c_i$ to the cluster center $\boldsymbol\mu_i$. 
They defined the separation $D(c_i)$ of cluster $c_i$ as the squared distance between the center $\boldsymbol\mu_i$ and the overall center $\boldsymbol\mu$ of the entire sample. 
Then the Calinski-Harabasz (CH) index~\cite{calinski1974dendrite} is defined as the ratio of the average cluster separation to the average within-cluster distance.
\begin{equation}
\Delta(c_i)  =  \sum\limits_{\mathbf x\in c_i} d^2(\mathbf x, \boldsymbol\mu_i), \ \ \ \ 
D(c_i)  =   |c_i| \  d^2( \boldsymbol\mu_i,  \boldsymbol\mu), \ \ \ \ 
\text{CH} = \frac{(n-k)\sum\limits_{c_i \in C}D(c_i)}{(k-1)\sum\limits_{c_i \in C}\Delta(c_i)}\ .
\end{equation}

A large value of the CH index denotes larger separation among clusters relative to the average within-cluster distance.
The CH index is limited to the interval $[0,+\infty]$ and should be maximized. 
The time needed to compute the CH index is $O(n+k)$, which is linear in the number of data points, and thus faster for large data matrices.

{\em  Dunn's Index:} 
Dunn defined the intra-cluster distance $\Delta(c_i)$ of a cluster $c_i$ as the maximum distance between pairs of data items $\mathbf x, \mathbf y\in c_i$ and the between-cluster distance $D(c_i,c_j)$ for a pair of clusters $c_i$ and $c_j$ as the minimum distance over all pairs of data items $\mathbf x\in c_i$ and $\mathbf y\in c_j$.
Dunn's index~\cite{dunn1974} then measures the ratio of the smallest between-cluster distance to the largest intra-cluster distance.
\begin{equation}
\label{eq:dunn}
\begin{array}{lcl}
\Delta(c_l)  =  \max\limits_{\mathbf x,\mathbf y\in c_l}  d(\mathbf x,\mathbf y),  \ \ \ \ \ \
D(c_i,c_j)  =  \min\limits_{\mathbf x\in c_i, \mathbf y\in c_j} d(\mathbf x,\mathbf y), \\ 
\text{Dunn} =\frac{1}{\max\limits_{c_l\in C} (\Delta(c_l)} \left(\min\limits_{\substack{c_i, c_j \in C, c_i\neq c_j}} D(c_i, c_j) \right).
\end{array}
\end{equation}
Dunn's index is limited to the interval $[0,+\infty]$ and should be maximized. 
The time complexity for computing Dunn's index is $O(n^2+k^2)$, which reduces to $O(n^2)$ for bounded $k$.
Therefore, Dunn's index is computationally expensive when the data matrix is large.

{\em  Average Silhouette Width (ASW):}  
Let $a(\mathbf x_i)$ be the average distance from the $i^{th}$ data item $\mathbf x_i$ to all other data items within the same cluster, and let $b(\mathbf x_i)$ be the lowest average distance from $\mathbf x_i$ to a cluster to which $\mathbf x_i$ does not belong. 
Then the \emph{silhouette width} of $\mathbf x_i$ is computed as follows:
\begin{equation}
S(\mathbf x_i) = \frac {b(\mathbf x_i) - a(\mathbf x_i)} {\max\{b(\mathbf x_i),a(\mathbf x_i)\}}.
\end{equation}
The Average Silhouette Width (AWS) of the partition $C$ is computed by averaging $S(\mathbf x_i)$ over all data items~\cite{rousseeuw1987silhouettes}.
The AWS is limited to the interval $[-1,1]$ and should be maximized.
The time to compute the AWS is $O(n^2)$, and similar to Dunn's index, it is computationally expensive when the data matrix is large.

{\em  Davies-Bouldin Index:} 
Davies and Bouldin~\cite{davies1979cluster} defined the intra-cluster distance $\Delta(c_i)$ of a cluster $c_i$ as the square root of the average squared distance from each data item $\mathbf x\in c_i$ to the cluster center $\boldsymbol\mu_i$. 
They defined the between-cluster distance $D(c_i,c_j)$ for a pair of clusters $c_i$ and $c_j$ as the distance between their clusters centers $\boldsymbol\mu_i$ and $\boldsymbol\mu_j$.
Then the Davies-Bouldin index is defined as follows:
\begin{equation}
\Delta(c_i)  =  \left\{\frac {1} {|c_i|} \sum\limits_{\mathbf x\in c_i} d^2(\boldsymbol\mu_i, \mathbf x)\right\}^{1/2}, \ \ \ 
\text{DB} = \frac 1 k \sum_{c_i\in C}  \ \max\limits_{\substack{c_j \in C \\ c_i\neq c_j} } \left\{ \frac {\Delta(c_i) + \Delta(c_j)} {d(\boldsymbol\mu_i, \boldsymbol\mu_j)} \right\}. 
\end{equation}
The Davies-Bouldin index is limited to the interval $[0,+\infty]$ and should be minimized. 
The time to compute the Davies-Bouldin index is $O(n+k^2)$, which is linear in the number of data points.
Therefore it is faster to compute for large data matrices.

{\em  S\_Dbw Index:}
Let $\boldsymbol\sigma^2(c_i)$ be a $p$-dimensional vector whose $q^{th}$ component $\sigma^2_q(c_i)$ contains the variance of data items $\mathbf x \in c_i$  in the $q^{th}$ dimension. 
Also let $\boldsymbol\sigma^2$ be a $p$-dimensional vector whose $q^{th}$ component $\sigma^2_q$ contains the variance of data items from all clusters in the $q^{th}$ dimension. 
Then the average intra-cluster distance $\Delta(C)$ of the partition $C$ is computed as follows:
\begin{equation}
\Delta(C)  =  \frac 1 k  \left\{\sum\limits_{c_i \in C}\lVert \boldsymbol\sigma^2(c_i)\rVert \right\}^{1/2},
\end{equation}
where $\lVert.\rVert$ is the $L^2$ norm.
Assume that the density $\rho(c_i)$ of the cluster $c_i$ is defined as the number of data points that lie within a hyper-sphere centered at $\boldsymbol\mu_i$ with $\Delta(C)$ as the radius. 
\begin{equation}
\rho(c_i) = \sum\limits_{\mathbf x\in c_i} f(\mathbf x, \boldsymbol\mu_i), \ \ 
\mbox{where } \ \ \ 
f(\mathbf x, \boldsymbol\mu_i) = \begin{cases} 0, & \mbox{if } d(\mathbf x, \boldsymbol\mu_i) > \Delta(C) \\ 
														1, & \mbox{otherwise } \end{cases}
\label{eq:density_sdbw}
\end{equation}
The average inter-cluster density can be computed as
\begin{equation}
\rho(C)  =  \frac 1 {k(k-1)} \sum\limits_{\substack{c_i, c_j \in C}}
\frac{\rho(c_i\cup c_j)} {\max\{\rho(c_i), \rho(c_j)\} } \ ,
\end{equation}
where $\rho(c_i\cup c_j)$ is the density computed after merging $c_i$ and $c_j$ and then using Eq.~\ref{eq:density_sdbw}.
The S\_Dbw index~\cite{halkidi2001sdbw} is then defined as the summation of the normalized intra-cluster distance and average inter-cluster density:
\begin{equation}
S\_Dbw =  \frac {(\Delta{C})^2} {\lVert \boldsymbol\sigma^2 \rVert} + \rho(C).
\end{equation}
The S\_Dbw index is limited to the interval $[0,+\infty]$, and should be minimized. 
The time to compute the S\_Dbw index is $O(nk^2)$, which is linear in the number of data points, and can be computed fast for large data matrices.

\begin{table}[!t]
   \centering
     \vspace{-.7 cm}
   \caption[List of internal cluster validation indices]{Summary of the cluster validation indices. In the complexity, $n$ is the number of data points in a sample and $k$ is the number of clusters. The dimension of a sample $p$ is omitted from the calculation of complexity.}
   \label{tab:cluster_val_indices}
   \begin{tabular}{@{} lccrr @{}} 
      \toprule
         Cluster validation methods    &  Range & Optimization & Time complexity \\
      \midrule
      Calinski-Harabasz~\cite{calinski1974dendrite}     &  [0, $\infty$] & maximize & $O(n+k)$ \\
      Dunn~\cite{dunn1974} & [0, $\infty$] & maximize   &  $O(n^2 + k^2)$  \\
      Average Silhouette Width~\cite{rousseeuw1987silhouettes}   &  [-1, 1] & maximize & $O(n^2)$ \\
      Davies-Bouldin~\cite{davies1979cluster} & [0, $\infty$] & minimize  &  $O(n + k^2)$  \\
      S\_Dbw~\cite{halkidi2001sdbw}  & [0, $\infty$] &  minimize & $O(nk^2)$ \\
      \bottomrule
   \end{tabular}
\end{table}

Table~\ref{tab:cluster_val_indices} gives a summary of these five internal cluster validation methods. 
Besides these five internal cluster validation methods, several other methods have also been proposed in the literature, such as Ball-Hall~\cite{ball1965isodata}, C~\cite{hubert1976quadratic},  Ray-Turi~\cite{ray1999determination}, Scott-Symons~\cite{scott1971clustering}, and other indices.
The papers \cite{halkidi2001clustering, gan2007clusteringBook, liu2010understanding} provide more details.
Different cluster validation methods are also implemented in several R packages such as {\tt clv}~\cite{manual_clv}, {\tt clValid}~\cite{manual_clvalid}, {\tt clusterCrit}~\cite{manual_clusterCrit}. 

\subsection{External cluster validation methods}
Unlike internal cluster validation methods, external methods evaluate a clustering result based on a prior knowledge of correct class labels or the ``gold standard".
In FC, we often seek completely unsupervised clustering because of the unavailability of ``gold standards".
However, occasionally prior human interpretation (e.g., manual gating ) is available for a subset of a larger FC dataset~\cite{aghaeepour2013critical}.
Then the external cluster validation methods can be applied to evaluate the consensus between the clustering solution provided by an algorithm and the known ``ground truth". 

In flow cytometry, F-measure~\cite{manning2008introduction} is a commonly used external method to evaluate the  closeness of a  partition to the ``gold standard" or manual gating.
Assume that $C=\{c_1, c_2, ..., c_k\}$ is a set of clusters computed by a clustering algorithm while $T=\{t_1, t_2, ..., t_l\}$ is a set of true clusters of the same sample.
Then, \emph{precision} is the number of data points in the computed cluster $c_j$ that belong to the correct cluster $t_i$ and \emph{recall} is the number of data points in the true cluster $t_i$ that are recovered by the computed cluster $c_j$.
The \emph{F-measure} $F(t_i, c_j)$ between $t_i$ and $c_j$ is the harmonic mean of \emph{precision} and \emph{recall}.
Let $|t_i|$ and $|c_j|$ be the number of data points in $t_i$ and $c_j$ respectively and $|t_i\cap c_j|$ be the number of points belonging to both $t_i$ and $c_j$ simultaneously.
Then the \emph{precision}, \emph{recall}, and \emph{F-measure} between $t_i$ and $c_j$ are defined as follows.
\begin{equation}
\begin{array}{lcl} 
\text{Precision}(t_i, c_j) = \frac{|t_i\cap c_j|}{|c_j|}, \ \ \ \ 
\text{Recall}(t_i, c_j) = \frac{|t_i\cap c_j|}{|t_i|} \\
F(t_i, c_j) =  \frac{2\ \cdot \ \text{Precision}(t_i, c_j) \ \cdot \ \text{Recall}(t_i, c_j)}{\text{Precision}(t_i, c_j)+\text{Recall}(t_i, c_j)}
\end{array}
\end{equation}
The overall F-measure for the partitioning $C$ with respect to $T$ is computed as
\begin{equation}
F(T,C) =  \sum_{t_i\in T} \frac{|t_i|}{n}\max_{c_j\in C} \{F(t_i, c_j)\},
\end{equation}
where $n$ is the number of data points in the sample.

Several other external cluster validation methods were also proposed in the literature, such as Rand index~\cite{rand1971objective}, Jaccard coefficient~\cite{jaccard1901}, Minkowski score~\cite{jardine1971mathematical}, Fowlkes-Mallows index~\cite{fowlkes1983method}, etc. 
External cluster validations based on combinatorial optimization have also been used, such the R-metric, transfer distance or partition distance that computes the minimum number of augmentations and removals of points needed to transform one partition into another~\cite{gusfield2002partition, konovalov2005partition, charon2006maximum, arabie1973multidimensional}.
Note that the external cluster validation approaches always compare two partitions of the same sample.
These methods are used to compare the quality of a clustering algorithm relative to the ``gold standard" and to compare the performance of multiple algorithms applied to the same sample.
However, they cannot be used to compute the dissimilarity between partitions from two different samples.
In Chapter~\ref{chap:matching}, I will discuss a combinatorial algorithm for computing dissimilarities between a pair of partitions from different samples.

\subsection{Selecting the number of clusters (cell populations)}
Most clustering algorithms take the number of clusters $k$ as an input and assign the data points into $k$ clusters. 
However, the optimum number of clusters $k^*$ is often not known in unsupervised clustering problems.
Especially in FC, the number of phenotypically distinct cell populations may not be known in advance.
In this situation, a clustering algorithm is run for different number of clusters in a range [$k_{min},k_{max}$]. 
The clustering quality for each $k$ is then computed by internal cluster validation methods and an optimum number of clusters $k^*$ is selected by maximizing/minimizing the validation indices.
For example, the values of avg. silhouette width (ASW), Calinski-Harabasz (C-H), and Dunn indices are maximized and Davies-Bouldin and S\_Dbw indices are minimized to obtain the optimum number of clusters.
However, if different cluster validation methods disagree with each other about $k^*$, a consensus (e.g., majority voting) of them is used to select $k^*$.

\subsection{Selecting the ``best" algorithm}
Given a collection of algorithms, which algorithm should we use to cluster a sample for the selected optimum number of clusters $k^*$?
To select an algorithm for clustering a particular sample, I consider the following three conditions.
First, if most clustering algorithms perform equally well under all validation indices, then  I select the fastest of the better performing algorithms. 
Second, if an algorithm consistently performs better than others under several validation methods, then I select the best-performing algorithm.
Finally, when no agreement can be reached about the best performing algorithm because different validation methods favor different algorithms, I compute a consensus of the clustering solutions provided by the algorithms. 

\section{Consensus clustering}
\label{sec:consensus_clust}
To describe how a consensus clustering is computed, we first need a dissimilarity measure between a pair of partitions.
Assume that a partition is defined as a binary membership matrix $P$ such that $P[i,j]=1$ if the $i^{th}$ object belongs to the $j^{th}$ cluster and otherwise, $P[i,j]=0$.  
Then the dissimilarity between two partitions $P$ and $Q$ is defined as 
\begin{equation}
d(P,Q) = \min_{\Pi} || P-Q\Pi ||,
\end{equation}
where the minimum is taken over all permutation matrices $\Pi$, and $||\cdot||$ is the Frobenius norm (so that $||Y||^2=\text{tr}(Y'Y)$).
This dissimilarity measure computes the minimum number of augmentations and removals of items needed to transform one partition into another and is called the transfer distance or partition distance \cite{gusfield2002partition, konovalov2005partition, charon2006maximum}.
The partition distance can be computed by formulating an edge-weighted matching problem in a bipartite graph and solving it by efficient combinatorial algorithms (e.g.,  the Hungarian method) in O($k^3$) time, where $k$ is the number of clusters in the partitions~\cite{gusfield2002partition, konovalov2005partition, papadimitriou1998combinatorial}.

Given a collection of partitions $E$ (possibly generated by several algorithms), a consensus clustering algorithm finds a new partition $P$ such that the total dissimilarity between $P$ and each partition in $E$ is minimized~\cite{hornik2008hard, gordon2001fuzzy, dimitriadou2002combination}.
The initial set of partitions $E$ is often called a \emph{cluster ensemble}~\cite{hornik2008hard, hornik2008clue}.
An optimum consensus partition $P^*$ minimizes the total dissimilarity between $P$ and each partition in the ensemble $E$:
\begin{equation}
P^* = \arg\!\min_{P} \sum_{Q\in E} \min_{\Pi} || P-Q\Pi ||.
\end{equation}
This optimization is an instance of multidimensional assignment problem, which is known to be NP-hard (by reducing it to the 3-dimensional matching problem \cite{gary1979computers}).
Hence, various heuristics have been developed to solve the consensus clustering problem.
Here I will discuss two heuristic approaches to approximately compute a consensus clustering.

The first heuristic approach is called {\tt Clue} developed by Hornik et al.~\cite{hornik2008hard, hornik2008clue}.
{\tt Clue} initializes the consensus partition $P$ with a randomly selected partition from the ensemble. 
In the $i^{th}$ iteration, the algorithm randomly selects a partition $Q$ not already included in the consensus clustering.
The algorithm then optimally matches the clusters between $Q$ and $P_{i-1}$ (the consensus partition created thus far) and creates an updated consensus partition $P_i$ by taking a weighted average of $P_{i-1}$ and $Q\Pi_i$, where $\Pi_i$ is the permutation matrix found by the optimum matching.

The second heuristic approach is {\tt flowMatch} developed by myself~\cite{azad2012matching}.
{\tt flowMatch} computes the dissimilarity between partitions by a mixed edge cover (MEC) algorithm that allows a cluster from a partition to be matched to zero or more clusters in another partition~\cite{azad2010identifying}.
Let $N$ be the number of partitions in the ensemble $E$.
In each iteration, the {\tt flowMatch} algorithm finds the most similar pair of partitions, merges them into an intermediate consensus partition and continues.
The merging is performed by matching clusters across two partitions and collapsing the matched clusters into meta-clusters.
Let $P$ be the consensus-partition with $K$ meta-clusters created by the above algorithm.
The collection of these meta-clusters then defines the consensus partition.
I will discuss this algorithm in more detail in Chapter~\ref{chap:matching} and Chapter~\ref{chap:template}.

\section{Results}
\label{sec:clustering_results}
\subsection{Identifying cell populations in a healthy sample}
\label{sec:clustering_hd}
{\em  Data description:} I cluster a representative sample from the HD dataset described in Section \ref{sec:hd_data_description}. 
Each sample in this dataset measures the abundance of  CD45, CD3, CD4, CD8, and CD19 protein markers in peripheral blood collected from a healthy individual.
I selected a representative sample from this dataset to demonstrate different aspects of clustering algorithms.
The selected sample is compensated to remove the effect of spectral overlap, transformed to stabilize variance, and gated on FS/SS channels to identify lymphocytes.
The preprocessed five-dimensional data is then clustered to identify different subsets of lymphocytes in the healthy immune system.

\begin{figure}[!t]
   \centering
   \includegraphics[scale=.62]{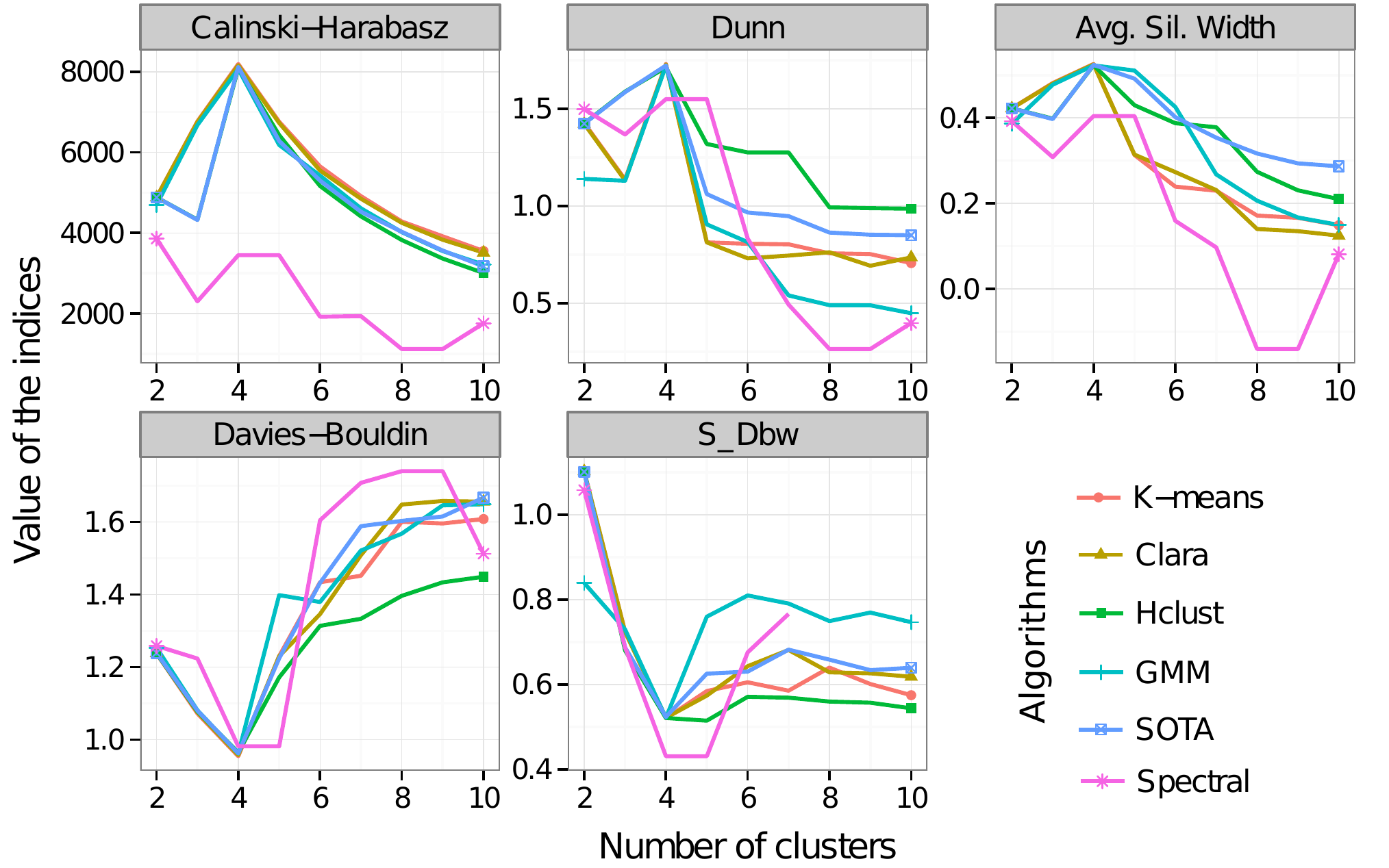} 
   \vspace{-.2 cm}
   \caption[The performance of different clustering algorithms on a healthy sample]{Evaluating the performance of six clustering algorithms (shown in different colors) by five internal cluster validation methods (shown in different panels). A representative 5-D sample from the HD dataset is clustered by the algorithms for a sequence of $k$. Each clustering solution is then evaluated by five cluster validation methods.
The optimum number of clusters, $k^*$ is selected by maxima in the top three methods and by minima in the bottom two methods.   
   }
   \label{fig:cluster_val_hd}
   \vspace{-.5 cm}
\end{figure}

{\em  Selecting the optimum number of clusters ($k^*$):}
I have applied six clustering (Table~\ref{tab:clustering_algorithms}) algorithms to the 5-D healthy sample for different number of clusters in the interval 2-10.
The range for $k$ can be increased if the optimum solution is not found in the selected range.
Each clustering solution is evaluated by five internal cluster validation methods described in Table~\ref{tab:cluster_val_indices}.
Fig.~\ref{fig:cluster_val_hd} shows the values of different cluster validation indices for different algorithms.
The optimum number of clusters $k^*$ is selected by maxima in the top three methods (Calinski-Harabasz, Dunn, and Average silhouette width) and by minima in the bottom two methods (Davies-Bouldin and S\_Dbw).
Form Fig.~\ref{fig:cluster_val_hd} we observe that all validation methods unanimously indicate $k=4$ as the optimum number of clusters across all clustering algorithms except for the spectral clustering.
The spectral clustering does not perform well because a user needs to pass a set of carefully tuned arguments to the algorithm for better performance~\cite{SamSPECTRAL}.
For this sample, I select $k^*=4$ as the optimum number of clusters as per the consensus of the five cluster validation methods.

\begin{figure}[!t]
   \centering
   \includegraphics[scale=.5]{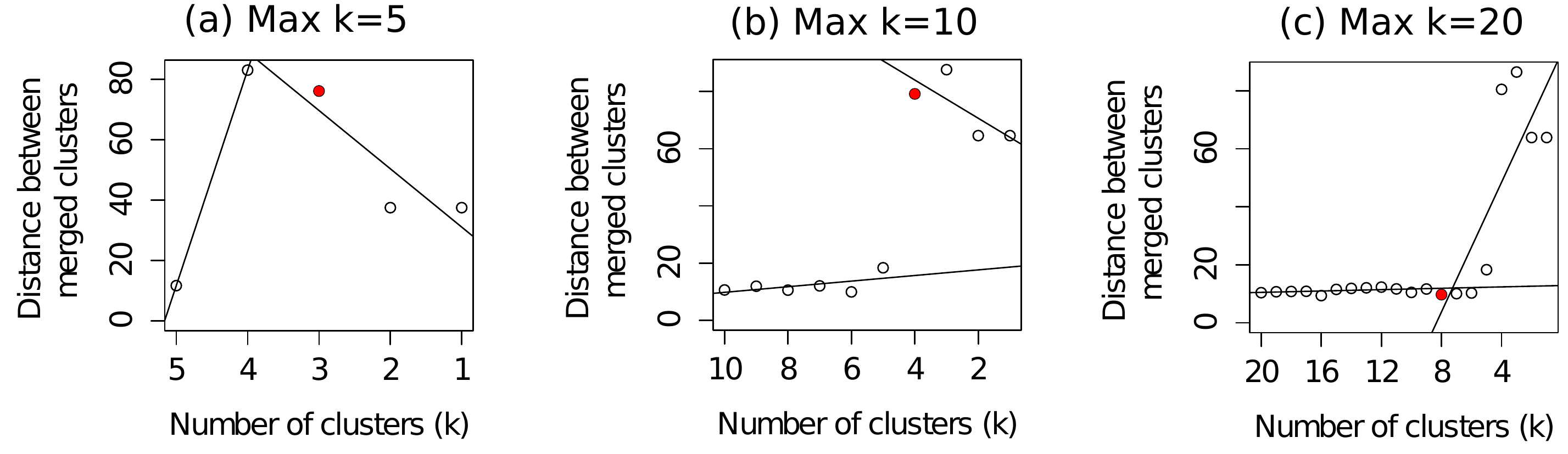} 
   \vspace{-.2 cm}
   \caption[Selecting the optimum number of clusters in a healthy sample]{ Selecting the optimum number of cell populations  in a sample from the HD dataset by the {\tt flowMeans} package~\cite{aghaeepour2011rapid, manual_flowMeans}. The maximum number of clusters ({\tt MaxN} parameter) is set to: (a) 5 clusters (automatically selected by algorithm), (b) 10 clusters, and (c) 20 clusters. The optimum number of clusters is  selected by detecting change point in the segmented regression lines and is shown with a red filled circle in each subfigure.}
   \vspace{-.5 cm}
   \label{fig:flowMeans_example}
\end{figure}

Researchers have proposed several \emph{ad hoc} criteria to select the optimum number of clusters $k^*$ for FC data. 
For example, the {\tt flowMeans} algorithm~\cite{aghaeepour2011rapid} developed by Aghaeepour et al. starts with a relatively large number of clusters (Max $k$) and merges two closest clusters in each iteration.
The algorithm then plots the distances between the merged clusters at each iteration and selects the value of $k^*$ where a sharp change in the segmented regression lines is observed.  
I cluster the same healthy sample used in the previous paragraph by the {\tt flowMeans} software package~\cite{manual_flowMeans} for three choices of Maximum $k$ (parameter {\tt MaxN} in  {\tt flowMeans} package) and plot the results in Fig.~\ref{fig:flowMeans_example}.
Observe that {\tt flowMeans} algorithm selected three different $k^*$ (shown by red circles) for three different choices of initial number of clusters.
Only Fig.~\ref{fig:flowMeans_example}(b) selects the correct $k^*$ as unanimously suggested by five well-established cluster validation methods in Fig.~\ref{fig:cluster_val_hd}.
Therefore, {\tt flowMeans} algorithm depends on a hidden parameter (Max $k$) when selecting $k^*$.
{\tt flowMeans} estimates the initial number of clusters to 5 when the user does not supply it; however, Fig.~\ref{fig:flowMeans_example}(a) shows that the auto tuning fails to identify the correct number of clusters for this sample.
Hence, I prefer using well-known cluster validation methods to select optimal number of clusters $k^*$.  

{\em  Selecting the ``best" algorithm:}
Which clustering algorithm should we use to cluster this sample?
Fig.~\ref{fig:cluster_val_hd} shows that except for spectral clustering, all clustering algorithms perform equally well since the validation indices have optima at $k=4$.
In this case, computing a consensus clustering does not improve the quality of the overall clustering solution because the individual clusterings from five algorithms (for $k=4$) have the same quality under different validation methods.
Therefore, I select the fastest k-means algorithm to cluster this sample.    

\begin{figure}[!t]
   \centering
   \includegraphics[scale=.65]{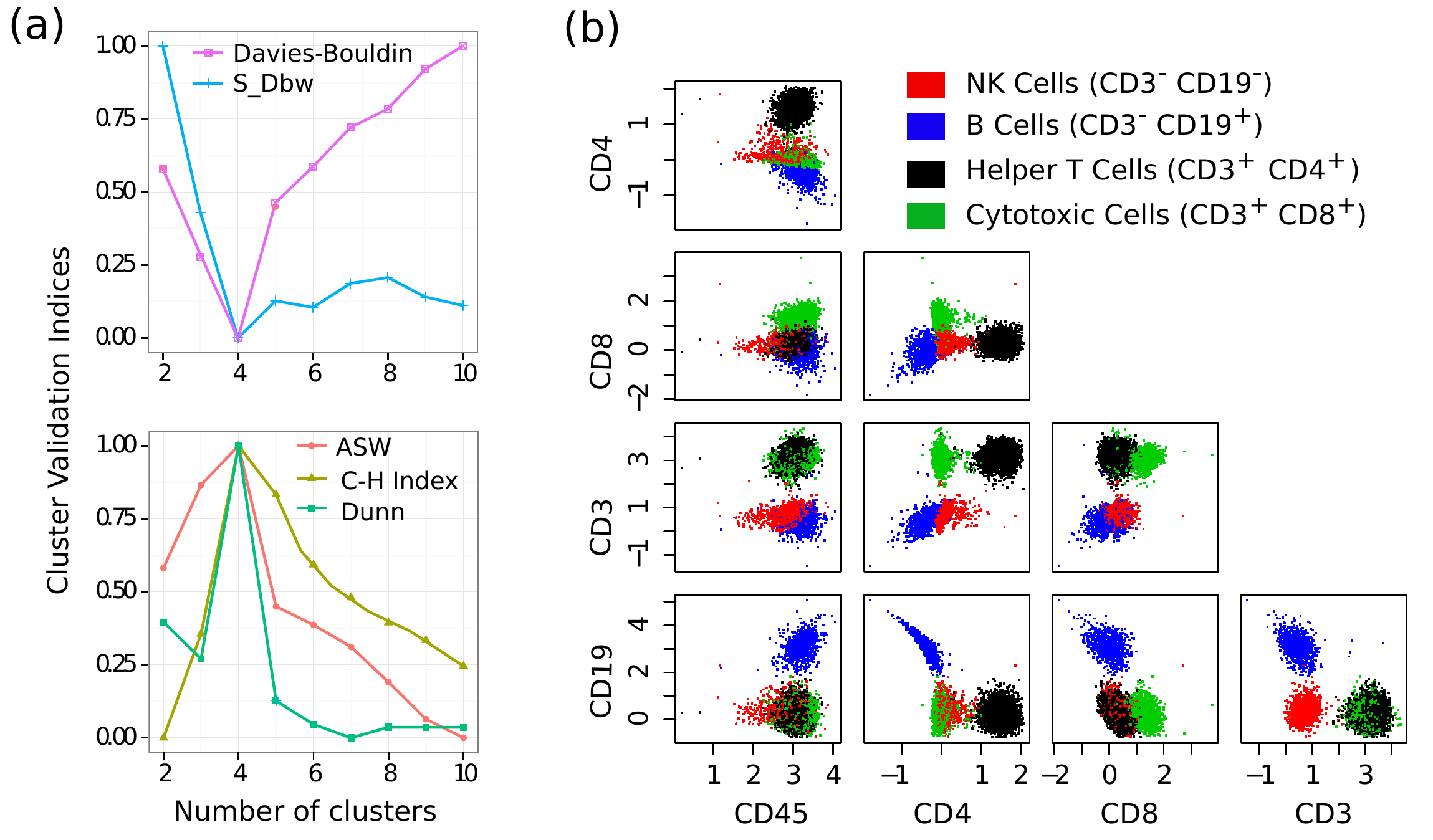} 
   \vspace{-.2 cm}
   \caption[Lymphocyte sub-populations in a healthy sample]{(a) Simultaneous optimization of five cluster validation criteria (scaled to [0,1]) suggests that four cell populations are present in the healthy sample.  Here three of the indices are maximized and two are minimized.  (b) Bivariate projections of lymphocyte sub-populations: red (natural killer cells), blue (B cells), black (helper T cells), and green (cytotoxic T cells). 
   }
   \label{fig:clustering_hd}
   \vspace{-.5 cm}
\end{figure}

{\em  Identifying lymphocyte sub-populations:}
The four clusters chosen by the k-means algorithm represent four biologically distinct lymphocyte sub-populations defined in the five dimensional marker space.
For visualization purposes, I show the cell populations by a collection of 2-D projections in Figure \ref{fig:clustering_hd}(b), where cell populations are shown in four different colors denoting (a) red:  natural killer cells (CD45$^+$CD3$^-$CD19$^-$), (b) blue: B cells (CD45$^+$CD3$^-$CD19$^+$), (c) black: helper T cells (CD45$^+$CD3$^+$CD4$^+$), and (d) green: cytotoxic T cells (CD45$^+$CD3$^+$CD8$^+$).
Here, every cell cluster is CD45$^+$ because CD45 is a common leukocyte antigen present in all lymphocytes and I pre-selected lymphocytes on the forward and side scatter channels.

\subsection{Identifying cell sub-populations in T cells}
\label{sec:clustering_tcp}
{\em  Data description:} As a second example, I cluster a pre-stimulation sample from the T cell phosphorylation (TCP) dataset described in Section~\ref{sec:tcp_data_description}.
The selected sample measures the abundance of four protein markers expressed on T cells: CD4, CD45RA, SLP-76 and ZAP-70.
The first two markers (CD4, CD45RA) are expressed on the surface of different T cell subsets and the last two (SLP-76 and ZAP-70) are highly expressed after T cells are phosphorylated~\cite{maier2007allelic}.
Since I selected a pre-stimulation sample (for simplicity), I consider only CD4, CD45RA to identify different subsets of T cells by using clustering algorithms.

\begin{figure}[!t]
   \centering
   \includegraphics[scale=.55]{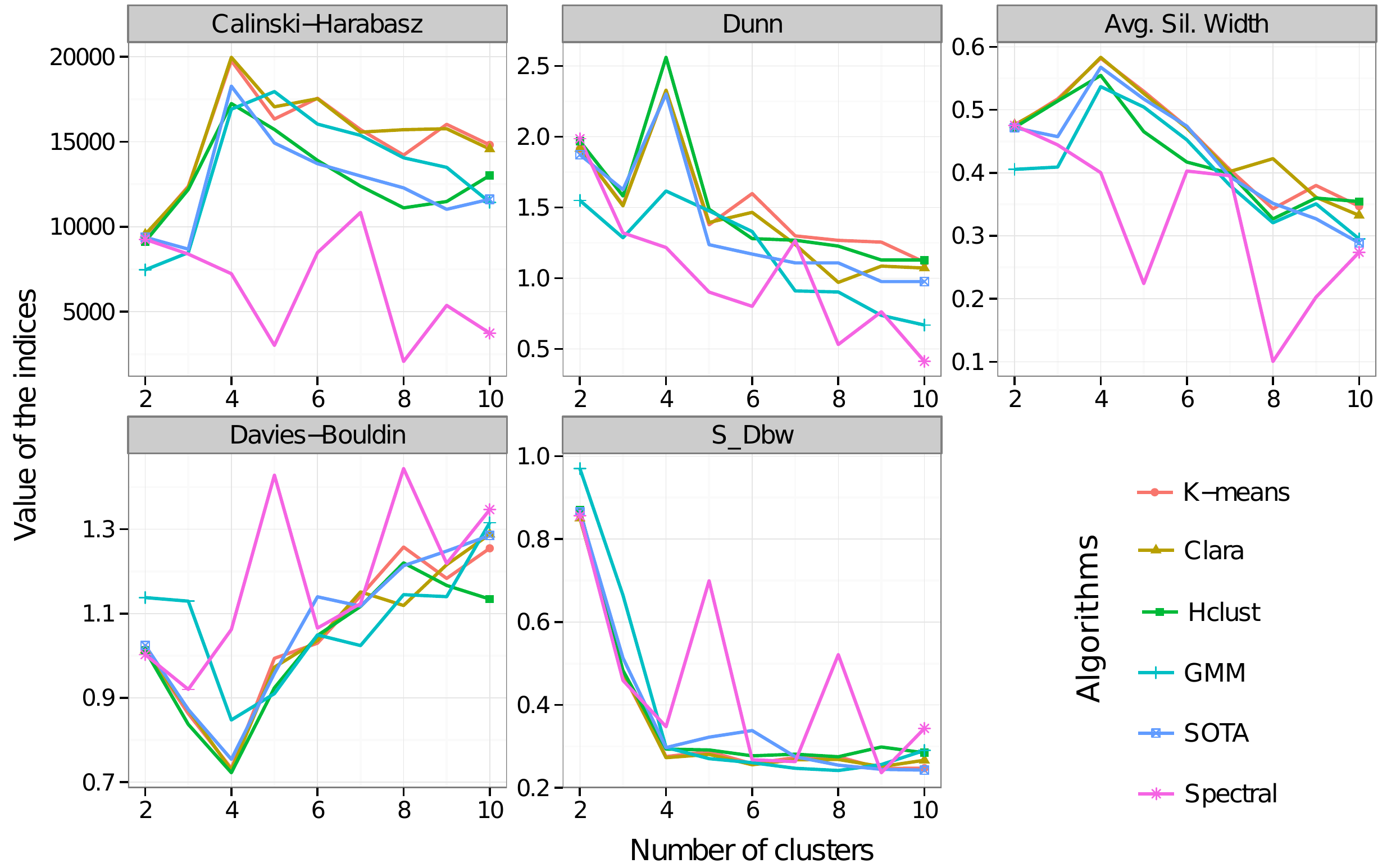} 
   \vspace{-.2 cm}
   \caption[The optimum number of clusters in a sample from the TCP dataset]{A representative 2-D sample from the TCP dataset is clustered using six clustering algorithms for a sequence of $k$ (number of clusters). Each clustering solution is evaluated by five cluster validation methods  (shown in different panels). 
The optimum number of clusters $k^*$ is selected by the maxima in the top three methods and by the minima in the bottom two methods. }
   \label{fig:cluster_val_phospho}
   \vspace{-.5 cm}
\end{figure}

{\em  Selecting the optimum number of clusters ($k^*$):}
I have applied six clustering algorithms to the selected 2-D sample.
The algorithms were run for different number of clusters and each clustering solution is evaluated by five cluster validation indices.
The values of the cluster validation indices are shown in Fig.~\ref{fig:cluster_val_phospho}.
The optimum number of clusters $k^*$ is selected by maxima in the top three methods (Calinski-Harabasz, Dunn, and Average silhouette width) and by minima in the bottom two methods (Davies-Bouldin and S\_Dbw).
Form Fig.~\ref{fig:cluster_val_phospho} we observe that all validation methods unanimously indicate $k=4$ as the optimum number of clusters across all clustering algorithms except for the spectral clustering.
Therefore, for this sample, I select $k^*=4$ as the optimum number of clusters.

{\em  Selecting the ``best" algorithm:}
Unlike the HD sample (Fig.~\ref{fig:cluster_val_hd}), the values of the cluster validation indices are different across the algorithms for $k^*=4$ as can be seen in Fig.~\ref{fig:cluster_val_phospho}.
Furthermore, different algorithms perform best under different validation indices.
For example, the hierarchical clustering is the best algorithm under Dunn's index, while K-means and Clara are the best performers under Calinski-Harabasz index.
Therefore, creating a consensus clustering is useful for this sample.

\begin{figure}[!t]
   \centering
   \includegraphics[scale=.48]{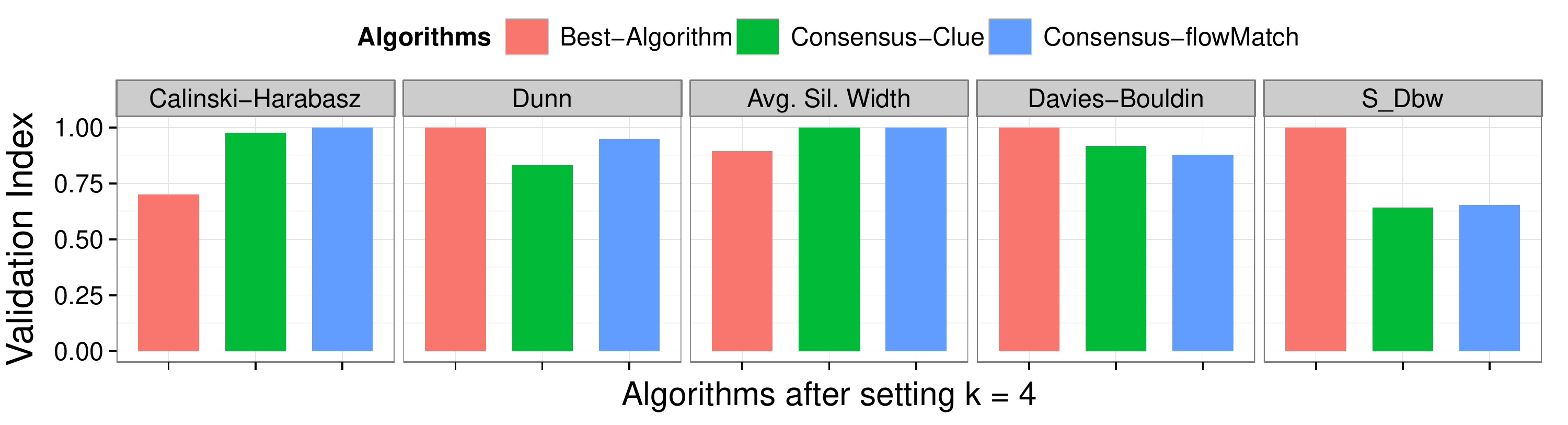} 
   \vspace{-.2 cm}
   \caption[Consensus clustering algorithms applied to a sample from the TCP dataset]{Five cluster validation methods are used to compare the quality of the ``best-performing" algorithms with two consensus clustering methods  ({\tt Clue} and {\tt flowMatch}) for a 2-D sample from the TCP dataset.
   A ``better" clustering solution is denoted by higher values of the first three indices and lower values of the last two indices.
   }
   \vspace{-.5 cm}
   \label{fig:consensus_compare}
\end{figure}

I compute consensus solutions by two heuristic algorithms, {\tt Clue} and {\tt flowMatch}, discussed in Section~\ref{sec:consensus_clust}.
The consensus clusterings are computed from the solutions of six clustering algorithms after the number of clusters is set to four ($k=4$).
Once again I evaluate the performance of the consensus clustering solutions by five cluster validation methods whose values are scaled to $[0,1]$ for comparison purpose. 
I show the values of the validation indices in five different panels in Fig.~\ref{fig:consensus_compare}.
In addition to the consensus partitions, I show the validation indices of the best-performing algorithm under each validation method in its respective panel.  
According to Table~\ref{tab:cluster_val_indices}, a ``better" clustering solution is denoted by higher values of the first three indices and lower values of the last two indices.
We observe in Fig.~\ref{fig:consensus_compare} that the consensus solutions perform better under four validation methods (except the Dunn's index) than the best performing algorithms. 
The consensus clustering based on {\tt flowMatch} performs slightly better than the solution based on {\tt Clue} under all validation methods.

The relatively poor performance of the consensus methods under Dunn's index can be explained if we look closely at the definition of the within-cluster and between-cluster distances used in Eq.~\ref{eq:dunn}.
Notice that the within-cluster distance is defined by the diameter of a cluster and the between-cluster distance is defined as the minimum distance between a pair of clusters.
These definitions are not robust especially in the presence of outlying data items that are far apart from their cluster centers.
Therefore, Dunn's index is less robust when evaluating the quality of robust consensus clustering solutions.

\begin{figure}[!t]
   \centering
   \includegraphics[scale=.7]{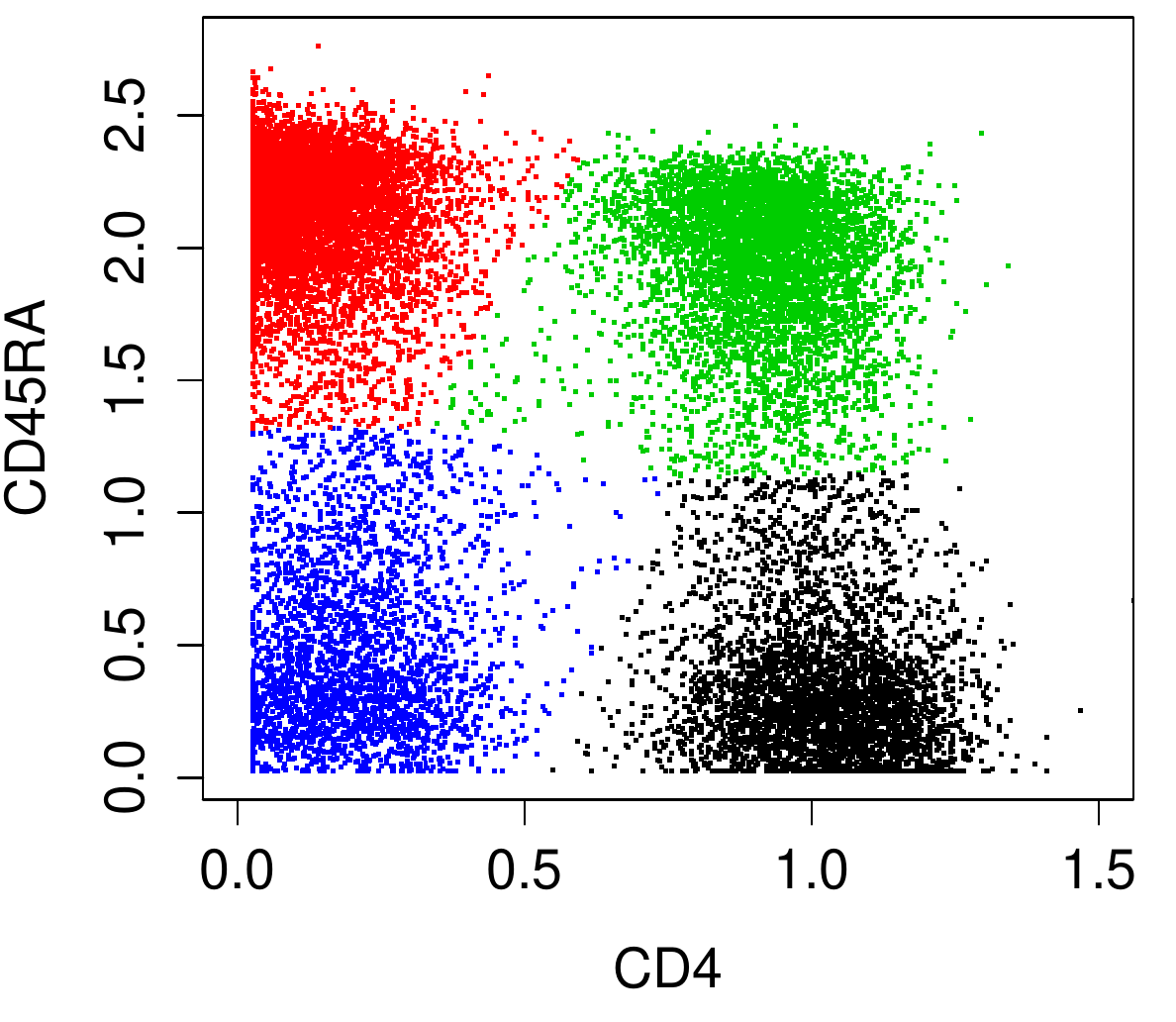} 
   \vspace{-.2 cm}
   \caption[T cell sub-populations within a sample from the TCP dataset]{The consensus clustering computed by the {\tt flowMatch} algorithm created from the solutions of six clustering algorithms with $k=4$.
   Four T cell sub-populations are shown in different colors; black: CD4$^+$ CD45RA$^\text{low}$ T cells, green: CD4$^+$ CD45RA$^\text{high}$ T cells, red: CD4$^-$ CD45RA$^\text{high}$ T cells, and blue: CD4$^-$ CD45RA$^\text{low}$ T cells.
   }
   \vspace{-.5 cm}
   \label{fig:clustering_phospho}
\end{figure}

{\em  Identifying T cell sub-populations:} I select the consensus clustering by \texttt{flowMatch} algorithm as the final clustering solution for the selected TCP sample.
The consensus clustering solution for this sample is shown in Fig.~\ref{fig:clustering_phospho}, where four clusters representing different T cell sub-populations are shown in different colors: (a) CD4$^+$ CD45RA$^\text{low}$ T cells in black, (b) CD4$^+$ CD45RA$^\text{high}$ T cells in green, (c) CD4$^-$ CD45RA$^\text{high}$ T cells in red, and (d) CD4$^-$ CD45RA$^\text{low}$ T cells in blue.
Here  `$+$' and  `high' indicate higher abundances  of a marker, and `$-$' and `low' indicate lower levels of it.
Two clusters with high CD4 expression, CD4$^+$ CD45RA$^\text{low}$ and CD4$^+$ CD45RA$^\text{high}$ cells, are called the memory and naive T cells respectively.
Few domains of the CD4$^+$ cell subsets are phosphorylated upon stimulated by anti-CD3 antibody.
To study the effect of this stimulation was the main purpose of the TCP dataset generated originally by Maier et al.~\cite{maier2007allelic}.
I will discuss algorithms to automatically detect phosphorylation changes across pre- and post-stimulated samples in Chapter~\ref{chap:matching} and Chapter~\ref{chap:template}.

\section{Conclusions}
\label{sec:clustering_conclusions}
We can achieve better confidence in a computed clustering when several algorithms produce similar clusterings.
I demonstrated that several simple, off-the-shelf clustering algorithms can be used to compute more accurate and robust partition of an FC sample than a specialized algorithm. 
Internal cluster validation methods can be used to select parameters of an algorithm such as the optimal number of clusters.
More than one validation method yet again provide more confidence and robustness in model selection.

I have discussed a set of five cluster validation methods -- Calinski-Harabasz, Dunn, Average silhouette width, Davies-Bouldin and S\_Dbw -- that can be simultaneously optimized to select algorithm parameters, as well as the best-performing algorithm for an FC sample. 
However, it is possible that different validation methods disagree about the best performing algorithm because different validation methods favor different algorithms.
Then a consensus of different clustering algorithm usually provides a ``better" solution.
I use the {\tt flowMatch} algorithm (originally developed for creating a template from similar samples) to create consensus clustering from a collection of partitions (cluster ensembles).
By comparing the consensus clustering with the ``best performing" algorithm, I showed that the consensus solution performs better when evaluated by different cluster validation methods.
The superior performance of consensus clustering was also observed in the clustering challenges organized by the FlowCAP consortium ({\tt http://flowcap.flowsite.org/}).
In these challenges, more than 15 clustering algorithms developed by different independent researchers were used to cluster same set of samples.
However, often the consensus clustering by {\tt Clue}~\cite{hornik2008clue} outperformed the ``best performing" algorithm based on their similarities with the manual (visual) gating~\cite{aghaeepour2013critical}.

Clustering (automated gating) is arguably the most researched topic in computational cytometry. 
More than twenty algorithms customized for FC data are available as software packages to be used by cytometrists.
However, the large number of options often confuses cytometrists due to the lack of validation methods needed to select a clustering algorithm for a particular dataset. 
I argue that several cluster validation methods should be used to evaluate the clustering quality and that a consensus clustering be constructed whenever necessary.
This approach produces a robust clustering solution and increases confidence on the quality of clustering in FC data.

%% file: ch4-matching.tex
 \chapter{Registering cell populations across FC samples}
 \label{chap:matching}
 
 \section{Introduction}
Cell populations with different phenotypes and functions often respond differently upon perturbation by stimuli or change of biological conditions.
To track these population-specific changes, we need to register corresponding cell clusters across samples.
The \emph{population registration} is a process where phenotypically similar cell clusters are matched to establish the correspondence of clusters across samples~\cite{pyne2009automated, azad2010identifying}.
Furthermore, the matched clusters are often labeled with abstract or biologically relevant descriptions, thus solving the population/cluster labeling  problem~\cite{spidlen2013genepattern}.

I solve the population registration problem with  a combinatorial \emph{mixed edge cover} (MEC) algorithm~\cite{azad2010identifying}.
Given a pair of clustered sample, I create a bipartite graph whose vertices are denoted with cell clusters from the samples.
An edge of the graph is created by joining a pair of clusters with the dissimilarity between the clusters as the edge-weight.
The MEC algorithm then matches vertices (clusters) in the bipartite graph by optimizing an appropriate objective function.
Unlike conventional matching algorithms, the MEC algorithm allows a cluster from one sample to be matched with zero or more clusters in another sample and thus solves the problem of missing or splitting biological populations.
In addition to solving the population registration problem, an optimal MEC solution provides a combinatorial dissimilarity measure between a pair of partitions from different FC samples.
This between-sample dissimilarity works as a building block of the meta-clustering and classification approaches discussed in Chapter \ref{chap:template} and Chapter \ref{chap:template-based-classification}.


The ability to register multi-dimensional populations across FC samples is useful in various applications of flow cytometry~\cite{pyne2009automated, azad2010identifying, azad2012matching, zeng2002matching, zeng2007feature}.
In a cluster labeling application, cell clusters in a collection of samples $\mathbb{S}$ are labeled based on the given labeling of another FC sample $A$.
This problem can be solved by matching clusters between $A$ and each sample $B\in \mathbb{S}$, and labeling each cluster in $B$ same as to the matched cluster from $A$.
I give an example of labeling different sub-populations of lymphocytes in samples from a healthy dataset.
In addition to labeling clusters,  the MEC algorithm allows an objective way of  evaluating the phenotypic differences between the matched clusters. 
By mapping clusters across two samples before and after stimulating T cells, I demonstrate how we can assess the population-specific effects of the stimulation experiment.
Furthermore, I register different cell types in a sample from an Acute Myeloid  Leukemia (AML) patient with a healthy sample and demonstrate a conventional procedure of AML diagnosis.

The MEC algorithm is similar -- only in principle -- to the FLAME approach proposed by Pyne et al. \cite{pyne2009automated}, while differing from it in significant ways.
First, FLAME does not match clusters directly across a pair of samples.
Instead, it constructs a template of a class of samples and matches generic meta-clusters from the template to the clusters from each sample; therefore, performs a global registration of clusters.
In contrast, MEC algorithm performs a direct, sample-to-sample cluster matching, thereby emphasizes the local patterns. 
In fact, the approach taken by FLAME is more similar to the meta-clustering algorithm discussed in Section \ref{chap:template}. 
Second, FLAME solves a relaxed min-cost flow problem \cite{schrijver2003combinatorial} by an integer-programming solver, whereas MEC uses a combinatorial algorithm to solve a relaxed minimum-weight matching in a bipartite graph.
Several other subtle differences also exist between these two approaches, such as the methods used to compute distance between clusters and the use of cluster sizes in matching.
The cluster labeling approaches \cite{finak2010optimizing, spidlen2013genepattern} discussed in FC literature are very different from the MEC algorithm.
These labeling algorithms run a second stage of clustering from the cluster-centers and hence perform a global labeling similar to the first phase of FLAME. 
I will discuss more about this approaches in the next Chapter \ref{chap:template}.

The rest of this chapter is organized as follows. 
At first, I discuss different methods to compute dissimilarity between a pair of clusters, which is used by population registration algorithms.
In Section~\ref{sec:mec_algorithm}, I describe the mixed edge cover algorithm with a proof of its optimality.
Section~\ref{sec:mec_propoerties} describes different properties of the MEC algorithm and their implications on the population registration problem.
In Section~\ref{sec:mec_results}, I demonstrate, with three representative datasets, that the MEC algorithm matches phenotypically similar populations across samples.
I conclude this chapter in Section~\ref{sec:mec_conclusions}.

 \section{Dissimilarity between a pair of cell clusters}
\label{sec:cluster_dist}
Computing distances between clusters is an integral part of any cluster-matching algorithm.
In the past, researchers have used various parametric and non-parametric methods in this purpose, such as the Euclidean distance \cite{pyne2009automated}, Mahalanobis distance \cite{finak2010optimizing}, the Kullback-Leibler (KL) divergence \cite{azad2010identifying, azad2012matching}, and the Earth Mover's distance (EMD) \cite{zimmerman2011}. 
To compute dissimilarity between two clusters $c_1$ and $c_2$, the parametric methods use distribution parameters of the clusters while non-parametric methods use statistical significance testing. 
I discuss very briefly several widely used parametric and non-parametric dissimilarity measures.

{\em  Parametric dissimilarity measures:}
To compute parametric distances, it is assumed that a cell cluster approximately follows a known probability distribution.
Dissimilarity between a pair of cell clusters is then computed by the dissimilarity between the corresponding multivariate probability distributions.
To model cell populations, researchers have used several well-characterized distribution families, such as normal, skew-normal, t, skew-t distributions \cite{pyne2009automated, lo2008automated, chan2008statistical, finak2010optimizing, azad2012matching}.
For simplicity, I assume normally distributed cell populations in the rest of my discussion, but other distributions are equally applicable to the dissimilarity methods discussed below.

Let $c_1(\boldsymbol {\mu_1}, \Sigma_1)$ and $c_2(\boldsymbol {\mu_2}, \Sigma_2)$ be two cell populations modeled by multivariate normal distributions with mean vectors $\boldsymbol {\mu_1}$, $\boldsymbol {\mu_2}$ and covariance matrices $\Sigma_1$, $\Sigma_2$ respectively.
The \emph{Euclidean distance} computes the distance between the centers ($\boldsymbol {\mu_1}$ and $\boldsymbol {\mu_2}$) of the distributions without considering the spreads of the distributions.
The symmetric version of the \emph{Kullback-Leibler (KL) divergence} \cite{ jeffreys1946invariant, kullback1951information} uses both centers and covariances of the distributions to compute dissimilarity $d_{KL}(c_1, c_2)$ between $c_1$ and $c_2$:
\begin{equation} d_{KL}(c_1, c_2)  =   \frac 1 2 (\boldsymbol {\mu_1} - \boldsymbol {\mu_2})^\top (\Sigma_1^{-1} + \Sigma_2^{-1}) (\boldsymbol {\mu_1} - \boldsymbol {\mu_2}) + \frac 1 2 \text{tr}(\Sigma_1^{-1}\Sigma_2 + \Sigma_2^{-1}\Sigma_1 - 2\boldsymbol{\text{I}}).
\label{eq:KL}
\end{equation}
The symmetrized KL divergence, however, is not a metric because it does not satisfy the triangle inequality. 
Mahalanobis distance \cite{mclachlan1999mahalanobis} uses the pooled estimate of common covariance $\Sigma_p$ from the two distributions to compute distance between clusters:
\begin{equation} \Sigma_p = \frac {(n_1-1) \Sigma_1 + (n_2-1) \Sigma_2} {n_1+n_2-2},  \ \ \ \ \ \ 
    d_{MD}(c_1, c_2)  =   \frac 1 2 (\boldsymbol {\mu_1} - \boldsymbol {\mu_2})^\top \Sigma_p^{-1} (\boldsymbol {\mu_1} - \boldsymbol {\mu_2}).
\label{eq:Mahalanobis}
\end{equation}
Note that, the above definition of Mahalanobis distance satisfies metric properties, whereas the Mahalanobis distance defined by Aghaeepour et. al. \cite{aghaeepour2011rapid} (Equation 3.7) is not a metric since it fails to satisfy triangle inequality \cite{zimmerman2011}.

{\em  Non-parametric dissimilarity measures}: The non-parametric measures use statistical significance testing to compute dissimilarity between cell populations. For example, the Kolmogorov-Smirnov (KS) statistic~\cite{smirnov1948table, massey1951kolmogorov} computes the maximum absolute vertical difference between two cumulative distribution functions (CDFs).   
Another non-parametric measure uses chi-squared statistics after applying adaptive probability binning \cite{zimmerman2011}
that divides a $p$-dimensional cell population into bins such that each bin contains the same number of cells. 
The normalized chi-squared statistics is then used to compute the dissimilarity between the populations. 
The \emph{earth mover's distance (EMD)} \cite{rubner2000earth, zimmerman2011} considers the histograms of two populations as two piles of dart in multivariate space and computes the minimum cost of moving one pile into another.
Here the cost of moving dart is defined as the amount of dirt moved times the distance by which it is moved.
EMD can be efficiently computed by first applying adaptive probability binning on the cell populations and then solving transportation problem on the bins across the clusters \cite{rubner2000earth}.

In the rest this chapter, I used Mahalanobis distance to compute dissimilarity between a pair of clusters because it is a metric and uses the centers and covariance matrices of the clusters.
However, every algorithm described in this thesis works with other dissimilarity measures as well. 

\section{The mixed edge cover (MEC) algorithm}
\label{sec:mec_algorithm}
\subsection{Overview of the algorithm}
\label{sec:mec_overview}
Given a dissimilarity measure between cell cluster, I develop an algorithm to optimally register clusters across a pair of FC samples.
I make the algorithm robust by allowing a cluster from one sample to be matched to zero or more clusters in another sample.
This approach covers possible circumstances when a cell cluster in one sample is absent from another sample, or when a cluster in one sample splits into two or more cell populations in a second sample, which can happen due to biological reasons or due to the limitations of clustering methods.

I characterize an FC sample by a mixture of cell clusters.
Consider two FC samples $A$ and $B$ containing of $k_a$ and $k_b$ cell clusters such that $A=\{a_1,a_2, ..., a_{k_a}\}$ and $B=\{b_1,b_2, ..., b_{k_b}\}$, where  $a_i$ is the $i^{th}$ cluster from $A$ and $b_j$ is the $j^{th}$ cluster from $B$. 
I developed a robust variant of matching algorithm called the Mixed Edge Cover (MEC) algorithm that matches a cluster from $A$ to zero or more clusters from $B$ \cite{azad2010identifying}.
Suppose {\tt mec} is a matching of clusters across  $A$ and $B$ such that ${\tt mec}(a_i) \in \mathcal P(B)$ and ${\tt mec}(b_j) \in \mathcal P(A)$, where $ \mathcal P(A)$ ($ \mathcal P(B)$) is a subset (possibly empty) of clusters in $A$ ($B$).
When a cluster $a_i$ (or $b_j$) remains unmatched under {\tt mec}, i.e., ${\tt mec}(a_i)=\emptyset$, I set $d(a_i,-)=\lambda$ where the fixed cost $\lambda$ is used as a penalty for leaving a cluster unmatched, and is set to a value such that the number of such clusters remains small.
I  select $\lambda$ empirically by plotting the total number of unmatched clusters  in all pair wise matchings of  samples
against $\lambda$, and  finding a ``knee'' in the curve~\cite{azad2010identifying}. 
The cost of {\tt mec} is therefore computed as the summation of the dissimilarities of all pairs of matched clusters and the penalties coming from the unmatched clusters.
A minimum-weight mixed edge cover is a mixed edge cover with the minimum cost: 
\begin{equation} \label{eq:mec}
\arg\!\min \left( {\sum_{i=1}^{k_a}\sum_{b_j\in{\tt mec}(a_i) }} d(a_i, b_j) \ + 
				{\sum_{i=1}^{k_b}\sum _{ a_j\in{\tt mec}(b_i)}} d(b_i, a_j)\right).
\end{equation}
Here $d(a_i, b_j)$ is the dissimilarity between clusters $a_i$ and $b_j$.
\subsection{Bipartite graph model}
I compute a minimum-weight mixed edge cover (MEC) algorithmically in a complete bipartite graph $G(A,B,E)$ created from a pair of sample, $A=\{a_1,a_2, ..., a_{k_a}\}$, and $B=\{b_1,b_2, ..., b_{k_b}\}$.
In the bipartite graph $G$, each cluster $a_i \in A$ represents a vertex in one part and each cluster $b_i\in B$ from represents a vertex in the other part.
The edge set $E$ is created by joining each pair  $(a_i,b_j) \in (A\times B)$ of vertices  with an edge of weight $d(a_i,b_j)$.
Here, $d(a_i,b_j)$ denotes the dissimilarity between  $a_i$ and $b_j$ and is computed by using Mahalanobis distance from Equation \ref{eq:Mahalanobis}.

Since low edge weight implies high similarity among vertices (clusters) I match vertices connected by edges with small weights and leave a vertex unmatched when it has no low-weight edge adjacent to it.
In the above graph model, a \emph{mixed edge cover (MEC)} is a collection of edges $EC$ and vertices  $V_{um}$ such that each vertex in $A\cup B \setminus V_{um}$ has {\em at least\/}  one edge incident on it and  $V_{um}$ remains unmatched with fixed penalty $\lambda$ for each vertex. 
Thus, a minimum-weight MEC is an MEC  that minimizes the following objective function:
\begin{equation} \label{eq:mec_obj}
	   min \left({\sum_{(a_i,b_j)\in EC} d(a_i, b_j) + \lambda * |V_{um}|  }\right).
\end{equation}
According to graph-theoretic terms, an MEC is a generalization of an {\em edge cover\/} represented by a subset of edges such that each vertex in the graph has {\em at least\/}  one edge incident on it. 
An {\em edge cover\/} again is a generalization of a {\em matching\/}  represented by a subset of edges such that each vertex in the graph has {\em at most\/} one edge incident on it.

\subsection{MEC algorithm on the bipartite graph}
A  mixed edge cover in $G$  can be computed  from an edge cover in a transformed  graph $G'$ obtained from $G$ by introducing two new distinguished vertices $a_0 \in A$ and $b_0 \in B$ representing two dummy clusters one in each sample.
In $G'$, I add an edge $\{a_0,b_0\}$ with $d(a_0,b_0)=0$, and edges $\{a_i,b_0\}$
for each $a_i \in A$, $\{b_i,a_0\}$ for each $b_j \in B$, with $d(a_i,b_0)= d(b_j,a_0) = \lambda$.
For each such edge $\{a_i,b_0\}$ (or  $\{b_j,a_0\}$) included in a minimum-weight edge cover computed on $G'$, I leave the $a_i$ ($b_j$) unmatched in a MEC computed on $G$, thereby paying a price of $\lambda$ for each unmatched vertex.
A  minimum-weight edge cover in $G'$ can be computed in polynomial time by making a copy of the graph and connecting
each vertex to its twin in the copy  by an edge with weight equal to twice the minimum weight  among 
original edges incident on it. A minimum-weight perfect matching in this graph 
can be used to compute  a minimum-weight edge cover in the original graph  \cite{schrijver2003combinatorial}.

Following the above discussion, a minimum-weight MEC in $G$ can be computed in the following step:
\begin{enumerate*}
\item \emph{Add dummy vertices:} Create a new augmented graph $G'$ from $G$ by adding two distinguished vertices $a_0 \in A$ and $b_0 \in B$ representing two dummy clusters one in each sample.
In $G'$ I add an edge $\{a_0,b_0\}$ with $d(a_0,b_0)=0$, and edges $\{a_i,b_0\}$
for each $a_i \in A$, $\{b_j,a_0\}$ for each $b_j \in B$, with $d(a_i,b_0)= d(b_j,a_0) = \lambda$.
Here $\lambda$ is the penalty for leaving a vertex unmatched.

\item \emph{Duplicate Graph:} Let $G''$ be a disjoint copy of $G'$. 
I create a new graph $\bar{G}$ by taking the union of $G'$ and $G''$
and adding an  edge $\{v',v''\}$ connecting every vertex $v'$ in $G'$ with its twin  $v''$ in $G''$.
The weight of edge $\{v',v''\}$ is set to $2\mu (v')$, where $\mu (v')$ is the minimum weight
of the edges of $G'$ incident on  $v'$.

\item \emph{Compute matching:} Compute a minimum-weight perfect matching $M$ in $\bar{G}$.

\item \emph{Compute edgecover from matching:} Obtain a  minimum-weight edge cover $EC'$ of $G'$ by replacing every  edge
$\{v',v''\} \in M$  by an edge of weight $\mu (v')$ in $G'$ incident on $v'$.

\item \emph{Compute MEC from edge cover:} Remove all edges $\{a_i,b_0\}$ and  $\{b_j,a_0\}$  from $EC'$ to obtain a reduced edge cover $EC$ in $G$.
Put all vertices in $G$ not covered by $EC$ to the set of unmatched vertices (clusters) $V_{um}$. 
The resulting edge cover $EC$ together with the set of unmatched vertices $V_{um}$ is a solution to the
mixed edge cover problem in $G$.
\end{enumerate*}

\begin{figure}[!t]
   \centering
   \includegraphics[scale=1]{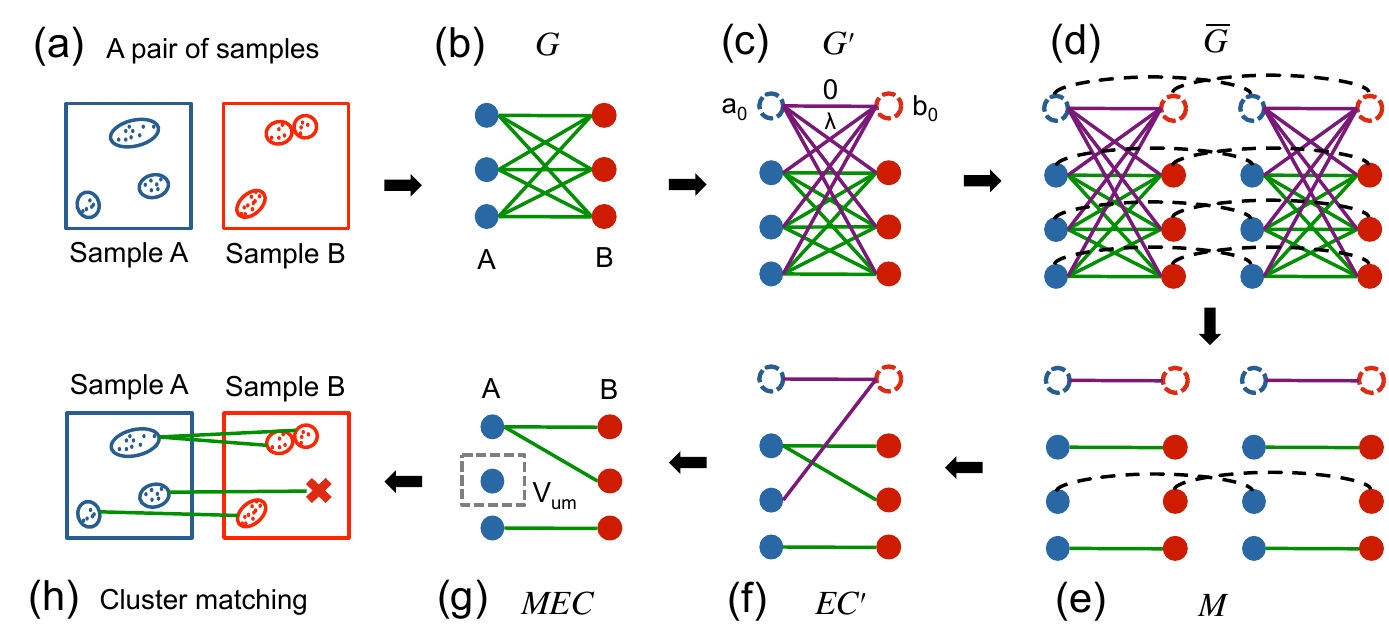} 
   \vspace{-.2cm}
   \caption[A working example of the mixed edge cover algorithm]{Steps of the MEC algorithm: (a) two FC samples $A$ and $B$, each with three cell clusters, (b) a complete bipartite graph $G$ is created from $A$ and $B$, 
   (c) an augmented graph $G'$ is obtained by adding two dummy clusters $a_0$ and $b_0$ as described in step 1 of the MEC algorithm,
   (d) an identical copy of $G'$ is combined with it to obtain $\bar{G}$, 
   (e) a minimum-weight perfect matching $M$ is computed in $\bar{G}$,
   (f) a minimum-weight edge cover $EC'$ in $G'$ is constructed  after replacing each edge of  $M$ connecting $G'$ and its copy with an edge of minimum weight within $G'$,
   (g) a minimum-weight MEC is computed from $EC'$ by removing dummy clusters and their adjacent edges, and finally 
   (h) the clusters are matched across $A$ and $B$.}
   \label{fig:mec_algorithm}
   \vspace{-.5cm}
\end{figure}

I show a workflow of the MEC algorithm with two hypothetical samples $A$ and $B$ each with three clusters in Figure \ref{fig:mec_algorithm}.
In this example, a single cell population is split into two clusters in sample $B$ requiring one-to-many matching of clusters. 
Additionally a cluster in sample $A$ remains unmatched because it does not have any corresponding cluster in sample $B$.
The MEC algorithm covers both cases and therefore provides a robust solution to the population registration problem.

\subsection{Proof of correctness}

\textbf{Lemma: } The Algorithm described above computes an optimum  mixed edge cover in $G$.

\textbf{Proof: }
The correctness of the  algorithm for computing the edge cover $EC'$ in the graph $G'$ is  shown in~\cite{schrijver2003combinatorial}.
I obtain a mixed edge cover in $G$ by deleting,  
the vertices $a_0$ and $b_0$ and the edges incident on these vertices in $EC'$.
Let $V_{um}$ be the set of the vertices adjacent to $a_0$ and $b_0$, which will be identified as unmatched vertices.
Let $EC$ be the edges remaining from the edge cover  $EC'$ in the modified graph $G\setminus V_{um}$.

I claim that  $EC$ together with $V_{um}$ is an optimal solution for the mixed edge cover problem in $G$.
Assume that there is an optimal solution in $G$ consisting of the set of unmatched vertices $V_{um}$ and an edge cover $EC^*$ in $G\setminus V_{um}$.
Clearly $\sum_{(a_i,b_j)\in EC} d(a_i, b_j) =  \sum_{(a_i,b_j)\in EC^*} d(a_i, b_j) $, for otherwise one of the solutions could be improved upon, thereby contradicting their minimality in $G\setminus V_{um}$ respectively. 
It remains to prove that there is no other MEC solution in $G$ with a different set of unmatched vertices and smaller cost.
Let $\tilde{EC}$ together  with $\tilde{V}_{um}$ gives a MEC solution in $G$ with a cost smaller than the MEC solution given by $EC$ and $V_{um}$ (Equation \ref{eq:mec_obj}).
Then $\tilde{EC}$ together with an edge $\{o,a_0\}$ or $\{o,b_0\}$ for every $o \in \tilde{V}_{um}$ and the edge $\{a_0,b_0\}$ is an  edge cover in $G'$ with cost smaller that the cost of $EC'$, contradicting the optimality of $EC'$.

\subsection{Complexity of the MEC algorithm}
For a graph with $n$ vertices and $m$ edges, a minimum-weight edge cover can be computed in time $O(n (n+m \log n))$~\cite{schrijver2003combinatorial}.
For a pair of sample $A=\{a_1,a_2, ..., a_{k_a}\}$, and $B=\{b_1,b_2, ..., b_{k_b}\}$, let $k=k_a+k_b$ be the total number of clusters.
By the construction of the graph model $n=O(k)$, $m=O(k^2)$.
Therefore the time complexity of a MEC algorithm is $O(k^3 \log k)$.
The number of cell clusters $k$ is usually small (less than 50 in typical experiments).
As such, the dissimilarity between a pair of samples can be computed in seconds on a desktop computer. 

\section{Properties of mixed edge cover} 
\label{sec:mec_propoerties}
To understand the conditions for an unmatched cluster, let $b_j$ be the nearest cluster in sample $B$ from a cluster  $a_i$ in sample $A$.
Then $a_i$ is always unmatched if $d(a_i, b_j) > 2 \lambda$ and is never unmatched if $d(a_i, b_j) <  \lambda$.
If $2 \lambda < d(a_i, b_j) < \lambda$, $a_i$ will be unmatched if and only if it is not matched to a vertex in duplicate graph during step $3$ of the algorithm.

The edges contained in a minimum-weight MEC do not create any path of length greater than two. 
Hence, MEC never creates a cycle in the optimum solution. 
To see this, consider an MEC solution {\tt mec} consisting of an edge cover $EC$ and a set of unmatched vertices $V_{um}$. 
 The graph induced by $EC$ in $G$ does not have a path of three edges since the middle edge can be removed from the optimal cover, thus reducing its weight further, contradicting the optimality of {\tt mec}.
 Therefore, {\tt mec} does not contain a path of length greater than two.
 In fact, {\tt mec}  is a union of  disjoint ``stars" (trees of height 1) where each star has no more than one vertex of degree greater than one, preventing cycle or long chain of matched vertices.
 This property allows a cluster from sample $A$ to be matched to the corresponding but spitted set of clusters in sample $B$ and vice versa.
 However, if a cell population is split up in both samples, the MEC algorithm is unable to match all parts across samples because this will require a path of length greater than 2 in the MEC solution, therefore violating the disjoint ``stars" constructions. 
This limitation, however, cannot be solved by an algorithm that considers only the between sample dissimilarities.

\section{Results}
\label{sec:mec_results}

\subsection{Registering populations across samples from two healthy subjects}
\label{sec:match_hd}
{\em  Data description:} I register cell populations across two five-dimensional samples collected from two healthy subjects.
These two samples are part of the HD dataset described in Section \ref{sec:hd_data_description} where blood was collected from five healthy individuals on different days to study natural and instrumental variations among samples.
For demonstration purpose, I randomly selected two samples from different subjects in the HD dataset.
Each sample is preprocessed and clustered independently to identify four cell clusters. 
Figure \ref{fig:clustering_hd} in Chapter~\ref{chap:clustering} displays these clusters with a collection of two dimensional projections.

\begin{figure}[!t]
   \centering
   \includegraphics[scale=.5]{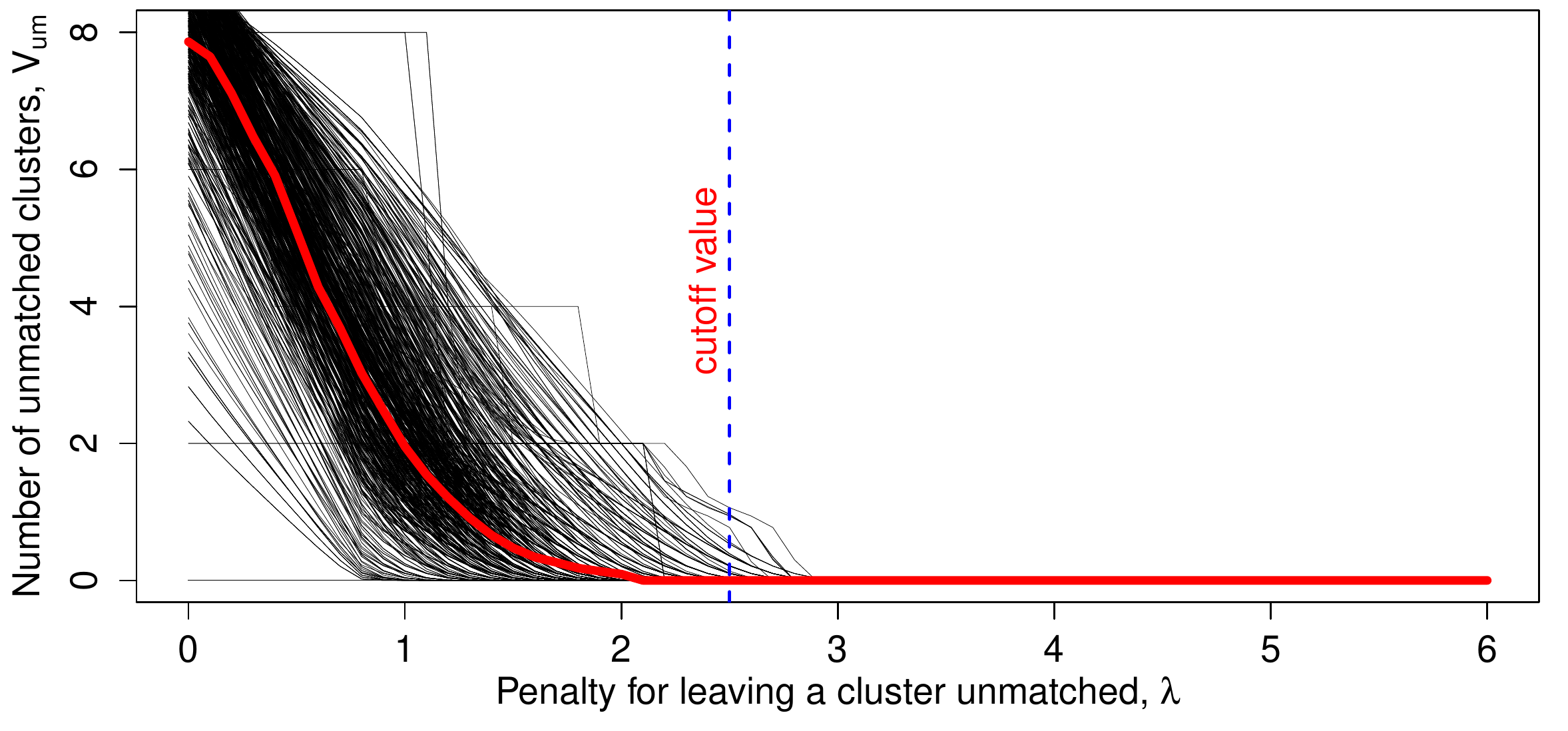} 
   \vspace{-.2cm}
   \caption[Selecting the unmatch-penalty $\lambda$ for the mixed edge cover algorithm]{ Selecting the unmatch-penalty, $\lambda$ for the HD dataset.
Each black curve plots the numbers of unmatched clusters, $V_{um}$ across a pair of HD samples for a sequence of unmatch-penalty $\lambda$ in the range [0,6].
  The red curve is a smooth-average of all individual curves and represents the general trend for the whole dataset.
From the average red curve, $\lambda = 3.5$ (shown as blue dashed line) is selected as a suitable unmatch-penalty because, at this value, the curve becomes stable and horizontal. 
   }
   \vspace{-.5cm}
   \label{fig:lambda_select}
\end{figure}

{\em  Selecting the unmatch-penalty, $\lambda$ :} 
To determine the unmatch-penalty $\lambda$ for the HD dataset, I compute the optimum mixed edge cover (MEC) between each pair of samples in this dataset for different choices of $\lambda$ in the range [0,6]. 
The number of unmatched clusters, $V_{um}$ obtained by each MEC solution is plotted against the corresponding $\lambda$ value in Figure \ref{fig:lambda_select}.
Each black curve in Figure \ref{fig:lambda_select} represents the the number of unmatched clusters for different choices of $\lambda$ when MEC algorithm is applied to a particular pair of samples.
I have applied local polynomial regression (loess) method to smoothen the curves.
The red curve is a smooth-average of all individual curves and represents the general trend for the whole dataset.
Observe that an increase of $\lambda$ quickly decreases the size of $V_{um}$, thus allowing more clusters to be matched ( from Equation \ref{eq:mec_obj}).
From the average red curve, $\lambda = 3.5$ (shown as blue dashed line in Figure \ref{fig:lambda_select}) is selected as a suitable unmatch-penalty because, at this value, the curve becomes stable and horizontal.
Since the samples are very similar (healthy) in this dataset, we do not expect many unmatched clusters and therefore $\lambda = 3.5$ looks like a reasonable choice.  
I use $\lambda = 3.5$ for any MEC computation between two samples in the HD dataset.

\begin{figure}[!t]
   \centering
   \includegraphics[scale=.56]{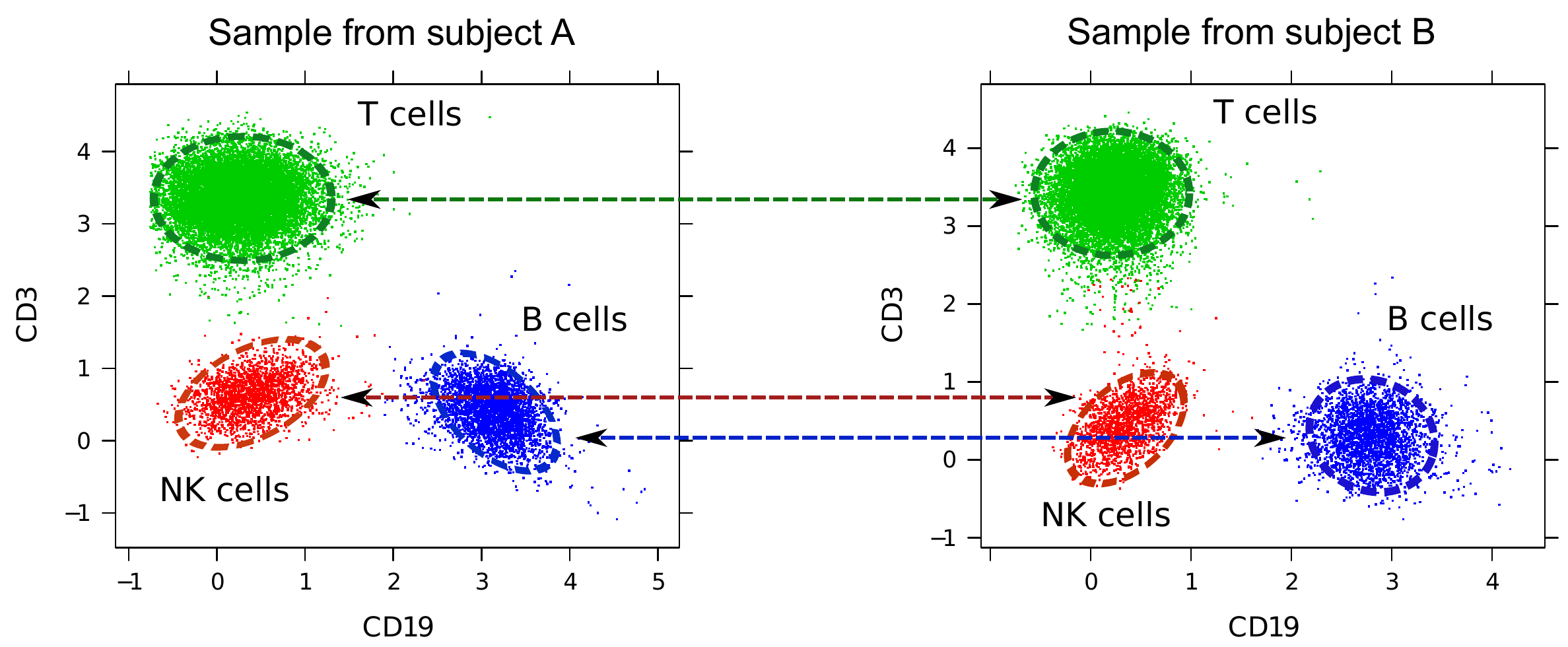} 
   \vspace{-.2cm}
   \caption[Registering cell populations across two representative samples from the HD dataset]{Cell populations are matched by the MEC algorithm across two representative samples in the HD dataset. 
 In addition to showing individual cells with dots, I draw the 95th quantile of each cell population with a dashed ellipse. I show a pair of matched clusters in same color and join them by a line with arrowheads. 
   Although the cell populations are defined in five dimensions,  I show a 2-D projection for visualization purpose. 
   }
   \label{fig:matching}
      \vspace{-.5cm}
\end{figure}

{\em  Registering populations:}
I now apply the MEC algorithm, with $\lambda = 3.5$, to match cell clusters across the selected pair of samples.
For visualization purpose, I project the matched populations onto two dimensions (CD3 and CD19) in Figure \ref{fig:matching}. 
I selected CD3 and CD19 markers for projection because they cleanly define three functional cell subsets -- T cells, B cells and Natural Killer (NK) cells.
The MEC algorithm, however, does not use these biological labels, but views each cluster as an unlabeled distribution of cells defined only by the distribution parameters.
In Figure \ref{fig:matching}, I show a pair of matched clusters in same color and join them by a line with arrowheads.
The matched clusters represent functionally equivalent cell subsets and therefore confirm that the algorithm is actually matching biologically equivalent cell populations across samples.
Additionally the proportion of lymphocytes in each clusters matches closely across the samples: T cell (71.1\%, 75.4\%), B cells (18.4\%, 15.1\%) and NK cells (10.5\%, 9.5\%).
The proportions closely follow the lymphocyte  proportions within a healthy immune system described by Berrington et al. \cite{berrington2005lymphocyte} and further validate the effectiveness of the clustering and matching algorithms. 
Note that, cell clusters are actually defined in five dimensions (Fig. \ref{fig:clustering_hd}), but I show a simplified 2-D projection for visualization purpose.

In Figure \ref{fig:matching}, it may seem trivial to match these three clusters across samples simply by visual inspection.
However, the clusters are defined in five dimensions and therefore it is impractical to visually match the clusters.
For example, we are unable to distinguish helper and cytotoxic T cell in the 2-D scatter plot shown in Figure \ref{fig:matching} because we need additional set of markers (CD4 and CD8) to differentiate them from one another.
Furthermore, the matching algorithm uses statistical parameters of clusters to compute their dissimilarity, which is not easy to perceive by visualization. 
The MEC algorithm is completely unsupervised and matches clusters solely based on their statistical parameters without using the biological labels (e.g., T cell, B cells etc.) of the clusters.  
However, if the cluster labels are available for a sample, then the algorithm has the ability to label another sample based on the matching pattern, thus solving the cluster labeling problem.

\subsection{Registering populations before and after stimulating T cells}
{\em  Data description:} As a second example, I use MEC algorithm to register cell populations across two four-dimensional samples collected before and after stimulating T cells in whole human blood.
These two samples are part of the TCP dataset described in Section~\ref{sec:tcp_data_description} where whole blood was collected from 29 subjects and stained using labeled antibodies against CD4, CD45RA, SLP-76, and ZAP-70 protein markers before and five minutes after stimulation with an anti-CD3 antibody.
The downstream phosphorylation event caused by the stimulation increases the levels of SLP-76, and ZAP-70 proteins in different T cell subsets \cite{maier2007allelic}.

I randomly selected a pair of samples from the same subject before and after the stimulation.
Each sample is clustered independently to obtain four cell clusters with the following expression profiles: (a) CD4$^+$ CD45RA$^\text{low}$, (b) CD4$^+$ CD45RA$^\text{high}$, (c) CD4$^-$ CD45RA$^\text{high}$, and (d) CD4$^-$ CD45RA$^\text{low}$.
(Recall that  `$+$' and  `high' indicate higher abundances  of a marker, and `$-$' and `low' indicate lower levels of it.)

\begin{figure}[!t]
   \centering
   \includegraphics[scale=.8]{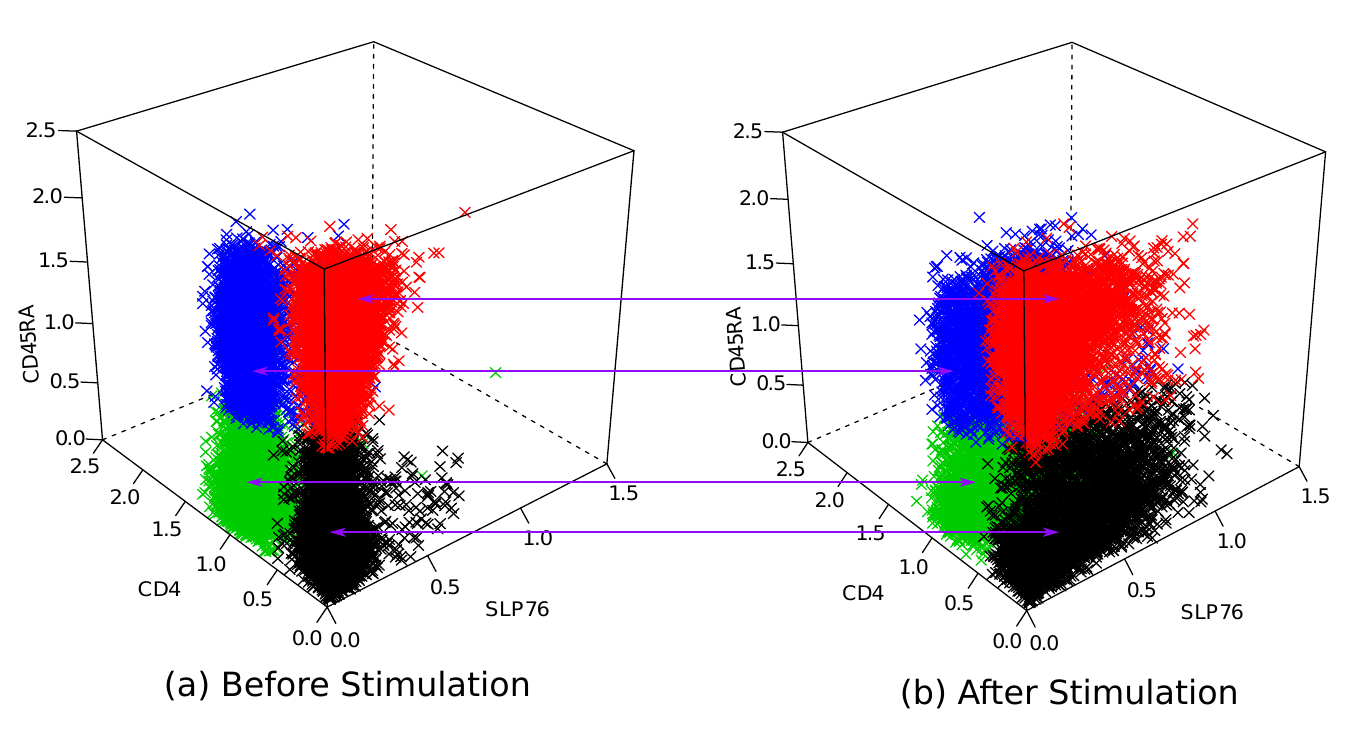}
   \vspace{-.2cm}
   \caption[Registering populations across a pair of samples before and after stimulating T cells with an anti-CD3 antibody]{Registering (matching) populations by the MEC algorithm across a pair of samples before and after stimulating T cells with an anti-CD3 antibody.
   Matched clusters are marked with same color and are joined them by a line with arrowheads. 
   The cell populations are defined in four dimensions, but I show a 3-D projection for visualization purpose.}
   \label{fig:matching_cluster_TCP}
   \vspace{-.5cm}
\end{figure}

{\em  Registering populations:}
I select unmatch-penalty $\lambda$ for the TCP dataset following the similar procedure discussed in Section~\ref{sec:match_hd}.
The MEC algorithm is then used to register cell population across the selected pair of samples.
For visualization purpose, I show the matching solution in three-dimensional projections (CD4, CD45RA and SLP-76) in Fig.~\ref{fig:matching_cluster_TCP}, where matched clusters are shown in same colors.
By comparing the matched clusters in Fig.~\ref{fig:matching_cluster_TCP}, we observe an increased level of SLP-76 proteins after the stimulation (and ZAP-70, although this is not included in the Fig.).
The mapping of clusters across pre- and post-stimulation samples can assess the population-specific effects of the stimulation experiment.

\begin{figure}[!t]
   \centering
   \includegraphics[scale=.55]{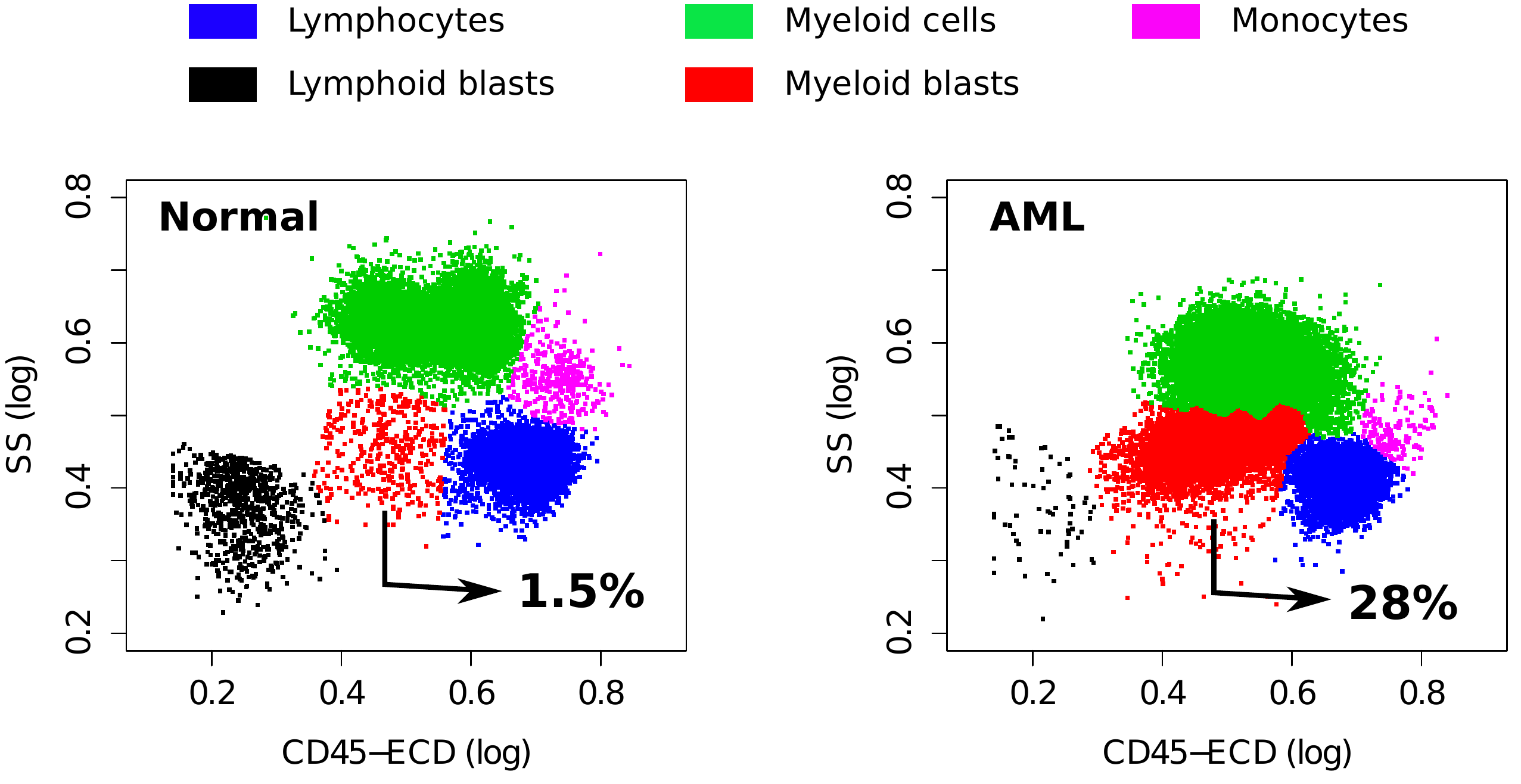}
   \vspace{-.2cm}
   \caption[Registering populations between normal and AML samples]{Registering different clusters between normal and AML samples. 
   Matched clusters are marked with same color. 
   The proportion of myeloid blasts (shown in red) increases significantly in the AML sample.}
   \label{fig:matching_cluster_AML}
   \vspace{-.5cm}
\end{figure}

\subsection{Registering population in diagnosing Acute Myeloid Leukemia (AML)}
Acute Myeloid Leukemia (AML) is often diagnosed by abnormal increase (greater than 20\% of the white blood cells) of premature myeloid blast cells~\cite{lacombe1997flow, kern2010role}.
In flow cytometry, the myeloid blasts can be  identified in SS vs. CD45 plot.
Here, the side scatter (SS) measures the granularity of cells and CD45 is a marker for the leukocytes.
To show the first step of  AML diagnosis, I use a normal and an AML samples from the AML dataset~\cite{aghaeepour2013critical}.
I only consider SS channel and CD45 marker for this analysis.

I cluster the healthy and the AML samples with the k-means algorithm and obtain five clusters in each sample.
The five clusters represent lymphocytes, lymphoid blasts, myeloid cells, myeloid blasts, and monocytes.
The myeloid blasts are identified by medium side scatter and medium CD45 expressions.
Next, I match cell clusters in order to detect same cell types between the samples.
The result is illustrated in Fig.~\ref{fig:matching_cluster_AML}, where a pair of matched clusters is shown in same color. 
After registering clusters, we observe that the fraction of myeloid blasts (shown in red) increases from 1.5\% in healthy sample to 28\% in the AML sample.
This observation confirms that the second sample has Acute Myeloid Leukemia.
The mixed edge cover algorithm is able to detect abnormally behaving cell populations indicative AML and therefore, can be used in automated diagnosis of this cancer. 

\section{Conclusions and future work}
\label{sec:mec_conclusions}
Population registration or labeling is a fundamental problem in flow cytometry since it is used to evaluate population level differences across biological conditions.
To solve this problem computationally, I have developed a robust mixed edge cover (MEC) algorithm that registers high-dimensional clusters across a pair of FC samples. 
The MEC algorithm uses a robust graph-theoretic framework to match a cluster from a sample to zero or more clusters in another sample and thus solves the problem of missing or splitting clusters.
I demonstrated, with three representative datasets, that this algorithm correctly matches phenotypically similar clusters despite heterogeneity across samples.
Given a sample with known cluster labels, the MEC algorithm labels clusters from another sample, thus solving the cluster labeling problem.

Beside matching and labeling populations, an optimal MEC solution computes a combinatorial dissimilarity between a pair of FC samples.
This between-sample dissimilarity works as a building block of the meta-clustering and classification approaches discussed in Chapter \ref{chap:template} and Chapter~\ref{chap:template-based-classification}.

MEC algorithm has three limitations.
First, it does not use the proportion of cells in the populations while matching them across samples.
It has been observed that samples from same biological status approximately preserve the proportion of cells in different populations (e.g., see the discussion in Section~\ref{sec:match_hd} and also by Pyne et al.~\cite{pyne2009automated}).
Hence, cell proportions can be used directly in Eq.~\ref{eq:mec_obj} to provide a second level of verification.
However, care must be taken when using cell proportion to match populations from different disease status.
For example, the proportion of CD4$^+$ cells reduces significantly after HIV infection, therefore this proportion of CD4$^+$ cells will not match between healthy subject and HIV patient.
Second, MEC algorithm can match populations more accurately if it uses the relative arrangement of populations within a sample.
In this approach, each sample is modeled by a network of populations and two such networks are aligned to match populations across samples.
This approach has been extensively used to align protein and gene regulatory networks \cite{sharan2006modeling, singh2008global}.
Aligning two networks, however, is an NP-complete problem; therefore, approximation algorithms are often used to align large networks.
Finally, the dissimilarity measure given by the MEC solution is not a metric since it fails to satisfy triangle inequality.
Distance metrics are especially useful for creating hierarchical organization of samples, as will be discussed in next Chapter~\ref{chap:template}. 

In current work, I am improving the MEC algorithm by addressing the limitations discussed in the previous paragraph.
In particular, I am developing a between-sample distance metric similar in spirit of the earth mover's distance~\cite{rubner2000earth, zimmerman2011} such that the combinatorial dissimilarity measure becomes a metric, as well as remains robust similar to the MEC.

%% file: ch5-template.tex
\chapter{Meta-clusters and class templates}
\label{chap:template}

\section{Introduction}
\begin{figure}[!t]
   \centering
   \includegraphics[scale=.75]{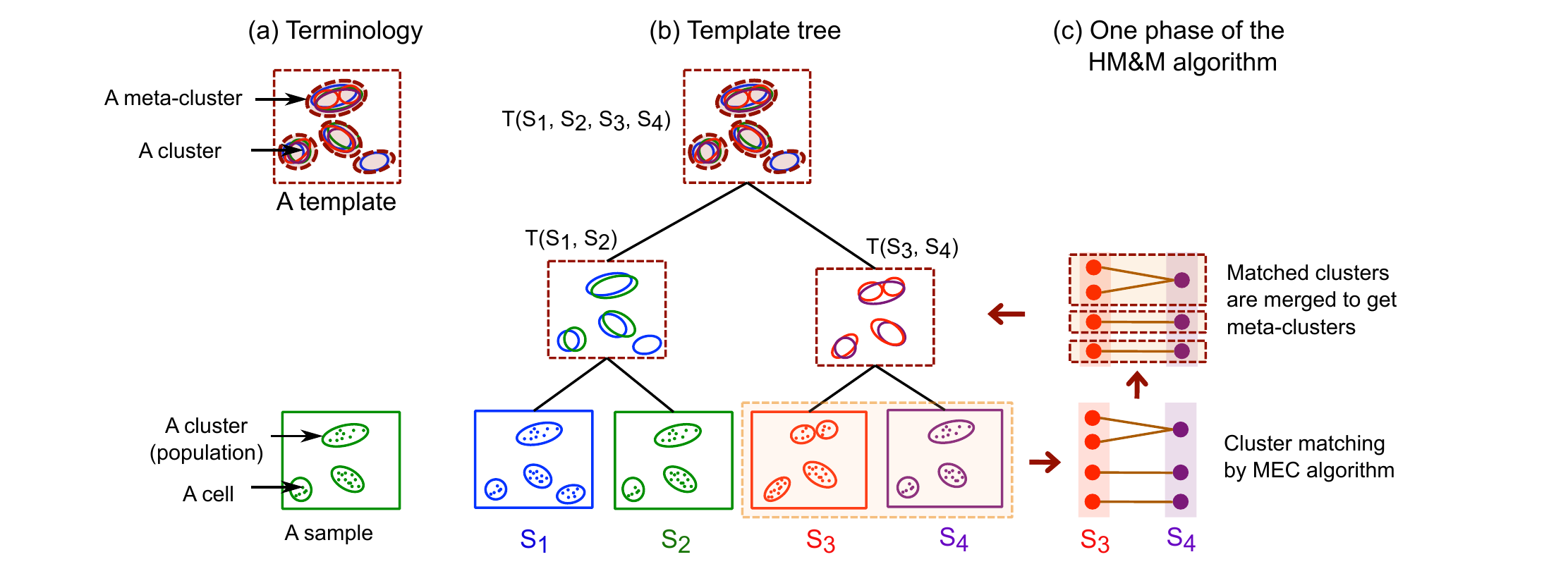}
     \vspace{-.2 cm}
   \caption[An example of a template-tree created from four hypothetical samples]{(a) The concepts of sample, cluster, template, and meta-cluster are illustrated graphically. Cells are denoted with dots, clusters with solid ellipses and meta-clusters with dashed ellipses. (b) An example of a hierarchical template-tree created from four hypothetical samples $S_1, S_2, S_3$ and $S_4$. A leaf node of the template-tree represents a sample and an internal (non-leaf) node represents a template created from its children in the tree.  (c) One step of the HM\&M algorithm creating a template $T(S_3,S_4)$ from a pair of samples $S_3$ and $S_4$. The algorithm first matches clusters across $S_3$ and $S_4$ by the mixed edge cover algorithm and then merges the matched clusters to construct new meta-clusters.}
   \label{fig:concepts}
     \vspace{-.5 cm}
\end{figure}

Consider a collection of flow cytometry samples belonging to a few representative classes with each class of samples indicating the same biological status (e.g., a disease).
I describe an algorithm that builds a \emph{template} for each class of samples by emphasizing the common properties of the class while omitting sample-specific details~\cite{azad2012matching, finak2010optimizing, pyne2009automated}. 
Samples within a class usually have shared populations expressing similar phenotypes, whereas samples from different classes contain some unrelated populations.
Cell populations with similar phenotypes in different samples of a class can be combined into a \emph{meta-population} (also called a  \emph{meta-cluster}) representing a generic phenotype shared by the populations.
A set of relatively homogeneous meta-clusters forms the core pattern of a class of samples and groups together into a prototypic template.
I summarize the concepts of meta-cluster and template in Table \ref{tab:terminology} and in Figure \ref{fig:concepts}(a).

I have developed a hierarchical matching-and-merging (HM\&M) algorithm for constructing templates from a collection of samples. 
The algorithm repeatedly merges the least dissimilar (most similar) pair of samples not already included in a template. 
The dissimilarity between a pair of samples is computed by the cost of an optimal mixed edge cover (MEC) solution discussed in Chapter~\ref{chap:matching}.
Towards this end, I assume that samples belonging to a particular class have smaller dissimilarity among themselves than samples from other classes. 
The HM\&M algorithm builds a binary \emph{template-tree} representing the hierarchical relationships among the samples.  
A leaf node of the template-tree represents a sample and an internal (non-leaf) node represents a template created from the samples.  
Figure  \ref{fig:concepts}(b) shows an example of a template-tree created from four hypothetical samples, $S_1, S_2, S_3$, and $S_4$.
An internal node in the template-tree is created by matching similar cell clusters across the left and right children and merging the matched clusters into meta-clusters. 
Fig.~\ref{fig:concepts}(c) illustrates the creations of the internal node $T(S_3,S_4)$ in Fig.~\ref{fig:concepts}(b) by matching clusters across samples $S_3$ and $S_4$.
The root of the template-tree defines a class-template when all samples in the leaf nodes belong to the same biological class, and otherwise, the tree can be cut at an appropriate height to discover multiple class templates.

\begin{table}[t]
   \centering
     \vspace{-.7 cm}
    \caption[Summary of terminology used in Chapter~\ref{chap:template}]{Summary of concepts -- cell population, sample, meta-cluster, and template -- used in this chapter.}
   \begin{tabular}{|p{0.18\linewidth}| p{0.75\linewidth}|} 
       \hline
      Terms    & \multicolumn{1}{c|}{Meaning}  \\
      \hline
      Cell population (cell cluster)      & a group of cells expressing similar features, e.g., helper T cells, B cells, etc. \\
             \hline
      Sample       & a collection of cell populations within a single biological sample \\
           \hline
      Meta-cluster       & a set of biologically similar cell clusters from different samples   \\
            \hline
      Template & a collection of meta-clusters from samples of same class  \\
      \hline
   \end{tabular}
   \label{tab:terminology}
\end{table}

Just as a sample is characterized by a mixture of cell populations, a template is characterized by a finite mixture of homogeneous meta-clusters.
The homogeneity of meta-clusters needs to be evaluated statistically since there is no accepted method for evaluating the biological homogeneity of clusters.
However, conventional methods for statistical significance testing (e.g., F-test or paired t-test) have a high probability of making a Type I error when used to evaluate the homogeneity of a meta-cluster because clusters in an FC sample are typically large~\cite{ekstrom2011introduction, steiger2004beyond}.
Instead, the ratio of  between-cluster to within-cluster variances, within a multivariate analysis of variance (MANOVA) model, can evaluate the homogeneity of a meta-cluster more effectively~\cite{cohen1988book}.
Based on a baseline experiment with a healthy dataset, I discuss a tolerable limit for the between-cluster variance in a homogeneous meta-cluster.

The ability to systematically characterize a class of multi-dimensional samples with a well-defined template is useful in various applications of flow cytometry. 
In addition to providing a cogent description of the core population structure that is shared among samples within a class, the templates also allow an objective way of assessing the overall differences in those structures across classes. 
I demonstrate the application of templates with a healthy donor (HD) dataset and a T cell phosphorylation (TCP) dataset. 
For the HD dataset, the algorithm is able to construct five well-separated templates for five healthy subjects, despite different sources of within-subject variations.
For the TCP dataset, the algorithm creates pre- and a post-stimulation templates from 29 pairs of samples collected before and after stimulating blood cells with an anti-CD3 antibody.  
By matching meta-clusters across templates, we can better assess the population-specific effects of the stimulation experiment.

{\em  Related work:} 
Several researchers have used the concepts of templates and meta-clusters to summarize collection of flow cytometry samples~\cite{pyne2009automated, finak2010optimizing, spidlen2013genepattern}. 
FLAME, proposed by Pyne et al. \cite{pyne2009automated}, pools cluster medoids from a class of samples and applies a second stage of clustering on the medoids to construct the global meta-clusters.
\emph{flowTrans}, developed by Finak et al. \cite{finak2010optimizing}, starts with seed meta-clusters and assigns each cluster to a meta-cluster to which it is most similar.
The HM\&M algorithm is significantly different from both FLAME and flowTrans in several ways. 
First, FLAME and \emph{flowTrans} both build a single template from samples of the same class.
Therefore, these algorithms need to know the class label of each sample, which is often unknown in practice.
In contrast, the HM\&M algorithm is able to identify templates as the roots of well-separated branches of a template-tree in an unsupervised manner.
This approach also permits multiple templates representing different sub-classes within a single class, and therefore, is more flexible to capture sample diversity.
Second, instead of clustering population centers, the HM\&M algorithm optimally matches populations across samples and then merges the matched clusters into meta-clusters (see Figure \ref{fig:concepts}(c) for an example).
Like FLAME, but unlike \emph{flowTrans}, this algorithm allows a cluster forming a self-contained meta-cluster by itself when the cluster is significantly different from all other clusters.
Finally, the hierarchical organization of samples shown in Fig.~\ref{fig:concepts}(b) provides additional flexibility in creating multi-layer templates, classifying samples, and updating templates dynamically.

The rest of the chapter is organized as follows.
In Section~\ref{sec:HMM}, I describe a hierarchical algorithm for creating templates from a collection of samples.
Section~\ref{sec:homogeneity_mc} discusses a statistical framework for analyzing the homogeneity of meta-clusters.
I demonstrate the application of templates with two representative FC datasets in Section~\ref{sec:template_results}.
I finish this chapter in Section~\ref{sec:template_conclusions} with concluding remarks.

\section {An algorithm for constructing templates}
\label{sec:HMM}
\subsection{Dissimilarity between samples or templates}
The template construction algorithm characterizes both a sample and a template with a finite mixture of multivariate  normal distributions each component of which is a cluster or a meta-cluster.
To compute the dissimilarity between a pair of samples or templates, I create a complete bipartite graph with the clusters (meta-clusters) in each sample (template) as vertices, and each edge is weighted by the Mahalanobis distance between its endpoints. 
The cost of an optimal mixed edge cover (MEC) in the bipartite graph gives the dissimilarity between the samples (templates).
This dissimilarity measure is similar in spirit with the R-metric, transfer distance or partition distance that computes the minimum number of augmentations and removals of cells needed to transform one partition into another \cite{gusfield2002partition, konovalov2005partition, charon2006maximum}.
However, the partition distance can only compare two partitions of the same sample, whereas MEC based dissimilarity can work with partitions from the same sample, or from two different samples.
Partition distance matches a cluster to at most one cluster, while MEC is able to match a cluster to zero or more clusters. 
Therefore, the latter approach is more robust in the presence of missing or splitting cell populations.

\subsection{The hierarchical matching-and-merging (HM\&M) algorithm}  
Consider a collection of $N$ flow cytometry samples \{$S_1, S_2, ..., S_N$\}.
Each sample is clustered independently to identify phenotypically distinct cell populations.
The HM\&M algorithm organizes the samples into a \emph{template-tree} data structure as was described in Fig.~\ref{fig:concepts}.
Each node $v_i$ in the template-tree represents either a sample (leaf node) or a template (internal node).
Let $v_i$ consist of $k_i$ clusters or meta-clusters,  $v_i=\{c^i_1$, $c^i_2$, $\ldots$, $c^i_{k_i}\}$.
$v_i$ is called an ``orphan" if it does not have a parent in the template-tree.
At each step, the HM\&M algorithm identifies the most similar pair of orphan nodes and merges them into a new node denoting a template of the merged nodes. 
This process continues until a single orphan node (root) remains. 
The HM\&M algorithm can be described with the following three steps.

1. \emph{Initialization}: Create a node $v_i$ for each of the $N$ samples  $S_i$. 
Define the set of orphan nodes, ${\tt Orphan}=\{v_1, v_2, ... ,v_N\}$. Repeat the following matching and merging steps until a single orphan node remains. 
 
2. {\em  Matching}: Compute the dissimilarity $D(v_i,v_j)$ between every pair of nodes  $v_i$ and $v_j$ in the current {\tt Orphan} set by the cost of an optimal mixed edge cover solution.

3. \emph{Merging:} Find a pair of orphan nodes $(v_i,v_j)$ with minimum dissimilarity $D(v_i, v_j)$ and merge them to create a new node $v_l$. 
Let ${\tt mec}$ be a function denoting the mapping of clusters from $v_i$ to $v_j$.
That is, if $c_x^i\in v_i$ is matched to $c_y^j\in v_j$, then $c_y^j = {\tt mec}(c_x^i)$, where  $1\leq x \leq k_i$ and $1\leq y \leq k_j$.
Create a new meta-cluster $c^l_z$ from each set of matched clusters, i.e., $c^l_z = \{c_x^i\cup  {\tt mec}(c^i_x)\}$.
Let $k_l$ be the number of new meta-clusters created above.
Then the new node $v_l$ is created as a collection of these newly created meta-clusters, i.e., $v_l=\{c^l_1,c^l_2, ..., c^l_{k_l}\}$.
The distribution parameters, $(\mu^l_z, \Sigma^l_z)$, of each of the newly formed meta-clusters $c^l_z$ are estimated by the Expectation-Maximization (EM) algorithm. 
The height of  $v_l$ is set to $D(v_i, v_j)$.
The node $v_l$ becomes the parent of $v_i$ and $v_j$, 
and the set of orphan nodes is updated to ${\tt orphan} = {\tt orphan} \setminus \{v_i, v_j\}\cup\{v_l\}$.  
If there are orphan nodes remaining, we return to the matching step, and otherwise, we terminate. 

\subsection{Creating templates from a template-tree}
Let the samples \{$S_1, S_2, ..., S_N$\} belong to $M$ disjoint classes where $M$ may or may not be known. 
Then, the following three cases are considered to create templates from the template-tree.

{\em  (a) Class label of each sample is known:} A template-forest with $M$ disjoint trees is constructed where each tree includes samples belonging to the same class.
The roots of the $M$ trees represent the templates of the $M$ classes of samples.

{\em  (b) $M$ is known but the class label of each sample is unknown:} A single template-tree is created from $N$ samples.
Then, the tree is cut at a suitable height so that $M$ disjoint subtrees are produced.
The root of each subtree represents a template of the samples placed in the leaves of that subtree. 

{\em  (c) Both $M$ and the class labels of the samples are unknown:} A single template-tree is created from $N$ samples.
The templates are created from the roots of the well-separated branches of the template-tree such that within-class variations (heights of the subtrees) are small relative to the between-class variations (heights of the ancestors of the subtrees). 
In this context, the HM\&M algorithm is similar to the spirit of the hierarchical clustering algorithm UPGMA \cite{gan2007clusteringBook, jain1999data}, with significant differences in the distance computation and management of the internal nodes.
Assume that $M$ well-separated templates are produced from the template tree.
Then, $N$ samples are predicted to be generated from these $M$ classes.
This approach leads to an unsupervised classification of samples discussed in the next chapter.

\subsection{Computational complexity}
The HM\&M algorithm requires $O(N^2)$ dissimilarity computations and $O(N)$ merge operations for creating a template from a collection of $N$ samples.
Let $k$ be the maximum number of clusters or meta-clusters in any of the nodes of the template-tree.
Then a dissimilarity computation takes $O(k^3\log k)$ time whereas the merge operation takes $O(k)$ time when distribution parameters of the meta-clusters are computed by maximum likelihood estimation. 
Hence, the time complexity of the algorithm is $O(N^2k^3 \log k)$, which reduces to $O(N^2)$ for bounded $k$.

\section{The homogeneity of meta-clusters and templates}
\label{sec:homogeneity_mc}
A template summarizes the common features of a group of similar samples.
These common features are captured by a collection of meta-clusters (meta-populations) formed by combining cell clusters expressing similar phenotypes in different samples.
If a template correctly represents a collection of samples with the same biological status, each meta-cluster would contain a homogeneous collection of cell clusters with similar phenotypes.
However, there is no standard method for computing biological homogeneity of a collection of cell populations.
Therefore, I consider a statistical procedure, in a Multivariate Analysis of Variance (MANOVA) model, to evaluate the homogeneity of meta-clusters. 
For simplicity, I first discuss the homogeneity of one-dimensional meta-clusters, followed by the multi-dimensional case.


 
\subsection{Analyzing homogeneity of a one-dimensional  meta-cluster}
\label{sec:homogeneity_one_dimension}
Let $C$ be a meta-cluster consisting of $k$ one-dimensional (1-D) clusters $c_1,c_2,..c_k$ defined within a single marker/channel.
A 1-D cluster usually represents a distribution of cells with a marker either being expressed or not being expressed.
For example, CD3$^+$ and CD3$^-$ represent two 1-D clusters in CD3 channel, with the former expressing high levels of  CD3 marker and the latter expressing low levels of it. 
CD3 is a marker of T lymphocytes, and therefore, a meta-cluster denoting T cells is biologically homogeneous if it contains only CD3$^+$ cell clusters.

Let the $i^{th}$ cluster $c_i \in C$ contain $n_i$ cells and be normally distributed with mean $\mu_i$ and variance $\sigma_i^2$.
The normality assumption was justified in Chapter~\ref{chap:variance_stabilization}, where I showed that cell clusters usually follow normal distribution after stabilizing their variances.
In spite of variance stabilization, I still use $\sigma_i^2$ to denote the variance of the $i^{th}$ cluster (instead of using a common variance) to emphasize that the variances can not be made exactly equal by the variance stabilization process.
The mean and variance of each cluster are estimated by the E-M algorithm (alternatively, with an unbiased maximum likelihood estimator) after the clusters are identified as was discussed in Chapter~\ref{chap:clustering}.

Let the entire meta-cluster $C$ contain $n$ cells in total with mean $\mu$. 
The separation of a cluster $c_i$ from the meta-cluster $C$ can be approximated by the squared deviation of their centers $(\mu_i - \mu)^2$.
In behavioral and biological sciences, $(\mu_i - \mu)^2$ is generally called the ``effect-size"  because it indicates the effect of the $i^{th}$ treatment~\cite{cohen1988book}.
In the meta-clustering context, it denotes the ``cluster separation" of the $i^{th}$ cluster from the meta-cluster center.
The between-cluster variation $\sigma_b^2$ of a meta-cluster is then computed with the average separation of the clusters: 
\begin{equation}
\sigma^2_b = \frac 1 {n-k} \sum_{i=1}^k (n_i-1)(\mu_i - \mu)^2.
\end{equation}
The within-cluster variance, $\sigma_w^2$, is estimated by a weighted average of all individual cluster variances (pooled variance): 
\begin{equation}
\sigma_w^2=\frac{1}{n-k}\sum_{i=1}^k {(n_i-1)\: \sigma^2_i}.
\label{eq:sigma_w}
\end{equation}
I calculate the ``Relative Cluster Separation" ($\phi$) after dividing $\sigma^2_b$ by $\sigma_w^2$ and taking a square root of the ratio:
\begin{equation}
\phi = \frac{\sigma_b} {\sigma_w}.
\label{eq:phi}
\end{equation}
Note that the square of the above ratio, $\phi^2$, is also known as the signal-to-noise ratio (SNR) \cite{fleishman1980confidence} or Cohen's $f^2$ statistics \cite{cohen1988book}.
$\phi$ is a dimensionless number and can take values between zero (when the clusters collapse into a single point) and an indefinitely large number (when the clusters become further apart relative to $\sigma_w$).
For example, $\phi=0.1$ conveys that the average separation among clusters ($\sigma_b$) is one-tenth of their (pooled) standard deviation ($\sigma_w$), indicating a significant overlap (homogeneity) among the clusters.
In contrasts, $\phi=2$ conveys a relatively inhomogeneous group of clusters because the average between-cluster separation is twice as large as the within-cluster standard deviation.
A meta-cluster is considered relatively homogeneous when $\phi$ is small relative to a predefined threshold.
For FC data, this threshold can be set to one, i.e., a collection of clusters is relatively homogeneous when $\phi<1$, and otherwise, inhomogeneous.
I justify this choice of threshold in Section \ref{sec:template_results_HD} with a healthy dataset. 
A guideline, although controversial, for setting the threshold at three levels --  $.1$ for small, $.25$ for medium and, $.4$ for large effect -- is suggested by Cohen for behavioral studies \cite{cohen1988book}.
However, I found these levels too small for relatively large cell clusters observed in FC samples.

I compute the uncertainty associated with $\phi$ by computing the confidence interval around it.
Assume, for simplicity, that the clusters within a meta-cluster have equal sizes $n'$ ( here, $n'=n/k$, where $n$ is the total size of the meta-cluster and $k$ is the number of cluster in the meta-cluster).
Then the confidence interval for $\phi$ is calculated from the confidence interval of the non-central F distribution whose non-centrality parameter $\lambda$ is calculated with:
\begin{equation}
\lambda = \frac {n'\sigma^2_b}{\sigma^2_w} = n'\phi^2.
\end{equation}
Let $[\lambda_L, \lambda_U]$ be the $100(1-\alpha)\%$ confidence interval for $\lambda$, then the following equations hold:
\begin{equation}
p[\lambda_L \leq \lambda \leq \lambda_U]=1-\alpha
\end{equation}
\begin{equation}
p\left[\sqrt{\frac{\lambda_L}{n'}} \leq \phi \leq\sqrt{\frac{\lambda_U}{n'}}\right]=1-\alpha.
\end{equation}
Hence, the confidence interval $[\phi_L, \phi_U]$ for $\phi$ is calculated with $\left[\sqrt{\frac{\lambda_L}{n'}} , \sqrt{\frac{\lambda_U}{n'}} \right]$ \cite{steiger2004beyond, kelley2007confidence}.

Note that I employ $\phi$ to evaluate the homogeneity of a meta-cluster obtained from any meta-clustering algorithm.
However, a meta-clustering algorithm can directly minimize $\phi$ and obtains the ``optimum" homogeneity of meta-clusters.
This approach is very similar to Fisher's linear discriminant analysis that maximizes the separation between clusters relative to within-cluster variations \cite{bishop2006pattern}.
I leave this optimization approach as a future work and will not discuss it any further.

\subsection{Comparison with null hypothesis based significance testing}
Consider a meta-cluster consisting of $k$ cell clusters, where the $i^{th}$ cluster contains $n_i$ cell with mean $\mu_i$.
To evaluate the homogeneity of this meta-cluster in an Analysis of Variance (ANOVA) model, we can consider a null hypothesis of no difference among cluster centers:
\begin{equation}
H_0: \mu_1=\mu_2=\cdots =\mu_k.
\end{equation}
The normality and homoskedasticity assumptions required for the ANOVA model are approximately satisfied by the variance stabilization step discussed in Chapter \ref{chap:variance_stabilization}. 
The validity of the above null hypothesis can be tested by an F-test that measures the following F-statistics:
\begin{equation}
F_{obs} = \frac{\frac{1} {k-1} {\sum \limits_{i=1}^k n_i(\mu_i - \mu)^2}} {\sigma_w^2}, 
\label{eq:F_obs}
\end{equation}
where, $\mu$ is the mean of the entire meta-cluster and $\sigma_w^2$ is the pooled variance of the clusters as defined in Eq.~\ref{eq:sigma_w}.
The $p$-value of the F-test is computed with the probability of obtaining a test statistic at least as extreme as
$F_{obs}$ assuming that the null hypothesis is true, i.e., we compute $P(F\geq F_{obs})$, where $F$ is a random variable following F-distribution with $k-1$ and $n-k$ degrees of freedom.
In a meta-cluster analysis, $p$-value evaluates whether the difference among cluster locations can occur by chance.
The null hypothesis $H_0$ is rejected when the $p$-value is less than a certain significance level (often 0.05 or 0.01). 
Hence, we can declare a meta-cluster to be inhomogeneous ($H_0$ is rejected) when the $p$-value of the F-test is less the significance level, and otherwise, declare it to be homogeneous ($H_0$ in not rejected).

The F-test depends on the size of clusters because the cluster size ($n_i$) appears on the numerator of $F_{obs}$ in Eq.~\ref{eq:F_obs}.
In the Null Hypothesis based Significance Testing (NHST) framework, large clusters produce high $F_{obs}$ (low $p$-value of the F-test) and therefore, tend to declare a meta-cluster to be inhomogeneous even though the average cluster-separation is relatively small.
As a matter of fact, given sufficiently large cluster sizes, an F-test always rejects $H_0$ unless the locations of the clusters collapse at a single point, which is extremely unlikely in practice.
In flow cytometry, the number of cells in a cluster is typically large, often in the range of tens of thousands cells.
Hence, NHST has a high probability of making a Type I error, even at a very low alpha such as at .001, and incorrectly declares a meta-cluster inhomogeneous when the cluster sizes are large.
In contrast to the F-test, the relative cluster separation ($\phi$) does not depend on the cluster sizes and is therefore more applicable to meta-cluster analysis in flow cytometry.

To see the relationship between $\phi$ and the paired t-test, consider a meta-cluster consisting of two equal-size ($n'$) clusters, with means $\mu_1$ and $\mu_2$ and a common variance $\sigma^2$.
In this case, Eq.~\ref{eq:phi} simplifies to 
\begin{equation}
\phi = \frac {\frac 1 2 |\mu_1-\mu_2|} {\sigma} = \frac d 2,
\label{}
\end{equation}
where $d$ is the standardized mean difference between two clusters, also known as Cohen's $d$~\cite{cohen1988book}.
In this settings, a two-group t-test statistics ($t_{obs}$) is computed by
\begin{equation}
t_{obs} = \sqrt{\frac {n'} 2} \cdot  \frac{\displaystyle{\mu_1 - \mu_2}}{\displaystyle \sigma}.
\label{eq:t_obs}
\end{equation}
Similar to $F_{obs}$ in Eq.~\ref{eq:F_obs}, $t_{obs}$ depends on the cluster sizes.
Consequently, the t-test can also report a small difference to be highly significant when the cluster sizes are large relative to the within-cluster variance.
Therefore, $\phi$ can also be used to assess the separation of a pair of clusters instead of traditional t-test.

\subsection{Analyzing homogeneity of high-dimensional meta-clusters}
\label{sec:manova}
I analyze the homogeneity of high-dimensional meta-clusters in a fixed-effect Multivariate Analysis of Variance (MANOVA) model~\cite{johnson2002applied}.
As before, let $C$ be a $p$-dimensional meta-cluster consisting of $k$ clusters, $c_1,c_2,..c_k$, with the $i^{th}$ cluster $c_i$ containing $n_i$ cells and the size of the entire meta-cluster to be $n=\sum_{i=1}^k n_i$.
Assume that the $i^{th}$ cluster  $c_i$ is modeled by a multivariate normal distribution with a $p$-dimensional mean vector $\boldsymbol{\mu}_i$ and a $p\times p$ covariance matrix $\Sigma_i$.
The mean vector $\boldsymbol{\mu}$ of the entire meta-cluster $C$ is estimated by $\frac 1 n\sum_{i=1}^{k}{n_i\boldsymbol{\mu}_i}$. 
Similar to the one-dimensional setting, I compute the within-cluster covariance matrix, $\Sigma_w$, and the between-cluster covariance matrix, $\Sigma_b$, as follows:
\begin{equation}
\Sigma_{w} = \frac {1} {n-k} \sum_{i=1}^{k} (n_i-1)\Sigma_i
\end{equation}
\begin{equation}
\Sigma_{b} =  \frac 1 {n-k} \sum_{i=1}^{k} (n_i-1)(\boldsymbol{\mu}_i - \boldsymbol{\mu}) (\boldsymbol{\mu}_i - \boldsymbol{\mu})^\intercal .
\end{equation}
The relative cluster separation, $\phi$, in multidimensional setting is then computed with:
 \begin{equation}
\phi = \sqrt{ \frac{1}{p} {\tt trace}(\Sigma_{w}^{-1} \Sigma_{b}) }.
\label{}
\end{equation}
In MANOVA, ${\tt trace}(\Sigma_{w}^{-1} \Sigma_{b})$ is known as  the Lawley-Hotelling statistics~\cite{hotelling1931generalization, pillai1959hotelling, johnson2002applied} and is computed by the summation of the eigenvalues of $\Sigma_{w}^{-1} \Sigma_{b}$.
I divide the Lawley-Hotelling statistics by $p$ so that $\phi$ becomes independent of the dimensionality and take a square root of it to keep it in the original unit of measurement. 
Similar to the univariate case,  $\phi < 1$ denotes that the between-cluster variation is less than naturally occurring within-cluster variation, and therefore, the meta-cluster can be considered homogeneous.
As pointed out by Ito et al. ~\cite[p. 75-78]{ito1964robustness}, $\phi$ is a robust estimator of the homogeneity of meta-clusters even in the presence of limited heteroskedasticity of covariances in large clusters.
Therefore, this measure can be applied to analyze homogeneity of heteroskedastic clusters as well.

In the multi-dimensional setting, computing an exact confidence interval around $\phi$  is difficult because the exact distribution of the Lawley-Hotelling statistics is not known for general $k$ and $p$.
According to the discussion in \cite{hughes1972approximating, pillai1959hotelling, steyn2009estimating}, I approximate
the distribution with a non-central high-dimensional F distribution with degrees of freedom $a$ and $b$ and the non-centrality parameter, $\lambda$, where $a=p(k-1)$, $b=p(n-k-p-1)+2$, and $\lambda = \phi^2 (a+b+1)$.
The confidence interval $[\phi_L, \phi_U]$ around $\phi$ is therefore computed with $[\sqrt{\frac{\lambda_L}{a+b+1}}, \sqrt{\frac{\lambda_U}{a+b+1}}]$, where $[\lambda_L, \lambda_U]$ is the confidence interval of the non-centrality parameter, $\lambda$.
I calculate the confidence interval for $\lambda$ by using the R package MBESS \cite{kelley2007methods}.
Note that $T^2_g = (n-k) {\tt trace}(\Sigma_{w}^{-1} \Sigma_{b})$ is Hotelling's generalized $T^2$ statistics~\cite{hotelling1931generalization, pillai1959hotelling, johnson2002applied} and is equivalent to the F statistics in the univariate case.
Similar to the univariate case, $T^2_g$ also depends on cluster sizes and has a high probability of making a Type I error by declaring trivial differences as statistically significant when cluster sizes are large.

\section{Results}
\label{sec:template_results}
\subsection{Creating templates from healthy samples }
\label{sec:template_results_HD}
{\em  Data description:}
The healthy donor (HD) dataset consists of 65 samples from five healthy individuals who donated blood on different days. 
Each sample was divided into five replicates and each replicate was stained using labeled antibodies against CD45, CD3, CD4, CD8, and CD19 protein markers.
Section~\ref{sec:hd_data_description} provides a detail description about this dataset. 
Each sample of the HD dataset was preprocessed, variance stabilized, and clustered to identify cell populations (see Fig.~\ref{fig:clustering_hd}). 

\begin{figure}[!t]
   \centering
   \includegraphics[scale=.6]{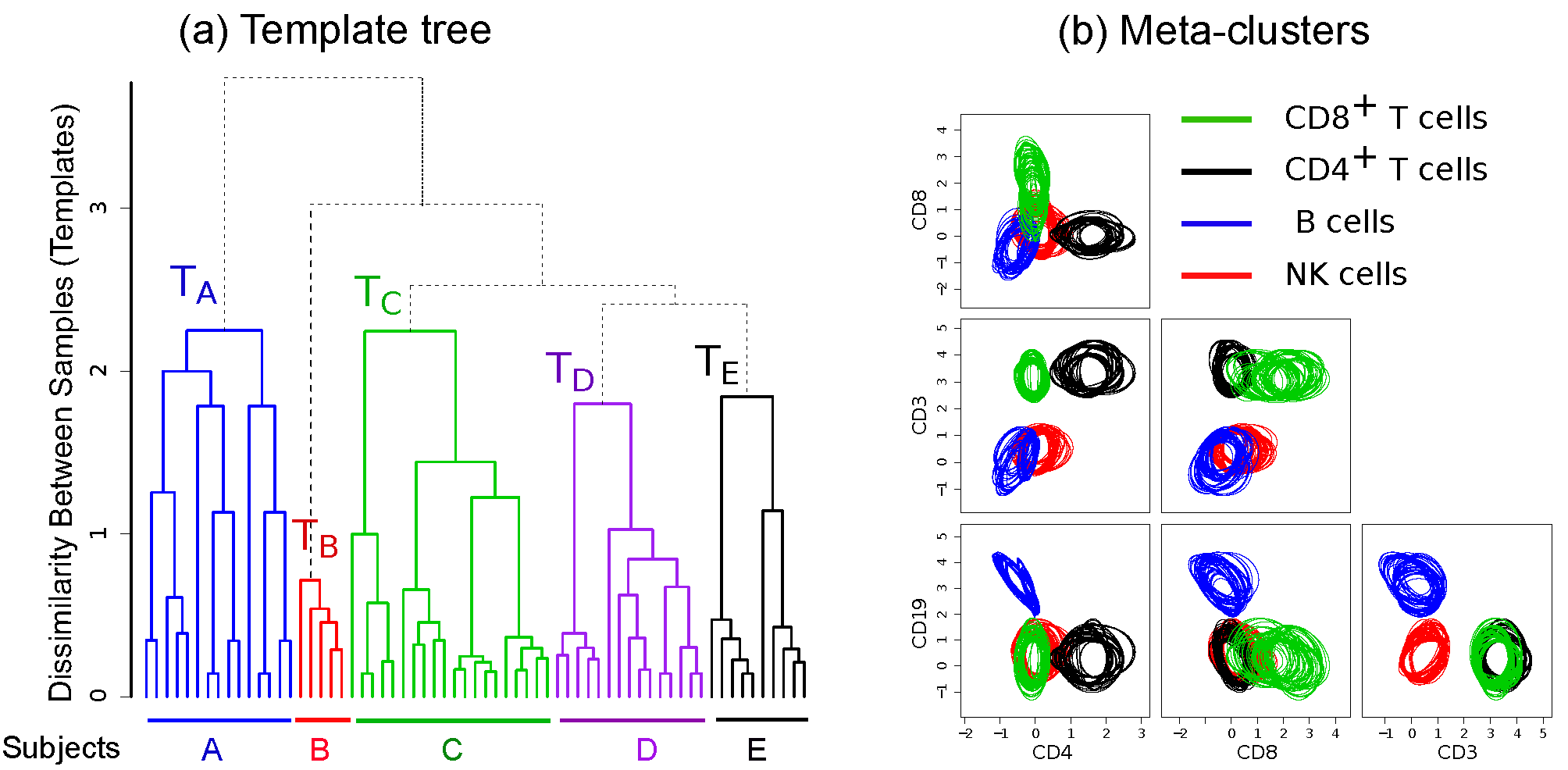} 
   \caption[The template-tree created from all samples of the HD dataset]{(a) The template-tree created by the HM\&M algorithm from all samples of the HD dataset. Leaves of the dendrogram denote samples from five healthy individuals. An internal node represents a template and the height of an internal node measures the dissimilarity between its left and right children. The sample-specific subtrees are drawn in different colors.  
   (b) Bivariate projections of the combined healthy template (the root of the tree in Subfig. (a)) are drawn in terms of the meta-clusters. Here, each meta-cluster is represented by a homogeneous collection of  cell clusters that are drawn with the 95th quantile contour lines. Clusters participating in a meta-cluster are drawn in same color. 
   }
    \vspace{-.5 cm}
   \label{fig:template_hd}
\end{figure}

{\em  Creating healthy templates:}
Figure~\ref{fig:template_hd}(a) shows the template-tree created from the HD dataset.
A leaf node of the tree denotes a healthy sample, and an internal node represents a template created from samples placed in the underlying subtree.
The height of an internal node measures the dissimilarity between its left and right children.
We observe that 65 samples from the five healthy subjects are organized into five well-separated branches shown in five different colors in Fig.~\ref{fig:template_hd}(a).
From the roots of these five subtrees, we can construct five subject-specific templates, e.g., $T_{A}$ represents a template created from 15 samples from subject A .   
In this dataset, the variations within a subject-specific template arise from the environmental impact on individual immune system on different days  and the technical variation in flow cytometry sample preparation and measurement. 
By contrast, the between-subject variations arise from the natural biological variations among healthy subjects. 
In this dataset, we observe more natural between-subject variations than the temporal and instrumental variations.
Hence, samples from the five subjects create concise and well separated templates representing immune profiles for different healthy individuals. 

The HD dataset includes three sources of variations -- the technical, day-to-day within-subject, and between-subject variations.
The template-tree captures these multi-level variations with different internal nodes that can be used as multi-layer templates. 
In the lower level, a day specific template is constructed from five replicates of a sample collected on a particular day from a subject.
In the middle level, samples from a subject collected on different days create a subject-specific template (the roots of colored subtrees in Fig.~\ref{fig:template_hd}(a)). 
Finally, in the top level, the root of the whole tree represents a combined template representing a healthy immune profile of these five subjects. 
This multi-level construction is especially useful to design a multi-level sample classification framework, which is the topic of the next chapter. 

{\em  Meta-clusters in the healthy template:}
The template created from the HD dataset (the root of the tree in Fig.~\ref{fig:template_hd}(a)) represents a healthy immune profile of the five subjects.
This template consists of four meta-clusters denoting four biologically distinct cell sub-types within lymphocytes (we have isolated lymphocytes on the scatter channels in the pre-processing steps). 
Each meta-cluster is a collection of biologically equivalent cell populations from different samples.
Figure \ref{fig:template_hd}(b) shows 2-D projections of these meta-clusters in terms of the participating clusters.
The 95th quantiles of the clusters within each meta-cluster are shown in same color denoting (1) green:  CD8$^+$ T cells (CD45$^+$CD3$^+$CD8$^+$), (2) black: CD4$^+$ T cells (CD45$^+$CD3$^+$CD4$^+$), (3) blue: B cells (CD45$^+$CD3$^-$CD19$^+$), and (4) red: natural killer cells (CD45$^+$CD3$^-$CD19$^-$).
These meta-clusters represents core population pattern of healthy lymphocytes excluding the subject-specific variations.

 \begin{figure}[!t]
   \centering
    \includegraphics[scale=.42]{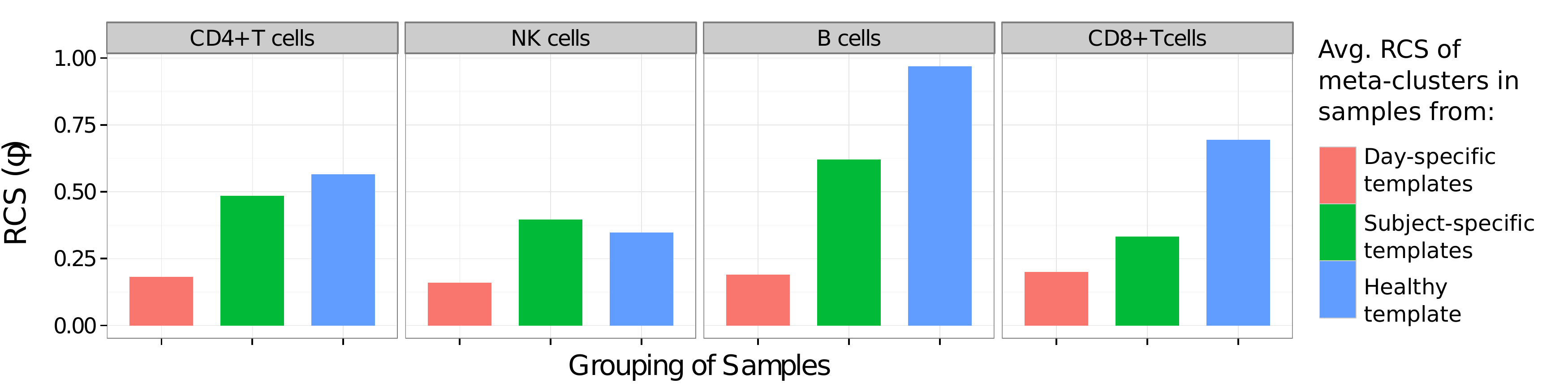}
    \vspace{-.3 cm}
   \caption[The homogeneity of meta-clusters for different groupings of samples in the HD dataset]{The homogeneity (Relative Cluster Separation, $\phi$) of four healthy meta-clusters shown in different panels.
The meta-clusters are created from three different groupings of the healthy samples shown in different colors.} 
   \vspace{-.5 cm}
    \label{fig:SCS_bar_hd}
\end{figure}

{\em  Homogeneity of meta-clusters:}
How homogeneous are the meta-clusters contained in different templates of the HD dataset?
Here, I discuss the homogeneity of three levels of templates: (a) the \emph{healthy template} created from all 65 samples, (b) 5 \emph{subject-specific templates}, each of which is created from samples from a subject, and (c) 13 \emph{day-specific templates}, each of which is created from five replicates of a sample collected on the same day from a subject.
Every template consists of four meta-clusters similar to the healthy template described in Fig.~\ref{fig:template_hd}(b).
We can evaluate the homogeneity of the high-dimensional meta-clusters in each template with the Relative Cluster Separation (RCS), $\phi$, as discussed in Section~\ref{sec:manova}.

Fig.~\ref{fig:SCS_bar_hd} shows the RCS ($\phi$) of four meta-clusters (different panels) denoting four subtypes of cells across three groups of templates (different colors).
For the subject-specific and day-specific templates, I computed the the average RCS of the meta-clusters in each group of templates.
Recall that a high value of $\phi$ indicates low homogeneity (high inhomogeneity) of a meta-cluster and vice versa. 
We observe, in each panel of Fig.~\ref{fig:SCS_bar_hd}, that a meta-cluster becomes increasingly inhomogeneous (i.e., $\phi$ decreases) as we increase variations from different sources.
This is expected because variations in data (even though from natural sources) reduce the uniformity of the cell populations.
However, we do not expect any biologically significant variation within a meta-cluster (denoting a cell-type) because the observed variations are originated from natural sources among healthy individual.
The highest value of $\phi$ is less than one for any meta-cluster in Fig.~\ref{fig:SCS_bar_hd}.
Therefore, we consider a meta-cluster to be biologically homogeneous when $\phi<1$.
That is, a meta-cluster is homogeneous when the between-cluster variation is less then the pooled within-cluster variation.
This assertion is satisfied by the healthy meta-clusters, as expected, because all samples are from the same biological status.

The threshold for $\phi$ can be adjusted for a biological experiment depending on the natural variability expected in the experimental condition.
Therefore, it is advisable to run a pilot study similar to the HD dataset to determine baseline/trivial homogeneity measurements, which can subsequently be used to assess biologically significant variations across different classes of samples.

\subsection{Creating templates from the TCP dataset}
{\em  Data description:}
T cell phosphorylation (TCP) dataset was originally generated by Maier et al.~\cite{maier2007allelic} to study the effect of phosphorylation upon stimulating blood with anti-CD3 antibody.
The abundance of four protein markers (CD4, CD45RA, SLP-76, and ZAP-70) was measured before and five minutes after stimulating whole blood with an anti-CD3 antibody~\cite{maier2007allelic}.
I reanalyzed 29 pairs of samples of this dataset to demonstrated the creation of templates and the detection of general effect of stimulation at the meta-cluster level.
Section~\ref{sec:tcp_data_description} provides a detail description about this dataset. 
Each of the 58 samples in the TCP dataset was preprocessed and clustered independently to identify cell populations (Section~\ref{sec:clustering_tcp}). 

{\em  General effects of stimulation on phosphorylation responses:}
During the stimulation, anti-CD3 antibody binds with T cell receptors (TCR) and activates the T cells, initiating the adaptive immune response. 
The binding with TCR induces  dramatic changes in  gene expression and cell morphology, 
and induces the formation of a phosphorylation-dependent signaling network via multi-protein complexes. 
ZAP-70 is a kinase that phosphorylates tyrosine in a trans-membrane protein called LAT,  and LAT and SLP-76 are part of a platform that assembles the signaling proteins~\cite{Brockmeyer+:phosphorylation}.  

Previous studies have shown that different T cell subsets (naive, memory, effector) display different phosphorylation responses upon stimulation~\cite{maier2007allelic, farber1997differential, ahmadzadeh2001heterogeneity, ahmadzadeh1999effector}. 
In these studies, each sample was gated to identify cell populations of interest, and each pair of samples -- before and after stimulation -- were compared to detect the phosphorylation responses. 
However, Maier et al.~\cite{maier2007allelic} reported that the autoimmune disease-associated allele at CTLA4 gene on chromosome 2q33 alters phosphorylation responses in naive and memory T cells. 
Thus, depending on their genetic profiles, different subjects might display different phosphorylation responses upon stimulation. 
It is therefore challenging to summarize the general effect of phosphorylation from observing the stimulation effect in individual samples. 
An alternative and robust approach is to create a pre-stimulation and a post-stimulation template. 
By matching meta-clusters across these templates we can better assess the population-specific effects of the stimulation experiment.

\begin{figure}[!t]
   \centering

\subfigure[CD4$^+$ CD45RA$^{low}$ memory T cells]{
    \includegraphics[scale=.4]{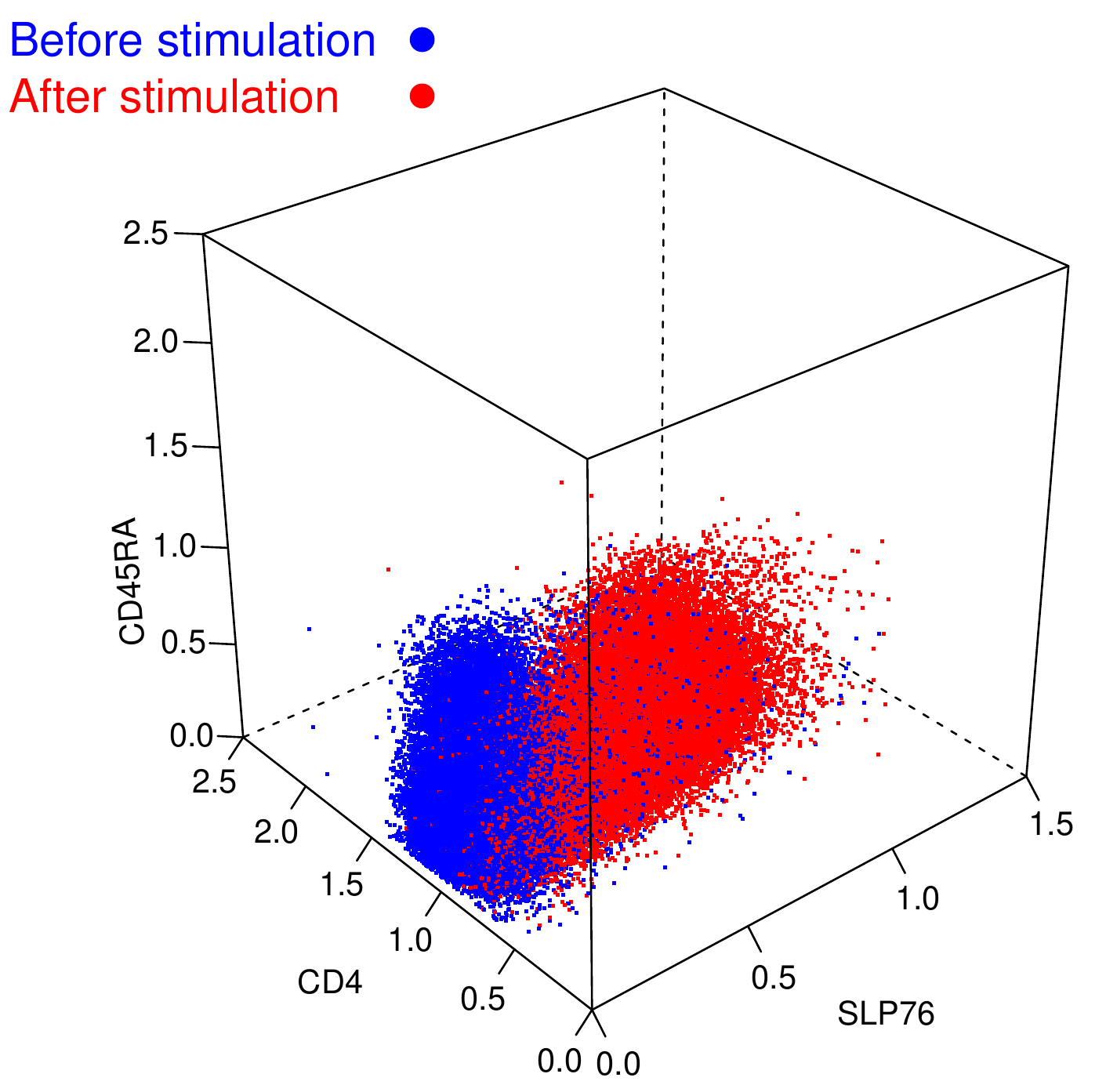}

   \label{fig:memory} 
	}
	\hspace{1.5 cm}
\subfigure[CD4$^+$ CD45RA$^{high}$ naive T cells]{
    \includegraphics[scale=.4]{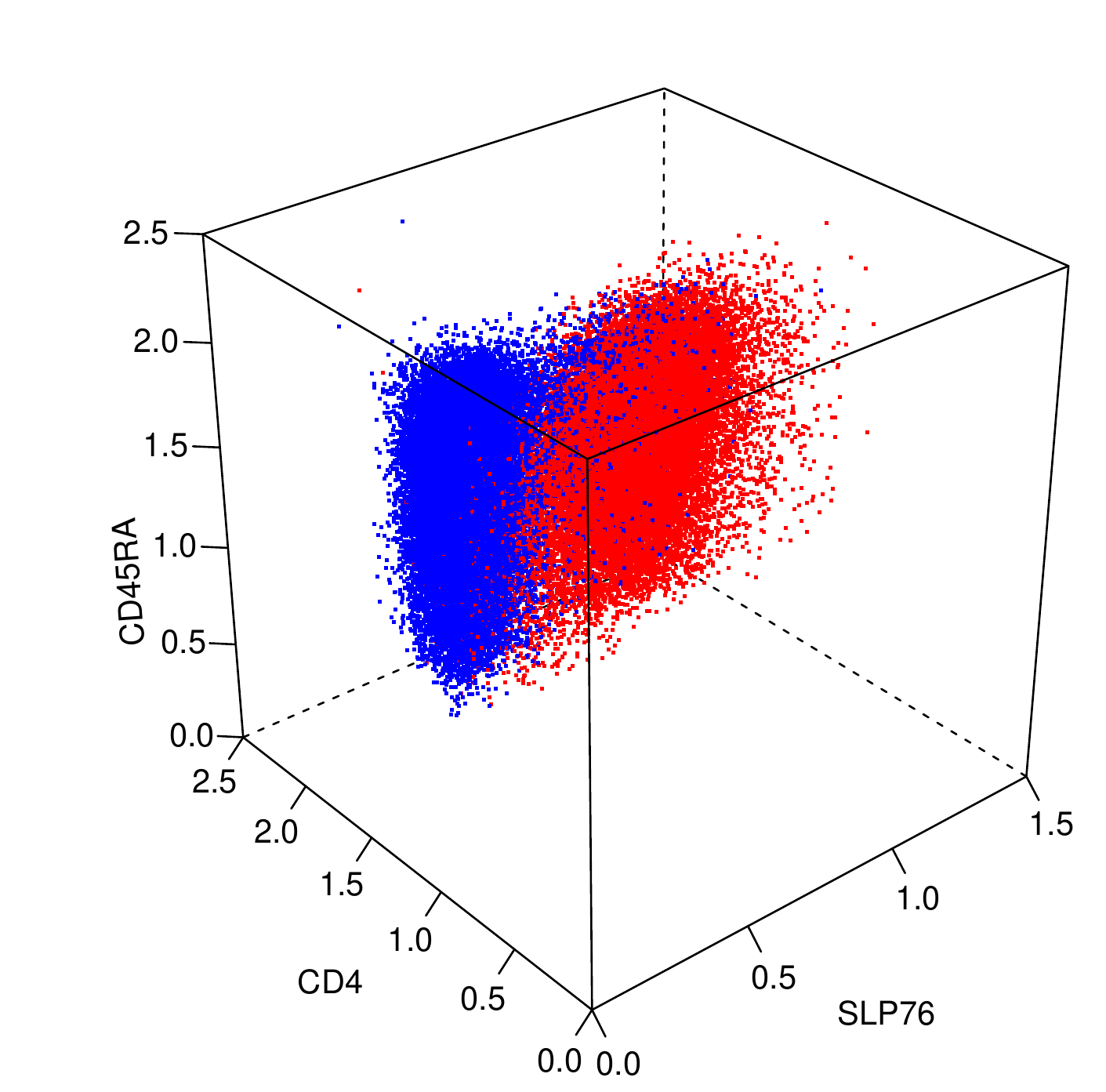}

    \label{fig:naive}	
    }
   \subfigure[CD4$^-$ CD45RA$^{high}$ cells]{
    \includegraphics[scale=.4]{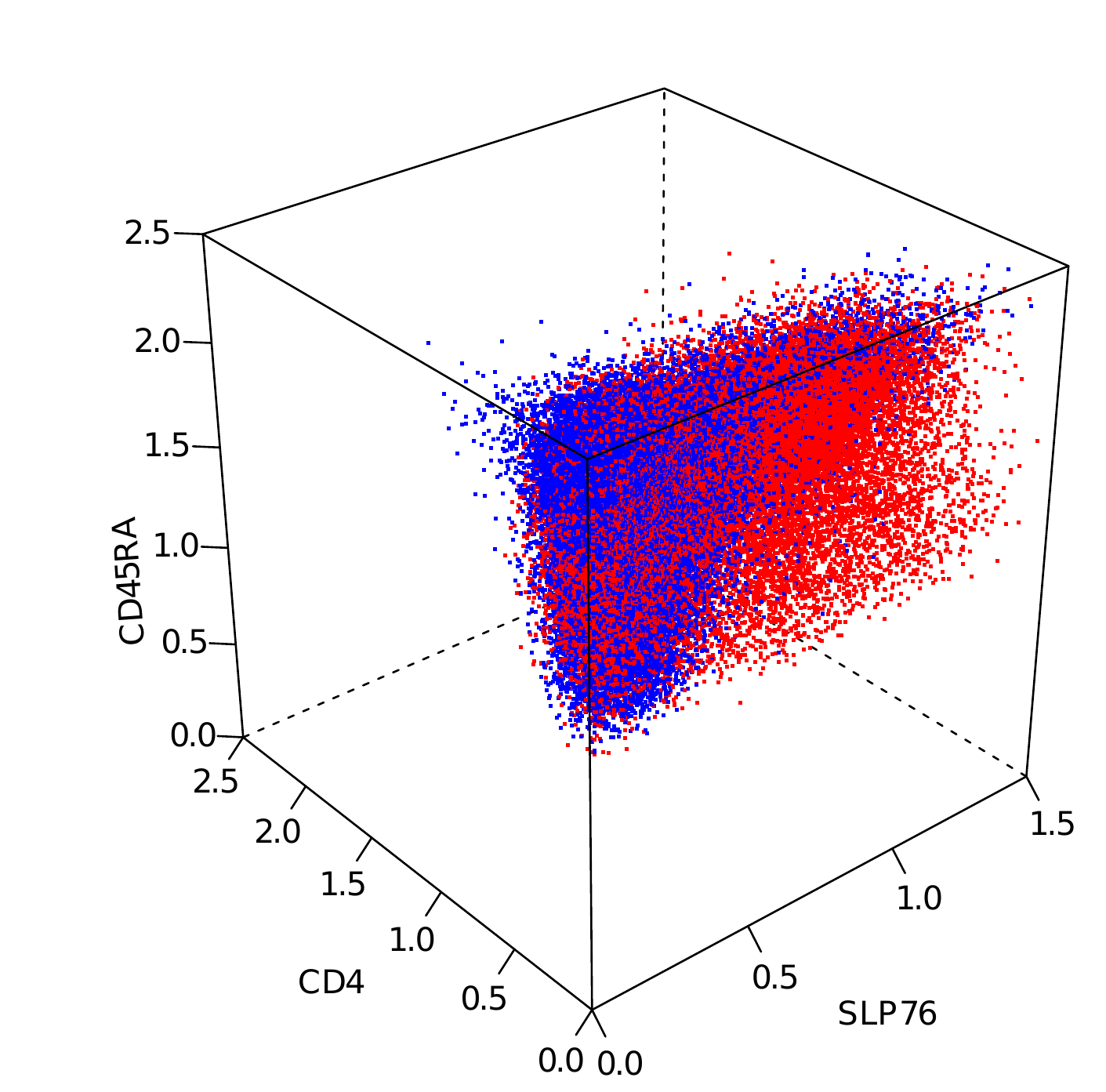}

   \label{fig_appendix:mc3_phospho}
	}
	\hspace{1.5 cm}
\subfigure[CD4$^-$ CD45RA$^{low}$ cells]{
    \includegraphics[scale=.4]{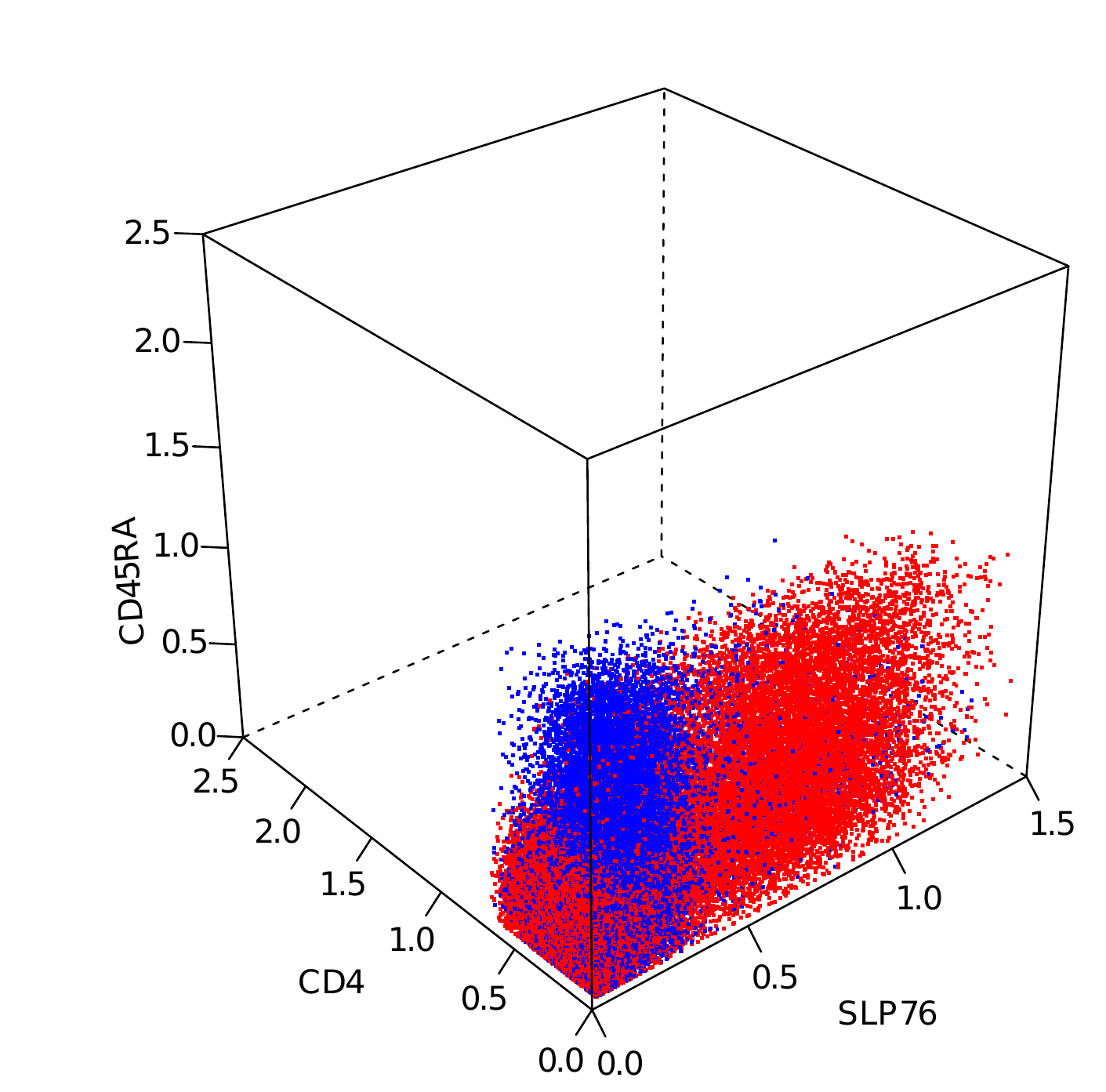}

    \label{fig_appendix:mc4_phospho}	
  }

  \label{fig:mc_TCP}
  \vspace{-.2 cm}
  \caption[Meta-clusters contained in the pre- and post- stimulation templates created from the TCP dataset]{Three-dimensional projections of the meta-clusters created from the TCP dataset: Pre- and post-stimulation meta-clusters are shown in blue and red, respectively. The four meta-clusters represent:  (a) CD4$^+$CD45RA$^{low}$ memory T cells, (b) CD4$^+$CD45RA$^{high}$ naive T cells, (c) CD4$^-$CD45RA$^{high}$ cells, and (d) CD4$^-$CD45RA$^{low}$ cells.}
    \vspace{-.5 cm}
 \end{figure}

{\em  Creating pre- and post-stimulation templates:}
 By applying the HM\&M algorithm, I create a  \emph{pre-stimulation template} from 29 samples before stimulation and a \emph{post-stimulation template} from 29 samples after the stimulation.
Both templates consist of four meta-clusters denoting: (a) CD4$^+$CD45RA$^{low}$ memory T cells, (b) CD4$^+$CD45RA$^{high}$ naive T cells, (c) CD4$^-$CD45RA$^{high}$ cells, and (d) CD4$^-$CD45RA$^{low}$ cells.
Fig.~\ref{fig:mc_TCP} shows three-dimensional projections of these meta-clusters: pre-stimulation meta-clusters in blue and post-stimulation meta-clusters in red. 
We observe an increased levels of SLP-76 protein in every meta-cluster after stimulation than its pre-stimulation state.
However, the increment is more clear in naive and memory T cells, as was observed in earlier publications~\cite{farber1997differential, ahmadzadeh1999effector, ahmadzadeh2001heterogeneity, maier2007allelic}.
Similar phosphorylation effect, although smaller than SLP-76, was observed for ZAP-70 protein as well.
Therefore, two templates can concisely represent stimulation states of 58 samples in terms of the generic meta-clusters, and comparing meta-clusters across templates evaluates the overall phosphorylation shifts across conditions.

{\em  Homogeneity of meta-clusters:}
How homogeneous are the pre- and post-stimulation meta-clusters?
To answer the question more rigorously, I create a \emph{combined template} from all 58 samples in the TCP dataset.
Table~\ref{tab:phi_TCP} shows the Relative Cluster Separation (RCS, $\phi$) with the 95\% confidence intervals for every four-dimensional meta-cluster from three templates (pre-stim, post-stim, and combined).
The combined meta-clusters in Table~\ref{tab:phi_TCP} are less homogeneous (i.e., larger values of $\phi$) than the pre- and post-stimulation meta-clusters.
As expected, the homogeneity of meta-clusters decreases with the increase of variations in the TCP dataset, as was also observed in the HD dataset.
In particular, the homogeneity of meta-clusters denoting the naive and memory T cells decreases by a factor of two (twofold increment of $\phi$) in the combined template.
For these two meta-clusters, the values of RCS are greater than one ($\phi > 1$) in the combined template indicating higher between-cluster variance than the within-cluster variance.
Therefore, the combined template is biologically inhomogeneous based on the homogeneity threshold ($\phi=1$) discussed earlier for the HD dataset. 
The inhomogeneity in the combined template implies that the combined template is created from a heterogeneous collection of samples from more than one biological classes, and one should not build a single template from these collection of samples.
Therefore, the RCS statistics provides an objective way of constructing and evaluating templates and prevents us from building templates from inhomogeneous collection of samples from multiple biologically distinct classes (such as the pre- and post-stimulation samples in the TCP dataset).

\begin{table}[!t]
   \centering
        \vspace{-.7 cm}
          \caption[The homogeneity of meta-clusters contained in the templates created from the TCP dataset]{Relative Cluster Separation ($\phi$) with 95\% confidence intervals (CI) for each of the four meta-clusters in the TCP dataset. Meta-clusters are computed from  (a) all 58 samples across before and after stimulation, (b) 29 samples before stimulation, and (c) 29 samples after stimulation.}
   \begin{tabular}{@{} cccc @{}} 
      \toprule

      Meta-cluster &  \multicolumn{3}{c}{$\phi$ [95\% CI]} \\
      \cline{2-4}
             & Before stimulation &  After stimulation &  Combined \\
           \midrule
         CD4$^+$CD45RA$^{low}$ & 0.541 [0.537, 0.545] & 0.495 [0.491, 0.499] & 1.023 [1.018, 1.026] \\
         CD4$^+$CD45RA$^{high}$ & 0.647 [0.642, 0.651] & 0.545 [0.540, 0.550]  & 1.056 [1.051, 1.060] \\
         CD4$^-$CD45RA$^{low}$  & 0.328 [0.324, 0.332] & 0.291 [0.287, 0.294] & 0.476 [0.472, 0.478] \\
         CD4$^-$CD45RA$^{high}$ & 0.565 [0.561, 0.568] & 0.364 [0.360, 0.367] & 0.400 [0.397, 0.403]  \\
         
       \bottomrule
   \end{tabular}
   \label{tab:phi_TCP}
\end{table}

\section{Conclusions and future work}
\label{sec:template_conclusions}
Templates and meta-clusters can succinctly summarize a large collection of samples belonging to a few biological classes.
With a collection of homogeneous meta-clusters, a template provides a concise description of a biological condition  and reduces the complexity of data by emphasizing the key characteristics of a class while masking statistical noises and low-level details.
I discuss a hierarchical matching-and-merging (HM\&M) algorithm for constructing templates that describe distinct biological states available in a collection of samples.
The HM\&M algorithm can operate both in supervised (class labels of samples are known) and unsupervised settings (class labels of samples are not known).
Templates created in an unsupervised setting can lead to a robust classification scheme, as will be discussed in the next chapter.

A meta-cluster within a template performs as a blueprint for a particular type of cells and represents the generic cell population with a statistical description while omitting sample-specific variations.
Meta-clusters can be employed for measuring  general effect on cell populations as biological conditions changes across templates.
As an example, the memory and naive T cells upon stimulated by anti-CD3 antibody display increased levels of SLP76 and ZAP70 proteins, an indication of increased phosphorylation in these T cell subpopulations.
Despite sample-specific variations, the T cell subpopulations from 29 pre-stimulation samples can be encapsulated with a set of homogeneous meta-clusters.
Therefore, we can evaluate the  population-specific effect of the stimulation by registering these meta-clusters with their corresponding meta-clusters in a post-stimulation template.

I discuss a statistical framework, in an multivariate analysis of variance (MANOVA) model, for  evaluating the homogeneity of meta-clusters.
For a collection of variance-stabilized clusters within a meta-cluster, this approach computes the ratio of  between-cluster to within-cluster variance and uses the ratio to evaluate the homogeneity of the meta-cluster.
Based on the natural variations among healthy individuals, a meta-cluster can be declared relatively homogeneous when the ratio of between- to within-cluster variance is less than one.

In continuing work, I plan to investigate the use of networks instead of trees to organize the templates, similar in spirit to the use of networks rather than trees in phylogenetics~\cite{Dress+:book}. 
Another issue is that the combinatorial dissimilarity measure between two samples is not a metric, and when the dissimilarity is extended to two templates, this value does not monotonically increase in the hierarchical matching and merging algorithm. 
I plan to investigate other dissimilarity measures in this regard.

%% file: ch6-template-based-classification.tex
\chapter{Classifying FC samples based on templates}
\label{chap:template-based-classification}

\section{Introduction}
Consider a  collection of flow cytometry samples, each sample consisting  of fluorescence measurements of protein markers made at the single-cell level from  hundreds of thousands of cells, indicating  different cell types present in each  sample. 
I describe an algorithm to dynamically classify such  samples into several classes based on phenotypically distinct templates. 
As described in Chapter~\ref{chap:template}, the class templates can be constructed by first organizing the samples into a template-tree data structure, and then identifying the templates by the roots of the well separated branches.
In this way, templates work as prototypes of different biological classes (e.g., disease status, time points, etc.)  and  emphasize the common properties of the class while omitting sample-specific details. 
In this chapter, I discuss classifying new samples by comparing them with templates and dynamically updating the templates (and the template-tree) as new samples are classified. 
The template-base classification is robust and efficient because it compares samples to cleaner and fewer class templates rather than the large number of noisy samples themselves.

The concepts of cluster, meta-cluster, sample, and template are explained in Chapter~\ref{chap:template}.
Briefly, a  \emph{template} is a collection of relatively homogeneous \emph{meta-clusters} commonly shared across samples of a given class, thus describing the key immune-phenotypes of an overall class of samples in a formal,  yet robust,  manner~\cite{azad2012matching, finak2010optimizing, pyne2009automated}. 
Here, a \emph{meta-clusters} or \emph{meta-population} is a generic (abstract) population formed by combining cell populations expressing similar phenotypes in different samples.
Clusters participating in a meta-cluster usually represent the same type of cells and thus have overlapping distributions in the marker space.
Therefore, a template is characterized by a mixture of (normally distributed) meta-clusters.

Besides their use in high-level visualization and between-class comparisons, templates can be employed to classify new samples with unknown status.
In this chapter, I use this approach to classify samples in terms of their expression of markers of the immune system. 
In the static classification approach, I build a fixed number of templates,  each representing samples from a particular class, and organize them into a template-tree data structure.
A new sample is then predicted to come from a class whose template it  is most similar to. 
In the dynamic classification approach, I update the templates and also  the template-tree, as new samples are classified. 

I demonstrate the use of template-based classification with two different datasets.
The first dataset measures the differences in phosphorylation events before and after stimulating T cells in human whole blood with an anti-CD3 antibody.
By creating pre-stimulation and post-stimulation templates, we can classify samples according to their stimulation status.
The second dataset studies the natural variations among different subsets of immune cells in five healthy individuals.
Blood was collected on up to four different days from each subject and five technical replicates were created from each subject.
Five templates could succinctly represent samples from five individuals despite the within-subject temporal and technical variations, demonstrating the fact that technical and day-to-day variations are smaller than the natural variation across individuals in this dataset.

Template-based classification has been used in several areas such as  face recognition, speech recognition, character recognition,  etc.
In face recognition \cite{brunelli1993face}, a template library is created with one or more digital images from each person.
An  unclassified image is compared to each database image by computing correlations of different features (eyes, nose, mouth etc.) and is classified as the one giving the highest cumulative score.
In speech recognition \cite{de2007template, deng2007structure}, a template is created for each speaker by a sequence of consecutive acoustic feature vectors and  an incoming signal is classified by comparing it with the templates using the Dynamic Time Warping algorithm.
In character recognition \cite{connell2001template}, representative prototypes for each character are created from different writing styles and an incoming character is classified by comparing it to existing prototypes using a feature matching algorithm.
The algorithms discussed in this chapter have similarities  to these methods in principle but differs from them significantly in how the templates are created, represented,  and compared  with incoming samples.
In contrast to these approaches, I maintain a hierarchy of the training samples in order to use their relationships in future classification.
A template is then represented with the shared features of all samples in a class whereas the methods discussed above use representatives from the training set.
Furthermore, the dynamic template algorithm continuously updates templates as new samples are classified, which improves the accuracy of future classification.

The rest of this chapter is organized as follows.
In Section~\ref{sec:static_template}, I describe the classification methods based on static templates created from a collection of training samples.
Section~\ref{sec:dynamic_template} outlines a robust dynamic classification algoritm.
The next Section~\ref{sec:template_classification_results} demonstrates the application of template-based classification with two representative datasets.
I conclude this chapter in Section~\ref{sec:classification_conclusions}.

\section{Classifying Samples with Static Templates} 
\label{sec:static_template}
\subsection{Creating static templates from a collection of FC samples}
Consider a collection of $N$ FC samples belonging to $M$ disjoint classes.
In Section~\ref{sec:HMM}, I described a hierarchical matching-and-merging (HM\&M) algorithm that organizes $N$ samples into a binary \emph{template-tree} data structure. 
A node in the tree represents either a sample (leaf node) or a template (internal node).
In both cases a node is characterized by a finite mixture of multivariate  normal distributions each component of which is a cluster or meta-cluster.
The height of an internal node in the template-tree is measured by the dissimilarity between its left and right children.
By recursion,  a template denoted by  a relatively lower internal node represents a relatively homogeneous collection of samples and vice versa.

After building a template-tree, we can cut the tree at a suitable height so that $M$ disjoint subtrees are produced.
The root of each subtree represents a template of the samples placed in the leaves of that subtree. 
The class (label) of a template is determined by the label of the majority of the samples in the subtree rooted at the template.
However, if the number of classes $M$ is not known {\em a priori\/},  $M$ is set to the number of well-separated branches based on the relative heights of  the subtrees. 
The roots of these well-separated subtrees represent  the class templates, 
where within-class variations (heights of the subtrees) are small relative to the 
between-class variations (heights of the ancestors of the subtrees).

\subsection{Classifying  new samples with  static templates}
Let $T_1$, $T_2$, $\ldots$, $T_M$ be $M$ templates created from a collection of $N$ samples, where the $i^{th}$ template $T_i$ summarizes samples of the  $i^{th}$ class.
The dissimilarity $D(S,T_i)$ between a sample $S$ and a template $T_i$ is computed by the mixed edge cover (MEC) algorithm described in Chapter~\ref{chap:matching}. 
(Note that both a sample and a template are characterized by finite mixtures of multivariate  normal distributions each component of which is a cluster or meta-cluster.)
The new sample $S$ is predicted to belong to the class whose template it is most similar (least dissimilar) to:
\begin{equation}
i^* = \arg\!\min_{1\leq i\leq M} D(S,T_i), \ \ \  \text{class}(S) = \text{class}(T_{i^*}).
\end{equation}
During classification, if the sample's dissimilarity with the closest template is above a threshold (i.e., it is not similar to any of the class templates), then it represents a new class not represented by a template.
Hence we need to create a new template for this sample. 
I address this issue in the next section. 

The template-based classification is  fast because we need to compare a new sample only with $M$ templates instead of with $N$ samples.
Let $K$ be the maximum number of clusters or meta-cluster present in a samples or template.
Then the time complexity of a classification is $O(MK^3 \log K)$, where $O(K^3 \log K)$ time is required to compute dissimilarity (mixed edge cover) between samples.
Typically the number of distinct classes ($M$) is much smaller than the number of samples $N$. 
Hence, computing $O(M)$ dissimilarity is faster than classifying the sample  from scratch, which requires $O(N)$ dissimilarity computations.

\section{Building dynamic templates}
\label{sec:dynamic_template}

\subsection{The algorithm}
The classification method based on static templates has two limitations.
First, the algorithm requires a significant number of samples from different biological classes to construct meaningful class templates. 
Hence, this approach is inadequate when samples arrive sequentially or in batches, as for instance in a 
longitudinal study of an epidemic.  
Second and more important,  the algorithm builds a fixed set of static templates and does not update 
templates as new samples are classified.
Therefore, future classification can not use the information gained from samples classified with the static template-tree. 

\begin{figure*}[!t]
  \begin{algorithmic}[1]
    \Procedure{\texttt{\tt insert}}{$u,v$}  \Comment{Insert leaf node $v$ in the subtree rooted at $u$}
    \If{$u$ is a empty} \Comment{Inserting in an empty tree}
     \State \textbf{return} $v$
      \EndIf
    \If{$u$ is a leaf} \Comment{Case 1}
   
    \State $w\gets$ empty node, \  $w_l \gets u$, \ $w_r \gets v$, \ $x\gets w$
    \Else 
    \State $D \gets \min\{D(u_l, u_r), D(u_l, v), D(u_r, v)\}$
    \If{$D(u_l, u_r) = D$} \Comment{Case 2}
    \State $w \gets$ empty node,  \  $w_l \gets u$, \ $w_r \gets v$, \ $x\gets w$
    \ElsIf{$D(u_l, v) = D$} \Comment{Case 3}
    \State $u_l \gets {\tt insert}(u_l, v)$, \ $x\gets u$
    \Else \Comment{when $D(u_r, v) = D$, \  Case 4}
    \State $u_r \gets \texttt{insert}(u_r, v)$, \ $x\gets u$
    \EndIf
    \EndIf
    \State update node $x$ by matching and merging meta-clusters from $x_l$ and $x_r$ 
     \State height($x$) = $D(x_l, x_r)$
    \State \textbf{return} $x$\Comment{Going up in the tree}
    \EndProcedure
  \end{algorithmic}
  \caption[An algorithm for inserting a sample into the template-tree]{Inserting a leaf node (sample) $v$ in a subtree rooted at $u$ (template or sample)}\label{insert}
  \label{fig:insert}
  \vspace{-.5cm}
\end{figure*}

I address both of these limitations with dynamic updates of templates as new samples are classified.
The update operation is performed by inserting the new sample into the current template-tree and updating necessary internal nodes (templates) in the tree.
Subsequent classification tasks are performed on the regularly updated templates.
This approach also works when we do not have any training dataset to begin with.
In that case, the template-tree is created from the scratch by repeated insertion of samples in a dynamic fashion starting with empty templates.
Consider an existing template-tree {\tt TT} (possibly empty) with $r$ as the root node.
Note that $r$ can be considered as the template of all samples 
in the leaves of the  tree. 
In order to insert a new sample $S$ in {\tt TT}, I first create a singleton node $v$ from $S$.
If {\tt TT} is an empty tree I make $v$ the root of the template-tree, and otherwise, I insert $v$ into the tree {\tt TT} by invoking the procedure {\tt insert} shown in Fig.~\ref{fig:insert} with $r$ and $v$ as the parameters.

\begin{figure*}[!t]
   \centering
   \includegraphics[scale=.42]{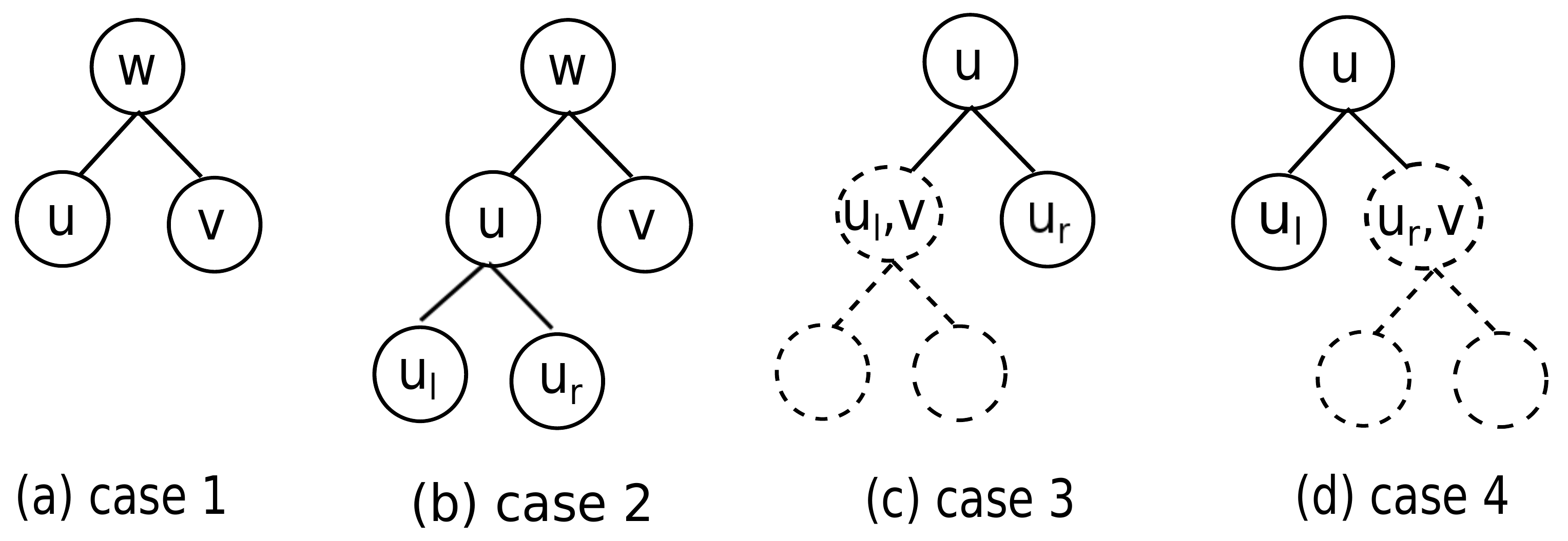} 
    \vspace{-.2cm}
   \caption[Four cases to consider when inserting a sample into the template-tree]{Four cases to consider when inserting a leaf node $v$ into the subtree rooted 
at $u$. (a) case 1 ($u$ is also a leaf): a new internal node $w$ is created and is 
made the parent of $u$ and $v$, (b) case 2 ($u$ is a non-leaf and the left and right children 
of $u$ are more similar to each other than to $v$): a new internal node $w$ is created 
and is made the parent of $u$ and $v$, (c) case 3 ($u$ is a non-leaf  and the left child $u_l$ 
of $u$ is more similar to $v$ than to the right child $u_r$): insert $v$ into the subtree 
rooted at $u_l$ by calling {\tt insert}$(u_l, v)$, and (d) case 4 ($u$ is a non-leaf  
and the right child $u_r$ of $u$ is more similar to $v$ than to the left child $u_l$): 
insert $v$ into the subtree rooted at $u_r$ by calling {\tt insert}$(u_r, v)$. 
The dotted parts in Subfigure (c) and (d) are determined by the {\tt insert} function in a recursive fashion.}
 \vspace{-.5cm}
   \label{fig:cases}
\end{figure*}

The procedure {\tt insert} works in a recursive fashion. 
It follows a path from the root to a node (a leaf or internal node),
to be identified by the algorithm,   where the new node $v$ is inserted.
The procedure then backtracks by updating the mixtures of the internal nodes found in the 
return path back to the root.
I consider four cases while inserting $v$ in a subtree rooted at $u$.
The cases are illustrated in Fig.~\ref{fig:cases}.
In the first case $u$ is a leaf node,  
and I create a new node $w$ and make $u$ and $v$ the children of $w$.
I create a template from the samples in the leaves $u$ and $v$ and save it in node $w$.
In the other cases $u$ is an internal node.
Let $u_l$ and $u_r$ be the left and right children of $u$, respectively. 
I compute dissimilarities $D(u_l, u_r)$, $D(u_l, v)$ and $D(u_r, v)$ between each pair of nodes from $u_l, u_r$, and $v$.
If $D(u_l, u_r)$ is the smallest among the three dissimilarities, 
then $v$ cannot be inserted in a subtree rooted at $u$.
Thus I create a new node $w$ and make $u$ and $v$ the children of $w$.
I create a new template from the template $u$ and sample $v$, save it in node $w$ and return.
When $D(u_l, v)$ is the smallest dissimilarity,  I insert $v$ in a subtree rooted at $u_l$ by calling the procedure {\tt insert} with $u_l, v$ as parameters. In this case the left subtree of $u$ gets updated.
Similarly, if $D(u_r, v)$ is the smallest then $v$ is inserted in the right subtree rooted at $u_r$.

\subsection{Computational complexity}
To insert a new sample, we need to traverse a path starting from the root to a leaf or an internal node in a template-tree.
In the worst case the length of the traversed path is the height of the template-tree.
Let $N$ be the number of samples and $h$ be the height of a template-tree where $(N-1)\leq h \leq \log_2(N)$.
The former equality holds when the tree is completely unbalanced (a chain) whereas the latter equality is satisfied when the tree is balanced. 
At each node in the traversed path we need to compute three dissimilarities and one update operation (when backtracking).  
Let $K$ be the maximum number of clusters or meta-clusters in a node of the template-tree and $p$ be the dimension (number of features) of the data. 
A dissimilarity computation takes $O(K^3\log K)$ time whereas an update operation takes $O(K^3\log K)$+$O(Kp)$ time when meta-cluster parameters are computed by the maximum likelihood estimation. 
Hence, the time complexity of inserting a sample in a template-tree is $O(hK^3\log K)$.

\subsection{Classifying a sample}
To classify a new sample $S$, it is inserted into the current template-tree using the procedure described in Fig. \ref{fig:insert}.
The class of $S$ is predicted to be the class of the template created from the subtree where $S$ is inserted.
At the time of insertion the template-tree is dynamically updated to reflect the information gained from the new sample. 
The dynamic template approach is especially useful in unsupervised classification where the class labels of the samples are not known in advance.
In that case, the class templates are created from the well-separated subtrees such that within-class variations (heights of the subtrees) are small relative to the between-class variations (heights of the ancestors of the subtrees).
In this context, the algorithm is similar to the spirit of the hierarchical clustering algorithm 
UPGMA, with significant differences in the distance computation and management of the internal nodes.
Furthermore, when a new sample is highly dissimilar to every existing template, the algorithm automatically creates a new branch in the tree indicating a new class.
This approach therefore has the ability to discover unknown classes from the incoming samples, which, for example, is very useful in detecting new strains of a disease.

The algorithm inserts samples into the template-tree in a particular order, which can determine the relative positions of nodes in the template-tree.
How sensitive is the template-based classification to the order in which the samples are inserted into the template-tree? 
If the between-class variation is significantly higher than the within-class variation (as is the case in the two datasets studied in this chapter), the classification accuracy is unaffected by the small differences in the subtrees of the template-tree. 
In current work, we are studying different rearrangement techniques for creating robust template-tree insensitive to the order of sample insertions.

\section{Results}
\label{sec:template_classification_results}
\subsection{Classifying stimulation status of T cells}
I apply the template-based classification algorithm to classify the stimulation status of 29 pairs of samples collected before and after stimulating blood with anti-CD3 antibody.
Each of the 29 pairs of samples in the T cell phosphorylation (TCP) dataset measures the abundance of four protein markers (CD4, CD45RA, SLP-76, and ZAP-70)~\cite{maier2007allelic}.
During the stimulation, anti-CD3 antibody binds with T cell receptors (TCR) and phosphorylate few domains of naive and memory T cells.
The effect of phosphorylation is revealed by the higher expressions of SLP-76 and ZAP-70 proteins in T cells after the stimulation.
Section~\ref{sec:tcp_data_description} describes this dataset in detail.  
 
For a pair of samples from the $i^{th}$ subject, I denote the unstimulated sample by $i-$ and the stimulated sample by $i+$.
Each of the 58 samples in the TCP dataset was preprocessed and clustered independently to identify four cell populations: (a) CD4$^+$CD45RA$^\text{low}$ T cells, (b) CD4$^+$CD45RA$^\text{high}$ T cells, (c) CD4$^-$CD45RA$^\text{high}$ T cells, and (d) CD4$^-$CD45RA$^\text{low}$ T cells (cf. Section~\ref{sec:clustering_tcp}).
(Recall that  `$+$' and  `high' indicate higher abundances  of a marker, and 
`$-$' and `low' indicate lower levels of it.)

\begin{figure}[!t]
   \centering
   \includegraphics[scale=.46]{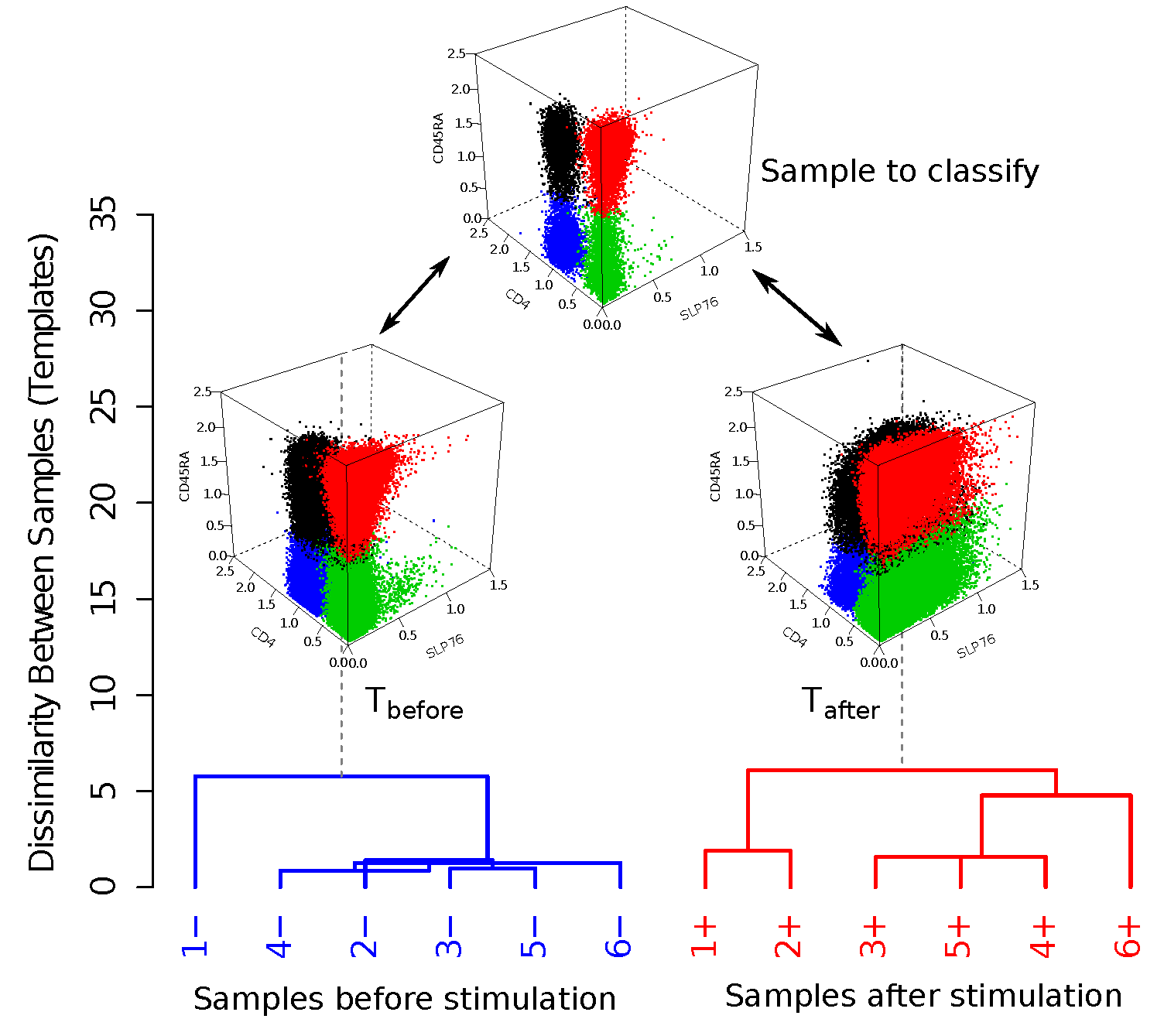} 
   \vspace{-.2cm}
   \caption[Sample classification based on static templates]{Illustration of sample classification based on static templates. 
The HM\&M algorithm creates two templates, $T_\text{before}$ for 
before-stimulation and $T_\text{after}$ for after stimulation classes, 
from six  pairs of samples in the TCP dataset.  
   A new sample is compared with the templates, 
and is classified with the template it is most similar to.}
   \label{fig:classify}
   \vspace{-.5cm}
\end{figure}

\emph{Classification with static templates}: 
I divide 29 pairs samples randomly into a training set and a test set.
From the training samples, the HM\&M algorithm constructs a template-tree where the left and right children of the root represent the pre- and post-stimulation templates.
Fig.~\ref{fig:classify} illustrates the classification process with a template tree created from six pairs of samples.
After constructing the tree, it is cut beneath the root to create the pre-stimulation ($T_{\text{before}}$) and post-stimulation ($T_{\text{after}}$) templates.
To predict the stimulation status of the  $i^{th}$ sample $S_i$ in the test set, I compute the dissimilarity $D$ between $S_i$ and each of the templates with the cost of the optimum mixed edge cover algorithm.
$S_i$ is predicted to come from the pre-stimulation class when $S_i$ is more similar to  $T_{\text{before}}$ than to $T_{\text{after}}$ (i.e., $D(S_i, T_{\text{before}}) < D(S_i, T_{\text{after}})$), and otherwise, from the post-stimulation class.
In Fig.~\ref{fig:classify}, I show a sample from the test set above the two templates.
The algorithm correctly classifies it as a pre-stimulation sample, and the correctness
of the classification can be verified visually as this sample does not show a phosphorylation shift in SLP-76.
(The 3-D scatter plots are shown for visualization only; classification is performed in 4-D.)

I study the accuracy of the template-based classifier with a cross-validation.
At each stage of the cross-validation, training and test sets are created from 10 and 48 samples respectively. 
The status of each test sample is predicted by comparing it with the templates created from the training set.
A sample is considered to be misclassified when its predicted class is different from the actual class.
This process is repeated 58 times for different combinations of training and test sets.  
I observed that three pre-stimulation samples, 9-, 10- and, 11-, 
were consistently classified with the after-stimulation class 
whenever they were  present in the test set. 
No other sample is classified into a wrong class in the cross-validation. 
I consider these three samples as outliers, show that they are likely to have been 
stimulated before the experiment, and discuss their properties further in the following section.

\begin{figure*}[!t]
   \centering
   \includegraphics[scale=.4]{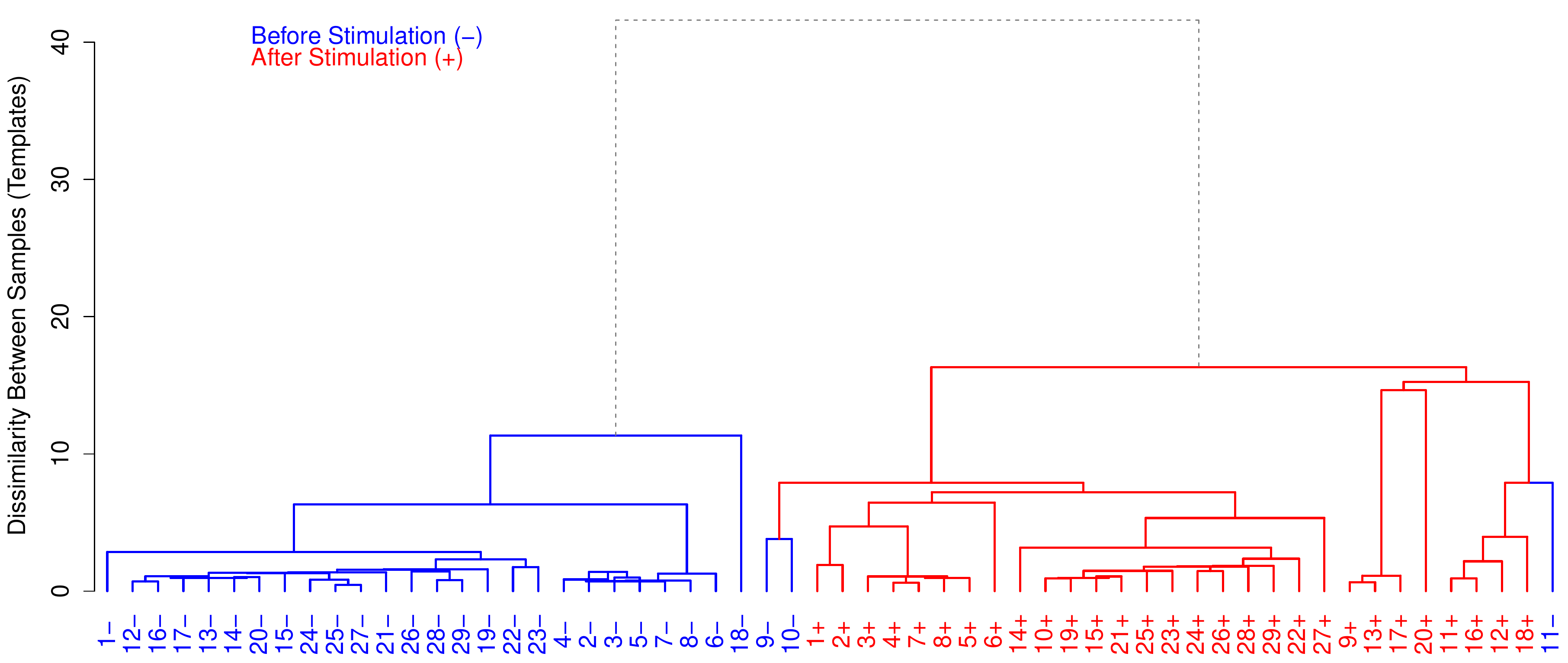} 
   \vspace{-.2cm}
   \caption[The dynamic template-tree created from the samples of the TCP dataset]{A dynamic template-tree created incrementally by adding the samples in the 
TCP dataset one after another.
   Minus and plus signs are appended to the subject number  to indicate  pre- and 
post-stimulation samples.
Pre-stimulation samples are in blue, and post-stimulation samples  are red. 
The height of a node measures the dissimilarity between its left and right children,
whereas the horizontal placement of a sample is arbitrary.}
   \label{fig:dynamic_tree}
   \vspace{-.5cm}
\end{figure*}

{\em  Classification with dynamic templates:}
In order to demonstrate the classification approach based on dynamic templates, 
I build a template-tree incrementally from the samples in the TCP dataset.
Starting with an empty tree, the algorithm described in Sec.~\ref{sec:dynamic_template} inserts the samples one after another into the current tree. 
Fig.~\ref{fig:dynamic_tree} shows the complete template-tree where pre-stimulation samples are shown in blue and post-stimulation samples are shown in red. 
Aside from three outlying samples, all samples create two well-separated branches of the root denoting the pre- and post-stimulation templates.
The height of an internal node in a template-tree is measured by the dissimilarity between the pair of samples (templates) denoted by the left and right children of the internal node.
The height of the root in Fig.~\ref{fig:dynamic_tree} is more than twice of the height of any other node.
Hence the algorithm successfully identifies two templates with small within-template deviation while maintaining a clear separation between them.

When inserting a new sample $S$, the template-tree is dynamically updated to reflect the information gained from the new sample.
After insertion, the position of $S$ in the tree determines its predicted class. 
$S$ is classified as a pre-stimulation sample when it is placed in the left (blue) subtree and otherwise, classified as a post-stimulation sample.
Similar to the classification with static templates, we observe in Fig.~\ref{fig:dynamic_tree} that all samples except 9-, 10- and 11-, are correctly classified. 

\begin{figure}[!t]
   \centering
   \includegraphics[scale=.33]{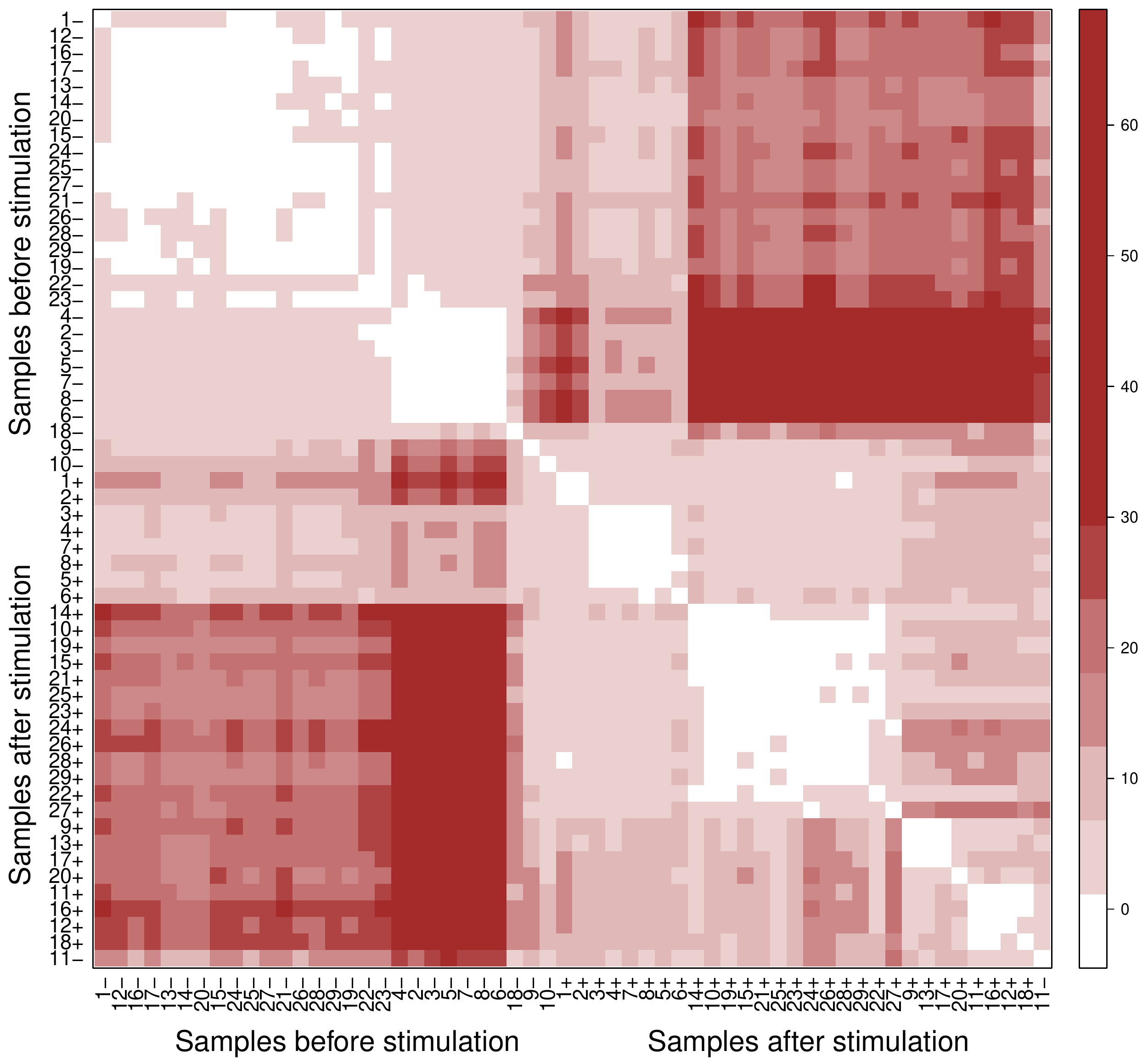} 
    \vspace{-.2cm}
   \caption[Heat plot showing the dissimilarity between pairs  of samples in the TCP dataset]{Heat plot showing the dissimilarity between pairs  of samples in the TCP dataset 
where the color of a square corresponds to the dissimilarity of a pair of samples.
A square is  drawn in a light shade  when the pair of samples is similar, 
and in a  dark shade  when the pair of samples is highly dissimilar.
The plot is symmetric about the main diagonal.
The samples are listed  as they are ordered horizontally in Fig.~\ref{fig:dynamic_tree}.
    }
   \label{fig:levelplot_phospho}
   \vspace{-.5cm}
\end{figure}

\begin{figure}[!t]
   \centering
   \includegraphics[scale=.65]{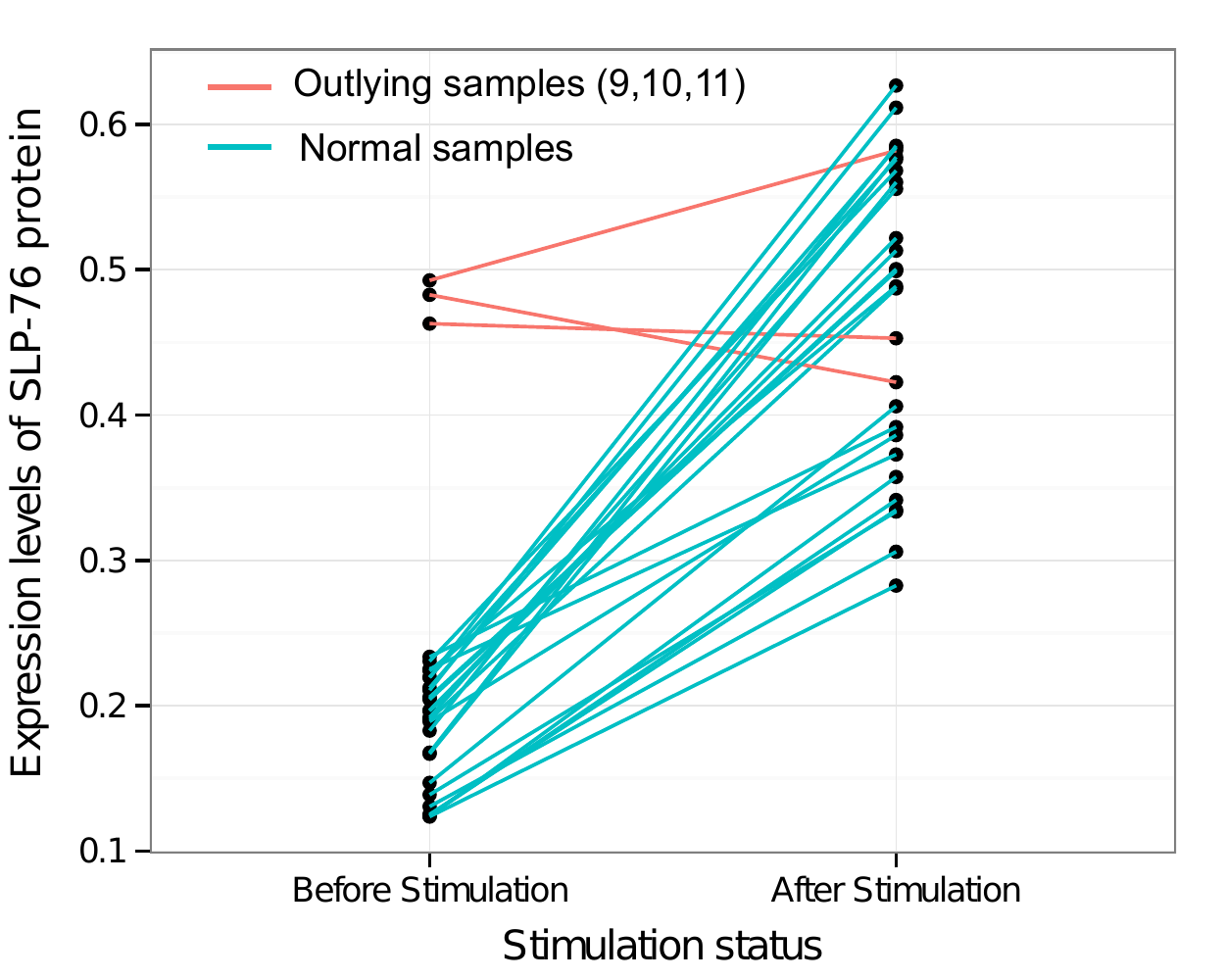}
    \vspace{-.2cm}
   \caption[Levels of SLP-76 expression in each pair of samples from the TCP dataset]{Levels of SLP-76 expression in each pair of samples (joined by a line) from the TCP dataset.
   For most samples, SLP-76 levels increase after the anti-CD3 stimulation.
   However, the three samples $\{9, 10, 11\}$   have high levels of SLP-76 protein in their 
pre-stimulation states,  and do not show the usual increase  after stimulation. 
    }
     \vspace{-.5cm}
   \label{fig:SLP76_stimulation}
\end{figure} 

{\em  Outlying samples:}
\label{sec:outliers}
Now I discuss the three outlying pre-stimulation samples, 9-, 10- and 11-, which 
are consistently classified with the post-stimulation samples by  both the static 
and dynamic classification algorithms.  
Fig.~\ref{fig:levelplot_phospho} shows the dissimilarity between every pair of samples in the dataset 
in a heatplot where a  square is drawn in a lighter shade  when a pair of samples is similar,
and in  darker shade  when a pair of samples is highly dissimilar. 
We observe that most squares in the top-left and bottom-right quadrants 
are in light colors indicating similarity among samples within pre- and post-stimulation classes.
However, three pre-stimulation samples, 9-, 10- and 11-, are more similar to the 
post-stimulation samples than to the pre-stimulation samples.
This anomaly can be explained by plotting the average expression of SLP-76 protein for each pair of samples in Figure \ref{fig:SLP76_stimulation}.
The three outlying samples $\{9, 10, 11\}$  have much higher levels of SLP-76 
than the remaining samples in their pre-stimulation states. 
On stimulation, two of these samples have decreased levels of SLP-76, while one shows an increase. 
Consequently these samples are classified with the post-stimulations samples by the 
templates-based classification algorithms presented here. 

\subsection{Classifying immune patterns in  healthy individuals}
I further validated the template-based classification with a healthy donor (HD) dataset described in Section~\ref{sec:hd_data_description}.
The HD dataset represents a ``biological simulation" where peripheral blood mononuclear cells (PBMC) were collected from five healthy subjects on different days, and each sample was divided into five parts and analyzed through a flow cytometer at Purdue's Bindley Biosciences Center.  
Each of the 65 samples is preprocessed, variance stabilized, and clustered to identify four cell populations: 
 (a) helper T cells (CD3$^+$CD4$^+$), (b) cytotoxic T cells (CD3$^+$CD8$^+$), (c) B cells (CD3$^-$CD19$^+$), and (d) natural killer cells (CD3$^-$CD19$^-$).
Detail discussion about the clustering step can be found in Section~\ref{sec:clustering_hd}.

\begin{figure}[!t]
   \centering
   \includegraphics[scale=.7]{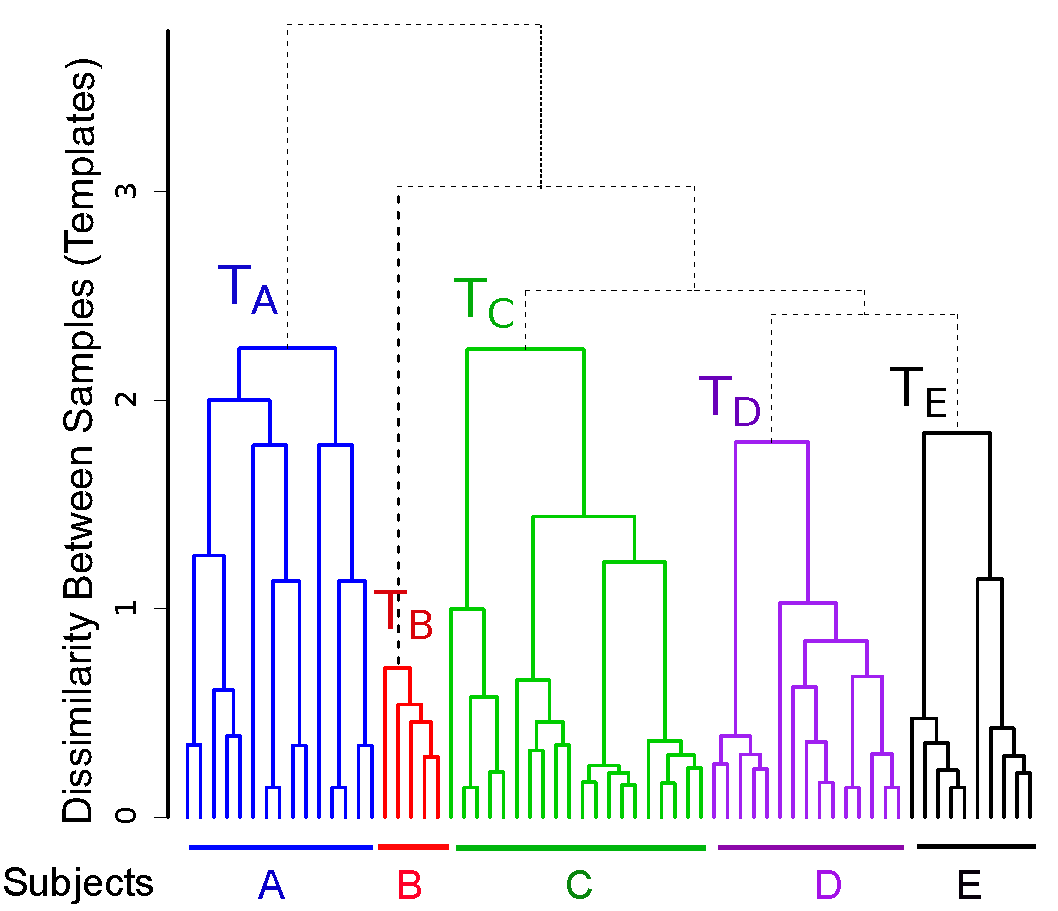} 
   \vspace{-.2cm}
   \caption[The template-tree created from the HD dataset]{The template-tree created from samples in the healthy donor dataset.
The algorithm identifies five well separated branches denoting templates for the five subjects.
A subtree consisting of samples from the same subject is shown in same color.
}
   \label{fig:hd_tree}
   \vspace{-.5cm}
\end{figure}

A template tree is created from the 65 samples in the HD dataset by repeated insertion of samples with the procedure described in Sec.~\ref{sec:dynamic_template}.
Fig.~\ref{fig:hd_tree} shows the complete template-tree where samples from five individuals are shown in five different colors.
We observe that samples from the five subjects create five well-separated branches and the roots of the five branches denote subject-specific templates.
Intuitively we expect samples from each subject to be classified together. 
Here, the within-subject variations of a subject come from day-to-day natural variations and technical variations in flow cytometry sample preparation and  measurement, whereas the between-subject variations come from the innate biological variability  in the healthy subjects.
In this dataset we observe more natural variation than the temporal and technical variations.

The observations from the healthy donor dataset confirm that we can build 
immune profiles for individuals  despite within-subject  variations from different sources. 
Additionally, the five templates from the five subjects create another level of hierarchy and the root of the tree in Fig.~\ref{fig:hd_tree} can be considered as a template for healthy individuals.
This combined template represents a healthy immune profile by preserving the common features of  healthy individuals and by removing between subject variations.
This  template can be compared against templates created from diseased samples in order to diagnose diseases and to perform comparative study of healthy and diseased immune profiles.


\section{Conclusions}
\label{sec:classification_conclusions}
I describe template-based classification methods for flow cytometry samples displaying a combination of different immunophenotypes.
A template built from samples of a class  provides a concise description of  the class 
by emphasizing the key characteristics while masking statistical noise and 
low-level details,  and  thus helps  to measure overall changes in cell populations 
across different conditions.
By moving beyond sample-specific variations, 
the templates act as the blueprints for different classes, and can be used to classify future 
samples to different classes  in a more relevant parameter space.
It is also more efficient to classify a sample using templates rather than all of the 
previously seen samples. 
I  maintain a hierarchy of the samples in a template-tree 
such that samples can be analyzed in higher resolution as  needed.

The major contribution in this chapter is a dynamic algorithm to construct 
and update the templates, and build and maintain the template-tree, 
when the samples arrive continuously over a period of time. 
As new samples come in, the  templates are dynamically updated to reflect the information 
gained from them. 
This is a desirable property in dynamic situations, as in the course of an epidemic, 
when new samples are being collected and analyzed. 
Another context where the dynamic classification approach is useful is when
the samples are collected at a large number of hospitals or labs; 
the data at each hospital can be analyzed {\em in situ\/}, 
and only the summaries 
need to be shared among the hospitals to create a global profile of the immune system, 
thus avoiding issues with privacy of clinical data. 


Dynamic classification  is a critical step towards characterizing 
diverse states of the human immune system from  big  datasets 
of samples collected at  geographically distributed laboratories, 
e.g., the Human Immunology  Project Consortium ({\tt www.immuneprofiling.org}). 
This work makes it possible to summarize the data from each laboratory 
using templates for each class, and then to merge the templates and template-trees across 
various laboratories, as the data is being continuously collected and 
analyzed.

%% file: aml-classification.tex
\chapter{Classification of Acute Myeloid Leukemia}
\label{chap:aml-classification}
\section{Introduction}
Can Acute Myeloid Leukemia (AML) samples be distinguished  from healthy ones  using 
flow cytometry data  from blood or bone marrow with a template-based classification method? 
This method builds a template for each class  to summarize the  samples belonging to the class,
and uses them to classify new samples as described in Chapter~\ref{chap:template-based-classification}. 
This question is interesting because AML is a heterogeneous disease with several subtypes
and hence it is not clear that a template can succinctly  describe all types  of AML. 
Furthermore, we wish to identify immunophenotypes (cell types in the bone marrow and blood) 
that are known to be characteristic of  subtypes of AML. 
Pathologists use these immunophenotypes to visualize AML and its subtypes,
and a computational procedure that can provide this information would be 
more helpful in clinical practice than a classification score that 
indicates if an individual is healthy or has AML. 

In Chapter~\ref{chap:template-based-classification}, I  described a template-based classification framework for analyzing flow cytometry (FC) data~\cite{azad2013classifying}. 
In this framework, each sample is characterized by means of the cell populations that it contains.
Similar samples belonging to the same class are  described by a template for the class. 
A  template consists of meta-clusters that characterize the cell populations present in  the samples that constitute the class. 
I described a hierarchical algorithm in Chapter~\ref{chap:template} that organizes the templates into a template tree. 
Given a sample to classify, we can compare it with the nodes in the template tree, and classify it to the template that it is closest to. 
A combinatorial measure for the dissimilarity of two samples or two templates,  computed by means of a mixed edge cover in a graph model (described in Chapter~\ref{chap:matching}), is at the heart of  this approach. 

The algorithmic pipeline for template-based classification has been applied to  various problems:
to distinguish the phosphorylation state of T cells; 
to study the biological, temporal, and technical variability of cell types in the 
blood of healthy individuals; to characterize changes in the immune cells of 
Multiple Sclerosis patients undergoing drug treatments; 
and to predict the vaccination status of HIV patients~\cite{azad2010identifying,azad2013classifying}. 
However, it is not clear if the AML data set can be successfully analyzed with 
this scheme, since AML is a heterogeneous disease at the morphologic, cytogenetic and molecular levels,
and a few  templates may not describe all of its  subtypes. 

AML is a disease of myeloid stem cells that differentiate to form several types of cells 
in the blood and marrow. It is characterized by the profusion of immature myeloid cells,
which are usually prevented from maturing due to the disease. 
The myeloid 
stem cell differentiates in several steps to form myeloblasts
and other cell types in a hierarchical process. 
This hierarchical differentiation process could be blocked at different cell types, 
leading to the multiple  subtypes of AML. 
Eight different subtypes of AML based on cell lineage are included in the French-American-British 
Cooperative Group (FAB)  classification scheme~\cite{bennett1985proposed}. 
(A different World Health Organization (WHO)  classification scheme has also 
been published.)
Since the prognosis  and treatment varies greatly among the subtypes of AML, 
accurate diagnosis is critical. 

In this chapter, I extend the template-based classification scheme presented in Chapter~\ref{chap:template-based-classification} by developing a scoring function that accounts for the subtleties of  FC data of AML samples~\cite{azad2014immunophenotypes}. 
Only a small number of the myeloid cell populations in AML samples
are specific to AML, and there are a larger number of cell populations that these samples
share with healthy samples. 
Furthermore, the scoring function needs to account for the diversity of the myeloid cell populations in the various subtypes of AML. 
This approach has the advantage of identifying immunophenotypes of clinical interest in AML from the templates. 
Earlier work on the AML dataset used in this chapter has classified AML samples using methods such as nearest neighbor classification, logistic regression, matrix relevance learning vector quantization, etc., but they have not identified these immunophenotypes; e.g.,~\cite{biehl2013analysis,manninen2013leukemia,qiu2012inferring}. 

Template-based classification has the advantage of being more robust than simple nearest neighbor classification since a template summarizes the characteristic properties of a class while ignoring small sample-to-sample variations. 
It is also  scalable to large numbers of samples, since it compares a sample to be classified only against a small number of templates rather than the much larger number of samples. 
The comparisons with the templates can  be performed efficiently using the structure of the template tree. 
It also reduces the data size by clustering the data to identify cell populations and then working with the statistical distributions characterizing the cell populations, in contrast to some of the earlier approaches that work with data sets 
even larger than the FC data by creating  multiple variables from a marker (reciprocal, powers, products and quotients of subsets of the markers, etc.~\cite{manninen2013leukemia}).

Template-based classification has been employed in other areas such as character, face, and image recognition, 
but its application to FC is relatively recent. 
In addition to our work, templates have been used for detecting the effects of phosphorylation~\cite{pyne2009automated},  evaluating the efficiency of data transformations~\cite{finak2010optimizing},  and labeling clusters across samples~\cite{spidlen2013genepattern}.

\section{Methods}

\begin{table}[!t]
\centering
\vspace{-.7 cm}
\caption{The fluorophore-conjugated antibodies contained in each of the 8 tubes in which the samples were incubated.}
\label{tab:AML_markers}
\begin{tabular}{cccccc}\toprule
Tubes   & FL1 & FL2 & FL3 & FL4 & FL5 \\
&   (FITC)\ \ \ \  & (PE)\ \ \ \  & (ECD)\ \ \ \  & (PC5)\ \ \ \  & (PC7)\ \  \\\toprule
	1 & IgG1 & IgG1 & CD45 & IgG1 & IgG1 \\ 
     	2 & Kappa & Lambda& CD45 & CD19 & CD20 \\  
     	3 & CD7 & CD4 & CD45 & CD8 & CD2 \\
     	4 & CD15 & CD13 & CD45 & CD16 & CD56 \\
     	5 & CD14 & CD11c & CD45 & CD64 & CD33 \\
    	6 & HLA-DR & CD117 & CD45& CD34& CD38 \\
     	7 & CD5 & CD19 & CD45 & CD3 & CD10 \\
     	8 & NA & NA & NA & NA & NA \\       \bottomrule
\end{tabular}
\end{table}

\subsection{The AML Dataset}
I have used an FC dataset on AML that  was included in the DREAM6/FlowCAP2 challenge of  2011~\cite{dream_AML}. 
The dataset consists of FC measurements of peripheral blood or bone marrow aspirate collected from 
$43$ AML positive patients and $316$ healthy donors over a one year period. 
Each patient sample was subdivided into eight aliquots (``tubes") and analyzed with different biomarker combinations, five markers per tube (most markers are proteins) as described in Table~\ref{tab:AML_markers}.
In addition to the markers, the forward scatter (FS) and side scatter (SS) of each sample was also measured in each tube. 
Hence, the dataset contains $ 359 \times8  = 2,872$ samples and each sample is seven-dimensional
(five markers and the two scatters).
Tube 1 is an isotype control used to detect non-specific antibody binding and Tube 8 is an unstained control for identifying background or autofluorescence of the system. 
Since the data has been compensated for autofluorescence and spectral overlap by experts, I omit these tubes from the analysis.
The disease status (AML/healthy) of $23$ AML patients and $156$ healthy donors are provided as training set,  and 
the challenge is to determine the disease status of the rest of the samples,  
$20$ AML and  $157$ healthy, based only on the information in the training set. 
The complete dataset is  available at {\tt http://flowrepository.org/}.

The side scatter (SS) and all of the  fluorescence channels are transformed logarithmically,  but the forward scatter (FS) is linearly transformed to the interval [0,1] so that all channels have values in the  same range.
This removes any bias towards FS channel in the multi-dimensional clustering phase.
After preprocessing, an FC  sample is stored as  an $n\times p$ matrix $A$, where the element  $A(i,j)$ quantifies the $j^{th}$ feature in the $i^{th}$ cell, and $p$ is the number of features measured in each of $n$ cells. 
In this dataset,  $p=7$ for each tube and $n$ varies among the samples.

\subsection{Identifying cell populations in each sample} 
\label{sec:clustering}
I employ a two-stage clustering approach for identifying phenotypically similar cell populations (homogeneous clusters of cells) in each sample.
At first, I apply the $k$-means clustering algorithm for a wide range of values for $k$, and select the optimum number of clusters $k^*$ by simultaneously optimizing the Calinski-Harabasz and S\_Dbw cluster validation methods~\cite{halkidi2001clustering}.
Next, the clusters identified by the $k$-means algorithm are modeled with a finite mixture model of multivariate normal distributions. 
In the mixture model, the $i^{th}$ cluster is  represented by two distribution parameters $\boldsymbol {\mu_i}$, 
the $p$-dimensional mean vector, and $\Sigma$, the $p\times p$ covariance matrix.
The distribution parameters for each cluster are then estimated using the Expectation-Maximization (EM) algorithm. 
The statistical parameters of a cluster are used to describe the corresponding cell population in the rest of the  analysis.
The population identification process is described in Chapter~\ref{chap:clustering}.

\subsection {Creating templates from a collection of samples}
\label{sec:templates}
The hierarchical matching-and-merging (HM\&M) algorithm described in Chapter~\ref{chap:template} is used to create templates representing distinct classes of samples. 
The HM\&M algorithm arranges a set of similar samples into a binary \emph{template tree} data structure \cite{azad2012matching}.
The dissimilarity between a pair of samples is computed by the cost of optimum mixed edge cover solution discussed in Chapter~\ref{chap:matching}.

\subsection{Classification score of a sample in AML dataset}
Consider a sample $X$ consisting of $k$ cell populations $S=\{c_1, c_2, ..., c_k\}$,
with the $i^{th}$ cluster $c_i$ containing $|c_i|$ cells.
Let $T^-$ and $T^+$ be the templates created from AML-negative (healthy) and AML-positive training samples,  respectively.
I describe how to compute a score $f(X)$ in order to classify the  sample $X$ to either the healthy class or the AML class. 

\begin{figure}[!tbp]
\centering
\includegraphics[scale=.4]{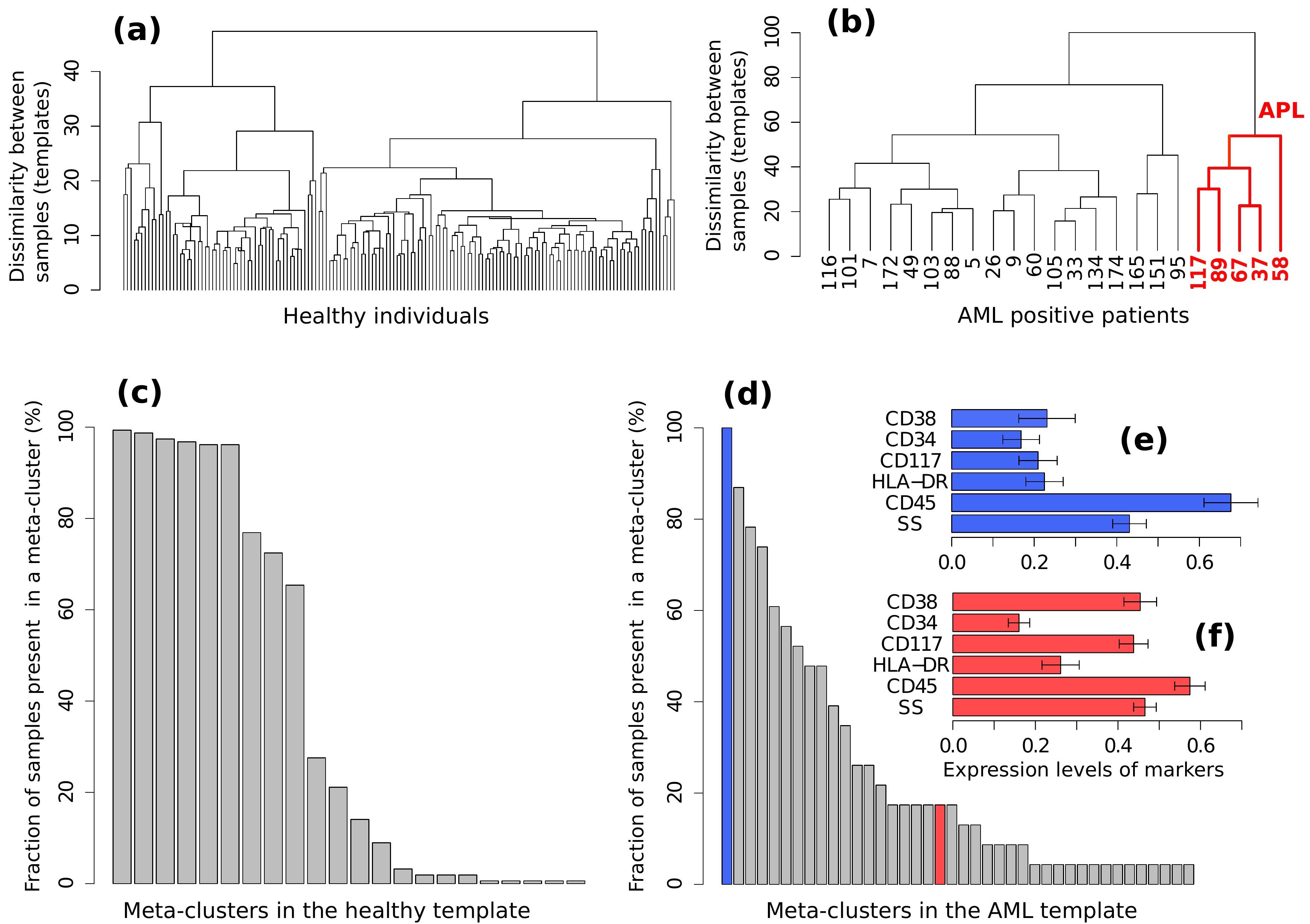}
\vspace{-.2cm}
\caption[The healthy and AML templates created from Tube 6 of AML dataset]{The healthy and AML templates created from Tube 6. 
{\bf (a)} The template-tree created from  156 healthy samples in the training set. 
{\bf (b)} The template-tree created from  23 AML samples in the training set. 
Samples in the red subtree  exhibit  the characteristics of Acute Promyelocytic Leukemia (APL) as shown in Subfigure (f).
{\bf (c)} Fraction of 156 healthy samples present in each of the 22 meta-clusters in the healthy template. 
Nine meta-clusters, each of them shared by at least 60\% of the healthy samples, form the core of the healthy template.
{\bf (d)} Fraction of 23 AML samples present in each of the 40 meta-clusters in the AML template.
The AML samples, unlike the healthy ones, are heterogeneously distributed over the meta-clusters.
{\bf (e)} The expression levels of markers in the meta-cluster shown with blue bar in 
Subfigure (d).
(Each horizontal bar in Subfigures (e) and (f) represents the average expression of a marker and the error bar shows its  standard deviation.) 
This meta-cluster represents lymphocytes denoted by medium SS and high CD45 expression 
and therefore does not express the AML-related markers measured in Tube 6.
{\bf (f)} Expression of markers in a meta-cluster shown with red bar in Subfigure (d).
This meta-cluster denotes myeloblast cells as defined by the SS and CD45 levels.  
This meta-cluster expresses HLA-DR$^-$CD117$^{+}$CD34$^-$CD38$^+$, a   characteristic immunophenotype of APL. 
Five AML samples sharing this meta-cluster are similar to each other as shown in the red 
subtree in Subfigure (b). 
}
\vspace{-.5cm}
\label{fig:templates_tube6}
\end{figure}

The intuition behind the score is as follows. 
An AML sample contains two kinds of cell populations: 
(1) AML-specific myeloblasts and myeloid cells, and 
(2) AML-unrelated cell populations, such as lymphocytes. 
The former cell populations correspond to the immunophenotypes of  AML-specific metaclusters in the AML template,  and hence when we compute  a mixed edge cover between the AML template and an AML  sample, these  clusters get matched to each other. 
(Such clusters in the sample do not match to any metacluster in the healthy template.) 
Hence a positive score is assigned to a cluster in sample when it satisfies this condition, signifying that it is indicative of AML. 
AML-unrelated cell populations in a sample could match to meta-clusters in the healthy template, and also to AML-unrelated meta-clusters in the AML template. 
When either of these conditions is satisfied, a cluster gets a negative score, signifying that it is not indicative of AML. 
Since AML affects only the myeloid cell line and its progenitors, it affects only a small number of AML-specific cell populations in an AML sample. 
Furthermore, different subtypes of AML affect different cell types in the myeloid cell line. 
Hence, there are many more clusters common to healthy samples than there are AML-specific clusters common to AML samples. 
(This is illustrated later in Fig.~\ref{fig:templates_tube6} (c) and (d).)
Thus the range of positive scores are made relatively higher  than the range of negative scores. 
This scoring system is designed to reduce the possibility of a false negative (an undetected AML-positive patient), since this is more serious in the diagnosis of AML. 
Additional data such as chromosomal translocations and images of bone marrow from microscopy could confirm an initial  diagnosis of AML from flow cytometry. 

In the light of the discussion above, we need to  identify AML-specific metaclusters initially.
Given the templates $T^+$ and $T^-$, we create a complete bipartite graph 
with the meta-clusters in each template as vertices, and with each  edge  weighted by 
the Mahalanobis distance between its endpoints. 
A minimum cost mixed edge cover solution in this graph will match meta-clusters common to both templates, and such  meta-clusters represent non-myeloid cell populations that are not AML-specific. 
On the other hand, meta-clusters in the AML template $T^+$ that are not matched to a meta-cluster in the healthy template $T^-$ correspond to AML-specific metaclusters. 
Such meta-clusters in the AML template $T^+$ are denoted by the set $M^+$. 

Now we can proceed to compare a sample against the template for healthy samples and the template for AML. 
I compute a minimum cost mixed edge cover between  a  sample $X$ and the healthy template 
$T^-$, and let ${\tt mec^-}(c_i)$ denote the set of meta-clusters in $T^-$ mapped to a cluster $c_i$ in the sample $X$. 
Similarly, compute a minimum cost mixed edge cover between $X$ and the AML template $T^+$, and let ${\tt mec^+}(c_i)$ denote the set of meta-clusters in $T^+$ mapped to a  cluster $c_i$. 
These sets could be empty if $c_i$ is unmatched in the mixed edge cover. 
I compute the average Mahalanobis distance between $c_i$ and the meta-clusters matched to it in the template $T^-$, and define this as the dissimilarity $d(c_i, {\tt mec^-}(c_i))$. 
From the formulation of the mixed edge cover in~\cite{azad2010identifying},  we have $d(c_i, {\tt mec^-}(c_i)) \leq 2\lambda$, where $\lambda$ is the cost of leaving a cluster unmatched in the mixed edge cover solution.
Hence we can define the {\em similarity\/} between $c_i$ and ${\tt mec^-}(c_i)$ as  $s(c_i, {\tt mec^-}(c_i)) = 2\lambda-d(c_i, {\tt mec^-}(c_i))$.
By analogous reasoning, the similarity between $c_i$ and ${\tt mec^+}(c_i)$ is defined as $s(c_i, {\tt mec^+}(c_i)) = 2\lambda-d(c_i, {\tt mec^+}(c_i))$.

The score of a sample is the sum of the scores of its clusters. 
I define the score of a cluster $c_i$,  $f(c_i)$,  as the sum of two functions $f^+(c_i)$ and $f^-(c_i)$ multiplied with suitable weights. 
A positive score indicates that the sample belongs to AML, and a negative score indicates that it is healthy. 

The function  $f^+(c_i)$ contributes  a positive score to the sum if $c_i$ is matched to an AML-specific meta-cluster in the mixed edge cover between the sample $X$ and the AML template $T^+$, and a non-positive score otherwise. 
For the latter case,  there  are two subcases: 
If $c_i$ is unmatched in the mixed edge cover, it corresponds to none of the meta-clusters in the template $T^+$, and gets a zero score. 
If $c_i$ is matched only to non-AML specific meta-clusters in the AML template $T^+$, then it is assigned a small negative score to indicate that it likely belongs to the healthy class 
(recall that $k$ is the number of clusters in sample $X$). 
Hence 
\begin{eqnarray*}
f^+(c_i) &=&   \begin{cases}
                   s\left(c_i, {\tt mec^+}(c_i)\right), &\text{if\ \ } {\tt mec^+}(c_i) \cap M^+ \neq  \emptyset,\\
     -\frac{1}{k} \left[s(c_i, {\tt mec^+}(c_i))\right], &\text{if\ \ } {\tt mec^+}(c_i) \cap M^+ =  \emptyset,
                   \ \  {\text{and\  \ } \tt mec^+}(c_i) \neq  \emptyset,\\
                   0,                              &\text{if\ \ } {\tt mec^+}(c_i) = \emptyset.\\
\end{cases}
\end{eqnarray*}

The function $f^-(c_i)$ contributes a negative score to a cluster $c_i$ in the sample $X$ 
if it is matched with some meta-cluster in the healthy template $T^-$, indicating that it likely belongs to the healthy class. 
If it is not matched to any  meta-cluster in $T^-$, then it is assigned a positive score $\lambda$. 
This latter subcase accounts for AML-specific clusters in the sample, or a cluster that is in neither template. 
In this last case, we acknowledge the diversity of cell populations  in AML samples. 
Hence we have 
\begin{eqnarray*} 
f^-(c_i) &=& \begin{cases}
             -\frac{1}{k} \left[s(c_i, {\tt mec^-}(c_i))\right], &\text{if\ \ } {\tt mec^-}(c_i) \neq \emptyset,\\
\lambda, &\text{if\ \ } {\tt mec^-}(c_i) = \emptyset.\\
\end{cases}   \\ 
\end{eqnarray*}

Finally, we define 
\begin{equation} 
f(X) = \sum_{c_i\in X} \frac{|c_i|}{|X|} \frac{1}{2} (f^+(c_i) + f^-(c_i)).
\label{eq:classif_score}
\end{equation}
Here $|X|$ is the number of cells in the sample $X$. The score of a cluster $c_i$ is weighted by
the fractional abundance of cells in it. 

\begin{figure}[!t]
   \centering
\includegraphics[scale=.4]{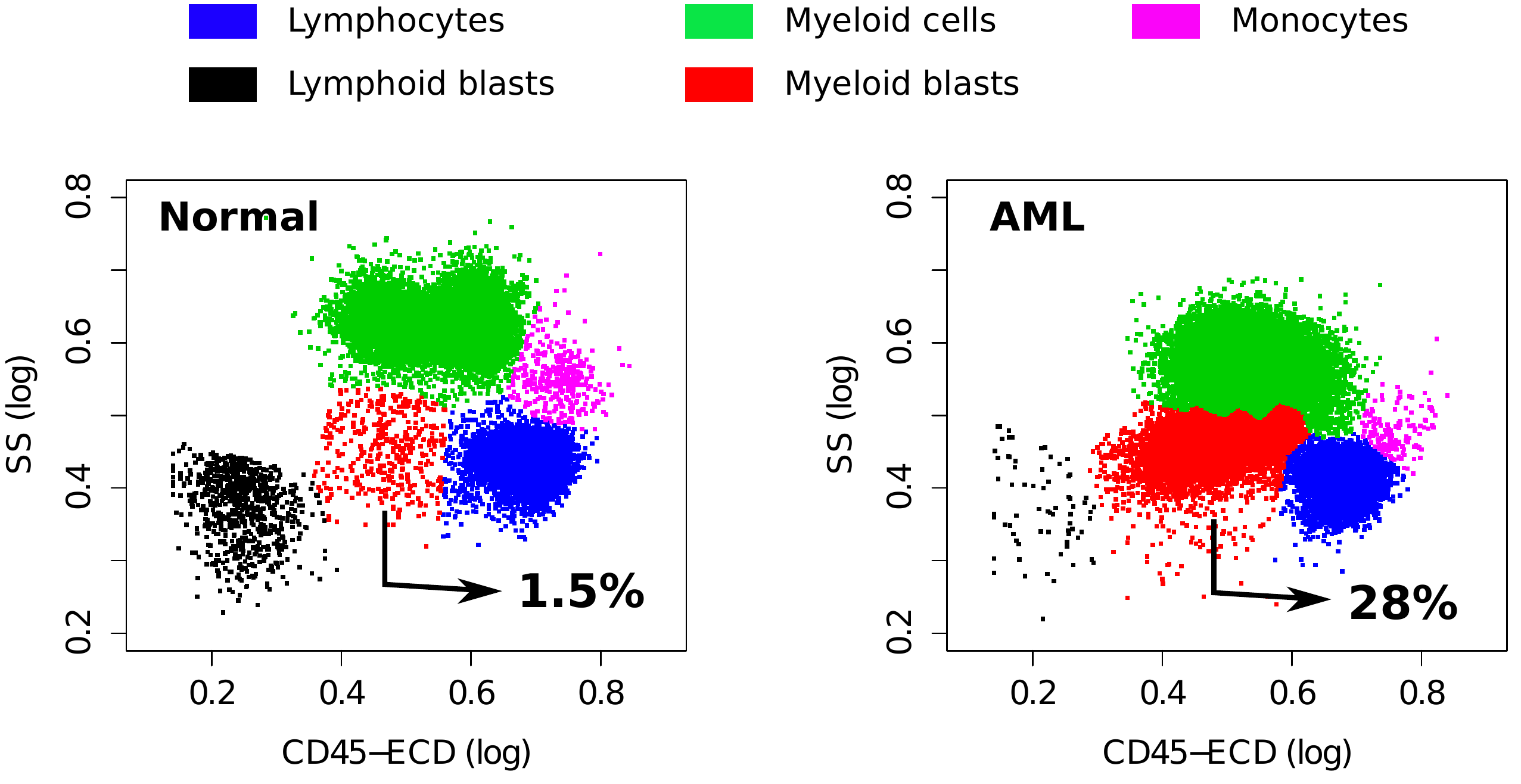}
\vspace{-.2cm}
\caption[Cell types identified on the side scatter (SS) and CD45 channels for a healthy and an AML positive sample]{Cell types identified on the side scatter (SS) and CD45 channels for a healthy and an AML positive sample.
Cell populations are discovered in the seven-dimensional samples with the clustering algorithm and then projected on these channels for visualization.
A pair of clusters denoting the same cell type is marked with the same color.
The proportion of myeloid blast cells (shown in red) increases significantly in the AML sample.
 }
 \vspace{-.5cm}
   \label{fig:AML_blasts}
\end{figure}

\section{Results}
\subsection{Cell populations in healthy and AML samples}
In each tube, I identify cell populations in the samples using  the clustering algorithm described in Section~\ref{sec:clustering}.
Each sample contains five major cell types that can be seen when cell clusters are projected on the side scatter (SS) and CD45 channels, as depicted in  Fig.~\ref{fig:AML_blasts}.
(Blast cells are immature progenitors of myeloid cells or lymphocytes.) 
The side scatter measures the granularity of cells, whereas CD45 is variably expressed by different white blood cells (leukocytes).
AML is initially diagnosed by  rapid growth of immature myeloid blast cells with medium SS and CD45 expressions~\cite{lacombe1997flow} marked in red in Fig.~\ref{fig:AML_blasts}.
According to the WHO guidelines, AML is initially confirmed when the sample contains more than 20\% blasts.
This is the  case for all, except one of the  AML samples in the DREAM6/FlowCAP2  training set, 
and the latter  will be discussed later.

\subsection{Healthy and AML templates} 
From  each tube of the AML dataset, using the training samples, we build two templates: one  for healthy samples, and one for AML. 
As described in Section~\ref{sec:templates},  the HM\&M algorithm organizes samples of the same class into a binary template tree whose root represents the class template.
The template trees created from the healthy and AML training samples in Tube 6 are shown in Subfigures \ref{fig:templates_tube6}(a) and \ref{fig:templates_tube6}(b) respectively.
The height of an internal node in the template tree measures the dissimilarity between its left and right children, whereas the horizontal placement of a sample is arbitrary.
In these trees, we observe twice as much heterogeneity in the AML samples than  among the healthy samples (in the dissimilarity measure), despite the number of healthy samples being five times as numerous as the  AML samples.
The larger heterogeneity among AML samples is observed in other tubes as well.
The  template-tree for AML partitions these samples into different subtrees that possibly  denote different subtypes of AML.
For example, the subtree in Fig.~\ref{fig:templates_tube6}(b) that is colored red includes samples (with subject ids 37, 58, 67, 89, and 117) with immunophenotypes of Acute Promyelocytic Leukemia (APL) (discussed later in this section).

Together, the meta-clusters in a healthy template  represent a healthy immune profile in the feature space of a tube from which the template is created. 
$22$ meta-clusters are obtained in the healthy template created from Tube 6.
The percentage of  samples from the training set participating in each of these meta-clusters is shown in Fig.~\ref{fig:templates_tube6}(c).
Observe that $60\%$ or more of the healthy samples participate in the nine most common meta-clusters (these constitute  the core of the healthy template).
The remaining thirteen meta-clusters include populations from a small fraction of samples. 
These populations could correspond to biological variability  in the healthy samples,variations in the FC experimental protocols,  and possibly also from the splitting of populations that could be an artifact of the clustering algorithm. 

The AML template created from Tube 6 consists of forty  meta-clusters (almost twice the number in the more numerous healthy samples).
Fig.~\ref{fig:templates_tube6}(d) shows that, unlike the healthy samples, the AML samples are heterogeneous with respect to  the meta-clusters they participate in:  
There are  $21$ meta-clusters that include cell populations from at least $20\%$ of the AML samples.
Some of the  meta-clusters common to a large number of AML samples  represent non-AML specific cell populations.
For example, Fig.~\ref{fig:templates_tube6}(e) shows the average marker expressions of the meta-cluster shown in the blue bar in Fig.~\ref{fig:templates_tube6}(d).
This meta-cluster has low to medium side scatter and high CD45 expression, and therefore represents lymphocytes  (Fig.~\ref{fig:AML_blasts}).
Since lymphocytes are not affected by AML, this meta-cluster does not express any AML-related markers, and hence can be described as HLA-DR$^-$CD117$^-$CD34$^-$CD38$^-$,  as expected.
Fig.~\ref{fig:templates_tube6}(f) shows the expression profile of another meta-cluster shown in the red bar in Fig.~\ref{fig:templates_tube6}(d).
This meta-cluster consists of five cell populations from five AML samples (with subject ids  37, 58, 67, 89, and 117) and exhibits medium side scatter and CD45 expression and therefore, represents myeloid blast cells.
Furthermore, this meta-cluster is HLA-DR$^-$CD117$^+$CD34$^-$CD38$^+$, 
and represents a profile known to be that of Acute Promyelocytic Leukemia (APL)~\cite{paietta2003expression}.
APL is subtype M3 in the FAB classification of AML~\cite{bennett1985proposed}) and is characterized by chromosomal translocation of retinoic acid receptor-alpha (RAR$\alpha$) gene on chromosome 17 with the promyelocytic leukemia gene (PML) on chromosome 15, a translocation denoted as t(15;17). 
In the feature space of Tube 6, these APL samples are similar to each other while significantly different from the other  AML samples.
The template-based classification algorithm groups these samples together in the subtree colored red in the AML template tree shown in Fig.~\ref{fig:templates_tube6}(b).

\subsection{Identifying meta-clusters symptomatic of AML}  
\label{sec:metaclusters}
In each tube, meta-clusters are registered across the AML and healthy templates using the mixed edge cover (MEC) algorithm.
The unmatch-penalty $\lambda$ is set to $\sqrt7$ so that a pair of meta-clusters get matched only if the average squared deviation across all dimensions is less than one.
Meta-clusters in the AML template that are not matched to any meta-clusters in the healthy template represent  the abnormal, AML-specific immunophenotypes while the matched meta-clusters represent healthy  or non-AML-relevant cell populations.         
Table~\ref{tab:unmatched_mc} lists several unmatched meta-clusters indicative of AML from different tubes. 
As expected, every unmatched meta-cluster displays medium side scatter and CD45 expression characteristic of myeloid blast cells, and therefore we can omit FS, SS, and CD45 values in Table~\ref{tab:unmatched_mc}.
I briefly discuss the immunophenotypes represented by each AML-specific meta-cluster in each tube,
omitting the isotype control Tube 1 and unstained Tube 8.

\begin{table}[!t]
\centering
  \vspace{-.7 cm}
\caption[Meta-clusters characteristic of AML for the 23 AML samples in the training set]{Some of the meta-clusters characteristic of AML for the 23 AML samples in the training set.
In the second column, `$-$',  `low', and  `$+$' denote very low, low and high,
abundance of a marker, respectively, and $\pm$ denotes a marker that is positively expressed by some samples 
and  negatively expressed by others. 
The number of samples participating in a meta-cluster is shown in the third column.
The average fraction of cells in a sample participating in a  meta-cluster, and the standard deviation,  are shown in the fourth column.}
\label{tab:unmatched_mc}
\begin{tabular}{cccc}\toprule
Tube   & Marker expression & \#Samples & Fraction of cells \\ \toprule
	2  \ \  & Kappa$^\text{low}$Lambda$^\text{low}$CD19$^+$CD20$^-$ & 5 &  $6.3\%(\pm 6.8)$ \\ 
	3  \ \  & CD7$^+$CD4$^-$CD8$^-$CD2$^-$ & 4 &  $18.0\%(\pm 4.8)$ \\ 
	4  \ \  & CD15$^-$CD13$^+$CD16$^-$CD56$^-$ & 17 &  $16.6\%(\pm 6.9)$ \\ 
	4  \ \  & CD15$^-$CD13$^+$CD16$^-$CD56$^+$ & 8 &  $11.1\%(\pm 5.7)$ \\ 
	5  \ \  & CD14$^-$CD11c$^-$CD64$^-$CD33$^+$ & 10 &  $13.5\%(\pm 5.2)$ \\ 
	5 \ \  & CD14$^-$CD11c$^+$CD64$^-$CD33$^+$ & 18 &  $10.8\%(\pm 3.8)$ \\ 
	5  \ \  & CD14$^\text{low}$CD11c$^+$CD64$^\text{low}$CD33$^+$ & 6 &  $13.8\%(\pm 4.3)$ \\ 
	6  \ \  & HLA-DR$^+$CD117$^+$CD34$^+$CD38$^+$ & 11 &  $13.3\%(\pm 2.6)$ \\ 
	6  \ \  & HLA-DR$^+$CD117$^{\pm}$CD34$^+$CD38$^+$ & 13 &  $17.3\%(\pm 6.6)$ \\ 
	6  \ \  & HLA-DR$^-$CD117$^{\pm}$CD34$^-$CD38$^+$ & 5 &  $12.9\%(\pm 4.7)$ \\ 
	7  \ \  & CD5$^-$CD19$^+$CD3$^-$CD10$^-$ & 3 &  $12.3\%(\pm 2.4)$ \\ 
	7  \ \  & CD5$^+$CD19$^-$CD3$^-$CD10$^-$ & 3 &  $10.0\%(\pm 8.5)$ \\ 
	7  \ \  & CD5$^-$CD19$^-$CD3$^-$CD10$^+$ & 1 &  $9.9\%$ \\ 
     	   \bottomrule
\end{tabular}
\end{table}

\begin{figure}[!t]
\centering
\includegraphics[scale=.35]{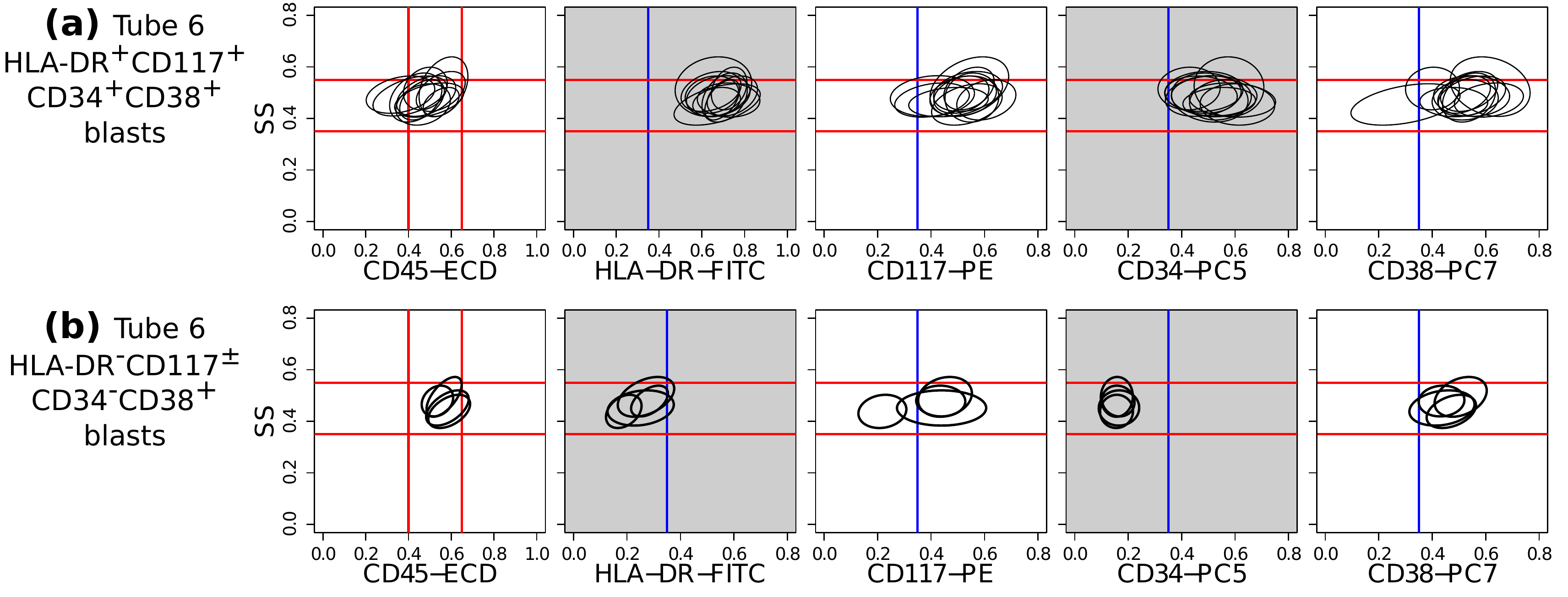}
\vspace{-.2cm}
\caption[Bivariate contour plots (side scatter vs. individual marker) for two meta-clusters indicative of AML]{Bivariate contour plots (side scatter vs. individual marker) for two meta-clusters (one in each row) indicative of AML.
The ellipses in a subplot denote the 95th quantile contour lines of cell populations included in the corresponding meta-cluster.
Myeloblast cells have medium side scatter (SS) and CD45 expressions.
The red lines indicate approximate myeloblast boundaries (located on the left-most subfigures in each row and extended horizontally to the  subfigures on the right) and confirm that these meta-clusters represent immunophenotypes of myeloblast cells. 
Blue vertical lines denote the +/- boundaries of a marker. 
Gray subplots show contour plots of dominant markers defining the meta-cluster in the same row. 
{\bf (a)} HLA-DR$^+$CD117$^+$CD34$^+$CD38$^+$ meta-cluster shared by 11 AML samples in Tube 6.
{\bf (b)}  HLA-DR$^-$CD117$^{\pm}$CD34$^-$CD38$^+$ meta-cluster shared by 5 AML samples in Tube 6. This meta-cluster is indicative of acute promyelocytic leukemia (APL).
These bivariate plots are shown for illustration only, 
since the populations of specific cell types are identified from seven-dimensional data. }
\vspace{-.5cm}
\label{fig:mc}
\end{figure}

{\bf Tube 6} is the most important panel for diagnosing AML since it includes several markers expressed by AML blasts.
HLA-DR is an MHC class II cell surface receptor complex that is expressed on 
antigen-presenting cells, e.g., B cells, dendritic cells, macrophages, and activated T cells.
It is expressed by myeloblasts in most subtypes of AML except M3 and M7~\cite{campana2000immunophenotyping}. 
CD117 is a tyrosine kinase receptor (c-KIT) expressed in blasts of some cases ($30-100\%$) of AML~\cite{campana2000immunophenotyping}.
CD34 is a cell adhesion molecule expressed on different stem cells and on the blast cells of many cases of AML ($40\% $) \cite{mason2006immunophenotype}.
CD38 is a glycoprotein found on the surface of blasts of several subtypes of AML but  usually not expressed in the 
M3 subtypes of AML~\cite{keyhani2000increased}.
In Tube 6, the matching algorithm has identified two AML-specific meta-clusters with high expressions of HLA-DR and CD34.
One of them also expresses CD117 and CD34, and Fig.~\ref{fig:mc}(c) shows the bivariate contour plots of the cell populations contained in this meta-cluster.
The second meta-cluster expresses positive but low levels of CD117 and CD34.
These two HLA-DR$^+$CD34$^+$ meta-clusters together are present in  $18$ out of the $23$  training AML samples.
The remaining five samples (subject id: 5, 7, 103, 165, 174) express HLA-DR$^-$CD117$^{\pm}$CD34$^-$CD38$^+$ myeloblasts, which is an immunophenotype of APL~\cite{paietta2003expression} as was discussed earlier.
Fig.~\ref{fig:mc}(d) shows the bivariate contour plots of this APL-specific meta-cluster.

{\bf Tube 5} contains several antigens typically expressed by AML blasts, of which CD33 is the most important. 
CD33 is a transmembrane receptor protein usually expressed on immature myeloid cells of the majority of cases of AML ($91\%$  reported in~\cite{legrand2000immunophenotype}).
The AML specific meta-clusters identified from markers in Tube 5 
(see Table~\ref{tab:unmatched_mc}) include CD33$^+$ myeloblasts from 
every sample in the training set.
Several of the CD33$^+$ populations also express CD11c, a type I transmembrane protein found on monocytes, macrophages and neutrophils.
CD11c is usually expressed by blast cells in acute myelomonocytic leukemia (M4 subclass of AML),
and acute monocytic leukemia (M5 subclass of AML)~\cite{campana2000immunophenotyping}.
Therefore CD14$^-$CD11c$^+$CD64$^-$CD33$^+$ meta-cluster could represent patients with M4 and M5 subclasses of AML.
I show the bivariate contour plots of this meta-cluster in Fig.~\ref{fig:mc}(b) .

{\bf Tube 4} includes several markers usually expressed by AML blasts, of which CD13 is the most important. 
CD13 is a zinc-metalloproteinase enzyme that binds to the cell membrane and degrades regulatory peptides~\cite{mason2006immunophenotype}.
CD13 is expressed on the blast cells of the majority of cases of AML (95\% as reported in \cite{legrand2000immunophenotype}).
Table~\ref{tab:unmatched_mc} shows two AML-specific  meta-clusters  detected from the blast cells 
in Tube 4.
Both of the meta-clusters are CD13$^+$ and they include populations from every AML sample in the training set.  
In addition to CD13, eight AML samples express CD56 glycoprotein that is naturally expressed on NK cells, a subset of CD4$^+$ T cells and a subset of CD8$^+$ T cells. 
Raspadori et al.~\cite{raspadori2001cd56} reported that CD56 was more often expressed by myeloblasts in FAB subclasses  M2 and M5, which covers about 42\% of AML cases in a study by Legrand et al.~\cite{legrand2000immunophenotype}.
In this dataset, we observe more AML samples expressing CD13$^+$CD56$^-$ blasts than expressing CD13$^+$CD56$^+$ blasts, which conforms to the findings of Raspadori et al.~\cite{raspadori2001cd56}.
Fig.~\ref{fig:mc}(a) shows the bivariate contour plots of the CD13$^+$CD56$^-$ meta-cluster.
CD15 is a carbohydrate adhesion molecule (not a protein) expressed on mature glycoproteins
and CD16 is an Fc (Fragment, crystallizable) receptor protein expressed on mature NK cells, macrophages, and neutrophils.
They are not usually expressed by myeloblasts and were absent in the AML-related meta-clusters.

{\bf Tube 2} is a B cell panel measuring B cell markers CD19 and CD20,  
and Kappa ($\kappa$) and Lambda ($\lambda$), immunoglobulin light chains present on the surface of antibodies produced by B lymphocytes.
This panel is used in diagnosing diseases  such as B-cell lymphomas and acute lymphoblastic leukemia (ALL)~\cite{campana2000immunophenotyping}.
B-cell specific markers are occasionally co-expressed with myeloid antigens especially in FAB M2 subtype of AML (with chromosomal translocation t(8;21)) \cite{campana2000immunophenotyping, walter2010aberrant}.
In Tube 2, the matching algorithm has identified a meta-cluster in the myeloblasts that  expresses high levels of CD19 and low levels of Kappa and Lambda.
The five samples with subject ids  5, 7, 103, 165, and 174 participating in this meta-cluster possibly belong to the FAB-M2 subtype of AML.

{\bf Tube 3} is a T cell panel measuring T cell specific markers CD4, CD8, CD2, and CD7.
This panel is frequently used in diagnosing T-lineage ALL~\cite{campana2000immunophenotyping, mason2006immunophenotype}.
In significant minority of cases of AML (e.g., 37\% of cases reported in~\cite{legrand2000immunophenotype}), CD7 is aberrantly expressed by myeloblasts~\cite{campana2000immunophenotyping}.
The matching algorithm identifies a meta-cluster in the blast region, which includes four populations expressing CD7$^+$CD4$^-$CD8$^-$CD2$^-$.
This finding confirms that T cell antigens are infrequently expressed by AML blasts, and therefore, they are less useful in AML diagnosis and classification.

{\bf Tube 7} is a lymphocyte panel with several markers expressed on T and B lymphocytes and is less important in detecting AML since they are infrequently expressed by AML blasts.
In Tube 7, we obtain three samples with CD19$^+$, three samples with CD5$^+$ and a sample with CD10$^+$ myeloblast cells.
The meta-clusters in rows 11-13 in Table~\ref{tab:unmatched_mc} confirm that lymphocyte antigens are infrequently expressed by AML blasts and are less useful in AML diagnosis and classification~\cite{campana2000immunophenotyping, walter2010aberrant}.

\begin{table}[!t]
   \centering
     \vspace{-.7 cm}
\caption[Four statistical measures  evaluating the performance of the template-based classification in the AML data]{Precision (P), Recall (R), Sensitivity (S) and F-measure (F) of the template-based classification in the training set and test set of the AML data.
The statistical measures are computed for each tube separately and two combinations of tubes.  
\label{tab:classification_stats}}
\begin{tabular}{c cccc p{.2cm}cccc}\toprule
{\bf Tubes} & \multicolumn{4}{c}{\bf Training set} & & \multicolumn{4}{c}{\bf Test set} \\ 
\cmidrule(lr){2-5} \cmidrule(lr){7-10}
   		& P  & R & S & F && P & R              & S & F\\ \toprule
1 		& 	 1.00  & 	0.26  	& 	1.00  	& 	0.41 	&&	1.00	&     0.15     &	1.00    & 	0.26 \\
2 		&  	 0.86  & 	0.26  	& 	0.99  	& 	0.40 	&&	0.83      &     0.25     &    	0.99    & 	0.38 \\
3 		& 	 1.00  & 	0.52  	& 	1.00  	& 	0.69	&&	1.00      &     0.35     &    	1.00    & 	0.52 \\
4 		&  	 0.94  & 	0.74  	& 	0.99  	& 	0.83 	&&	1.00      &     0.75     &  	1.00    &  	0.86\\
5 		&  	 0.75  & 	0.91  	& 	0.96  	& 	0.82 	&&	0.65      &     0.85     &   	0.94    & 	0.74\\
6 		&  	 1.00  & 	0.70  	& 	1.00  	& 	0.82 	&&	1.00      &     0.80     &   	1.00    & 	0.89\\
7 		&  	 0.52  & 	0.48  	& 	0.94  	& 	0.50 	&&	0.48      &     0.60     &    	0.92    & 	0.53\\ \midrule
All (2-7)     &  	 1.00  & 	0.74  	& 	1.00 	& 	0.85	&&	1.00      &     0.85     &   	1.00    & 	0.92 \\
4,5,6 	& 	 1.00  & 	0.96  	& 	1.00  	& 	0.98	&&	1.00      &     1.00     &   	1.00    &	1.00 \\
     	   \bottomrule
\end{tabular}
\end{table}

\subsection{Impact of each tube in the classification}
As discussed in the methods section, I build six independent classifiers based on the healthy and AML templates created from Tubes 2-7 of the AML dataset. 
Tubes 1 and 8 are omitted since the former is an isotope control and the latter is unstained, and therefore, they are uninformative for classification.
A sample is classified as an AML sample if the classification score is positive, and as a healthy sample otherwise. 
Let true positives (TP) be the number of AML samples correctly classified, true negatives (TN) be the number of healthy samples correctly classified, false positives (FP) be the number of healthy samples incorrectly classified as AML, and false negatives (FN) be the number of AML samples incorrectly classified as healthy.
Then, I evaluate the performance of each template-based classifier with the well-known four statistical measures: Precision, Recall(Sensitivity), Specificity, and F-value, defined as 
$\text{Precision}=\frac{\text{TP}}{\text{TP}+\text{FP}}$,  $\text{Recall(Sensitivity)}=\frac{\text{TP}}{\text{TP}+\text{FN}}$,
$\text{Specificity}=\frac{\text{TN}}{\text{FP}+\text{TN}}$, and  
$\text{F-value}= \frac{2(\text{Precision} \times \text{Recall})}{\text{Precision} + \text{Recall}}$.
These four measures take values in the interval [0,1], and the higher the values the better the classifier.

\begin{figure}[!t]
\centering
\includegraphics[scale=.44]{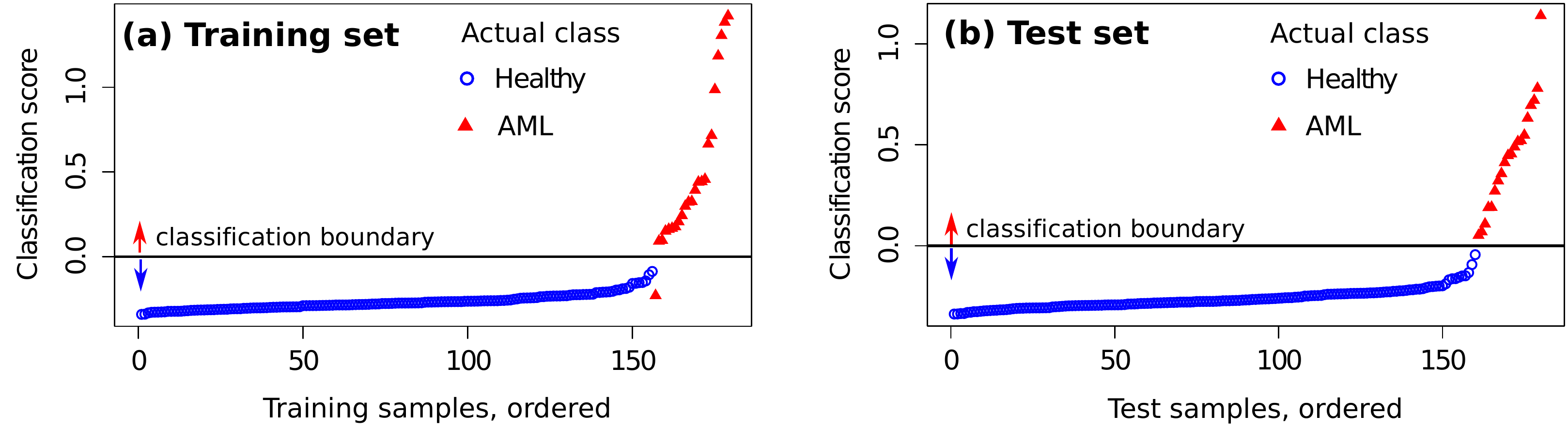}
\vspace{-.2cm}
\caption[Classification of the AML and healthy samples]{Average classification score from Tubes 4,5,6 for each sample in the (a) training set and (b) test set. 
Samples with scores above the horizontal line are classified as  AML, and as  healthy otherwise.
The actual class of each sample is also shown. 
An AML sample (subject id 116) is always misclassified in the training set, and this is discussed
in the text.
}
 \vspace{-.5cm}
\label{fig:classificationTrainTest}
\end{figure}

First, I evaluate the impact of each tube in the classification of the training samples.
For a training sample $X$, the classification score is computed by comparing it with the healthy and AML templates created from the training set after removing $X$.
The predicted status of $X$ is then compared against true status to evaluate the classification accuracy.
Table \ref{tab:classification_stats} (left panel) shows various statistical measures for the classifiers defined in Tubes~2-7 of the training set.
The classifiers based on Tubes 4, 5, and 6 have the highest sensitivity because  these tubes include several markers relevant to AML diagnosis~\cite{campana2000immunophenotyping, paietta2003expression}.
The number of true negatives TN is high in every tube since the identification of healthy samples does not depend on the detection of AML-specific markers. 
Hence specificity is close to one for all tubes.
Analogously, FP is low for most tubes, and we observe high precision for most tubes. 
The F-value is a harmonic mean of precision and recall,  and denotes the superior 
classification ability of markers in Tubes 4-6.
Averaging scores from all tubes does not improve the sensitivity and F-value dramatically.
However, combining Tubes 4-6 gives almost perfect classification  
with one misclassification for the training set.  
I plot the average classification scores from Tubes 4-6 for the training samples in Fig.~\ref{fig:classificationTrainTest}(a).
On the x-axis, the healthy samples are placed first followed by the AML samples, and samples in each group  are placed in the ascending order of the average classification score.  
The class labels of samples are also shown (blue circles for healthy and red triangles for AML samples).

\begin{figure}[!t]
\centering
\includegraphics[scale=.5]{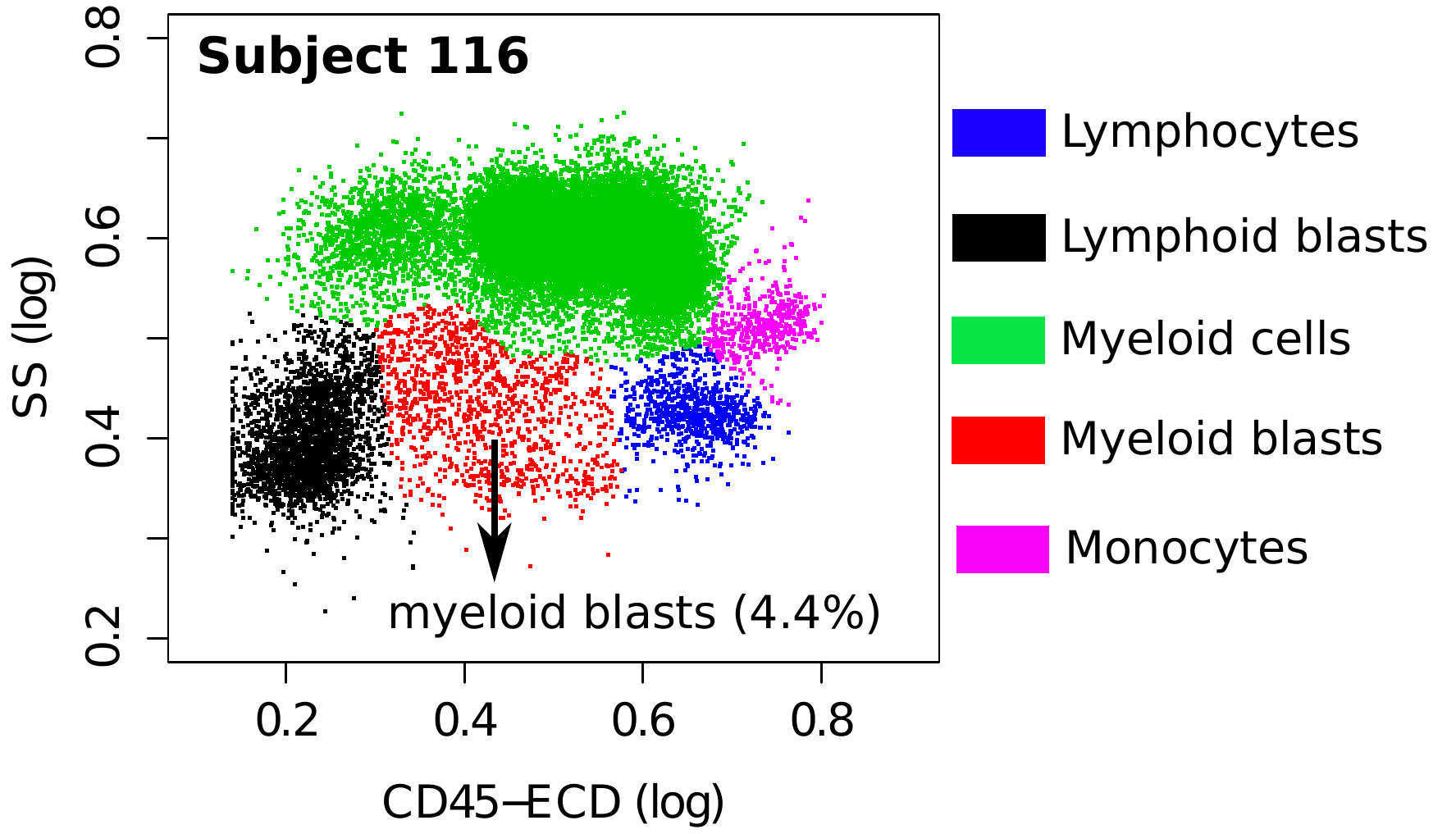}
\caption[Cell populations in a misclassified AML sample]{Cell populations in a sample from subject 116.
Even though this sample is marked as AML by the organizers of DREAM6/FlowCAP2 competition, it contains only 4.4\% myeloid blast cells (shown in red).
According to the World Health Organization (WHO) guidelines, AML is only confirmed when peripheral blood contains at least 20\%  immature myeloblasts.
Hence, this subject is either a rare AML subtype, or has been incorrectly labeled. 
}
 \label{fig:subject116}
\end{figure}

In Fig.~\ref{fig:classificationTrainTest}(a), we observe an AML sample (subject id 116) with score below the classification boundary.
Fig.~\ref{fig:subject116} shows the cell populations present in subject 116 projected on the SS and CD45 channels.
In this subject, the proportion of myeloid blasts is 4.4\%, which is lower than the minimum 20\% AML blasts necessary to recognize a patient to be AML-positive according to the WHO guidelines~\cite{estey2006acute} (the FAB threshold is even higher,  at $30\%$). 
(Recall  Fig.~\ref{fig:AML_blasts} for a plot of the cell types in healthy and AML samples.)
Hence this is either a rare case of AML,  or  one with minimal residual disease after therapy, or perhaps it was incorrectly labeled as AML  in the training set. 
Since the template-based classifier weighs individual clusters with their cell proportions, the abnormal myelobasts contribute insignificantly in the final score computed by Eq.~\ref{eq:classif_score}.
As a result, Subject 116 is placed with the healthy samples in Fig.~\ref{fig:classificationTrainTest}. 
Subject 116 was classified with the healthy samples by methods in other published work~\cite{biehl2013analysis}.

\subsection{Classifying test samples}
Now we turn to the test samples. 
For each tube, I compute the classification score for each sample in the test set using templates created from the training set and applying Eq.~\ref{eq:classif_score}.
Since the average classification score from Tubes 4-6 performs best for the training set, it is used as a classifier for the test set as well.
Since the status of  test samples was  released after the DREAM6/FlowCAP2 challenge, we can determine the classification accuracy of the test samples. 
Fig.~\ref{fig:classificationTrainTest}(b) shows the classification scores of the test samples, where samples are placed in ascending order of classification scores.
In Fig.~\ref{fig:classificationTrainTest}(b), we observe  perfect classification in the test set.
Similar to the training set, I tabulate statistical measures for  the classifiers  in Table~\ref{tab:classification_stats}.

When classifying a sample $X$, we can assume the null hypothesis: $X$ is healthy (non-leukemic).
The sample $X$ receives a positive score  if it contains  AML-specific immunophenotypes, and  the higher the score,  the stronger the evidence against the null hypothesis.
Since Tube 1 (isotype control) does not include any AML-specific markers, it can provide a background distribution for the  classification scores. 
In Tube 1,  174 out of 179 training samples have negative classification scores, but  five samples have positive scores, with values  less than $0.2$. 
In the best classifier designed from Tubes 4, 5, 6, we observe that two AML-positive samples in the training set and three AML-positive samples in the test set have scores between 0 and 0.2. 
The classifier is relatively less confident about these samples;  nevertheless, the p-values of these five samples (computed from the distribution in Tube 1) are still small ($<0.05$),  so that they can be classified as AML-positive. 
The rest of the AML samples in the training and test sets have scores greater than $0.2$ and the classifier is quite confident about their status (p-value zero).

Four AML samples in the test set (ids 239, 262, 285, and 326) were subclassified as APL by comparing against distinct  template trees for APL and the other AML  samples in the training set (cf. Fig.~\ref{fig:templates_tube6} (b)). 

Finally, I state the computational times required on an iMac with four 2.7 GHz cores and 8 GB memory. 
The code is written in C++ and R. 
Consider a single tube with $359$ samples in it. 
The $k$-means clustering of all samples took one hour, primarily because we need to 
run the algorithm multiple times (about ten on the average) to  find the optimal value of the 
number of clusters. 
Creating the healthy template from $156$ samples in the training set required $10$ seconds (s)  on one core, and the AML template for $23$ AML samples took $0.5$s on one core. 
Cross validation (leave one out) of the training set took $30$ minutes, and computing the classification score for the $180$ test samples took $15$s, both  on four cores. 
I could have reduced  the running time by executing the code in parallel on more cores. 
In an ongoing project, we have made the dominant step, the $k$-means clustering of all the samples with an optimal number of clusters, faster using a GPU, reducing the total time to a few minutes. 

\section{Conclusions}
In this chapter, I have demonstrated that an algorithmic pipeline for template-based classification can successfully
identify  immunophenotypes of clinical interest in AML. 
These could be used to differentiate the subtypes of AML, which is advantageous since prognosis and treatment depends on the subtype. 
The templates enable us to classify AML samples in spite of their  heterogeneity. 
This was accomplished by creating a scoring function that accounts for the 
subtleties in cell populations within AML samples.
We are currently applying  this approach to a larger AML data set,  and intend to analyze  other heterogeneous data sets.  

%% file: ch7-conclusions.tex
\chapter{Conclusions and future work}
\label{chap:conclusions-future-work}
\section{Conclusions}
In my dissertation, I have explored, designed and developed a collection of algorithms 
for analyzing multi-parametric flow cytometry data.
These algorithms solve sample preprocessing, variance stabilization, clustering, population registration, meta-clustering, sample classification, and other related problems arising in FC data analysis. 
I have assembled the algorithms into a data analysis pipeline, \emph{flowMatch}\cite{manual_flowMatch}, that is made available as an open-source R package in Bioconductor ({\tt http://www.bioconductor.org/}).  
The pipeline has been employed to analyze several practical datasets and the results are published in peer-reviewed journals and conferences~\cite{azad2010identifying, azad2012matching, azad2013classifying}.

Flow cytometry (FC) is a widely used platform for measuring phenotypes of individual cells from millions of cells in biological samples. 
Modern FC technology gives rise to high-dimensional and high-content data that challenges the ability to manually investigate the data and interpret the results.
To address the limitations of manual analysis, researchers have developed a new branch of research called the ``computational cytometry".  
Along with a number of contemporary software tools~\cite{aghaeepour2013critical, pyne2009automated, spidlen2013genepattern, Bagwell1993, finak2010optimizing, ray2012computational, lo2008automated, aghaeepour2011rapid, robinson2012computational, hahne2009flowcore}, \emph{flowMatch} automates different analysis steps of FC data with an aim to prevent the data analysis from being the bottleneck in scientific discovery based on cytometry.

The \emph{flowMatch} pipeline consists of six well defined algorithmic modules for
(1) unmixing of spectra (compensation) to remove the effect of overlapping fluorescence channels, 
(2) transforming data  in order to stabilize variance and normalize data, 
(3) identifying cell population by automated gating or clustering algorithm in the high-dimensional marker space,
(4) registering cell populations (cluster labeling or matching) across samples,
(5) representing a class of samples with high-level templates, and 
(6) classifying  samples based on the phenotypic pattern of the cell clusters.
Each module of \emph{flowMatch} is designed to perform a specific task independent of other modules of the pipeline.
However, they can also be employed sequentially in the order described in Fig.~\ref{fig:pipeline} to perform the complete data analysis.

I have developed a variance stabilization (VS) method called \emph{flowVS} for removing the mean-variance correlation arisen in fluorescence based flow cytometry.
\emph{flowVS} stabilizes variance by transforming FC samples with the inverse hyperbolic sine function whose parameters are estimated by minimizing Bartlett's likelihood-ratio statistics. 
For population identification, I demonstrate that several simple, off-the-shelf clustering algorithms provide more accurate and robust partition of an FC sample. 
I describe several cluster validation methods that can be used to select the optimum parameters for a clustering algorithm (e.g., the optimum number of clusters), as well as choosing the best algorithm for a dataset.
Furthermore, I describe an algorithm for computing the consensus of multiple clustering solutions, which performs better than any individual algorithm.

For population registration~\cite{pyne2009automated, azad2012matching}, I developed a robust mixed edge cover (MEC) algorithm that matches cell populations across a pair of FC samples. 
The MEC algorithm uses a robust graph-theoretic framework to match a cluster from a sample to zero or more clusters in another sample and thus solves the problem of missing or splitting populations.
Next, I describe a hierarchical algorithm for encapsulating a collection of sample belonging to the same class with a \emph{template}.
A  \emph{template} is a collection of relatively homogeneous \emph{meta-clusters} commonly shared across samples of a given class, thus describing the key immunophenotypes of an overall class of samples. 
Finally, I demonstrate that the use of templates leads to efficient classification algorithms.
By comparing a sample with class templates, the sample is predicted to come from a class whose template it  is most similar to. 
The template-base classification is robust and efficient because it compares samples to cleaner and fewer class templates rather than the large number of noisy samples themselves.

I have employed different components of \emph{flowMatch} for classifying leukemia samples, evaluating the phosphorylation effects on T cells, classifying healthy immune profiles, comparing the impact of two treatments against Multiple Sclerosis, and classifying the HIV vaccination status. 
In these analyses, \emph{flowMatch} is able to reach biologically meaningful conclusions with the automated algorithms.
The algorithms included in \emph{flowMatch}  can also be applied to problems outside of flow cytometry such as from microarray data analysis and image recognitions.

\section{Future work}
{\em  Application to larger and higher-dimensional datasets:}
Thus far, I have  employed \emph{flowMatch} to five FC datasets, the largest of which is an AML dataset with nearly 3000 seven-dimensional samples. 
More rigorous testing with other datasets will certainly increase the confidence about the correctness and efficiency of the pipeline.  
I am in the process of getting access to large cytomics data from Immune Tolerance Network ({\tt http://www.immunetolerance.org/}), FlowRepository ({\tt http://flowrepository.org/}), Cytobank ({\tt http://www.cytobank.org}), and from other public repositories. 
I have tested the \emph{flowMatch} pipeline with up to eight dimensional samples. 
However, the newly developed mass cytometry technology~\cite{bendall2011single} can investigate more than 40 parameters of a single cell.
\emph{flowMatch} needs to be tested and enhanced (if needed) for this high dimensional data.

{\em  Enhancing the modules of the \emph{flowMatch} pipeline:}
The \emph{flowMatch} pipeline is still under development.
I plan to enhance different components of the pipeline with additional functionalities from which the users can choose depending on the objective of the data analysis. 
For variance stabilization, in addition to the inverse hyperbolic sine (asinh) function, several other transformations can be tested.
It also remains a future work to design and implement a multi-dimensional variance stabilization algorithm. 
In current work, I am incorporating the proportion of cells into the population registration algorithm, which will lead to a between-sample distance metric similar in spirit of the earth mover's distance~\cite{rubner2000earth, zimmerman2011}.
For the meta-clustering algorithm, I investigate the  use of networks instead of trees to organize the templates, similar in spirit to the use of  networks rather than trees in phylogenetics~\cite{Dress+:book}. 
Finally, I am considering other classification algorithms, such as the logistic regression~\cite{hosmer2013applied}, for complementing the template-based classification methods.

{\em  Parallelizing algorithms for faster processing:}
At present, the modules of \emph{flowMatch} are implemented as serial algorithms.
However, given the ever increasing volume of data and the availability of high-performance computers, the algorithms of the pipeline can be parallelized to analyze a dataset faster without incurring any additional cost.
I consider parallelizing \emph{flowMatch} at different levels.
The variance stabilization algorithm is embarrassingly parallel where density peaks on different channels can be identified and their likelihood-ratio test can be evaluated on different processing units.  
A number of parallel clustering algorithms have been discussed in the literature~\cite{zhao2009parallel, foti2000scalable, li1989parallel} and they can be integrated into the \emph{flowMatch} pipeline.
The quality of partitions can be evaluated in parallel for different choices of clustering parameters. 
The population registration problem can be solved faster by parallelizing the mixed edge cover algorithm. 
I have already developed algorithms for parallel cardinality matching on bipartite graphs \cite{azad2012multithreaded, azad2012ipdps, azad2011computing}.
Developing parallel algorithms for weighted matching problem remains a future work.
However, the number of populations is generally small (tens to hundreds) in typical FC samples, and therefore parallel population registration might not decrease the processing time significantly.
The samples are processed in a sequential order by the meta-clustering algorithm, which is difficult to parallelize. 
However, we can relax the strict order of sample processing, and then parallelize the hierarchical algorithm.
In the dynamic classification algorithm, we can insert multiple samples into the template tree simultaneously whenever the inserted samples update different parts of the template-tree.

{\em  Applying the pipeline to problem outside of the flow cytometry:}
Stabilizing variance, clustering, matching, creating templates are general concepts with applications to different areas. 
Therefore, the algorithms developed in this dissertation can be applied -- possibly with simple modifications-- to problems outside of flow cytometry.
I have already applied the variance stabilization framework to microarray data and compared the results with a state-of-the-art software developed for microarrays in Chapter~\ref{chap:variance_stabilization}.
Likewise, other algorithms have applications to problems from microarrays,  ChIp-Seq, etc. 

FC data projected on a lower dimension has considerable similarities with images  from tradition photography and from bio-imaging technologies such as imaging cytometry, MRI, etc.
Therefore, the algorithms in \emph{flowMatch} have direct application in analyzing images from these technologies.
For example, consider an image recognition application where the clustering algorithm segments images into different features (eyes, nose, etc.), the matching algorithm registers different features across images, and the classification algorithm recognizes the images with templates created from existing image library.
This image recognition problem can be solved by the clustering, matching and classification algorithms from the \emph{flowMatch} pipeline.

I expect this dissertation to be a strong algorithmic contribution with application to flow cytometry, as well as to domains outside of flow cytometry.

%% file: vita.tex
\begin{vita}

Ariful Azad was born in Ullapara, in the district of Sirajganj, Bangladesh on September 1, 1982.
He is the youngest son of Anwar Hossain and Ayesha Begum.
He finished high school from Pabna Cadet College in 2000 and completed his Bachelors in Computer Science and Engineering from Bangladesh University of Engineering and Technology (BUET) in 2006. 
Ariful started his Ph.D. in Computer Science at Purdue University in August 2008 under the supervision of Prof. Alex Pothen. 
His research interests include bioinformatics, graph algorithms, and parallel computing.
His Ph.D. research has been published to several peer reviewed journals and conferences.

Ariful is a recipient of an IBM PhD fellowship (2013), outstanding poster award (Cyto 2013), fellowship incentive grant (2010), Purdue research fellowship (2010) and several travel awards from NSF, NIH and ISAC.
He worked as a summer intern at Pacific Northwest National Laboratory (PNNL) in summer 2010, 2011.
During his study at Purdue, he served as the president of Bangladesh students organization (2012-13) and the webmaster of Purdue graduate student government (2009-11).

Ariful is married to Rubaya Pervin who is also a graduate of Bangladesh University of Engineering and Technology (BUET).
His favorite sports are Cricket and Tennis.
He enjoys to travel and watch movies in his leisure time. 

\end{vita}

%% file: thesis.bbl
\begin{thebibliography}{100}

\bibitem{coico2009immunology}
R.~Coico and G.~Sunshine, {\em Immunology: A Short Course}.
\newblock Wiley-Blackwell, 2009.

\bibitem{mellors1997plasma}
J.~W. Mellors, A.~Munoz, J.~V. Giorgi, J.~B. Margolick, C.~J. Tassoni,
  P.~Gupta, L.~A. Kingsley, J.~A. Todd, A.~J. Saah, R.~Detels, {\em et~al.},
  ``Plasma viral load and {CD4}$^+$ lymphocytes as prognostic markers of
  {HIV}-1 infection,'' {\em Annals of Internal Medicine}, vol.~126, no.~12,
  pp.~946--954, 1997.

\bibitem{maecker2012standardizing}
H.~T. Maecker, J.~P. McCoy, and R.~Nussenblatt, ``Standardizing
  immunophenotyping for the human immunology project,'' {\em Nature Reviews
  Immunology}, vol.~12, no.~3, pp.~191--200, 2012.

\bibitem{van2012euroflow}
J.~Van~Dongen, L.~Lhermitte, S.~B{\"o}ttcher, J.~Almeida, V.~Van~der Velden,
  J.~Flores-Montero, A.~Rawstron, V.~Asnafi, Q.~L{\'e}crevisse, P.~Lucio, {\em
  et~al.}, ``Euroflow antibody panels for standardized n-dimensional flow
  cytometric immunophenotyping of normal, reactive and malignant leukocytes,''
  {\em Leukemia}, vol.~26, no.~9, pp.~1908--1975, 2012.

\bibitem{shapiro2005practical}
H.~M. Shapiro, {\em Practical Flow Cytometry}.
\newblock Wiley-Liss, 2005.

\bibitem{lugli2010data}
E.~Lugli, M.~Roederer, and A.~Cossarizza, ``{Data analysis in flow cytometry:
  The future just started},'' {\em Cytometry Part A}, vol.~77, pp.~705--713,
  2010.

\bibitem{bendall2011single}
S.~C. Bendall, E.~F. Simonds, P.~Qiu, D.~A. El-ad, P.~O. Krutzik, R.~Finck,
  R.~V. Bruggner, R.~Melamed, A.~Trejo, O.~I. Ornatsky, {\em et~al.},
  ``Single-cell mass cytometry of differential immune and drug responses across
  a human hematopoietic continuum,'' {\em Science}, vol.~332, no.~6030,
  pp.~687--696, 2011.

\bibitem{perfetto2004seventeen}
S.~P. Perfetto, P.~K. Chattopadhyay, and M.~Roederer, ``Seventeen-colour flow
  cytometry: Unravelling the immune system,'' {\em Nature Reviews Immunology},
  vol.~4, no.~8, pp.~648--655, 2004.

\bibitem{peters2011leukemia}
J.~M. Peters and M.~Q. Ansari, ``Multiparameter flow cytometry in the diagnosis
  and management of acute leukemia,'' {\em Archives of Pathology \& Laboratory
  Medicine}, vol.~135, no.~1, pp.~44--54, 2011.

\bibitem{seder2008t}
R.~A. Seder, P.~A. Darrah, and M.~Roederer, ``T-cell quality in memory and
  protection: Implications for vaccine design,'' {\em Nature Reviews
  Immunology}, vol.~8, no.~4, pp.~247--258, 2008.

\bibitem{perfetto2006amine}
S.~P. Perfetto, P.~K. Chattopadhyay, L.~Lamoreaux, R.~Nguyen, D.~Ambrozak,
  R.~A. Koup, and M.~Roederer, ``Amine reactive dyes: An effective tool to
  discriminate live and dead cells in polychromatic flow cytometry,'' {\em
  Journal of Immunological Methods}, vol.~313, no.~1, pp.~199--208, 2006.

\bibitem{le2007data}
N.~Le~Meur, A.~Rossini, M.~Gasparetto, C.~Smith, R.~R. Brinkman, and
  R.~Gentleman, ``Data quality assessment of ungated flow cytometry data in
  high throughput experiments,'' {\em Cytometry Part A}, vol.~71, no.~6,
  pp.~393--403, 2007.

\bibitem{roederer2001spectral}
M.~Roederer, ``Spectral compensation for flow cytometry: Visualization
  artifacts, limitations, and caveats,'' {\em Cytometry}, vol.~45, no.~3,
  pp.~194--205, 2001.

\bibitem{Bagwell1993}
C.~B. Bagwell and E.~G. Adams, ``Fluorescence spectral overlap compensation for
  any number of flow cytometry parameters,'' {\em Annals of the New York
  Academy of Sciences}, vol.~677, no.~1, pp.~167--184, 1993.

\bibitem{Novo2013}
D.~Novo, G.~Gr\'egori, and B.~Rajwa, ``Generalized unmixing model for
  multispectral flow cytometry utilizing nonsquare compensation matrices,''
  {\em Cytometry Part A}, vol.~83, pp.~508--520, May 2013.

\bibitem{snow2004flow}
C.~Snow, ``Flow cytometer electronics,'' {\em Cytometry Part A}, vol.~57,
  no.~2, pp.~63--69, 2004.

\bibitem{bagwell2005hyperlog}
C.~Bagwell, ``Hyperlog -- {A} flexible log-like transform for negative, zero,
  and positive valued data,'' {\em Cytometry Part A}, vol.~64, no.~1,
  pp.~34--42, 2005.

\bibitem{parks2006new}
D.~Parks, M.~Roederer, and W.~Moore, ``A new logicle display method avoids
  deceptive effects of logarithmic scaling for low signals and compensated
  data,'' {\em Cytometry Part A}, vol.~69, no.~6, pp.~541--551, 2006.

\bibitem{dvorak2005modified}
J.~Dvorak and S.~Banks, ``Modified {B}ox-{C}ox transform for modulating the
  dynamic range of flow cytometry data,'' {\em Cytometry}, vol.~10, no.~6,
  pp.~811--813, 2005.

\bibitem{novo2008flow}
D.~Novo and J.~Wood, ``Flow cytometry histograms: Transformations, resolution,
  and display,'' {\em Cytometry Part A}, vol.~73, no.~8, pp.~685--692, 2008.

\bibitem{efron1982transformation}
B.~Efron, ``Transformation theory: How normal is a family of distributions?,''
  {\em The Annals of Statistics}, vol.~10, no.~2, pp.~323--339, 1982.

\bibitem{huber2002variance}
W.~Huber, A.~Von~Heydebreck, H.~S{\"u}ltmann, A.~Poustka, and M.~Vingron,
  ``Variance stabilization applied to microarray data calibration and to the
  quantification of differential expression,'' {\em Bioinformatics}, vol.~18,
  no.~suppl 1, pp.~S96--S104, 2002.

\bibitem{finak2010optimizing}
G.~Finak, J.~Perez, A.~Weng, and R.~Gottardo, ``Optimizing transformations for
  automated, high throughput analysis of flow cytometry data,'' {\em {BMC}
  Bioinformatics}, vol.~11, no.~1, p.~546, 2010.

\bibitem{ray2012computational}
S.~Ray and S.~Pyne, ``A computational framework to emulate the human
  perspective in flow cytometric data analysis,'' {\em PLoS One}, vol.~7,
  no.~5, p.~e35693, 2012.

\bibitem{chan2008statistical}
C.~Chan, F.~Feng, J.~Ottinger, D.~Foster, M.~West, and T.~B. Kepler,
  ``Statistical mixture modeling for cell subtype identification in flow
  cytometry,'' {\em Cytometry Part A}, vol.~73, no.~8, pp.~693--701, 2008.

\bibitem{lo2008automated}
K.~Lo, R.~Brinkman, and R.~Gottardo, ``Automated gating of flow cytometry data
  via robust model-based clustering,'' {\em Cytometry Part A}, vol.~73, no.~4,
  pp.~321--332, 2008.

\bibitem{pyne2009automated}
S.~Pyne, X.~Hu, K.~Wang, E.~Rossin, T.~Lin, L.~Maier, C.~Baecher-Allan,
  G.~McLachlan, P.~Tamayo, D.~Hafler, {\em et~al.}, ``Automated
  high-dimensional flow cytometric data analysis,'' {\em Proceedings of the
  National Academy of Sciences}, vol.~106, no.~21, pp.~8519--8524, 2009.

\bibitem{qian2010elucidation}
Y.~Qian, C.~Wei, F.~Eun-Hyung~Lee, J.~Campbell, J.~Halliley, J.~A. Lee, J.~Cai,
  Y.~M. Kong, E.~Sadat, E.~Thomson, {\em et~al.}, ``Elucidation of seventeen
  human peripheral blood {B}-cell subsets and quantification of the tetanus
  response using a density-based method for the automated identification of
  cell populations in multidimensional flow cytometry data,'' {\em Cytometry
  Part B: Clinical Cytometry}, vol.~78, no.~S1, pp.~S69--S82, 2010.

\bibitem{walther2009automatic}
G.~Walther, N.~Zimmerman, W.~Moore, D.~Parks, S.~Meehan, I.~Belitskaya, J.~Pan,
  and L.~Herzenberg, ``Automatic clustering of flow cytometry data with
  density-based merging,'' {\em Advances in Bioinformatics}, vol.~2009, 2009.

\bibitem{zare2010data}
H.~Zare, P.~Shooshtari, A.~Gupta, and R.~Brinkman, ``Data reduction for
  spectral clustering to analyze high throughput flow cytometry data,'' {\em
  {BMC} Bioinformatics}, vol.~11, no.~1, p.~403, 2010.

\bibitem{aghaeepour2011rapid}
N.~Aghaeepour, R.~Nikolic, H.~H. Hoos, and R.~R. Brinkman, ``Rapid cell
  population identification in flow cytometry data,'' {\em Cytometry Part A},
  vol.~79, no.~1, pp.~6--13, 2011.

\bibitem{aghaeepour2013critical}
N.~Aghaeepour, G.~Finak, H.~Hoos, T.~R. Mosmann, R.~Brinkman, R.~Gottardo,
  R.~H. Scheuermann, {\em et~al.}, ``Critical assessment of automated flow
  cytometry data analysis techniques,'' {\em Nature Methods}, vol.~10, no.~3,
  pp.~228--238, 2013.

\bibitem{jain1999data}
A.~K. Jain, M.~N. Murty, and P.~J. Flynn, ``Data clustering: A review,'' {\em
  {ACM} Computing Surveys ({CSUR})}, vol.~31, no.~3, pp.~264--323, 1999.

\bibitem{halkidi2001clustering}
M.~Halkidi, Y.~Batistakis, and M.~Vazirgiannis, ``On clustering validation
  techniques,'' {\em Journal of Intelligent Information Systems}, vol.~17,
  no.~2-3, pp.~107--145, 2001.

\bibitem{hornik2008hard}
K.~Hornik and W.~B{\"o}hm, ``Hard and soft euclidean consensus partitions,'' in
  {\em Data Analysis, Machine Learning and Applications}, pp.~147--154,
  Springer, 2008.

\bibitem{azad2012matching}
A.~Azad, S.~Pyne, and A.~Pothen, ``Matching phosphorylation response patterns
  of antigen-receptor-stimulated {T} cells via flow cytometry,'' {\em {BMC}
  Bioinformatics}, vol.~13, no.~Suppl 2, p.~S10, 2012.

\bibitem{azad2010identifying}
A.~Azad, J.~Langguth, Y.~Fang, A.~Qi, and A.~Pothen, ``Identifying rare cell
  populations in comparative flow cytometry,'' {\em Lecture Notes in Computer
  Science}, vol.~6293, pp.~162--175, 2010.

\bibitem{allison2006microarray}
D.~B. Allison, X.~Cui, G.~P. Page, and M.~Sabripour, ``Microarray data
  analysis: From disarray to consolidation and consensus,'' {\em Nature Reviews
  Genetics}, vol.~7, no.~1, pp.~55--65, 2006.

\bibitem{spidlen2013genepattern}
J.~Spidlen, A.~Barsky, K.~Breuer, P.~Carr, M.-D. Nazaire, B.~A. Hill, Y.~Qian,
  T.~Liefeld, M.~Reich, J.~P. Mesirov, P.~Wilkinson, R.~H. Scheuermann, R.-P.
  Sekaly, and R.~R. Brinkman, ``Genepattern flow cytometry suite,'' {\em Source
  Code for Biology and Medicine}, vol.~8, no.~1, p.~14, 2013.

\bibitem{kotecha2010web}
N.~Kotecha, P.~O. Krutzik, and J.~M. Irish, ``Web-based analysis and
  publication of flow cytometry experiments,'' {\em Current Protocols in
  Cytometry}, vol.~10, 2010.

\bibitem{azad2014flowMatch}
A.~Azad, B.~Rajwa, and A.~Pothen, ``{flowMatch}: {A}n algorithmic pipeline for
  flow cytometry data analysis,'' {\em Submitted}, 2014.

\bibitem{manual_flowMatch}
A.~Azad, {\em flowMatch: Matching and meta-clustering in flow cytometry}, 2013.
\newblock R package version 0.99.2.

\bibitem{gentleman2004bioconductor}
R.~C. Gentleman, V.~J. Carey, D.~M. Bates, B.~Bolstad, M.~Dettling, S.~Dudoit,
  B.~Ellis, L.~Gautier, Y.~Ge, J.~Gentry, {\em et~al.}, ``Bioconductor: Open
  software development for computational biology and bioinformatics,'' {\em
  Genome Biology}, vol.~5, no.~10, p.~R80, 2004.

\bibitem{maier2007allelic}
L.~Maier, D.~Anderson, P.~De~Jager, L.~Wicker, and D.~Hafler, ``Allelic variant
  in {CTLA4} alters {T} cell phosphorylation patterns,'' {\em Proceedings of
  the National Academy of Sciences}, vol.~104, no.~47, p.~18607, 2007.

\bibitem{azad2014immunophenotypes}
A.~Azad, B.~Rajwa, and A.~Pothen, ``Immunophenotypes of acute myeloid leukemia
  from flow cytometry data using templates,'' {\em ArXiv}, 2014.

\bibitem{azad2013classifying}
A.~Azad, A.~Khan, B.~Rajwa, S.~Pyne, and A.~Pothen, ``Classifying
  immunophenotypes with templates from flow cytometry,'' in {\em Proceedings of
  the International Conference on Bioinformatics, Computational Biology and
  Biomedical Informatics ({ACM BCB})}, p.~256, ACM, 2013.

\bibitem{azad2012ipdps}
A.~Azad, M.~Halappanavar, S.~Rajamanickam, E.~G. Boman, A.~Khan, and A.~Pothen,
  ``Multithreaded algorithms for maximum matching in bipartite graphs,'' in
  {\em {IEEE} Parallel \& Distributed Processing Symposium ({IPDPS})},
  pp.~860--872, IEEE, 2012.

\bibitem{azad2011computing}
A.~Azad, M.~Halappanavar, F.~Dobrian, and A.~Pothen, ``Computing maximum
  matching in parallel on bipartite graphs: Worth the effort?,'' in {\em
  Proceedings of the First Workshop on Irregular Applications: Architectures
  and Algorithm}, pp.~11--14, ACM, 2011.

\bibitem{azad2012multithreaded}
A.~Azad and A.~Pothen, ``Multithreaded algorithms for matching in graphs with
  application to data analysis in flow cytometry,'' in {\em {IEEE} Parallel and
  Distributed Processing Symposium Workshops \& PhD Forum ({IPDPSW})},
  pp.~2494--2497, IEEE, 2012.

\bibitem{waddell2009resampling}
P.~J. Waddell and A.~Azad, ``Resampling residuals: Robust estimators of error
  and fit for evolutionary trees and phylogenomics,'' {\em arXiv preprint
  arXiv:0912.5288}, 2009.

\bibitem{waddell2010resampling}
P.~J. Waddell, A.~Azad, and I.~Khan, ``Resampling residuals on phylogenetic
  trees: Extended results,'' {\em arXiv preprint arXiv:1101.0020}, 2010.

\bibitem{hahne2010per}
F.~Hahne, A.~Khodabakhshi, A.~Bashashati, C.~Wong, R.~Gascoyne, A.~Weng,
  V.~Seyfert-Margolis, K.~Bourcier, A.~Asare, T.~Lumley, {\em et~al.},
  ``Per-channel basis normalization methods for flow cytometry data,'' {\em
  Cytometry Part A}, vol.~77, no.~2, pp.~121--131, 2010.

\bibitem{Brockmeyer+:phosphorylation}
C.~Brockmeyer, W.~Paster, D.~Pepper, {\em et~al.}, ``{T Cell Receptor
  ({TCR})-induced tyrosine phosphorylation dynamics identifies {THEMIS} as a
  new {TCR} signalosome component},'' {\em Journal of Biological Chemistry},
  vol.~286, no.~9, pp.~7535--7547, 2011.

\bibitem{aghaeepour2012early}
N.~Aghaeepour, P.~K. Chattopadhyay, A.~Ganesan, K.~O'Neill, H.~Zare, A.~Jalali,
  H.~H. Hoos, M.~Roederer, and R.~R. Brinkman, ``Early immunologic correlates
  of {HIV} protection can be identified from computational analysis of complex
  multivariate {T}-cell flow cytometry assays,'' {\em Bioinformatics}, vol.~28,
  no.~7, pp.~1009--1016, 2012.

\bibitem{qiu2011extracting}
P.~Qiu, E.~F. Simonds, S.~C. Bendall, K.~D. Gibbs~Jr, R.~V. Bruggner, M.~D.
  Linderman, K.~Sachs, G.~P. Nolan, and S.~K. Plevritis, ``Extracting a
  cellular hierarchy from high-dimensional cytometry data with {SPADE},'' {\em
  Nature Biotechnology}, vol.~29, no.~10, pp.~886--891, 2011.

\bibitem{robinson2012computational}
J.~P. Robinson, B.~Rajwa, V.~Patsekin, and V.~J. Davisson, ``Computational
  analysis of high-throughput flow cytometry data,'' {\em Expert Opinion on
  Drug Discovery}, vol.~7, no.~8, pp.~679--693, 2012.

\bibitem{maecker2012new}
H.~T. Maecker, T.~M. Lindstrom, W.~H. Robinson, P.~J. Utz, M.~Hale, S.~D. Boyd,
  S.~S. Shen-Orr, and C.~G. Fathman, ``New tools for classification and
  monitoring of autoimmune diseases,'' {\em Nature Reviews Rheumatology},
  vol.~8, no.~6, pp.~317--328, 2012.

\bibitem{bashashati2009survey}
A.~Bashashati and R.~R. Brinkman, ``A survey of flow cytometry data analysis
  methods,'' {\em Advances in Bioinformatics}, vol.~2009, 2009.

\bibitem{hahne2009flowcore}
F.~Hahne, N.~LeMeur, R.~Brinkman, B.~Ellis, P.~Haaland, D.~Sarkar, J.~Spidlen,
  E.~Strain, and R.~Gentleman, ``flowcore: A bioconductor package for high
  throughput flow cytometry,'' {\em {BMC} Bioinformatics}, vol.~10, no.~1,
  p.~106, 2009.

\bibitem{sarkar2008using}
D.~Sarkar, N.~Le~Meur, and R.~Gentleman, ``Using flowviz to visualize flow
  cytometry data,'' {\em Bioinformatics}, vol.~24, no.~6, pp.~878--879, 2008.

\bibitem{lo2009flowclust}
K.~Lo, F.~Hahne, R.~Brinkman, and R.~Gottardo, ``flowclust: A bioconductor
  package for automated gating of flow cytometry data,'' {\em {BMC}
  Bioinformatics}, vol.~10, no.~1, p.~145, 2009.

\bibitem{hahne2010flowstats}
F.~Hahne, N.~Gopalakrishnan, A.~Khodabakhshi, and C.~Wong, {\em {flowStats}:
  Statistical methods for the analysis of flow cytometry data}.
\newblock R package version 3.20, available at http://www.bioconductor.org.

\bibitem{scheuermann2009immport}
R.~Scheuermann, Y.~Qian, C.~Wei, and I.~Sanz, ``{ImmPortFLOCK}: Automated cell
  population identification in high dimensional flow cytometry data,'' {\em The
  Journal of Immunology}, vol.~182, no.~Meeting Abstracts 1, pp.~42--17, 2009.

\bibitem{schena1995quantitative}
M.~Schena, D.~Shalon, R.~W. Davis, and P.~O. Brown, ``Quantitative monitoring
  of gene expression patterns with a complementary {DNA} microarray,'' {\em
  Science}, vol.~270, no.~5235, pp.~467--470, 1995.

\bibitem{chen1997ratio}
Y.~Chen, E.~R. Dougherty, and M.~L. Bittner, ``Ratio-based decisions and the
  quantitative analysis of {cDNA} microarray images,'' {\em Journal of
  Biomedical optics}, vol.~2, no.~4, pp.~364--374, 1997.

\bibitem{durbin2002variance}
B.~P. Durbin, J.~S. Hardin, D.~M. Hawkins, and D.~M. Rocke, ``A
  variance-stabilizing transformation for gene-expression microarray data,''
  {\em Bioinformatics}, vol.~18, no.~suppl 1, pp.~S105--S110, 2002.

\bibitem{Bartlett1937}
M.~Bartlett, ``Properties of sufficiency and statistical tests,'' {\em
  Proceedings of the Royal Society of London. Series A: Mathematical and
  Physical Sciences}, vol.~160, no.~901, pp.~268--282, 1937.

\bibitem{anscombe1948transformation}
F.~J. Anscombe, ``The transformation of poisson, binomial and negative-binomial
  data,'' {\em Biometrika}, vol.~35, no.~3/4, pp.~246--254, 1948.

\bibitem{bar1988classical}
S.~K. Bar-Lev and P.~Enis, ``On the classical choice of variance stabilizing
  transformations and an application for a poisson variate,'' {\em Biometrika},
  vol.~75, no.~4, pp.~803--804, 1988.

\bibitem{bartlett1936square}
M.~Bartlett, ``The square root transformation in analysis of variance,'' {\em
  Supplement to the Journal of the Royal Statistical Society}, vol.~3, no.~1,
  pp.~68--78, 1936.

\bibitem{tibshirani1988estimating}
R.~Tibshirani, ``Estimating transformations for regression via additivity and
  variance stabilization,'' {\em Journal of the American Statistical
  Association}, vol.~83, no.~402, pp.~394--405, 1988.

\bibitem{zhang2008wavelets}
B.~Zhang, J.~M. Fadili, and J.-L. Starck, ``Wavelets, ridgelets, and curvelets
  for poisson noise removal,'' {\em IEEE Transactions on Image Processing},
  vol.~17, no.~7, pp.~1093--1108, 2008.

\bibitem{qian2012fcstrans}
Y.~Qian, Y.~Liu, J.~Campbell, E.~Thomson, Y.~M. Kong, and R.~H. Scheuermann,
  ``{FCSTrans}: An open source software system for fcs file conversion and data
  transformation,'' {\em Cytometry Part A}, vol.~81, no.~5, pp.~353--356, 2012.

\bibitem{wilk1968probability}
M.~B. Wilk and R.~Gnanadesikan, ``Probability plotting methods for the analysis
  for the analysis of data,'' {\em Biometrika}, vol.~55, no.~1, pp.~1--17,
  1968.

\bibitem{motakis2006variance}
E.~Motakis, G.~P. Nason, P.~Fryzlewicz, and G.~Rutter, ``Variance stabilization
  and normalization for one-color microarray data using a data-driven
  multiscale approach,'' {\em Bioinformatics}, vol.~22, no.~20, pp.~2547--2553,
  2006.

\bibitem{fryzlewicz2005data}
P.~Fryzlewicz and V.~Delouille, ``A data-driven haar-fisz transform for
  multiscale variance stabilization,'' in {\em IEEE/SP 13th Workshop on
  Statistical Signal Processing}, pp.~539--544, IEEE, 2005.

\bibitem{levene1960robust}
H.~Levene, ``Robust tests for equality of variances1,'' {\em Contributions to
  Probability and Statistics: Essays in Honor of Harold Hotelling}, vol.~2,
  p.~278, 1960.

\bibitem{brown1974robust}
M.~B. Brown and A.~B. Forsythe, ``Robust tests for the equality of variances,''
  {\em Journal of the American Statistical Association}, vol.~69, no.~346,
  pp.~364--367, 1974.

\bibitem{finak2009merging}
G.~Finak, A.~Bashashati, R.~Brinkman, and R.~Gottardo, ``Merging mixture
  components for cell population identification in flow cytometry,'' {\em
  Advances in Bioinformatics}, vol.~2009, 2009.

\bibitem{sugar2010misty}
I.~P. Sug{\'a}r and S.~C. Sealfon, ``Misty mountain clustering: Application to
  fast unsupervised flow cytometry gating,'' {\em {BMC} Bioinformatics},
  vol.~11, no.~1, p.~502, 2010.

\bibitem{naumann2010curvhdr}
U.~Naumann, G.~Luta, and M.~Wand, ``The {curvHDR} method for gating flow
  cytometry samples,'' {\em {BMC} Bioinformatics}, vol.~11, no.~1, p.~44, 2010.

\bibitem{naim2010swift}
I.~Naim, S.~Datta, G.~Sharma, J.~S. Cavenaugh, and T.~R. Mosmann, ``{SWIFT}:
  Scalable weighted iterative sampling for flow cytometry clustering,'' in {\em
  IEEE International Conference on Acoustics Speech and Signal Processing
  (ICASSP)}, pp.~509--512, IEEE, 2010.

\bibitem{quinn2007statistical}
J.~Quinn, P.~W. Fisher, R.~J. Capocasale, R.~Achuthanandam, M.~Kam, P.~J.
  Bugelski, and L.~Hrebien, ``A statistical pattern recognition approach for
  determining cellular viability and lineage phenotype in cultured cells and
  murine bone marrow,'' {\em Cytometry Part A}, vol.~71, no.~8, pp.~612--624,
  2007.

\bibitem{hornik2008clue}
K.~Hornik, ``A {CLUE} for {CLUster Ensembles},'' {\em Journal of Statistical
  Software}, vol.~14, September 2005.

\bibitem{SamSPECTRAL}
H.~Zare, P.~Shooshtari, A.~Gupta, and R.~Brinkman, ``Data reduction for
  spectral clustering to analyze high throughput flow cytometry data,'' {\em
  {BMC} Bioinformatics}, vol.~11, no.~1, p.~403, 2010.

\bibitem{hartigan1979algorithm}
J.~A. Hartigan and M.~A. Wong, ``Algorithm {AS} 136: A k-means clustering
  algorithm,'' {\em Journal of the Royal Statistical Society. Series C (Applied
  Statistics)}, vol.~28, no.~1, pp.~100--108, 1979.

\bibitem{lloyd1982least}
S.~Lloyd, ``Least squares quantization in {PCM},'' {\em IEEE Transactions on
  Information Theory}, vol.~28, no.~2, pp.~129--137, 1982.

\bibitem{kaufman2009finding}
L.~Kaufman and P.~J. Rousseeuw, {\em Finding Groups in Data: An Introduction to
  Cluster Analysis}, vol.~344.
\newblock Wiley, 2009.

\bibitem{fastcluster}
D.~M\"ullner, ``{fastcluster}: Fast hierarchical, agglomerative clustering
  routines for {R} and {Python},'' {\em Journal of Statistical Software},
  vol.~53, no.~9, pp.~1--18, 2013.

\bibitem{bishop2006pattern}
C.~M. Bishop and N.~M. Nasrabadi, {\em Pattern Recognition and Machine
  Learning}, vol.~1.
\newblock Springer New York, 2006.

\bibitem{lebret2012rmixmod}
R.~Lebret, S.~Iovleff, and F.~Langrognet, {\em Rmixmod: An interface of
  MIXMOD}.
\newblock R package version 2.0.1, available at http://cran.r-project.org.

\bibitem{herrero2001sota}
J.~Herrero, A.~Valencia, and J.~Dopazo, ``A hierarchical unsupervised growing
  neural network for clustering gene expression patterns,'' {\em
  Bioinformatics}, vol.~17, no.~2, pp.~126--136, 2001.

\bibitem{von2007tutorial}
U.~Von~Luxburg, ``A tutorial on spectral clustering,'' {\em Statistics and
  Computing}, vol.~17, no.~4, pp.~395--416, 2007.

\bibitem{yan2009fastSpec}
D.~Yan, L.~Huang, and M.~Jordan, ``Fast approximate spectral clustering,'' in
  {\em Proceedings of the International Conference on Knowledge Discovery and
  Data Mining ({ACM SIGKDD})}, pp.~907--916, ACM, 2009.

\bibitem{gan2007clusteringBook}
G.~Gan, C.~Ma, and J.~Wu, {\em Data Clustering: Theory, Algorithms, and
  Applications}, vol.~20.
\newblock SIAM, 2007.

\bibitem{R_kmeans}
{R Development Core Team}, {\em R: A language and environment for statistical
  computing}.
\newblock R Foundation for Statistical Computing, Vienna, Austria, 2012.

\bibitem{cluster_manual}
M.~Maechler, P.~Rousseeuw, A.~Struyf, M.~Hubert, and K.~Hornik, {\em Cluster:
  Cluster analysis basics and extensions}.
\newblock R package version 1.14.4, available at http://cran.r-project.org.

\bibitem{brock2011clvalid}
G.~Brock, V.~Pihur, S.~Datta, and S.~Datta, ``{clValid}, an {R} package for
  cluster validation,'' {\em Journal of Statistical Software}, vol.~25, no.~4,
  2008.

\bibitem{karatzoglou2004kernlab}
A.~Zeileis, K.~Hornik, A.~Smola, and A.~Karatzoglou, ``Kernlab--an {S4} package
  for kernel methods in {R},'' {\em Journal of Statistical Software}, vol.~11,
  no.~9, pp.~1--20, 2004.

\bibitem{ramze1998new}
M.~Ramze~Rezaee, B.~P. Lelieveldt, and J.~H. Reiber, ``A new cluster validity
  index for the fuzzy c-mean,'' {\em Pattern Recognition Letters}, vol.~19,
  no.~3, pp.~237--246, 1998.

\bibitem{pal1995cluster}
N.~R. Pal and J.~C. Bezdek, ``On cluster validity for the fuzzy c-means
  model,'' {\em IEEE Transactions on Fuzzy Systems}, vol.~3, no.~3,
  pp.~370--379, 1995.

\bibitem{xie1991validity}
X.~L. Xie and G.~Beni, ``A validity measure for fuzzy clustering,'' {\em IEEE
  Transactions on Pattern Analysis and Machine Intelligence}, vol.~13, no.~8,
  pp.~841--847, 1991.

\bibitem{calinski1974dendrite}
T.~Cali{\'n}ski and J.~Harabasz, ``A dendrite method for cluster analysis,''
  {\em Communications in Statistics-theory and Methods}, vol.~3, no.~1,
  pp.~1--27, 1974.

\bibitem{dunn1974}
J.~Dunn, ``Well-separated clusters and optimal fuzzy partitions,'' {\em Journal
  of Cybernetics}, vol.~4, no.~1, pp.~95--104, 1974.

\bibitem{rousseeuw1987silhouettes}
P.~Rousseeuw, ``Silhouettes: A graphical aid to the interpretation and
  validation of cluster analysis,'' {\em Journal of Computational and Applied
  Mathematics}, vol.~20, pp.~53--65, 1987.

\bibitem{davies1979cluster}
D.~Davies and D.~Bouldin, ``A cluster separation measure,'' {\em IEEE
  Transactions on Pattern Analysis and Machine Intelligence}, no.~2,
  pp.~224--227, 1979.

\bibitem{halkidi2001sdbw}
M.~Halkidi and M.~Vazirgiannis, ``Clustering validity assessment: Finding the
  optimal partitioning of a data set,'' in {\em IEEE International Conference
  on Data Mining ({ICDM})}, pp.~187--194, IEEE, 2001.

\bibitem{ball1965isodata}
G.~H. Ball and D.~J. Hall, ``{ISODATA}, a novel method of data analysis and
  pattern classification,'' {\em Technical Report, {DTIC} Document}, 1965.

\bibitem{hubert1976quadratic}
L.~Hubert and J.~Schultz, ``Quadratic assignment as a general data analysis
  strategy,'' {\em British Journal of Mathematical and Statistical Psychology},
  vol.~29, no.~2, pp.~190--241, 1976.

\bibitem{ray1999determination}
S.~Ray and R.~H. Turi, ``Determination of number of clusters in k-means
  clustering and application in colour image segmentation,'' in {\em
  Proceedings of the 4th International Conference on Advances in Pattern
  Recognition and Digital Techniques}, pp.~137--143, 1999.

\bibitem{scott1971clustering}
A.~Scott and M.~J. Symons, ``Clustering methods based on likelihood ratio
  criteria,'' {\em Biometrics}, pp.~387--397, 1971.

\bibitem{liu2010understanding}
Y.~Liu, Z.~Li, H.~Xiong, X.~Gao, and J.~Wu, ``Understanding of internal
  clustering validation measures,'' in {\em IEEE International Conference on
  Data Mining ({ICDM})}, pp.~911--916, IEEE, 2010.

\bibitem{manual_clv}
L.~Nieweglowski., {\em clv: Cluster validation techniques}.
\newblock R package version 0.3.2, available at http://cran.r-project.org.

\bibitem{manual_clvalid}
G.~Brock, V.~Pihur, S.~Datta, and S.~Datta, {\em clValid: Validation of
  clustering results}, 2011.
\newblock R package version 0.6-4.

\bibitem{manual_clusterCrit}
B.~Desgraupes, {\em clusterCrit: Clustering indices}.
\newblock R package version 1.2.3, available at http://cran.r-project.org.

\bibitem{manning2008introduction}
C.~D. Manning, P.~Raghavan, and H.~Sch{\"u}tze, {\em Introduction to
  Information Retrieval}, vol.~1.
\newblock Cambridge University Press Cambridge, 2008.

\bibitem{rand1971objective}
W.~M. Rand, ``Objective criteria for the evaluation of clustering methods,''
  {\em Journal of the American Statistical Association}, vol.~66, no.~336,
  pp.~846--850, 1971.

\bibitem{jaccard1901}
P.~Jaccard, ``Distribution de la flore alpine dans le bassin des drouces et
  dans quelques regions voisines,'' {\em Bulletin de la Soci{\'e}t{\'e}
  Vaudoise des Sciences Naturelles}, vol.~37, no.~140, pp.~241--272, 1901.

\bibitem{jardine1971mathematical}
N.~Jardine and R.~Sibson, {\em Mathematical Taxonomy}.
\newblock Wiley, New York, 1971.

\bibitem{fowlkes1983method}
E.~B. Fowlkes and C.~L. Mallows, ``A method for comparing two hierarchical
  clusterings,'' {\em Journal of the American Statistical Association},
  vol.~78, no.~383, pp.~553--569, 1983.

\bibitem{gusfield2002partition}
D.~Gusfield, ``Partition-distance: A problem and class of perfect graphs
  arising in clustering,'' {\em Information Processing Letters}, vol.~82,
  no.~3, pp.~159--164, 2002.

\bibitem{konovalov2005partition}
D.~A. Konovalov, B.~Litow, and N.~Bajema, ``Partition-distance via the
  assignment problem,'' {\em Bioinformatics}, vol.~21, no.~10, pp.~2463--2468,
  2005.

\bibitem{charon2006maximum}
I.~Charon, L.~Denoeud, A.~Gu{\'e}noche, and O.~Hudry, ``Maximum transfer
  distance between partitions,'' {\em Journal of Classification}, vol.~23,
  no.~1, pp.~103--121, 2006.

\bibitem{arabie1973multidimensional}
P.~Arabie and S.~A. Boorman, ``Multidimensional scaling of measures of distance
  between partitions,'' {\em Journal of Mathematical Psychology}, vol.~10,
  no.~2, pp.~148--203, 1973.

\bibitem{papadimitriou1998combinatorial}
C.~H. Papadimitriou and K.~Steiglitz, {\em Combinatorial Optimization:
  Algorithms and Complexity}.
\newblock Courier Dover Publications, 1998.

\bibitem{gordon2001fuzzy}
A.~Gordon and M.~Vichi, ``Fuzzy partition models for fitting a set of
  partitions,'' {\em Psychometrika}, vol.~66, no.~2, pp.~229--247, 2001.

\bibitem{dimitriadou2002combination}
E.~Dimitriadou, A.~Weingessel, and K.~Hornik, ``A combination scheme for fuzzy
  clustering,'' {\em International Journal of Pattern Recognition and
  Artificial Intelligence}, vol.~16, no.~07, pp.~901--912, 2002.

\bibitem{gary1979computers}
M.~R. Gary and D.~S. Johnson, {\em Computers and Intractability: A Guide to the
  Theory of NP-completeness}.
\newblock WH Freeman and Company, New York, 1979.

\bibitem{manual_flowMeans}
N.~Aghaeepour, {\em flowMeans: Non-parametric flow cytometry data gating}.
\newblock R package version 1.15, available at http://www.bioconductor.org.

\bibitem{zeng2002matching}
Q.~Zeng, M.~Wand, A.~J. Young, J.~Rawn, E.~L. Milford, S.~J. Mentzer, and R.~A.
  Greenes, ``Matching of flow-cytometry histograms using information theory in
  feature space,'' in {\em Proceedings of the AMIA Symposium}, p.~929, American
  Medical Informatics Association, 2002.

\bibitem{zeng2007feature}
Q.~T. Zeng, J.~P. Pratt, J.~Pak, D.~Ravnic, H.~Huss, and S.~J. Mentzer,
  ``Feature-guided clustering of multi-dimensional flow cytometry datasets,''
  {\em Journal of Biomedical Informatics}, vol.~40, no.~3, pp.~325--331, 2007.

\bibitem{schrijver2003combinatorial}
A.~Schrijver, {\em Combinatorial Optimization: Polyhedra and Efficiency},
  vol.~24.
\newblock Springer, 2003.

\bibitem{zimmerman2011}
N.~Zimmerman, ``A computational approach to identification and comparison of
  cell subsets in flow cytometry data,'' {\em Doctoral Dissertation}, 2011.

\bibitem{jeffreys1946invariant}
H.~Jeffreys, ``An invariant form for the prior probability in estimation
  problems,'' {\em Proceedings of the Royal Society of London. Series A.
  Mathematical and Physical Sciences}, vol.~186, no.~1007, pp.~453--461, 1946.

\bibitem{kullback1951information}
S.~Kullback and R.~Leibler, ``On information and sufficiency,'' {\em The Annals
  of Mathematical Statistics}, vol.~22, no.~1, pp.~79--86, 1951.

\bibitem{mclachlan1999mahalanobis}
G.~McLachlan, ``Mahalanobis distance,'' {\em Resonance}, vol.~4, no.~6,
  pp.~20--26, 1999.

\bibitem{smirnov1948table}
N.~Smirnov, ``Table for estimating the goodness of fit of empirical
  distributions,'' {\em The Annals of Mathematical Statistics}, vol.~19, no.~2,
  pp.~279--281, 1948.

\bibitem{massey1951kolmogorov}
F.~J. Massey~Jr, ``The kolmogorov-smirnov test for goodness of fit,'' {\em
  Journal of the American Statistical Association}, vol.~46, no.~253,
  pp.~68--78, 1951.

\bibitem{rubner2000earth}
Y.~Rubner, C.~Tomasi, and L.~J. Guibas, ``The earth mover's distance as a
  metric for image retrieval,'' {\em International Journal of Computer Vision},
  vol.~40, no.~2, pp.~99--121, 2000.

\bibitem{berrington2005lymphocyte}
J.~Berrington, D.~Barge, A.~Fenton, A.~Cant, and G.~Spickett, ``Lymphocyte
  subsets in term and significantly preterm {UK} infants in the first year of
  life analyzed by single platform flow cytometry,'' {\em Clinical \&
  Experimental Immunology}, vol.~140, no.~2, pp.~289--292, 2005.

\bibitem{lacombe1997flow}
F.~Lacombe, F.~Durrieu, A.~Briais, P.~Dumain, F.~Belloc, E.~Bascans,
  J.~Reiffers, M.~Boisseau, and P.~Bernard, ``Flow cytometry {CD45} gating for
  immunophenotyping of acute myeloid leukemia,'' {\em Leukemia}, vol.~11,
  no.~11, pp.~1878--1886, 1997.

\bibitem{kern2010role}
W.~Kern, U.~Bacher, C.~Haferlach, S.~Schnittger, and T.~Haferlach, ``The role
  of multiparameter flow cytometry for disease monitoring in {AML},'' {\em Best
  Practice \& Research Clinical Haematology}, vol.~23, no.~3, pp.~379--390,
  2010.

\bibitem{sharan2006modeling}
R.~Sharan and T.~Ideker, ``Modeling cellular machinery through biological
  network comparison,'' {\em Nature Biotechnology}, vol.~24, no.~4,
  pp.~427--433, 2006.

\bibitem{singh2008global}
R.~Singh, J.~Xu, and B.~Berger, ``Global alignment of multiple protein
  interaction networks with application to functional orthology detection,''
  {\em Proceedings of the National Academy of Sciences}, vol.~105, no.~35,
  pp.~12763--12768, 2008.

\bibitem{ekstrom2011introduction}
C.~T. Ekstr{\o}m and H.~S{\o}rensen, {\em Introduction to Statistical Data
  Analysis for the Life Sciences}.
\newblock CRC Press, 2011.

\bibitem{steiger2004beyond}
J.~Steiger, ``Beyond the f test: Effect size confidence intervals and tests of
  close fit in the analysis of variance and contrast analysis.,'' {\em
  Psychological Methods}, vol.~9, no.~2, p.~164, 2004.

\bibitem{cohen1988book}
J.~Cohen, {\em Statistical {P}ower {A}nalysis for the {B}ehavioral {S}ciences}.
\newblock Lawrence Erlbaum, 1988.

\bibitem{fleishman1980confidence}
A.~Fleishman, ``Confidence intervals for correlation ratios,'' {\em Educational
  and Psychological Measurement}, vol.~40, no.~3, pp.~659--670, 1980.

\bibitem{kelley2007confidence}
K.~Kelley, ``Confidence intervals for standardized effect sizes: Theory,
  application, and implementation,'' {\em Journal of Statistical Software},
  vol.~20, no.~8, pp.~1--24, 2007.

\bibitem{johnson2002applied}
R.~A. Johnson and D.~W. Wichern, {\em Applied Multivariate Statistical
  Analysis}, vol.~5.
\newblock Prentice hall Upper Saddle River, NJ, 2002.

\bibitem{hotelling1931generalization}
H.~Hotelling, ``The generalization of student's ratio,'' {\em The Annals of
  Mathematical Statistics}, pp.~360--378, 1931.

\bibitem{pillai1959hotelling}
K.~S. Pillai and P.~Samson~Jr, ``On {H}otelling's generalization of {T}$^2$,''
  {\em Biometrika}, pp.~160--168, 1959.

\bibitem{ito1964robustness}
K.~Ito and W.~J. Schull, ``On the robustness of the {T}$_0^2$ test in
  multivariate analysis of variance when variance-covariance matrices are not
  equal,'' {\em Biometrika}, vol.~51, no.~1-2, pp.~71--82, 1964.

\bibitem{hughes1972approximating}
D.~T. Hughes and J.~G. Saw, ``Approximating the percentage points of
  {H}otelling's generalized {T}$^2_0$ statistic,'' {\em Biometrika}, vol.~59,
  no.~1, pp.~224--226, 1972.

\bibitem{steyn2009estimating}
H.~Steyn~Jr and S.~Ellis, ``Estimating an effect size in one-way multivariate
  analysis of variance ({MANOVA}),'' {\em Multivariate Behavioral Research},
  vol.~44, no.~1, pp.~106--129, 2009.

\bibitem{kelley2007methods}
K.~Kelley, ``Methods for the behavioral, educational, and social sciences
  ({MBESS}): An {R} package,'' {\em Behavior Research Methods}, vol.~39, no.~4,
  pp.~979--984, 2007.

\bibitem{farber1997differential}
D.~Farber, O.~Acuto, and K.~Bottomly, ``Differential {T} cell receptor-mediated
  signaling in naive and memory {CD4} {T} cells,'' {\em European Journal of
  Immunology}, vol.~27, no.~8, pp.~2094--2101, 1997.

\bibitem{ahmadzadeh2001heterogeneity}
M.~Ahmadzadeh, S.~Hussain, and D.~Farber, ``Heterogeneity of the memory {CD4}
  {T} cell response: Persisting effectors and resting memory {T} cells,'' {\em
  The Journal of Immunology}, vol.~166, no.~2, p.~926, 2001.

\bibitem{ahmadzadeh1999effector}
M.~Ahmadzadeh, S.~Hussain, and D.~Farber, ``Effector {CD4} {T} cells are
  biochemically distinct from the memory subset: Evidence for long-term
  persistence of effectors in vivo,'' {\em The Journal of Immunology},
  vol.~163, no.~6, p.~3053, 1999.

\bibitem{Dress+:book}
A.~Dress, K.~T. Huber, J.~Koolen, V.~Moulton, and A.~Spillner, {\em Basic
  Phylogenetic Combinatorics}.
\newblock Cambridge University Press, 2012.

\bibitem{brunelli1993face}
R.~Brunelli and T.~Poggio, ``Face recognition: Features versus templates,''
  {\em IEEE Transactions on Pattern Analysis and Machine Intelligence},
  vol.~15, no.~10, pp.~1042--1052, 1993.

\bibitem{de2007template}
M.~De~Wachter, M.~Matton, K.~Demuynck, P.~Wambacq, R.~Cools, and
  D.~Van~Compernolle, ``Template-based continuous speech recognition,'' {\em
  IEEE Transactions on Audio, Speech, and Language Processing}, vol.~15, no.~4,
  pp.~1377--1390, 2007.

\bibitem{deng2007structure}
L.~Deng, H.~Strik, {\em et~al.}, ``Structure-based and template-based automatic
  speech recognition-comparing parametric and nonparametric approaches,'' in
  {\em Proceedings of Interspeech}, pp.~898--901, 2007.

\bibitem{connell2001template}
S.~D. Connell and A.~K. Jain, ``Template-based online character recognition,''
  {\em Pattern Recognition}, vol.~34, no.~1, pp.~1--14, 2001.

\bibitem{bennett1985proposed}
J.~M. Bennett, D.~Catovsky, M.~T. Daniel, G.~Flandrin, D.~A. Galton, H.~R.
  Gralnick, and C.~Sultan, ``Proposed revised criteria for the classification
  of acute myeloid leukemia: A report of the {French-American-British
  Cooperative Group},'' {\em Annals of Internal Medicine}, vol.~103, no.~4,
  pp.~620--625, 1985.

\bibitem{biehl2013analysis}
M.~Biehl, K.~Bunte, and P.~Schneider, ``Analysis of flow cytometry data by
  matrix relevance learning vector quantization,'' {\em {PLoS} One}, vol.~8,
  no.~3, p.~e59401, 2013.

\bibitem{manninen2013leukemia}
T.~Manninen, H.~Huttunen, P.~Ruusuvuori, and M.~Nykter, ``Leukemia prediction
  using sparse logistic regression,'' {\em {PLoS} One}, vol.~8, no.~8,
  p.~e72932, 2013.

\bibitem{qiu2012inferring}
P.~Qiu, ``Inferring phenotypic properties from single-cell characteristics,''
  {\em {PLoS} One}, vol.~7, no.~5, p.~e37038, 2012.

\bibitem{dream_AML}
``{DREAM6/FlowCAP2} molecular classification of acute myeloid leukaemia
  challenge.'' http://www.the-dream-project.org/.
\newblock Accessed: 2013-12-25.

\bibitem{paietta2003expression}
E.~Paietta, ``Expression of cell-surface antigens in acute promyelocytic
  leukaemia,'' {\em Best Practice \& Research Clinical Haematology}, vol.~16,
  no.~3, pp.~369--385, 2003.

\bibitem{campana2000immunophenotyping}
D.~Campana and F.~G. Behm, ``Immunophenotyping of leukemia,'' {\em Journal of
  Immunological Methods}, vol.~243, no.~1, pp.~59--75, 2000.

\bibitem{mason2006immunophenotype}
K.~D. Mason, S.~K. Juneja, and J.~Szer, ``The immunophenotype of acute myeloid
  leukemia: is there a relationship with prognosis?,'' {\em Blood Reviews},
  vol.~20, no.~2, pp.~71--82, 2006.

\bibitem{keyhani2000increased}
A.~Keyhani, Y.~O. Huh, D.~Jendiroba, L.~Pagliaro, J.~Cortez, S.~Pierce,
  M.~Pearlman, E.~Estey, H.~Kantarjian, and E.~J. Freireich, ``Increased {CD38}
  expression is associated with favorable prognosis in adult acute leukemia,''
  {\em Leukemia Research}, vol.~24, no.~2, pp.~153--159, 2000.

\bibitem{legrand2000immunophenotype}
O.~Legrand, J.-Y. Perrot, M.~Baudard, A.~Cordier, R.~Lautier, G.~Simonin,
  R.~Zittoun, N.~Casadevall, and J.-P. Marie, ``The immunophenotype of 177
  adults with acute myeloid leukemia: proposal of a prognostic score,'' {\em
  Blood}, vol.~96, no.~3, pp.~870--877, 2000.

\bibitem{raspadori2001cd56}
D.~Raspadori, D.~Damiani, M.~Lenoci, D.~Rondelli, N.~Testoni, G.~Nardi,
  C.~Sestigiani, C.~Mariotti, S.~Birtolo, M.~Tozzi, {\em et~al.}, ``{CD56}
  antigenic expression in acute myeloid leukemia identifies patients with poor
  clinical prognosis,'' {\em Leukemia}, vol.~15, no.~8, pp.~1161--1164, 2001.

\bibitem{walter2010aberrant}
K.~Walter, P.~Cockerill, R.~Barlow, D.~Clarke, M.~Hoogenkamp, G.~Follows,
  S.~Richards, M.~Cullen, C.~Bonifer, and H.~Tagoh, ``Aberrant expression of
  {CD19} in {AML} with t (8; 21) involves a poised chromatin structure and
  pax5,'' {\em Oncogene}, vol.~29, no.~20, pp.~2927--2937, 2010.

\bibitem{estey2006acute}
E.~Estey and H.~D{\"o}hner, ``Acute myeloid leukaemia,'' {\em The Lancet},
  vol.~368, no.~9550, pp.~1894--1907, 2006.

\bibitem{hosmer2013applied}
D.~W. Hosmer~Jr, S.~Lemeshow, and R.~X. Sturdivant, {\em Applied Logistic
  Regression}.
\newblock Wiley. com, 2013.

\bibitem{zhao2009parallel}
W.~Zhao, H.~Ma, and Q.~He, ``Parallel k-means clustering based on
  {MapReduce},'' in {\em Cloud Computing}, pp.~674--679, Springer, 2009.

\bibitem{foti2000scalable}
D.~Foti, D.~Lipari, C.~Pizzuti, and D.~Talia, ``Scalable parallel clustering
  for data mining on multicomputers,'' in {\em Parallel and Distributed
  Processing}, pp.~390--398, Springer, 2000.

\bibitem{li1989parallel}
X.~Li and Z.~Fang, ``Parallel clustering algorithms,'' {\em Parallel
  Computing}, vol.~11, no.~3, pp.~275--290, 1989.

\end{thebibliography}
